\documentclass[aps,prb,reprint,twocolumn,superscriptaddress,floatfix,nofootinbib,longbibliography]{revtex4-1}
\usepackage{amsmath,amssymb,color,comment,physics}
\usepackage[makeroom]{cancel}
\usepackage[caption=false]{subfig}
\usepackage{mathrsfs}
\usepackage{graphicx}
\usepackage{subfig}
\usepackage[countmax]{subfloat}
\usepackage[english]{babel}
\usepackage[bookmarks=true,colorlinks,linkcolor=OrangeRed,urlcolor=NavyBlue,citecolor=RoyalBlue]{hyperref}
\usepackage[dvipsnames]{xcolor}
\usepackage{ulem} 
\usepackage{dsfont}

\definecolor{mygreen}{rgb}{0,0.5,0}
\definecolor{myblue}{rgb}{0,0,0.75}
\definecolor{mymagenta}{cmyk}{0,1,0,0.12}
\definecolor{mygray}{rgb}{0.5,0.5,0.5}

\definecolor{mypink1}{rgb}{0.858, 0.188, 0.478}
\definecolor{mypurple}{rgb}{0.49,0.18,0.56}
\definecolor{mygold}{rgb}{0.93,0.59,0.13}
\definecolor{mygreen}{rgb}{0,0.5,0}
\definecolor{myblue}{rgb}{0,0,0.75}
\definecolor{mymagenta}{cmyk}{0,1,0,0.12}
\definecolor{mygray}{rgb}{0.5,0.5,0.5}

\newcommand{\canc}[1]{}

\begin{document}
\title{Diffusive-to-ballistic crossover of symmetry violation in open many-body systems}
\author{Jad C.~Halimeh}
\affiliation{INO-CNR BEC Center and Department of Physics, University of Trento, Via Sommarive 14, 38123 Povo (TN), Italy}

\author{Philipp Hauke}
\affiliation{INO-CNR BEC Center and Department of Physics, University of Trento, Via Sommarive 14, 38123 Povo (TN), Italy}

\begin{abstract}
Conservation laws in a quantum many-body system play a direct role in its dynamic behavior. Understanding the effect of weakly breaking a conservation law due to coherent and incoherent errors is thus crucial, e.g., in the realization of reliable quantum simulators. In this work, we perform exact numerics and time-dependent perturbation theory to study the dynamics of \textit{symmetry violation} in quantum many-body systems with slight coherent (at strength $\lambda$) or incoherent (at strength $\gamma$) breaking of their local and global symmetries. We rigorously prove the symmetry violation to be a divergence measure in Hilbert space. Based on this, we show that symmetry breaking generically leads to a crossover in the divergence growth from diffusive behavior at onset times to ballistic or hyperballistic scaling at intermediate times, before diffusion dominates at long times. More precisely, we show that for local errors the leading coherent contribution to the symmetry violation cannot be of order lower than $\propto\lambda t^2$ while its leading-order incoherent counterpart is typically of order $\propto\gamma t$. This remarkable interplay between unitary and incoherent gauge-breaking scalings is also observed at higher orders in projectors onto symmetry (super)sectors. Due to its occurrence at short times, the diffusive-to-ballistic crossover is expected to be readily accessible in modern ultracold-atom and NISQ-device experiments.
\end{abstract}

\date{\today}
\maketitle
\tableofcontents

\section{Introduction}
Symmetries play a quintessential role in nature, from being the progenitors of different phases of matter\cite{Sachdev_book,Balents_NatureReview,Savary2016} to manifesting  conservation laws in and out of equilibrium.\cite{Brading_book} The group-theoretic notion of symmetries, i.e., defining symmetry as invariance under a group transformation, has become a cornerstone in the development of modern physics.\cite{Zee_book} Symmetries are also used to simplify a physical problem by reducing its effective Hilbert space based on what symmetry sectors the relevant physics takes place in (see glossary in Sec.~\ref{sec:glossary} for our nomenclature). These sectors involve global or local, as well as both continuous or discrete symmetries, and are manifestations of conservation laws. In the case of continuous symmetries, the associated conservation laws are formalized in Noether's theorem, which asserts an equivalence between them and the continuous symmetries in a physical system. As an example, the Fermi-- and Bose--Hubbard models\cite{Tasaki1998,Gersch1963} host a global $\mathrm{U}(1)$ symmetry, which translates into the conservation of particle number. Nevertheless, this equivalence is not restricted to continuous symmetries, as their discrete counterparts can also endow the system with conserved quantities. For example, coordinate inversion (P), charge conjugation (C), and time reversal (T) are discrete symmetries equivalent to the conservation of spatial, charge, and time parities, respectively.\cite{Zee_book,Peskin2016} Whereas (continuous and discrete) global symmetries give rise to the conservation of \textit{global charge}, local symmetries lead to \textit{local charges} being conserved. A prime example is gauge invariance,\cite{Jackson_review,Cheng_book} which is known as Gauss's law in quantum electrodynamics, and forms the principal property of gauge theories. By default, conservation of local charges leads to conservation of the global charge, but the converse is not necessarily true.

Given the richness afforded to physics by symmetries, a fundamental question is the effect of weakly breaking an underlying symmetry of a model on its subsequent dynamics. Even though traditionally studied in systems with weak integrability breaking, prethermalization has recently been generalized to nonintegrable systems with perturbative coherent breaking of global \cite{Mallayya2019,Ray2020} and local symmetries.\cite{Halimeh2020b,Halimeh2020c} Interestingly, in the case of weak breaking of the local gauge symmetry in a lattice gauge theory of $N$ matter sites, a \textit{prethermalization staircase} arises in the dynamics of the gauge violation (see Sec.~\ref{sec:definition} for definition), composed of prethermal plateaus at timescales $\propto\lambda^{-s}$, with $s=0,1,2,\ldots,N/2$ and $\lambda$ the strength of  gauge invariance-breaking unitary errors. This rich prethermal behavior has the property of delaying the timescale of full gauge violation exponentially in system size, and it can also be observed in other local observables.\cite{Halimeh2020b,Halimeh2020c} 
Also in case of slight breaking of global symmetries, the equilibrating dynamics is strongly affected. This can occur when conservation laws due to, e.g., integrability are perturbatively broken.\cite{Moeckel2008,Eckstein2009,Kollar2011,Tavora2013,Nessi2014,Essler2014,Bertini2016,Fagotti2015,Reimann2019} In this case, the initial equilibration to a generalized Gibbs ensemble\cite{Rigol2007,Rigol2009a,Rigol2009b,Vidmar2016,Essler2016,Cazalilla2016,Caux2016,Mallayya2018,Mallayya2019} (GGE) steady state is replaced by thermal equilibrium at a later timescale $\propto\lambda^{-2}$, with $\lambda$ the strength of the perturbative integrability-breaking term, as can be shown by kinetic Boltzmann-like equations that can be derived through employing time-dependent perturbation theory starting from the GGE steady state.\cite{Stark2013,DAlessio_review} This behavior is not restricted to quenched systems, but can also be seen in weakly interacting driven models.\cite{Lazarides2014,Canovi2016} As these examples illustrate, coherent errors can drastically change the dynamical properties of a quantum many-body system.

Going beyond unitary closed-system dynamics, decoherence has been a central topic of research in quantum many-body physics.\cite{Zeh1970,Schlosshauer2005} Its mitigation is a necessary capability to achieve reliable quantum computers, because these devices rely on the principles of superposition and entanglement, and are therefore particularly sensitive to interactions with the environment. Examples abound such as $1/f$-noise in superconducting quantum interference devices (SQUIDs) that constitutes a dominant adverse effect on superconducting qubits,\cite{Yoshihara2006,Kakuyanagi2007,Bialczak2007,Bylander2011,Wang2015,Kumar2016} CMB-photon noise in superconducting cryogenic detectors,\cite{Day2003} and thermomechanical motion in microwave cavity interferometers.\cite{Regal2008} In open quantum many-body systems, the effects of decoherence have been studied on light-cone dynamics and the spread of correlations,\cite{Poulin2010,Marino2012,Descamps2013,Bernier2013,Halimeh2018,Bernier2018} and in a recent experiment\cite{Maier2019} a ballistic-to-diffusive crossover in quantum transport has been observed due to environmental noise in a $10$-qubit network of interacting spins. Moreover, the effect of decoherence on weakly driven quantum many-body systems have also been studied in the context of long-time steady states in the presence of approximately conserved quantities, where it is shown that a GGE state can also arise.\cite{Lenarcic2018,Lange2018} Therefore, like their coherent counterparts, incoherent symmetry-breaking errors due to decoherence can fundamentally change the properties and behavior of a physical system.

From a technological point of view, in modern ultracold-atom experiments that aim to quantum-simulate a given target model,\cite{Gross_review,Lewenstein_book,Hauke2012} it is of critical importance to reliably implement certain conservation laws due to both local and global symmetries. As global symmetries define the fixed points of renormalization-group flow, they decisively influence quantum phase diagrams. For example, particle-number conservation in the form of a global $\mathrm{U}(1)$ symmetry is crucial in the transition between the superfluid and Mott-insulator phases in the Bose--Hubbard model,\cite{Zhou2006,Maruyama2007,Roy2019,Donatella2020} a paradigm phase transition of cold-atom experiments.\cite{Greiner2002} Similarly, local gauge symmetries have fundamental consequences such as massless photons and a long-ranged Coulomb law,\cite{Melnikov2000,Tu2004} but their realization in quantum simulators requires careful engineering---in contrast to fundamental theories of nature such as quantum electrodynamics or quantum chromodynamics, in quantum devices they are not given by fundamental laws. This has recently generated a surge of research investigating unitary errors in lattice gauge theories that compromise gauge invariance, including ways of protecting against them.\cite{Zohar2011,Zohar2012,Banerjee2012,Zohar2013,Hauke2013,Stannigel2014,Kuehn2014,Kuno2015,Kuno2017,Negretti2017,Barros2019,Schweizer2019,Halimeh2020a,Yang2020,Halimeh2020d,Halimeh2020e,Mathis2020,Lamm2020,Tran2020}
As these examples highlight, the existence of coherent and incoherent symmetry-breaking errors in realistic setups necessitates a rigorous understanding of their influence on the dynamics of quantum many-body systems. Particularly relevant is to see if such errors are \textit{controlled} insomuch that one may extract the ideal theory dynamics despite their presence.

In this paper, we investigate the effect of experimentally motivated incoherent global and local symmety-breaking errors on the dynamics of quantum many-body systems. We demonstrate that the symmetry violation---the expectation value of the symmetry generator or its square, which has been often used in the past to estimate the effects of errors---is a rigorous divergence measure in Hilbert space that quantifies the deviation of the state from the target symmetry sector.
We exploit this insight to show, using exact numerics and rigorous proofs in time-dependent perturbation theory, the existence of a diffusive-to-ballistic crossover in the dynamics of the symmetry violation as a result of competition of these errors with their coherent counterparts. These results extend and generalize upon the findings presented in Ref.~\onlinecite{Halimeh2020f}, which considered quenches from a separable gauge-invariant initial state in a $\mathrm{Z}_2$ gauge theory. By presenting a thorough analysis of various sources of errors and various different model scenarios, our results provide a guideline for quantum-simulation experiments on noisy intermediate-scale quantum (NISQ) devices that aim to realize target models with a given symmetry.

The rest of the paper is organized as follows. In Sec.~\ref{sec:preamble}, we define the symmetry violation, clarify our nomenclature in a short glossary, and provide a summary of our main results. In Sec.~\ref{sec:meaning}, we rigorously explain the physical meaning of the symmetry violation by equating it with a divergence measure in Hilbert space and by relating it to the decrease of the overlap between states that differ only by a symmetry transformation. Our findings and conclusions are then illustrated on three main models: the extended Bose--Hubbard model in Sec.~\ref{sec:eBHM}, the $\mathrm{Z}_2$ lattice gauge theory in Sec.~\ref{sec:Z2LGT}, and the $\mathrm{U}(1)$ quantum link model in Sec.~\ref{sec:U1QLM}. We conclude and discuss possible future directions in Sec.~\ref{sec:conclusion}. Appendix~\ref{sec:TDPT} contains our detailed derivations in time-dependent perturbation theory that explain the various scalings seen in our exact diagonalization results. Appendix~\ref{sec:NumSpec} provides details on our numerical implementation.

\section{Preamble}\label{sec:preamble}
Before entering in the details of our work, we define in this section our main figure of merit, the symmetry violation, we provide a short glossary for nomenclature clarity, and we present a concise summary of our main findings.

\subsection{Definition of symmetry violation}\label{sec:definition}
The motivation behind our work is the assessment of quantum many-body systems in modern experimental settings where realistically coherent and incoherent errors will always be present at least to a perturbative degree. These errors may break target local and global symmetries, which may or may not be desired in the experiment, but where a thorough understanding of their effects on the dynamics is nevertheless advantageous. These effects can be qualitatively and quantitatively studied by calculating dynamics of the symmetry violation and other relevant observables.

The symmetry violation $\varepsilon$ is defined as the expectation value of a (local or global) symmetry generator\cite{footnote} with respect to a target symmetry sector (see glossary in Sec.~\ref{sec:glossary}), or the square of this expectation value, and is often used to estimate the effect of slightly breaking a symmetry.\cite{Zohar2011,Zohar2012,Banerjee2012,Zohar2013,Hauke2013,Stannigel2014,Kuehn2014,Kuno2015,Kuno2017,Negretti2017,Barros2019,Schweizer2019,Halimeh2020a,Yang2020,Halimeh2020d,Halimeh2020e,Mathis2020,Lamm2020,Tran2020} To formalize its definition, let us consider a many-body model described by the Hamiltonian $H_0$ with a local symmetry generated by the operators $G_j$, with $j$ a lattice-site index (for the local gauge symmetries considered below, matter particles reside on the lattice sites and gauge fields on the links in between). The eigenvalues of $G_j$ are the local charges $g_j$, and a given combination of them on the lattice classifies a gauge-invariant sector $\mathbf{g}=\{g_1,g_2,\ldots,g_N\}$. We select a target gauge-invariant sector $\mathbf{g}_\text{tar}=\{g_1^\text{tar},g_2^\text{tar},\ldots,g_N^\text{tar}\}$, and call a given state gauge-invariant or symmetric iff $G_j\rho_0=g_j\rho_0,\,\forall j$. We prepare the system in an initial state $\rho_0$, which may be gauge-invariant or not. Gauge invariance restricts dynamics within a gauge sector, and so in case $\rho_0$ lies in a given gauge-invariant sector $\mathbf{g}$, then the system will remain in this sector for all times if the dynamics is solely due to $H_0$, because due to the gauge invariance of the latter, $[H_0,G_j]=0,\,\forall j$. In the presence of unitary or incoherent gauge-breaking errors, the gauge violation generically will spread across various gauge sectors $\mathbf{g}$, and can in general be quantified as
\begin{align}\label{eq:MeasureGeneral}
\varepsilon(t)=\frac{1}{N}\sum_j\Tr\Big\{\rho(t)\big[G_j-g_j^\text{tar}\big]^2\Big\},
\end{align}
where $\rho(t)$ is the density matrix of the time-evolved system at time $t$. The motivation behind this measure lies in the assumption that the system is desired to reside within the target gauge-invariant sector $\mathbf{g}_\text{tar}$. Thus, any coherent or incoherent errors during the preparation of $\rho_0$ or the subsequent dynamics that take the system away from $\mathbf{g}_\text{tar}$ will make $\varepsilon$ as defined by  Eq.~\eqref{eq:MeasureGeneral} nonzero.

Much the same way, this definition can be extended to the case of global-symmetry models, with the only caveat being that there the deviation is across global-symmetry sectors, each of which consists of all states with a given fixed value of the global charge (see glossary in Sec.~\ref{sec:glossary}). We select a target global-symmetry sector defined by the total global charge $g_\text{tar}$. The system is prepared in an initial state $\rho_0$ which may be in the target sector  $g_\text{tar}$ or not. The global symmetry is generated by the operator $G$, and we denote the Hamiltonian of the global-symmetry model as $H_0$, i.e., $[H_0,G]=0$. The initial state $\rho_0$ is said to be symmetric iff $G\rho_0=g\rho_0$. Consequently, the symmetry violation becomes
\begin{align}
\label{eq:violationglobalsymmetry}
\varepsilon(t)=\frac{1}{N^2}\Tr\Big\{\rho(t)\big[G-g_\text{tar}\big]^2\Big\}.
\end{align}
The normalization with $N^2$ is chosen since $G$ typically is an extensive quantity such as the total particle number, see, e.g., Eq.~\eqref{eq:viol_eBHM} below.

\subsection{Glossary}\label{sec:glossary}
\textbf{\textit{Symmetry sectors and symmetric states.}} In case of a local symmetry, a state $\rho$ is said to be gauge-invariant or symmetric iff $G_j\rho=\rho G_j=g_j\rho,\,\forall j$ where $G_j$ are the local-symmetry generators of the gauge group at matter sites $j$ with eigenvalues $g_j$ that depend on the gauge symmetry of the model. A given set of values $\mathbf{g}=\{g_1,g_2,\ldots,g_N\}$ constitute a gauge-invariant sector. In the case of a global symmetry, a state $\rho_0$ is symmetric iff $G\rho=\rho G=g\rho$, where $G$ is the generator of the global symmetry, and its eigenvalue $g$ denotes the global charge of the corresponding symmetry sector. As a concrete example, in the Bose--Hubbard model $G$ can be chosen as the total particle number. In this case, a given symmetry sector $g$ would consist of all Fock states $\ket{n_1,n_2,\ldots,n_N}$ where the individual on-site particle numbers $n_j$ sum to $\sum_{j=1}^Nn_j=g$.

\medskip

\textbf{\textit{Target symmetry sector.}} In an experiment, it is often desired to prepare the system and restrict its dynamics within a given symmetry sector in a local- or global-symmetry model. This is called the target symmetry sector. The symmetry violation measures how far off a state $\rho$ is from this target symmetry sector.

\medskip

\textbf{\textit{Gauge-invariant supersector.}} Whereas both local and global symmetries have sectors, in our nomenclature only local symmetries have supersectors. A given supersector $\alpha$ is the set of all gauge-invariant sectors $\mathbf{g}=\{g_1,g_2,\ldots,g_N\}$ that have a total violation $\alpha$ with respect to the target gauge-invariant sector $\mathbf{g}_\text{tar}$. This definition can be formalized as $\sum_{j=1}^N(G_j-g_j^\text{tar})^p\rho=\alpha\rho$, with $p=1$ for the $\mathrm{Z}_2$ LGT and $p=2$ for the $\mathrm{U}(1)$ QLM. In the case of the $\mathrm{Z}_2$ LGT discussed in Sec.~\ref{sec:Z2LGT} and a target gauge-invariant sector $\mathbf{g}_\text{tar}=\mathbf{0}$, a gauge-invariant supersector can be defined by the set of gauge-invariant sectors $\mathbf{g}$ each carrying $M$ violations with respect to $\mathbf{g}_\text{tar}=\mathbf{0}$. For notational brevity, these supersectors are then denoted by $M$ and the projectors onto them by $\mathcal{P}_M$, with each then being a sum of all projectors onto the constituent sectors; cf.~Eq.~\eqref{eq:SupProj}.

\subsection{Summary of main findings}\label{sec:summary}
The main result of our work is the crossover in the short-time dynamics of the symmetry violation from a diffusive spread through symmetry sectors caused by incoherent errors to a ballistic spread caused by coherent errors. Formally, we consider a system ideally described by a Hamiltonian $H_0$, which has either a local or global symmetry generated by the local or global operators $G_j$ or $G$, respectively, with $j$ indicating a site in the associated lattice. In practice, these symmetries will be compromised in an experiment without unrealistic fine-tuning and isolation from the environment. We represent the coherent errors by the Hamiltonian $\lambda H_1$, with $\lambda$ their strength, while we model decoherence (dissipation and dephasing) using a Lindblad master equation with the coupling strength to the environment denoted by $\gamma$ (see further below for the specific terms used).

We prepare the system in an initial state $\rho_0$ and quench at $t=0$ with $H_0+\lambda H_1$ in the presence of decoherence. We demonstrate that the symmetry violation yields a divergence measure for the quantum state across symmetry sectors, enabling us to identify an increase as $t^a$ with diffusive ($a=1$), ballistic ($a=2$), and hyperballistic scaling ($a>2$). As we elaborate in detail below, the leading order of incoherent contributions to the symmetry violation is $\propto\gamma t$. The leading order of the coherent contribution depends on the structure of $H_1$ and $\rho_0$. If $\rho_0$ is symmetric or is a generic eigenstate of $H_0$, then the coherent contribution cannot be of an order lower than $\propto\lambda^2 t^2$. If $\rho_0$ is neither symmetric nor a generic eigenstate of $H_0$, then theoretically the leading order of the coherent contribution can be $\propto\lambda t$, but we find this to happen only in two rather pathological cases and one engineered (artificial) case. The first pathological scenario is when $\rho_0$ is the ground state of $H_0+\lambda_\text{i} H'_1$ with $\lambda_\text{i}\neq0$ and $H'_1$ a highly nonlocal Hamiltonian, and the second is when $H'_1$ is local but $H_1$ is highly nonlocal. If $H_1=H'_1$, the contribution $\propto\lambda t$ vanishes identically to zero, and then the order of the coherent contribution cannot be lower than $\propto\lambda t^2$. We find this also to be the case if $H'_1$ and $H_1$ are both local, or mildly nonlocal, but not equal. One may also get the scaling $\propto\lambda t$ by artificially engineering the initial state in a common eigenbasis of $H_0$ and the symmetry generator to be an arbitrary superposition of eigenstates degenerate with respect to $H_0$ but not with respect to the symmetry generator. Conversely, one can also construct specific combinations of initial state and coherent gauge breaking that result in hyperballistic expansion across symmetry sectors.

Therefore, it is shown that in quantum many-body systems with small coherent and incoherent symmetry breaking, a crossover from diffusive to ballistic spread takes place in the short-time dynamics of the symmetry violation and certain other observables. The crossover time is $t\propto\gamma/\lambda^2$ (or $t\propto\gamma^{\frac{1}{3}}/\lambda^{\frac{2}{3}}$ in case of hyperballistic spread) for initial states that reside in a symmetry sector or are generic (possibly unsymmetric) eigenstates of $H_0$, and $t\propto\gamma/\lambda$ for unsymmetric initial states that are not eigenstates of $H_0$. Our work presents an extension to and generalization of Ref.~\onlinecite{Halimeh2020f}, which has studied this crossover in a $\mathrm{Z}_2$ lattice gauge theory starting in a gauge-invariant initial product state. Qualitatively, we find that our main findings are similar for systems with breakings of local or of global symmetries.

\section{Meaning of the symmetry violation}\label{sec:meaning}
In this section, we imbue the symmetry violation with mathematical and physical meaning, as a divergence measure across symmetry sectors in case of local or global symmetries, as well as the overlap between states that differ only by a symmetry transformation. This section generalizes the discussion of Ref.~\onlinecite{Halimeh2020f}, which considered local gauge symmetries, and provides further details.

\subsection{Symmetry violation as a divergence measure in Hilbert space}
\label{sec:divergencemeasure}
We follow a similar reasoning as in Ref.~\onlinecite{Hauke2015}, where a measure for divergence in Hilbert space has been defined in the context of many-body localization.\cite{Abanin2019} We first consider gauge-symmetry models, and start by defining the mean-square displacement across the gauge sectors, 
\begin{align}
	D^{(2)}_{\mathbf{g}_\text{tar}}(t)=\sum_{\mathbf{g}} d^2(\mathbf{g},\mathbf{g}_\text{tar})\Tr\big\{ \rho(t)P_\mathbf{g}\big\}, 
\end{align}
where $P_\mathbf{g}$ is the projector onto gauge-invariant sector $\mathbf{g}$, and $\mathbf{g}_\text{tar}$ is the target gauge-invariant sector (see Sec.~\ref{sec:definition}). Using the definition $d^2(\mathbf{g},\mathbf{g}_\text{tar})=[\mathbf{g}-\mathbf{g}_\text{tar}]^2$, we can rewrite this expression as 
\begin{align}
\label{eq:D2vsepsilon}
D^{(2)}_{\mathbf{g}_\text{tar}}(t)&=\mathrm{Tr}\Big\{\rho(t) \sum_{\mathbf{g}} P_\mathbf{g} \big[\mathbf{g}-\mathbf{g}_\text{tar}\big]^2  \Big\}\nonumber \\
&=\sum_j\mathrm{Tr}\Big\{\rho(t)\big [G_j-g_j^\text{tar}\big]^2\Big\}\\
&=\sum_j\langle \big[G_j-g_j^\text{tar}\big]^2\rangle. \nonumber
\end{align}
By comparison with Eq.~\eqref{eq:MeasureGeneral}, we obtain 
\begin{align}
\varepsilon(t)=\frac{1}{N}D^{(2)}_{\mathbf{g}_\text{tar}}(t).
\end{align}

Thus, commonly used measures of gauge violation, in $\mathrm{U}(1)$ as well as $\mathrm{Z}_2$ gauge theories,\cite{Stannigel2014,Halimeh2020a,Halimeh2020b,Halimeh2020c} rigorously define a mean-square displacement across gauge sectors as encapsulated in Eq.~\eqref{eq:D2vsepsilon}. 
One can immediately apply exactly the same reasoning to global symmetries, where only $\mathbf{g}$ needs to be replaced by the scalar $g$ representing the global-charge sector (and analogously for $\mathbf{g}_\text{tar}$ and $g_\text{tar}$).

In this sense, $\varepsilon(t)\propto \gamma t$ is associated with a diffusive spreading of the wavefunction across gauge (global-charge) sectors in local- (global-) symmetry models, with diffusion constant $\propto \gamma$, while $\varepsilon(t)\propto \lambda^2 t^2$ is indicative of a ballistic spreading, a characteristic property of coherent quantum dynamics. 

\subsection{Symmetry violation as overlap between gauge-transformed states}
We can further understand the gauge violation as quantifying how fast the overlap diminishes between states that differ only by a gauge transformation. Let us again begin by considering the case of local symmetry. By definition, a gauge-invariant state should be oblivious to a local gauge transformation generated by a unitary $U(\boldsymbol{\phi}) = \prod_{j=1}^N e^{i\phi_j G_j}$,  with $\boldsymbol{\phi}=\{\phi_1,\dots,\phi_N\}$ and $\phi_{j}$ arbitrary and independent angles. 
To quantify by how much gauge-breaking errors compromise gauge invariance, one can use the overlap between the state of interest and the transformed state, which for a pure state reads $\mathcal{C}=|\bra{\psi(t)}U(\boldsymbol{\phi})\ket{\psi(t)}|^2$. 
The rate with which the overlap reduces under a gauge transformation is 
\begin{align}
\eta_j=-\frac{1}{2}\frac{\partial^2 \mathcal{C}}{\partial \phi_j^2} = \langle{G_j^2}\rangle-\langle{G_j}\rangle^2,
\end{align}
which is nothing else but the variance of the local Gauss's-law generator $G_j$. 

For a homogeneous system, and assuming without restriction of generality $\mathbf{g}_\text{tar}=\mathbf{0}$, definition \eqref{eq:MeasureGeneral} yields $\eta_j=\varepsilon-\langle{G_j}\rangle^2$. For the $\mathrm{Z}_2$ gauge theory considered below, where $G_j^2=2G_j$, we further have $\eta_j=\varepsilon(2-\varepsilon)$ [upon normalizing $\varepsilon$ by adding an inconsequential factor of $1/2$ in Eq.~\eqref{eq:MeasureGeneral}].

Thus, the sensitivity of a state towards a gauge transformation is related to the gauge violation $\varepsilon$. For small gauge violations in a homogeneous $\mathrm{Z}_2$ gauge theory, both even coincide (apart from an inconsequential factor of 2), as they do in a homogeneous $\mathrm{U}(1)$ gauge theory with $\langle{G_j}\rangle=0$. This insight lends further physical motivation for $\varepsilon$ as a good measure of gauge violation. As a side note, for mixed states, $\eta_j$ is given by the quantum Fisher information of the Gauss's-law generator, which for thermal states, e.g., can be measured through linear-response susceptibilities \cite{Hauke2016} or engineered quench dynamics.\cite{Almeida2020} 

Again, these considerations can be immediately adapted to global symmetries. In this case, the unitary transformation is defined as $U({\phi}) = e^{i\phi G}$ and the rate of change is $\eta=-\frac{1}{2}\frac{\partial^2 \mathcal{C}}{\partial \phi^2} = \langle{G^2}\rangle-\langle{G}\rangle^2$. For translationally-invariant systems with $\mathbf{g}_\text{tar}=0$ and $\langle{G}\rangle=0$, using definition \eqref{eq:violationglobalsymmetry} this relation becomes $\eta=N^2\varepsilon$.

\section{Extended Bose--Hubbard model}\label{sec:eBHM}

In this section, we numerically evaluate the interplay of coherent and incoherent breakings of a global symmetry. To this end, we consider the paradigmatic example of the extended Bose--Hubbard model\cite{Kuehner1998,Kuehner2000,Dutta2015} (eBHM).

\subsection{Model and quench protocol}
\begin{figure}[htp]
	\centering
	\includegraphics[width=.48\textwidth]{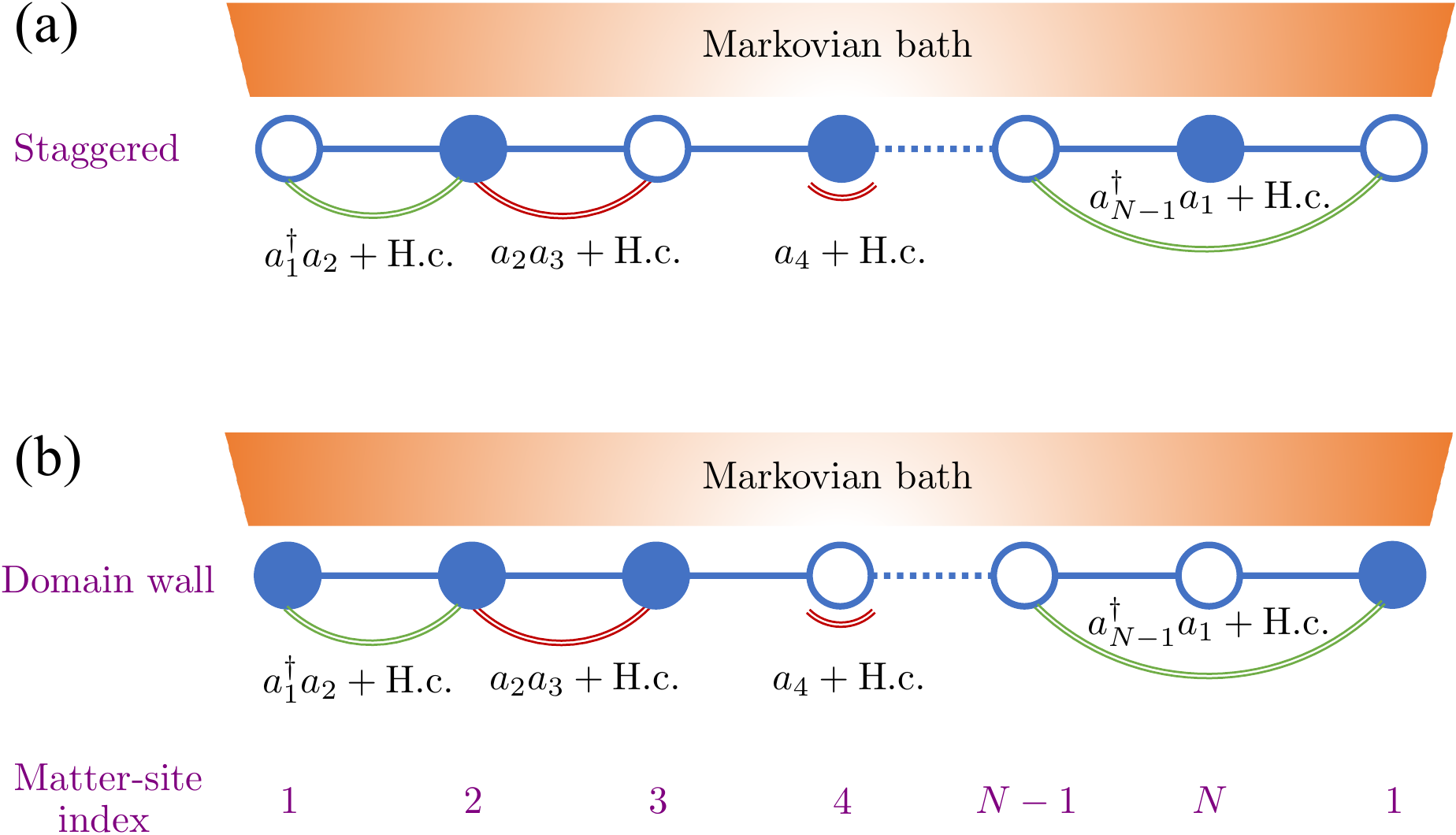}
	\caption{(Color online). Initial product states prepared at half-filling ($g_\text{tar}=N/2$), as used in the dynamics of the extended Bose--Hubbard model in the presence of coherent and incoherent errors. Green (red) arcs exemplify coherent processes preserving (breaking) global $\mathrm{U}(1)$ symmetry. In this work, we use chains of $N=6$ sites with periodic boundary conditions for this model. (a) A staggered product state where odd sites are empty and each even site contains one hard-cose boson. (b) A domain-wall product state where the left half of the lattice has a hard-core boson at each site and the right half of the chain is empty.
	}
	\label{fig:eBHM_InitialStates} 
\end{figure} 

The eBHM considered here is defined on a one-dimensional spatial lattice with $N$ sites and assuming periodic boundary conditions, and is described by the Hamiltonian
\begin{align}\nonumber
\label{eq:HeBHM}
H_0=&-\sum_{j=1}^N\big(J_1a_j^\dagger a_{j+1}+J_2a_j^\dagger a_{j+2}+\text{H.c.}\big)\\
&+\frac{U}{2}\sum_{j=1}^Nn_j(n_j-1)+W\sum_{j=1}^Nn_jn_{j+1},
\end{align}
where $a_j$ is the bosonic annihilation operator at site $j$ satisfying the canonical commutation relations $[a_j,a_l]=0$ and $[a_j,a_l^\dagger]=\delta_{j,l}$. The eBHM is nonintegrable at finite nonzero values of $J_1$, $J_2$, $U$, and $W$. Generically, it has two integrable points: the atomic limit of $J_1=J_2=0$ and the free-boson limit of $U=W=0$.\cite{Kollath2010} In our case, we additionally impose a hard-core constraint on the bosons, through which the term $\propto U$ does not play any role in the dynamics, so we can remove it from the Hamiltonian in Eq.~\eqref{eq:HeBHM} (formally, the hard-core constraint amounts to setting $U=\infty$). This leads to another integrable point at $J_2=0$, where the eBHM becomes equivalent to the XXZ model.\cite{Kollath2010} To avoid any effects due to integrability breaking, we therefore set $J_1=1$, $J_2=0.83$, and $W=0.11$ in our numerics, although we have checked that other generic values of these parameters yield the same conclusions. In all our results for the eBHM we use a periodic chain with $N=6$ sites.

\begin{figure}[htp]
	\centering
	\includegraphics[width=.48\textwidth]{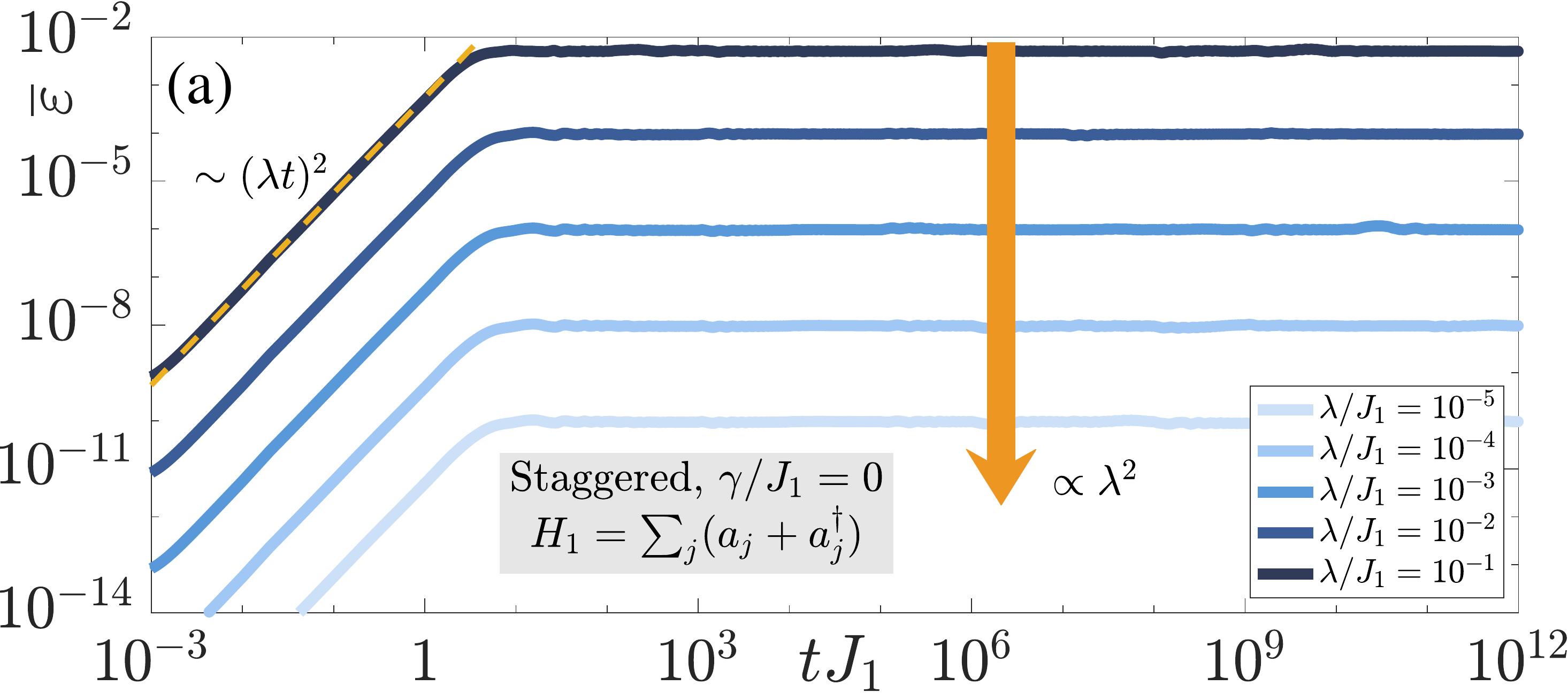}\\
	\includegraphics[width=.48\textwidth]{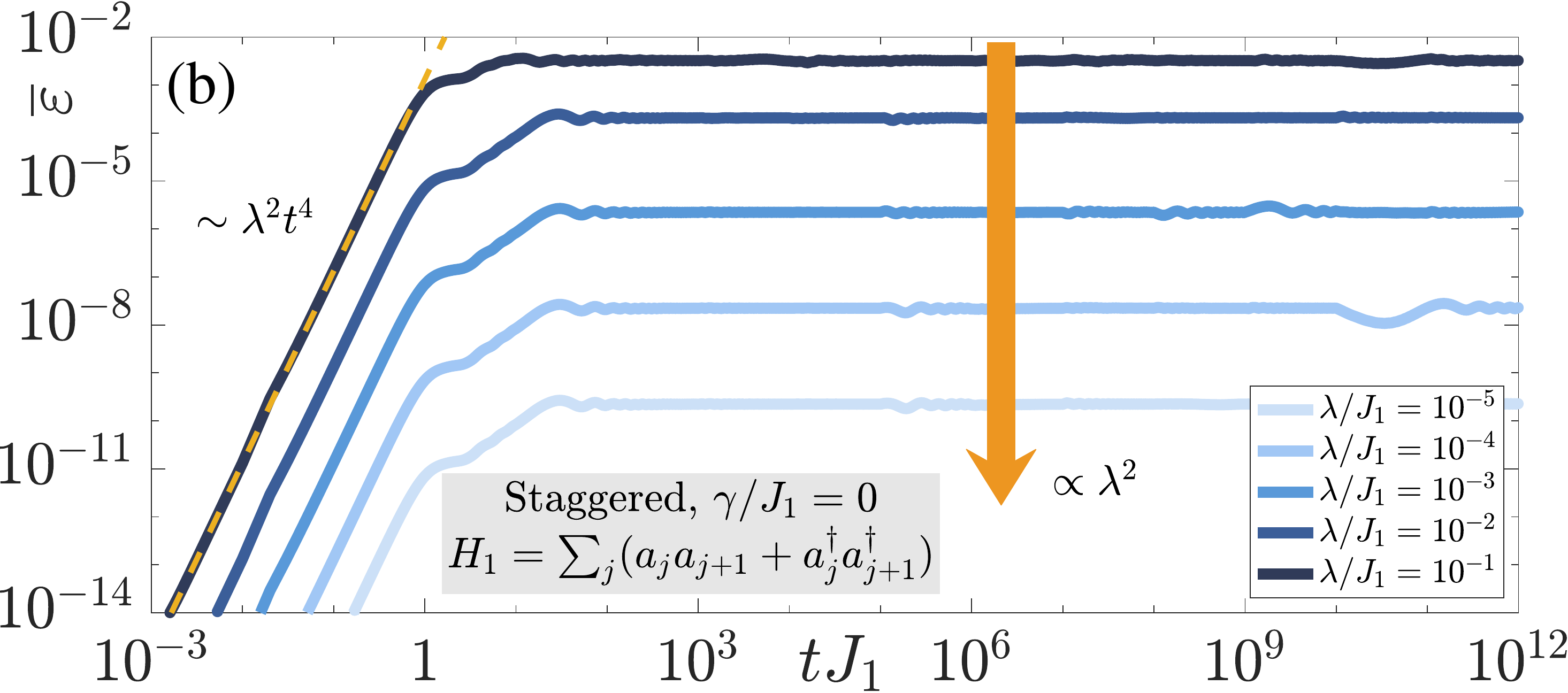}\\
	\includegraphics[width=.48\textwidth]{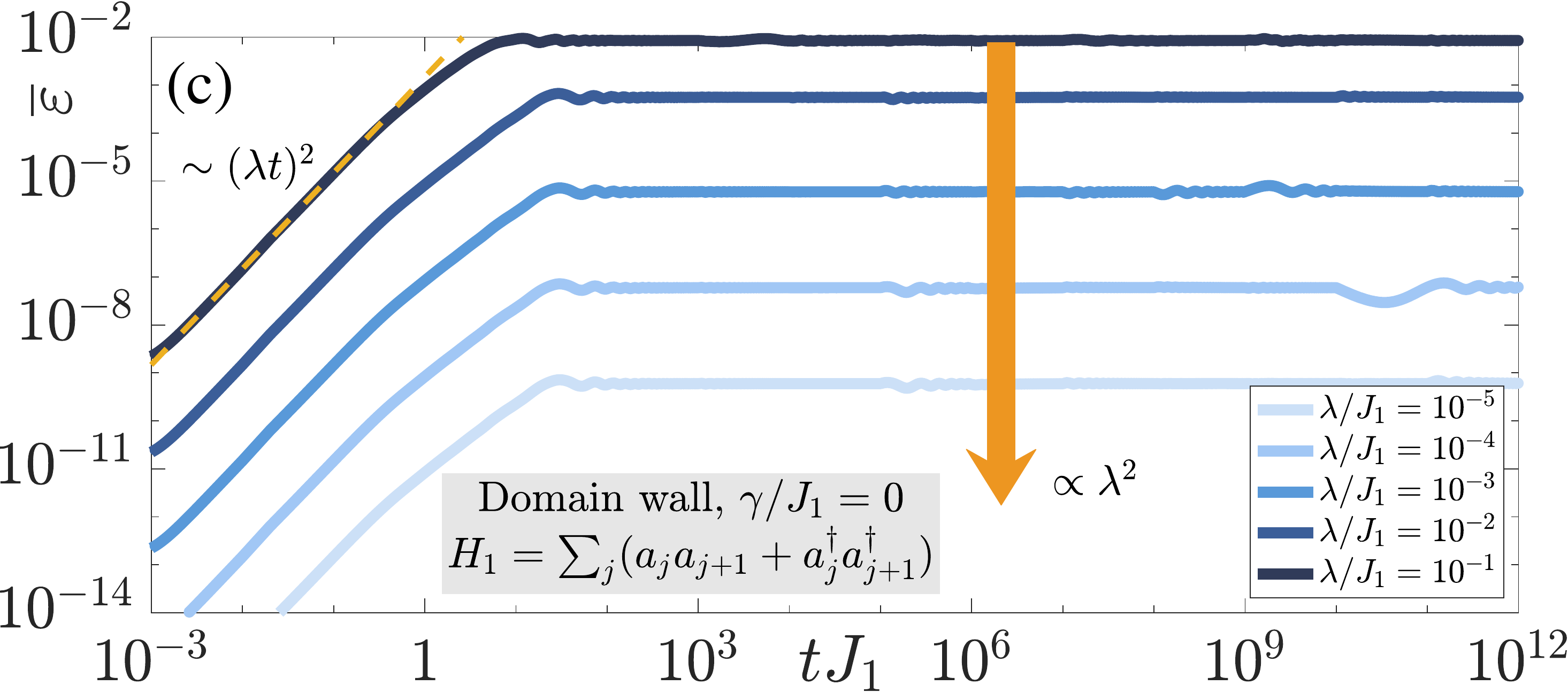}
	\caption{(Color online). Quench dynamics of the symmetry violation in the closed ($\gamma=0$) extended Bose--Hubbard model in the presence of unitary symmetry-breaking errors $\lambda H_1$ and starting in a symmetric initial state. The staggered initial product state shown in Fig.~\ref{fig:eBHM_InitialStates}(a) is used for (a) and (b), while the domain-wall product state in Fig.~\ref{fig:eBHM_InitialStates}(b) is used as the initial state in (c). System size is $N=6$ sites, and periodic boundary conditions are assumed. $H_1$ is composed of single-body terms in (a), while it is a two-body error in (b) and (c). Generically, the initial increase of symmetry violation is $\propto (\lambda t)^2$ (a,c), while specific combinations of initial state and error terms can increase the order of the short-time behavior, to $\lambda^2 t^4$ in the case of panel (b). Within our simulation times, the maximal value of the violation in all cases is $\propto\lambda^2$, which is the same order in $\lambda$ at which the violation grows at short times.}
	\label{fig:eBHM_closed} 
\end{figure}

The eBHM conserves the total particle number because $\sum_j[H_0,a_j^\dagger a_j]=0$. This translates to the eBHM hosting a global $\mathrm{U}(1)$ symmetry. In a realistic experiment, the implementation of this model will suffer from coherent and incoherent errors that in the best case slightly break this symmetry. The coherent errors are described by a Hamiltonian $\lambda H_1$, where $\lambda$ denotes the strength of these errors, and dynamics in the presence of decoherence is modeled by the Lindblad master equation\cite{Breuer_book,Manzano2020}
\begin{align}\nonumber
\dot{\rho}=&-i[H_0+\lambda H_1,\rho]\\\label{eq:EOM_eBHM}
&+\gamma\sum_j\Big(L_j\rho L_j^\dagger-\frac{1}{2}\big\{L_j^\dagger L_j,\rho\big\}\Big),
\end{align}
where $L_j=a_j$ are the jump operators describing the dissipation of our system with the environment, and $\gamma$ is the environment-coupling strength. Below, we test various examples of initial state and terms for $H_1$.

In the following, we prepare our system in an initial state $\rho_0$ and solve Eq.~\eqref{eq:EOM_eBHM} for the dynamics of the symmetry violation
\begin{align}\label{eq:viol_eBHM}
\varepsilon(t)&=\Tr\big\{\mathcal{G}\rho(t)\big\},\,\,\,\mathcal{G}=\frac{1}{N^2}\bigg[\sum_{j=1}^Na_j^\dagger a_j -g_\text{tar}\bigg]^2,
\end{align}
where $g_\text{tar}=N/2$ indicates the target half-filling sector.

\subsection{Symmetric initial state}
Let us first prepare our system in the staggered initial product state shown in Fig.~\ref{fig:eBHM_InitialStates}(a). This state is symmetric because it lies in the half-filling sector: $\mathcal{G}\rho_0=0$. The subsequent time evolution of the symmetry violation in Eq.~\eqref{eq:viol_eBHM} \textit{without decoherence} under the unitary errors
\begin{align}\label{eq:H1OneBody}
\lambda H_1=\lambda\sum_{j=1}^N\big(a_j+a_j^\dagger\big),
\end{align}
is shown in Fig.~\ref{fig:eBHM_closed}(a). The error term in Eq.~\eqref{eq:H1OneBody} would describe, e.g., the coupling of the system to another internal state (call its associated annihilation operator $b$), which is occupied by a condensate that we assume to be evenly spread across the entire lattice.\cite{Maruyama2007} Then, a Rabi flop between the two states gives $H_1=\sum_j\big(b^\dagger a_j+a_j^\dagger b\big)\approx \sum_j\big(\langle b\rangle^* a_j+\langle b\rangle a_j^\dagger\big)$.
Without restriction of generality, we set the phase of the condensate to $0$. Thus, we get $H_1 = \langle b\rangle  \sum_j\big(a_j+a_j^\dagger\big)$, thereby achieving Eq.~\eqref{eq:H1OneBody} by absorbing $\langle b\rangle$ in the definition of the unitary-error strength $\lambda$. The short-time scaling in Fig.~\ref{fig:eBHM_closed}(a) is $\sim(\lambda t)^2$. As we show in Sec.~\ref{sec:TDPT_coherent} through time-dependent perturbation theory (TDPT), this is the lowest order of coherent contributions to the symmetry violation in the case of a symmetric initial state. 

\begin{figure}[htp]
	\centering
	\includegraphics[width=.48\textwidth]{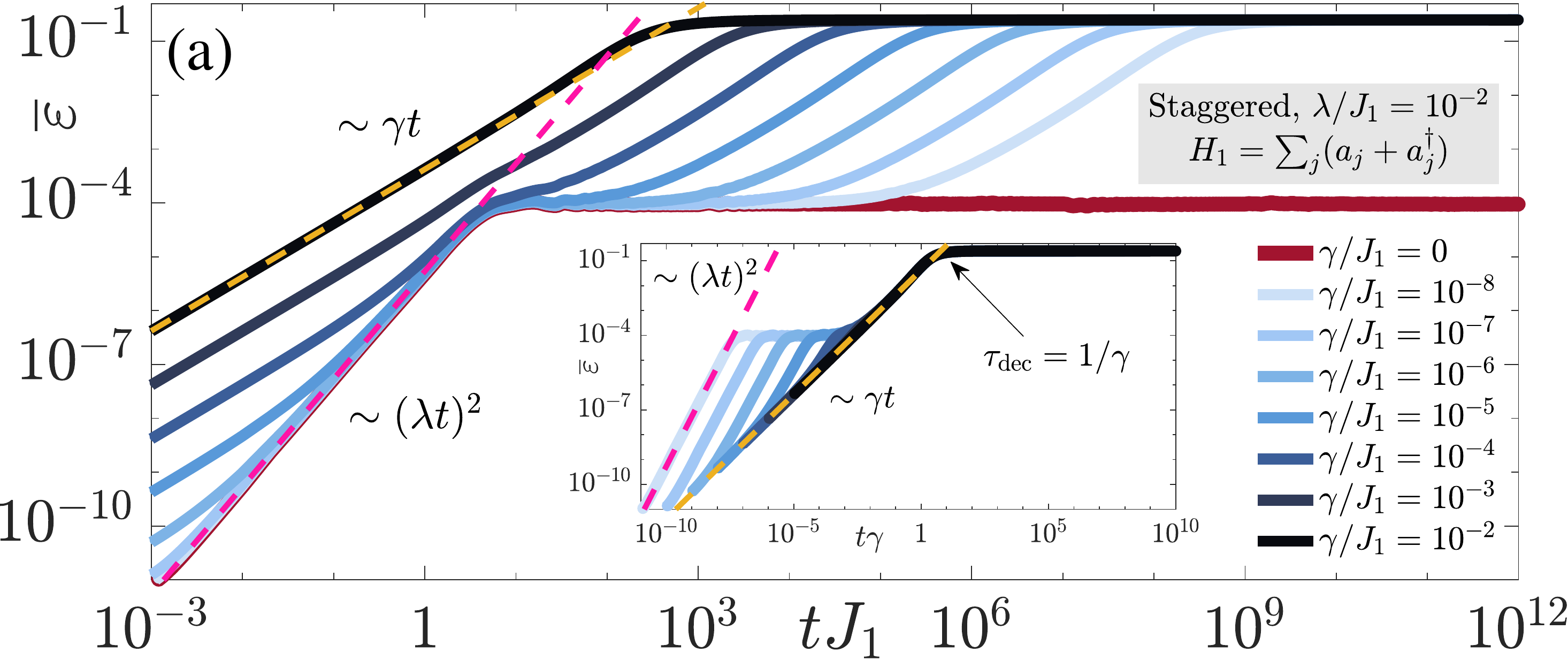}\\
	\includegraphics[width=.48\textwidth]{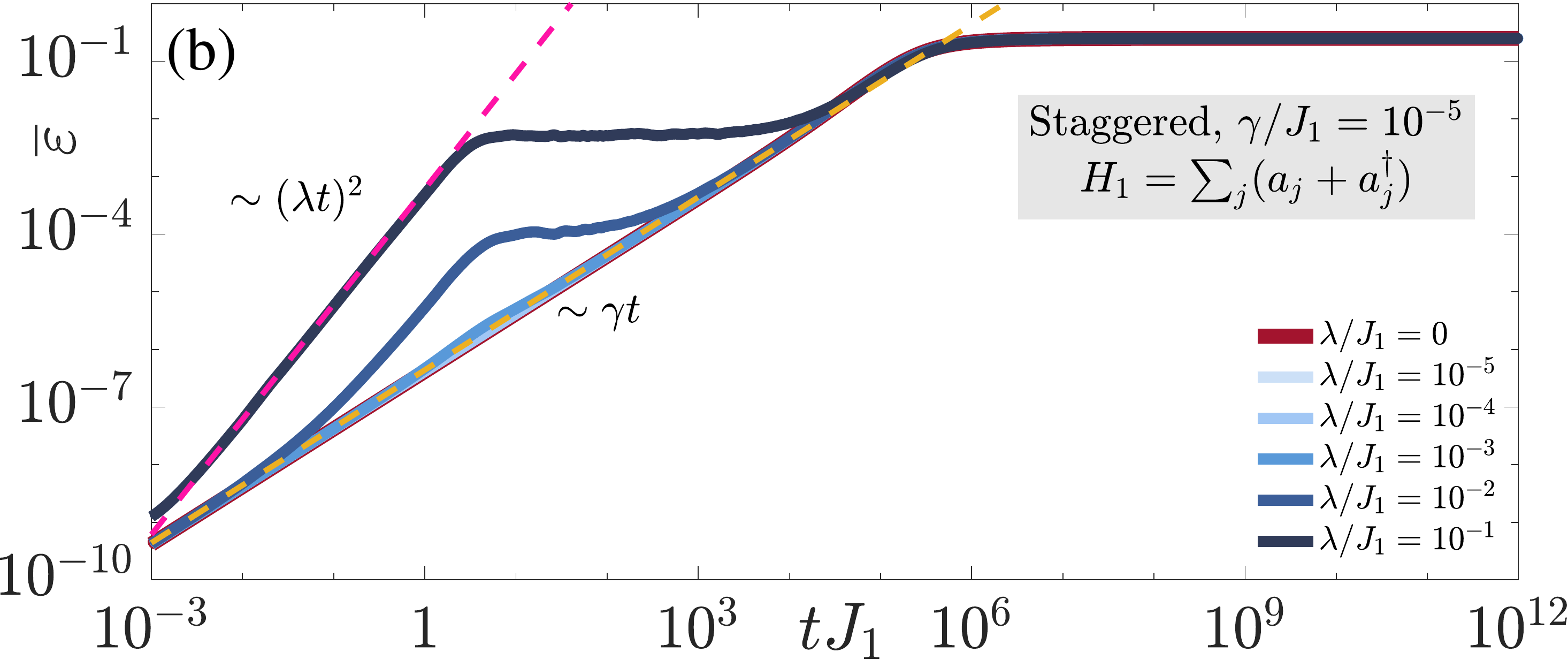}
	\caption{(Color online). Quench dynamics of the symmetry violation in the $N=6$-site open eBHM starting in a staggered initial product state and in the presence of the single-body unitary errors of Eq.~\eqref{eq:H1OneBody} at strength $\lambda$. The coupling to the environment is at strength $\gamma$ with the jump operator $L_j=a_j$ at each site. (a) Symmetry violation over time with fixed $\lambda$ at various values of $\gamma$. (b) Symmetry violation over time with fixed $\gamma$ at various values of $\lambda$. Note how decoherence, regardless the value of $\lambda$, takes the symmetry violation to its maximal value $g_\text{tar}^2/N^2=1/4$ at long-enough times. A diffusive-to-ballistic crossover occurs at $t\propto\gamma/\lambda^2$ where the symmetry violation goes from a diffusive spread $\sim\gamma t$ for $t\lesssim\gamma/\lambda^2$ to ballistic spread $\sim\lambda^2t^2$ for $t\gtrsim\gamma/\lambda^2$.}
	\label{fig:eBHM_OneBodyError_open} 
\end{figure}

\begin{figure}[htp]
	\centering
	\includegraphics[width=.48\textwidth]{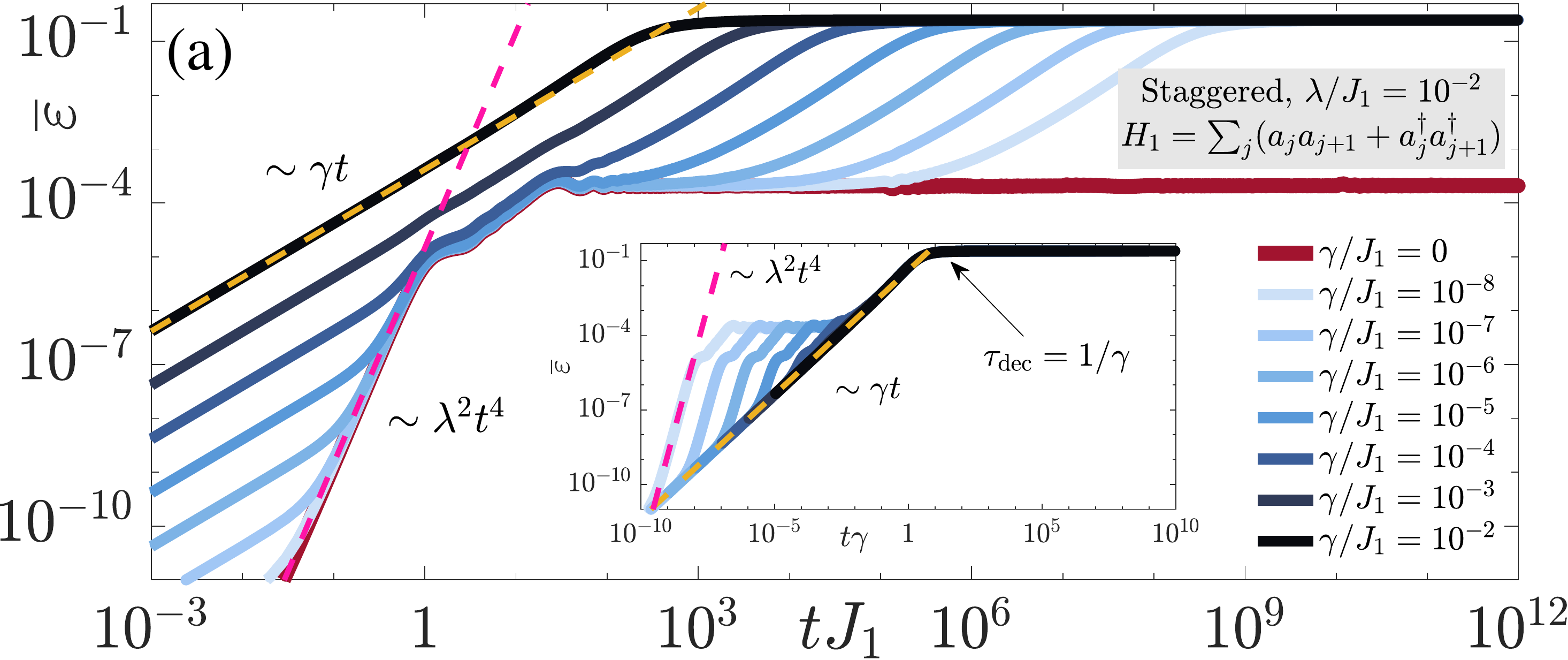}\\
	\includegraphics[width=.48\textwidth]{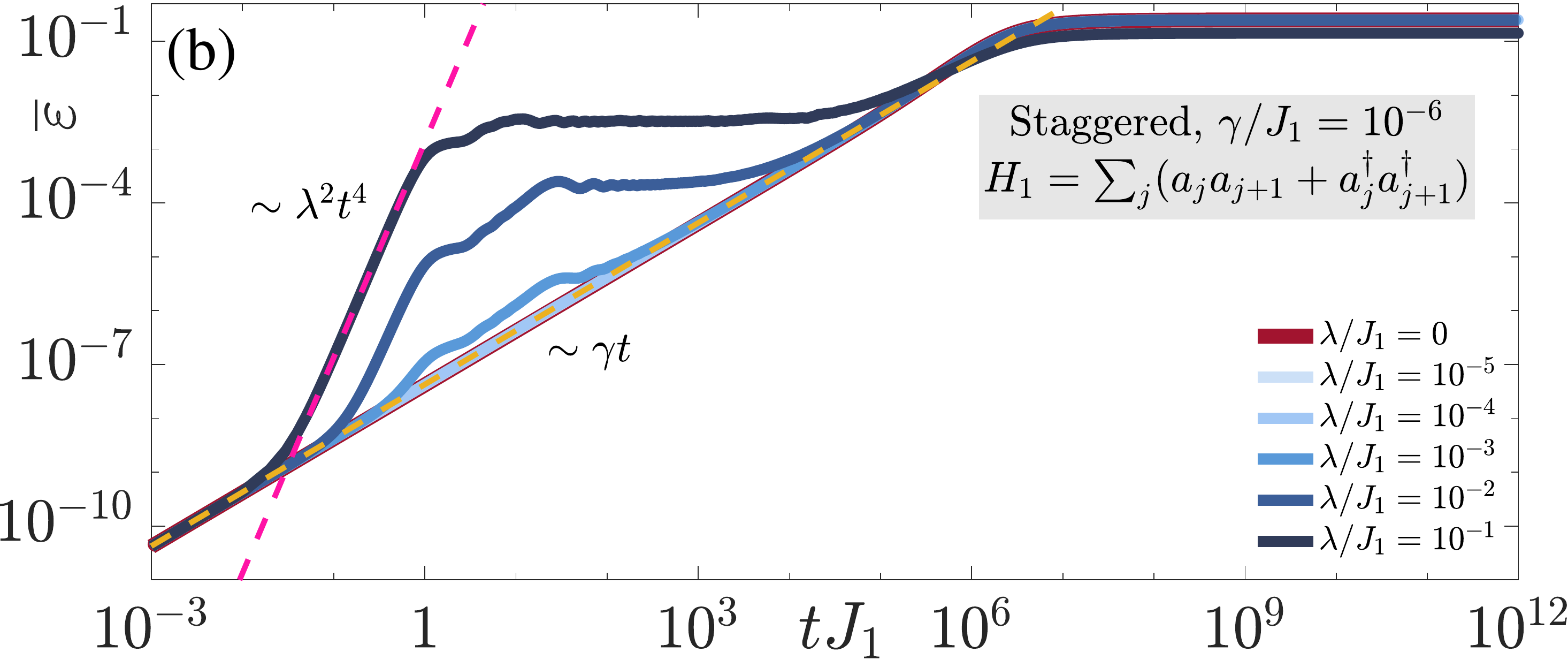}
	\caption{(Color online). Same as Fig.~\ref{fig:eBHM_OneBodyError_open} but for the two-body error of Eq.~\eqref{eq:H1TwoBody}. The initial state can play a nontrivial role in the short-time dynamics of the symmetry violation. In this case, the crossover is from diffusive to \textit{hyperballistic} spread and occurs at $t\propto(\gamma/\lambda^2)^{\frac{1}{3}}$. This behavior is qualitatively different from the case of the single-body unitary errors in Eq.~\eqref{eq:H1OneBody}, and is due to the fact that $H_1$ in Eq.~\eqref{eq:H1TwoBody} annihilates $\rho_0$ when it acts on it: $H_1\rho_0=0$, which leads to the contribution $\propto\lambda^2t^2$ vanishing, and thus the next leading order $\propto\lambda^2t^4$ sets in for $t\gtrsim(\gamma/\lambda^2)^{\frac{1}{3}}$. Decoherence allows the symmetry violation to reach its maximal value of $1/4$ that unitary errors alone cannot achieve.}
	\label{fig:eBHM_TwoBodyError_open} 
\end{figure}

However, one can engineer the initial state or coherent error to make higher orders the leading contributions. As an example, we consider the same initial state shown in Fig.~\ref{fig:eBHM_InitialStates}(a), but consider the two-body unitary errors
\begin{align}\label{eq:H1TwoBody}
\lambda H_1=\lambda\sum_{j=1}^N\big(a_ja_{j+1}+a_j^\dagger a_{j+1}^\dagger\big).
\end{align}
Similarly to the motivation behind the error term of Eq.~\eqref{eq:H1OneBody}, one can imagine immersing an optical lattice into a condensate of biatomic molecules, where bonding $\propto a_j a_{j+1}$ gives a molecular annihilation operator $m$.\cite{Zhou2006} 
Then, a term that drives the coupling to the molecular condensate is $H_1=\sum_j\big(m^\dagger a_ja_{j+1}+a_j^\dagger a_{j+1}^\dagger m\big)
\approx \langle m\rangle \sum_j\big(a_ja_{j+1}+a_j^\dagger a_{j+1}^\dagger\big)$, which by absorbing $\langle m\rangle$ into the definition of $\lambda$, we achieve the error term in Eq.~\eqref{eq:H1TwoBody}. 

When $\rho_0$ is the staggered initial state of Fig.~\ref{fig:eBHM_InitialStates}(a), the coherent contribution $\lambda^2t^2\Tr\{\mathcal{G}H_1\rho_0H_1\}$ to the symmetry violation is finite in case of the unitary errors in Eq.~\eqref{eq:H1OneBody}, but vanishes in case of the unitary errors of Eq.~\eqref{eq:H1TwoBody}. The reason is that the staggered initial occupation yields $H_1\rho_0=0$. As shown in Sec.~\ref{sec:TDPT_coherent}, the next leading contribution $\propto\lambda^2t^4\Tr\{\mathcal{G}H_1H_0\rho_0H_0H_1\}$ now dominates, where it does not vanish because $H_0$ induces tunneling processes that can bring bosons on adjacent sites, allowing $H_1$ to act on $\rho_0$ without destroying it. This is exactly what we see in Fig.~\ref{fig:eBHM_closed}(b). Such an increase of the mean-square displacement with a power of $t$ larger than $2$ is a hallmark of hyperballistic expansion, and sets the crossover timescale to $t\propto(\gamma/\lambda^2)^{\frac{1}{3}}$ in the presence of decoherence (see below).

Nevertheless, the short-time scaling $\sim\lambda^2t^4$ is not a generic feature of the unitary errors of Eq.~\eqref{eq:H1TwoBody}, but is rather a combination of such errors and the fact that we start in the staggered initial state. Indeed, if we consider these same \textit{pairing errors} but instead start in, say, the domain-wall initial state shown in Fig.~\ref{fig:eBHM_InitialStates}(b), then the leading coherent contribution to the symmetry violation is again $\propto\lambda^2t^2$, as shown in Fig.~\ref{fig:eBHM_closed}(c). The reason is that a domain-wall initial state already has bosons on adjacent sites, and thus $H_1$ can act on it nontrivially.

In what follows, we take $\rho_0$ as the staggered initial state and include decoherence through jump operators $L_j^\dagger$. We consider first in Fig.~\ref{fig:eBHM_OneBodyError_open} the unitary errors of Eq.~\eqref{eq:H1OneBody}. As can be shown in TDPT (see Appendix~\ref{sec:TDPT_leadingIncoherent}), the leading incoherent contribution to the symmetry violation in case of a symmetric initial state is $\propto\gamma t\sum_j\Tr\{\mathcal{G}L_j \rho_0L_j^\dagger\}$. As such, at times $t\lesssim\gamma/\lambda^2$, the symmetry violation shown in Fig.~\ref{fig:eBHM_OneBodyError_open} scales diffusively $\sim\gamma t$, before being overtaken by the ballistic spread $\sim\lambda^2t^2$ due to the leading coherent contribution. Note that the maximal value $\approx1/4$ reached by the symmetry violation in the steady-state due to decoherence is larger than that due to purely coherent symmetry breaking.

We now consider the same staggered initial state but use the unitary error terms of Eq.~\eqref{eq:H1TwoBody}. The only qualitative difference here is that the diffusive scaling $\sim\gamma t$ due to incoherent errors is overtaken by hyperballistic spread $\sim\lambda^2 t^4$, due to the leading coherent contribution, at a crossover time $t\propto(\gamma/\lambda^2)^{\frac{1}{3}}$. Again here in the presence of decoherence the symmetry violation attains its maximal value of $1/4$, even though in the purely unitary case it does not.

\begin{figure}[htp]
	\centering
	\includegraphics[width=.48\textwidth]{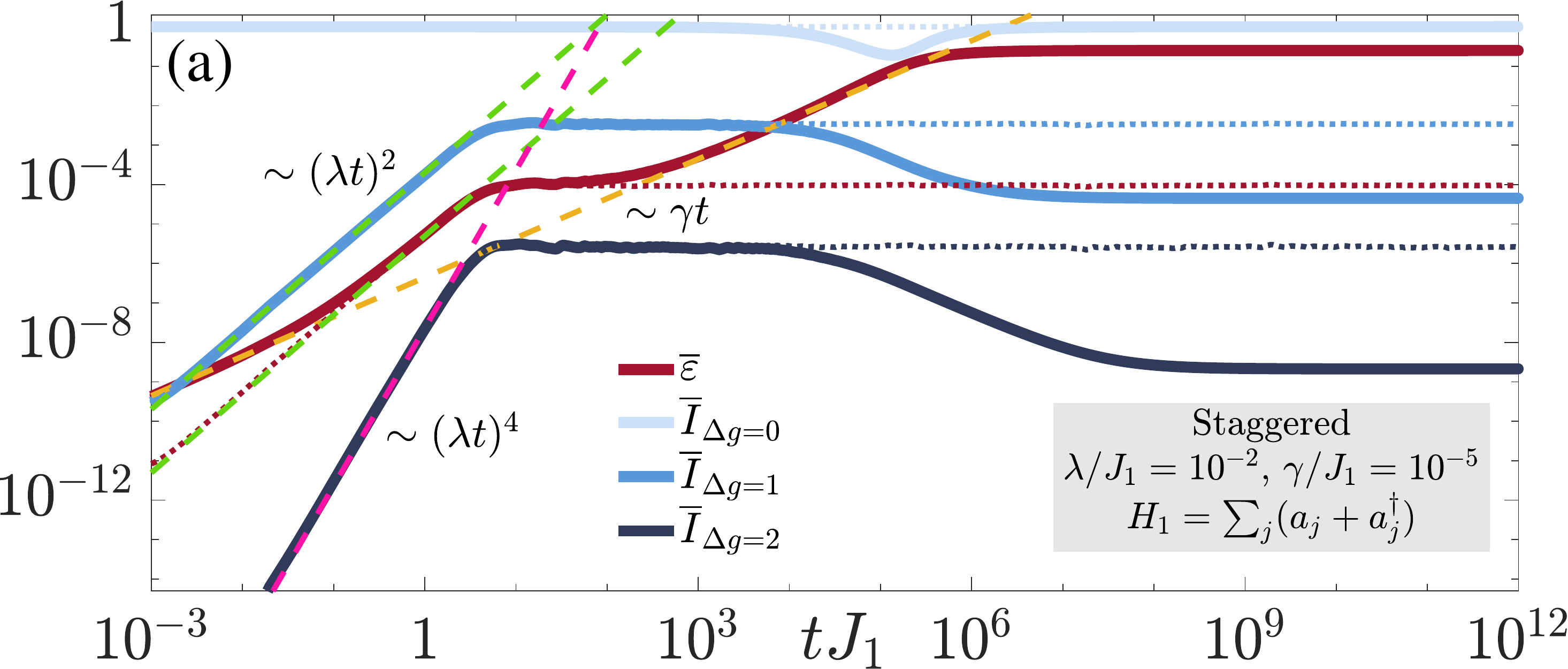}\\
	\includegraphics[width=.48\textwidth]{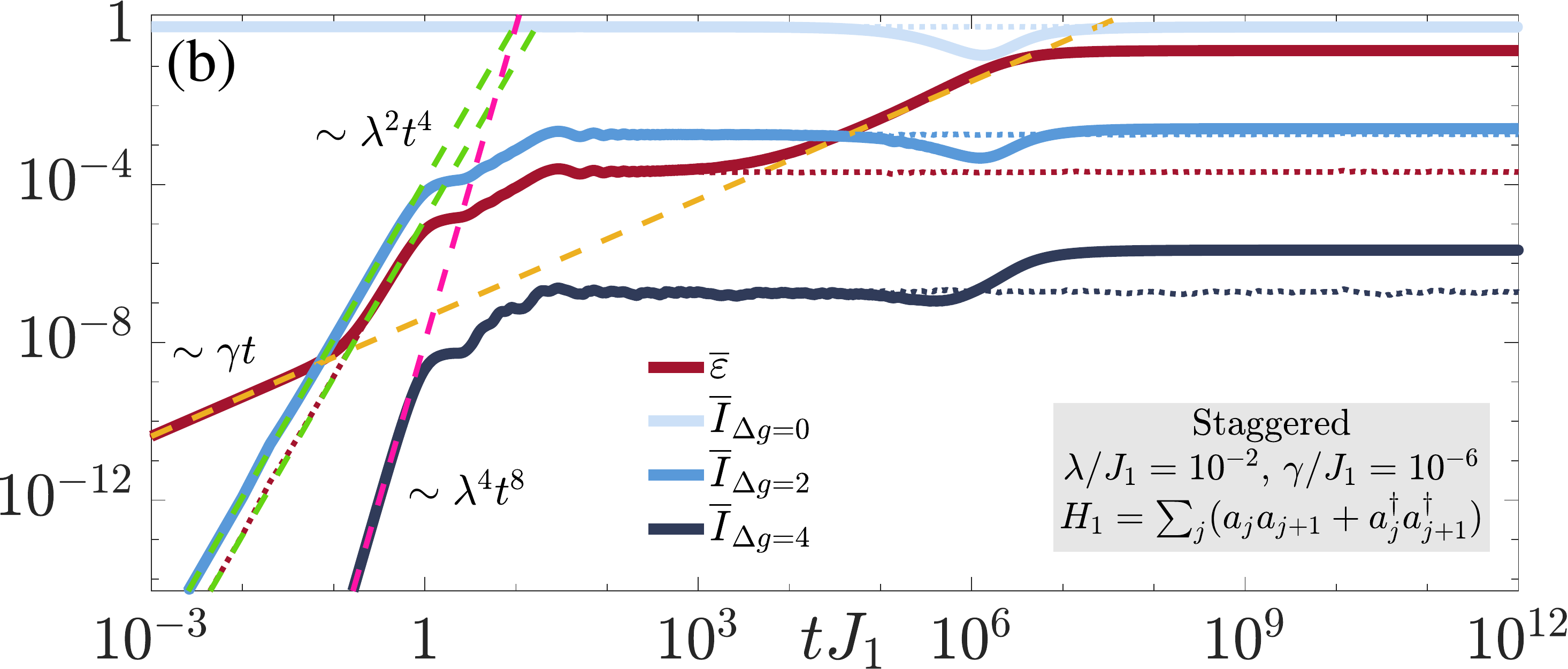}
	\caption{(Color online). Multiple quantum coherences $I_{\Delta g}$ (solid blue curves) at fixed coherent- and incoherent-error strengths $\lambda$ and $\gamma$ (see gray boxes for exact values), respectively, in the quench dynamics of the extended Bose-Hubbard model starting in the staggered initial state of Fig.~\ref{fig:eBHM_InitialStates}(a). For reference, we also show the symmetry violation (solid red curve) as well as the MQCs in the purely coherent case ($\gamma=0$; same color but dotted lines). In (a) the coherent errors are single-body terms given by Eq.~\eqref{eq:H1OneBody}. Decoherence compromises all $I_{\Delta g>0}$, causing them to settle into steady-state values lower than those in the purely coherent case. However, $I_0$ does not deviate much from its coherent steady-state value, suggesting that decoherence does not affect quantum coherences within each sector much. In (b) the coherent errors are the two-body terms given by Eq.~\eqref{eq:H1TwoBody}. Even though $I_0$ behaves the same as in the case of single-body errors, $I_{\Delta g>0}$ here are nonzero only in the case of even $\Delta g$ and settle into steady-state values \textit{larger} than those in the purely coherent case. }
	\label{fig:eBHM_MQC} 
\end{figure}

\begin{figure}[htp]
	\centering
	\includegraphics[width=.48\textwidth]{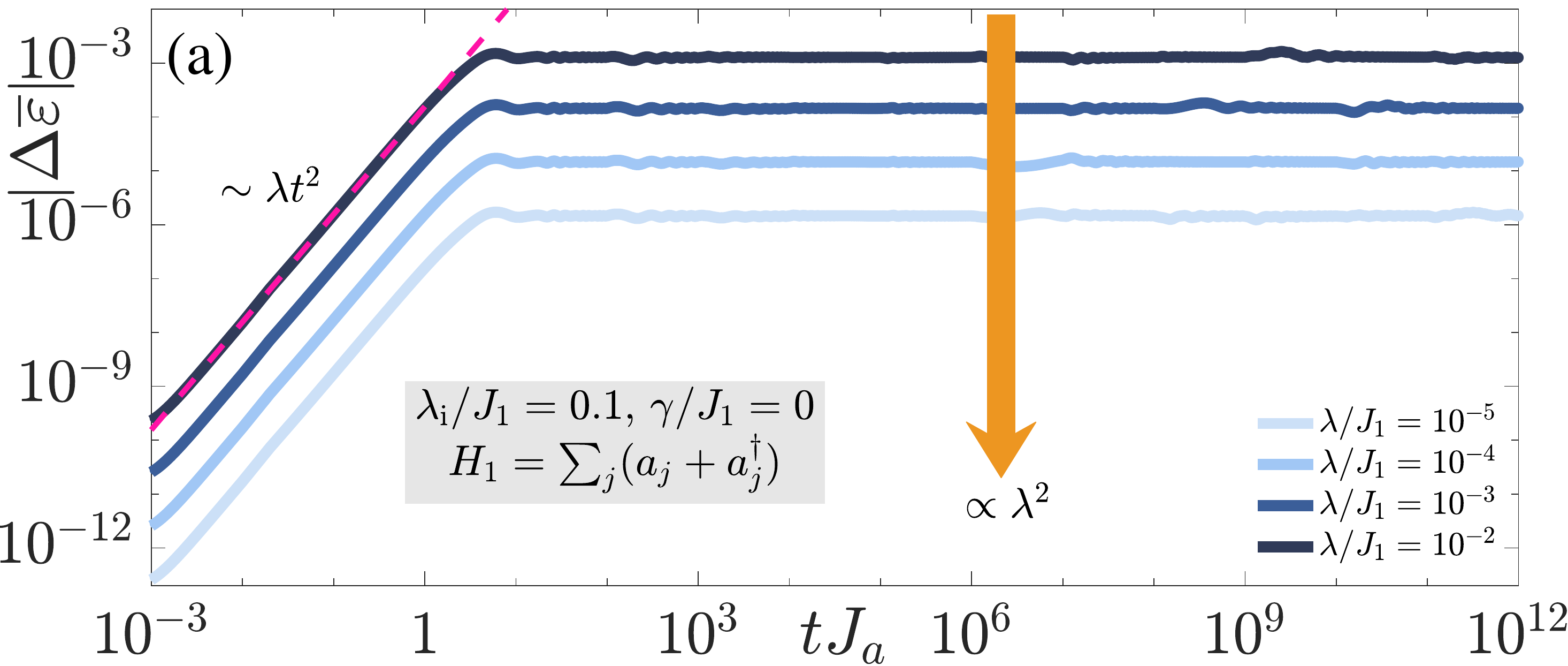}\\
	\includegraphics[width=.48\textwidth]{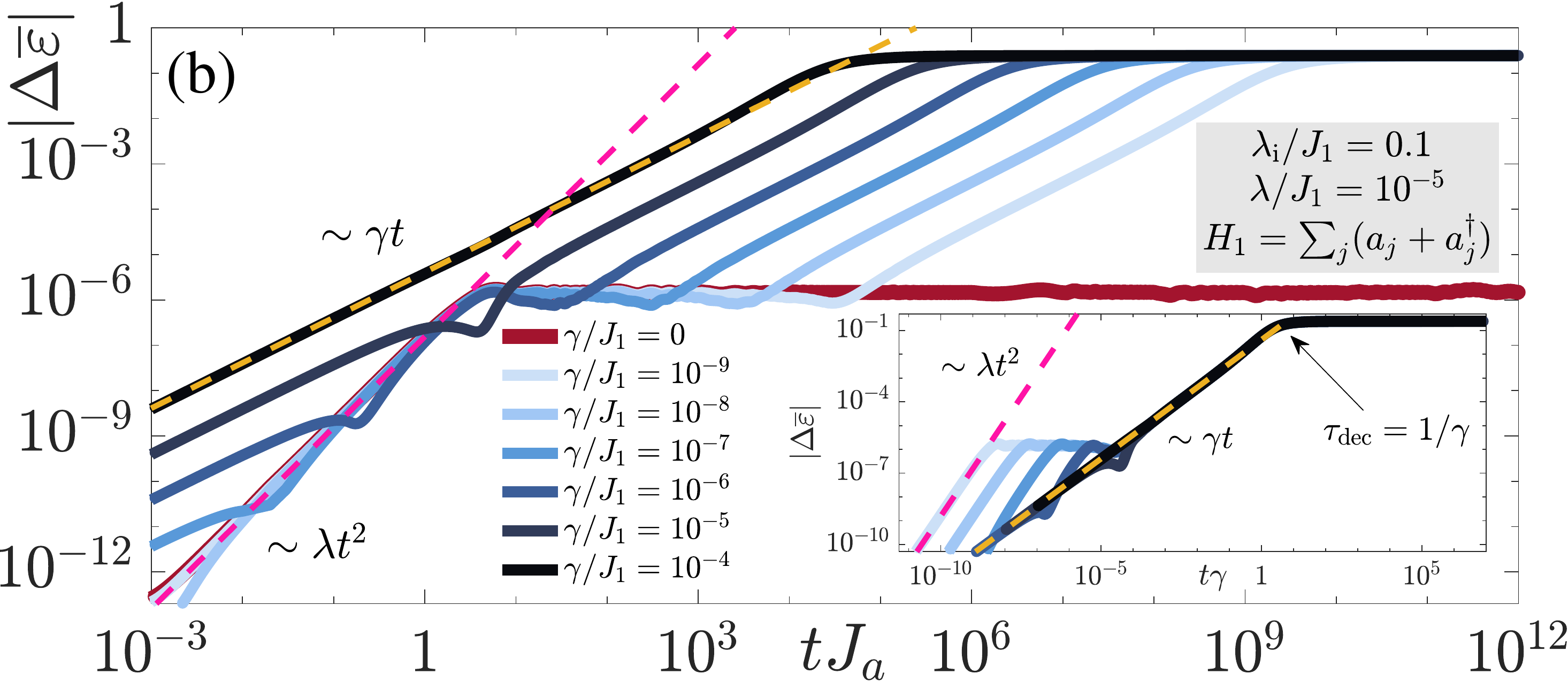}
	\caption{(Color online). Starting in the unsymmetric ground state $\rho_0$ of $H_0+\lambda_\text{i} H_1$ with $H_1$ given in Eq.~\eqref{eq:H1OneBody}. (a) Quench dynamics of the symmetry-violation change in the case of a closed eBHM chain with $N=6$ matter sites. Since $\rho_0$ is unsymmetric, the leading coherent contribution to the symmetry violation is $\propto\lambda t^2$. Note here that the steady-state value is $\propto\lambda$ rather than $\propto\lambda^2$ as in Fig.~\ref{fig:eBHM_closed} when the initial state is symmetric. (b) Decoherence brings about a diffusive-to-ballistic crossover at an earlier timescale $\gamma/\lambda<\gamma/\lambda^2$ than the case of a symmetric initial state.}
	\label{fig:eBHM_unsymmetric} 
\end{figure}

\subsection{Multiple quantum coherences}\label{sec:eBHM_MQC}
In order to further quantify the effects of decoherence especially in the late-time dynamics of our quenches, we analyse multiple quantum coherences (MQC). These are experimentally accessible quantities that have been used to study quantum coherences in nuclear magnetic resonance imaging,\cite{Pines1985,Pines1986,Pines1986} as well as decoherence effects on correlated spins\cite{SpinDecoherence2014} and many-body localization.\cite{NMRLocalizationPRL2010,NMRLocalizationScience2015} Moreover, they have also been connected to multipartite entanglement\cite{Gaerttner2018} and out-of-time-ordered correlators,\cite{Gaerttner2018,LewisSwan2020} and they have been measured in trapped-ion experiments.\cite{Garttner2017}

Let us call $\Delta g$ the difference in global charge between two global-symmetry sectors. The associated MQC is then defined as
\begin{subequations}\label{eq:eBHM_MQC}
\begin{align}
I_{\Delta g}&=\Tr\big\{\rho_{\Delta g}^\dagger\rho_{\Delta g}\big\},\\
\rho_{\Delta g}&=\sum_g P_{g+\Delta g}\rho P_g,
\end{align}
\end{subequations}
where $P_g$ is the projector onto the sector of global charge $g$. 

In Fig.~\ref{fig:eBHM_MQC}, we show the MQC for various $\Delta g$ for both unitary errors in Eqs.~\eqref{eq:H1OneBody} and~\eqref{eq:H1TwoBody} at fixed $\lambda$ and $\gamma$. For the single-body coherent errors of Eq.~\eqref{eq:H1OneBody}, we see that odd values of $\Delta g$ give rise to a finite MQC as shown in Fig.~\ref{fig:eBHM_MQC}(a). This behavior is plausible, as in this case a first-order process in $H_1$ removes or adds a single boson on a given site $j$. We see that the MQC is dominated by processes within the same sector (corresponding to $\Delta g=0$). Whereas the symmetry violation (red curve) starts at $t=0$ to grow $\sim\gamma t$, the MQCs, being measures of quantum \textit{coherences} between global-symmetry sectors, do not show any scaling related to incoherent processes at early times. Rather, their growth is $\sim(\lambda^2 t^2)^{\Delta g}$. As can be expected in case of decoherence,\cite{Halimeh2020f} the $I_{\Delta g}$ with $\Delta g>0$ show a decrease at $t\approx1/\gamma$ as compared to the steady-state value of the purely coherent case. In contrast, $I_0$ exhibits an increase in its values after an initial decrease, finally settling at roughly the coherent steady-state value. This behavior suggests that even though decoherence compromises coherences between different sectors, those within the same sector are not affected much by the decoherence studied here.

By changing $H_1$ to Eq.~\eqref{eq:H1TwoBody}, the above picture changes, as shown in Fig.~\ref{fig:eBHM_MQC}(b). First, MQCs due to odd $\Delta g$ are identically zero, as there can be no coherent processes between sectors differing by an odd number of bosons---$H_1$ as per Eq.~\eqref{eq:H1TwoBody} removes or adds two neighboring bosons simultaneously. Furthermore, due to the staggered occupation of $\rho_0$, the MQCs scale $\sim(\lambda t^2)^{\Delta g}$.
 Whereas $I_0$ behaves the same as in the case of the single-body error, $I_{\Delta g>0}$ behave fundamentally differently in case of the two-body error in Eq.~\eqref{eq:H1TwoBody}. Interestingly, at $t\approx1/\gamma$ they begin to decrease in value, but at later times they increase again and settle at a value larger than that of their purely coherent dynamics, represented by dotted lines in Fig.~\ref{fig:eBHM_MQC}. This suggests that decoherence populates sectors that cannot be populated by $H_1$ alone. Indeed, in the case of purely unitary dynamics with only coherent errors as in Eq.~\eqref{eq:H1TwoBody}, we have checked that $\langle P_g\rangle=0$ identically for odd $g$. In the case of decoherence, however,  $\langle P_g\rangle$ acquire nonzero values, and this allows then, through $H_1$, coherence within and between sectors with odd global charges, but separated by an even $\Delta g$. This is quite counterintuitive in that decoherence here seems to help in building up quantum coherences by allowing access to previously inaccessible sectors.

\subsection{Unsymmetric initial state}
We have so far considered only initial states lying in the half-filling sector. Let us now consider an initial state $\rho_0$ that is \textit{unsymmetric}, i.e., $\big[\mathcal{G},\rho_0\big]\neq 0$. In particular, we consider $\rho_0$ to be the ground state of the Hamiltonian $H_0+\lambda_\text{i} H_1$, where $\lambda_\text{i}=0.1$ is the prequench strength of the error term and $H_1$ is given by Eq.~\eqref{eq:H1OneBody}. Such a scenario may arise when aiming at adiabatically preparing the ground state of $H_0$ in the presence of errors. As has been done until now, we quench $\rho_0$ with $H_0+\lambda H_1$ in the presence of decoherence under the jump operators $L_j=a_j$. 

The ensuing dynamics of the change in symmetry violation $|\Delta\varepsilon|$ is shown in Fig.~\ref{fig:eBHM_unsymmetric}. In the case of no decoherence ($\gamma=0$), shown in Fig.~\ref{fig:eBHM_unsymmetric}(a), $|\Delta\varepsilon|\sim\lambda t^2$ at short times rather than $\lambda^2t^n$ (even $n\geq2$) as in the case of a symmetric initial state; cf.~Figs.~\ref{fig:eBHM_closed}--\ref{fig:eBHM_TwoBodyError_open}. As explained in Sec.~\ref{sec:TDPT_coherent} through TDPT, the coherent contribution to the symmetry violation $\propto\lambda t^2$, given by Eq.~\eqref{eq:coherent_last} always vanishes when the initial state is symmetric or an eigenstate of $H_0$, but not when $\rho_0$ is unsymmetric yet not an eigenstate of $H_0$ as in the case of Fig.~\ref{fig:eBHM_unsymmetric}. Moreover, the contribution $\propto\lambda t$ in Eq.~\eqref{eq:lamt} completely vanishes in this case, since the errors $H_1$ in preparing $\rho_0$ are the same as those in the subsequent dynamics (see Appendix~\ref{sec:TDPT_coherent}). Note how,  consequently, the steady-state value at which $|\Delta\varepsilon|$ settles is $\propto\lambda$. Upon introducing decoherence ($\gamma>0$) in Fig.~\ref{fig:eBHM_unsymmetric}(b), a diffusive-to-ballistic crossover at $t\propto\gamma/\lambda$ occurs taking the spread of the symmetric violation from diffusive $\sim\gamma t$ to ballistic $\sim\lambda t^2$. Note that the diffusive-to-ballistic crossover here occurs at an earlier timescale than that in the case of a symmetric initial state. Finally, as in the case of a symmetric initial state, the symmetry violation reaches its maximal value of $g_\text{tar}^2/N^2=1/4$ also when $\rho_0$ is unsymmetric.

\section{$\mathrm{Z}_2$ lattice gauge theory}\label{sec:Z2LGT}
In this Section, we present results that supplement those on a $\mathrm{Z}_2$ gauge theory presented in Ref.~\onlinecite{Halimeh2020f} in various ways: by studying the effect of decoherence under different Lindblad operators including those for particle loss; by starting in various initial states including gauge-noninvariant ones; by investigating the effect of dissipation and dephasing set at different environment-coupling strengths; by analyzing the addition of energy-penalty terms on the dynamics; and by adding further results on MQCs under decoherence.

\begin{figure}[htp]
	\centering
	\includegraphics[width=.48\textwidth]{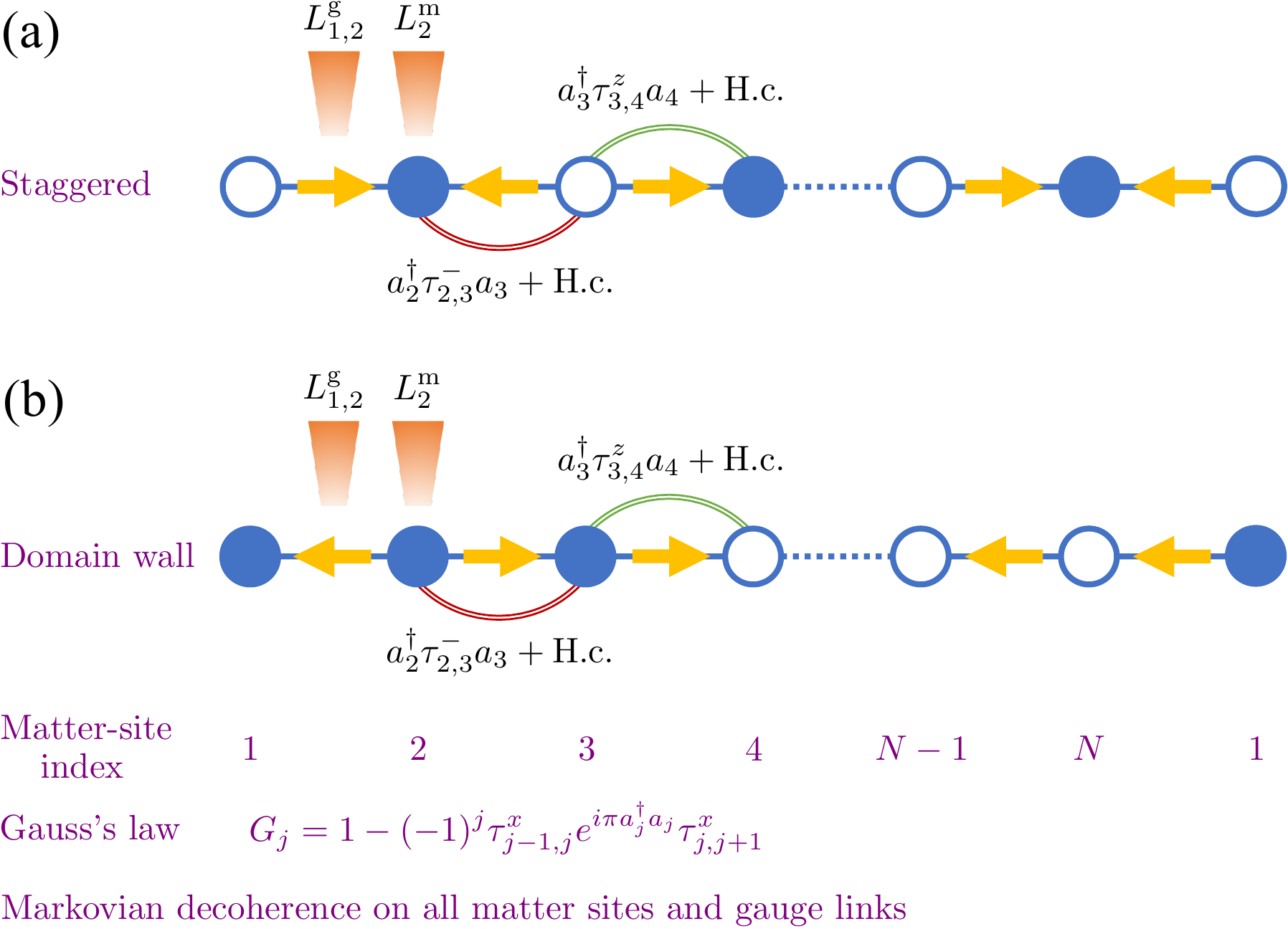}
	\caption{(Color online). Symmetric initial product states used in the dynamics of the $\mathrm{Z}_2$ LGT in the presence of coherent errors. (a) A staggered product state where odd matter sites are empty and each even matter site contains one hard-cose boson, with the electric fields along links pointing from odd to even matter sites, thereby satisfying Gauss's law at each local constraint. (b) A ``domain-wall'' product state where the left half of the lattice has a hard-cose boson at each site and the right half of the chain is empty. The electric fields along the links are oriented such that Gauss's law is satisfied at each local constraint.}
	\label{fig:Z2LGT_InitialStates} 
\end{figure} 

\subsection{Model and quench protocol}
A recent experiment\cite{Schweizer2019} has employed a Floquet setup to successfully implement a building block of the $\mathrm{Z}_2$ LGT described by the Hamiltonian\cite{Zohar2017,Barbiero2019,Borla2019}
\begin{align}\label{eq:H0}
H_0=&\,\sum_{j=1}^N\big[J_a\big(a^\dagger_j\tau^z_{j,j+1}a_{j+1}+\text{H.c.}\big)-J_f\tau^x_{j,j+1}\big], 
\end{align}
where $N$ is the number of matter sites, $a_j$ is the annihilation operator of a hard-core boson at site $j$ obeying the canonical commutation relations $[a_j,a_l]=0$ and $[a_j,a^\dagger_l]=\delta_{j,l}(1-2a_j^\dagger a_j)$, and the Pauli matrix $\tau^{x(z)}_{j,j+1}$ represents the electric (gauge) field at the link between matter sites $j$ and $j+1$. The first term of Eq.~\eqref{eq:H0} represents assisted matter tunneling and gauge flipping at strength $J_a$, which, e.g., forms the essence of Gauss's law in quantum electrodynamics. The electric field's energy is given by $J_f$. In this work, we adopt periodic boundary conditions, which means our effective system size is $2N$, and we set $J_a=1$ and $J_f=0.54$ throughout our paper, even though we have checked that our conclusions are not restricted to these values. 

Gauge invariance is embodied in local conservation laws, the generators of which are
\begin{align}
G_j=1-(-1)^j\tau^x_{j-1,j}(1-2a_j^\dagger a_j)\tau^x_{j,j+1},
\end{align}
where $[H_0,G_j]=0,\,\forall j$. As discussed above, the eigenvalues $g_j$ of $G_j$ are known as \textit{local charges}, and a set of their values $\mathbf{g}=\{g_1,g_2,\ldots,g_N\}$ defines a gauge-invariant sector (see Sec.~\ref{sec:definition}). A gauge-invariant supersector $M$ is defined as the set of gauge-invariant sector that satisfy $\sum_jg_j=2M$ (see Sec.~\ref{sec:glossary}).
\begin{figure*}[htp]
	\centering
	\includegraphics[width=.48\textwidth]{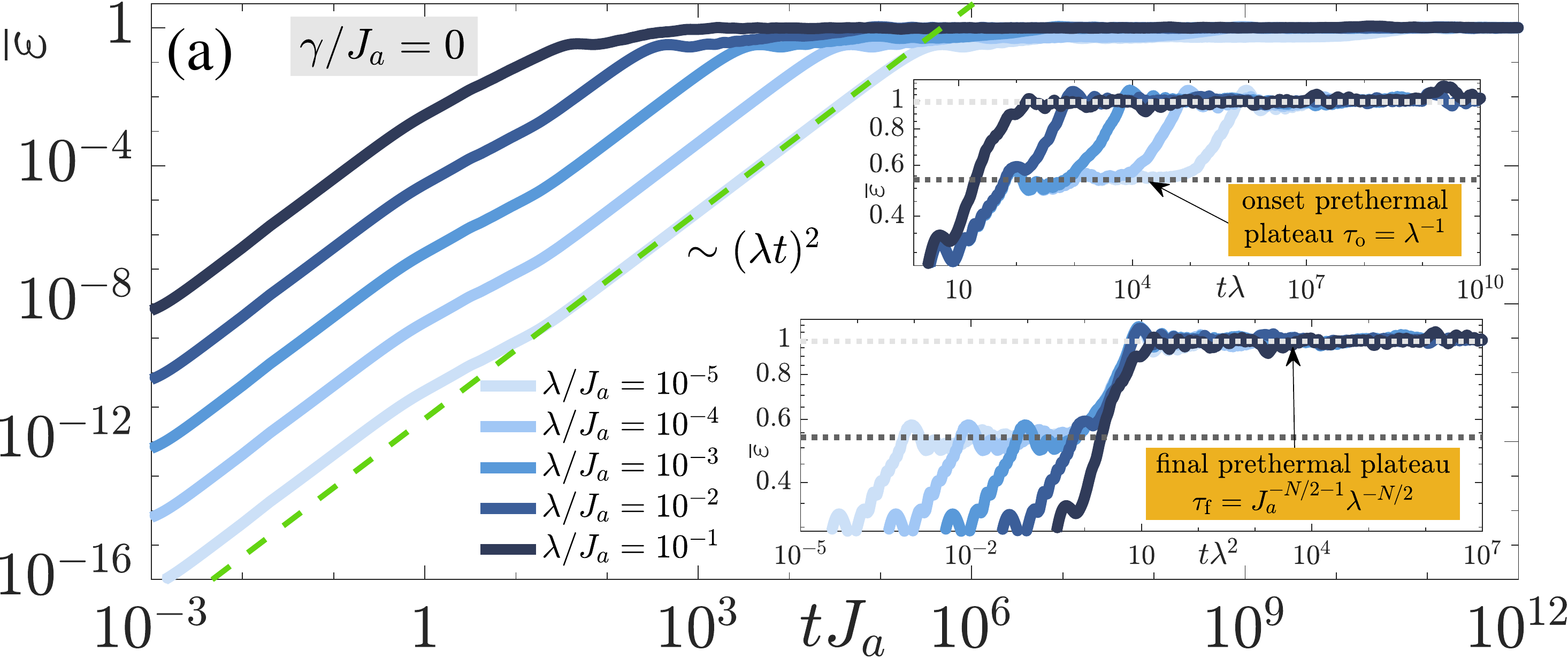}\quad
	\includegraphics[width=.48\textwidth]{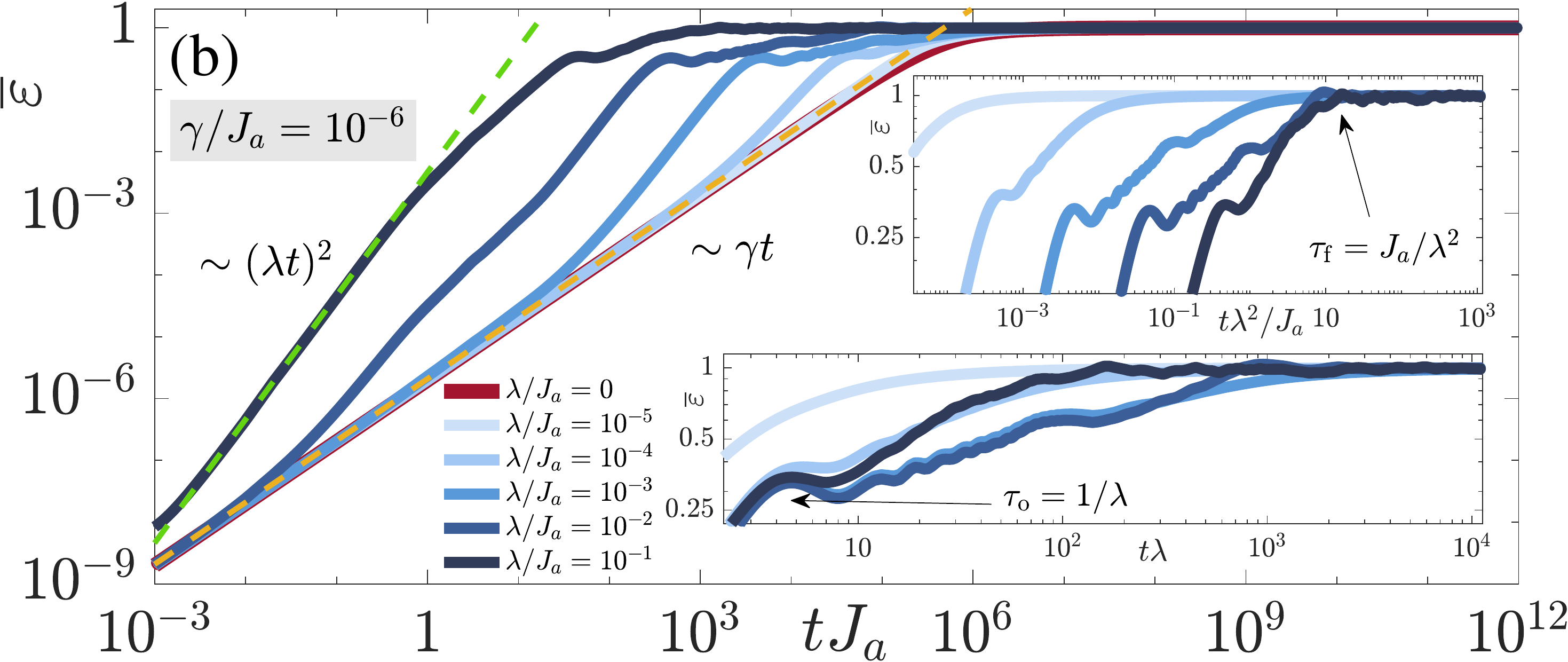}\\
	\includegraphics[width=.48\textwidth]{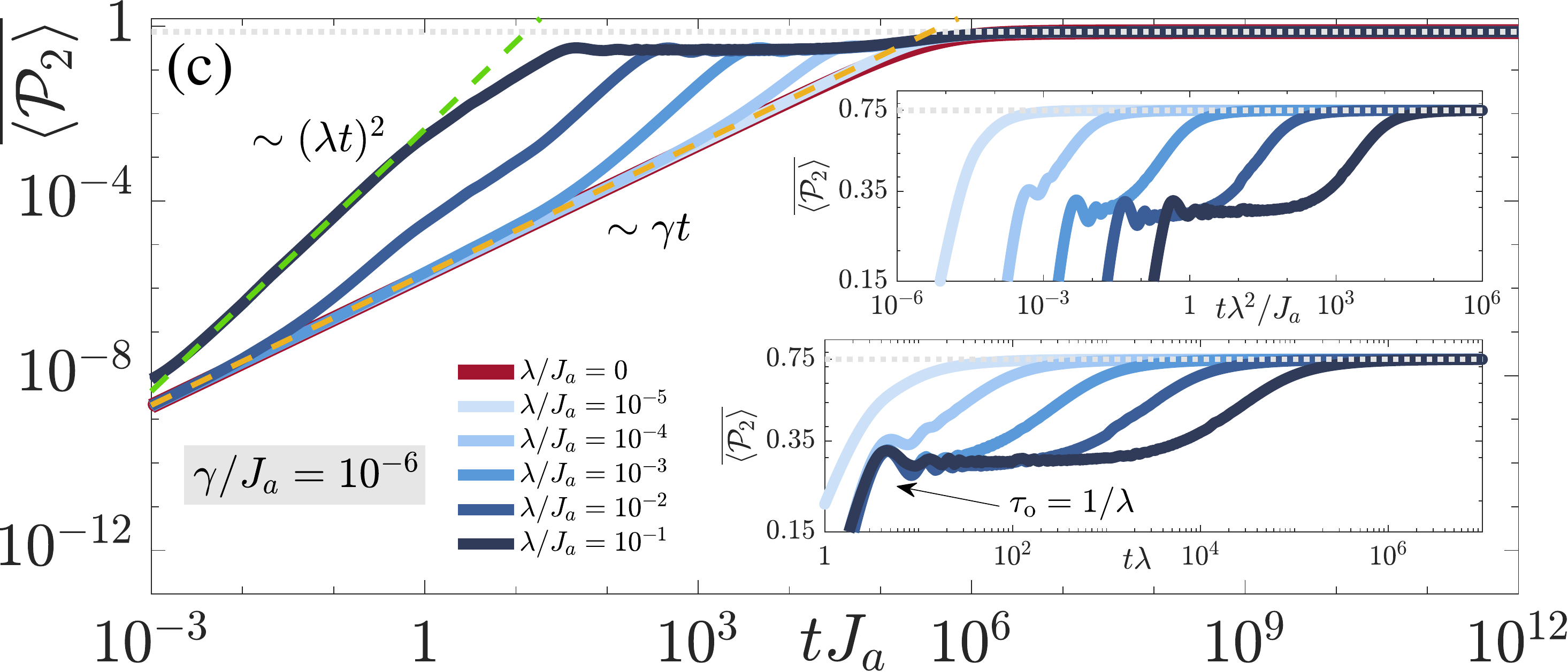}\quad
	\includegraphics[width=.48\textwidth]{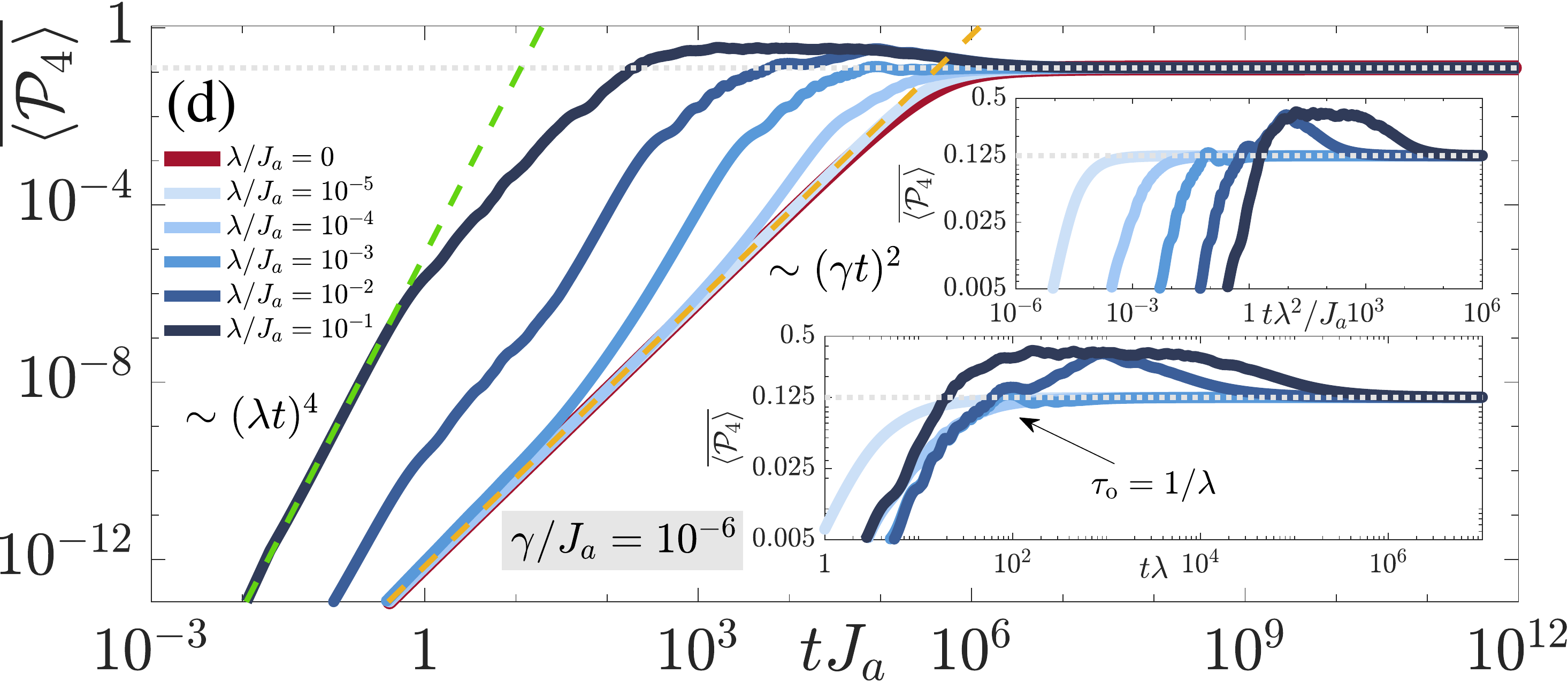}\\
	\includegraphics[width=.48\textwidth]{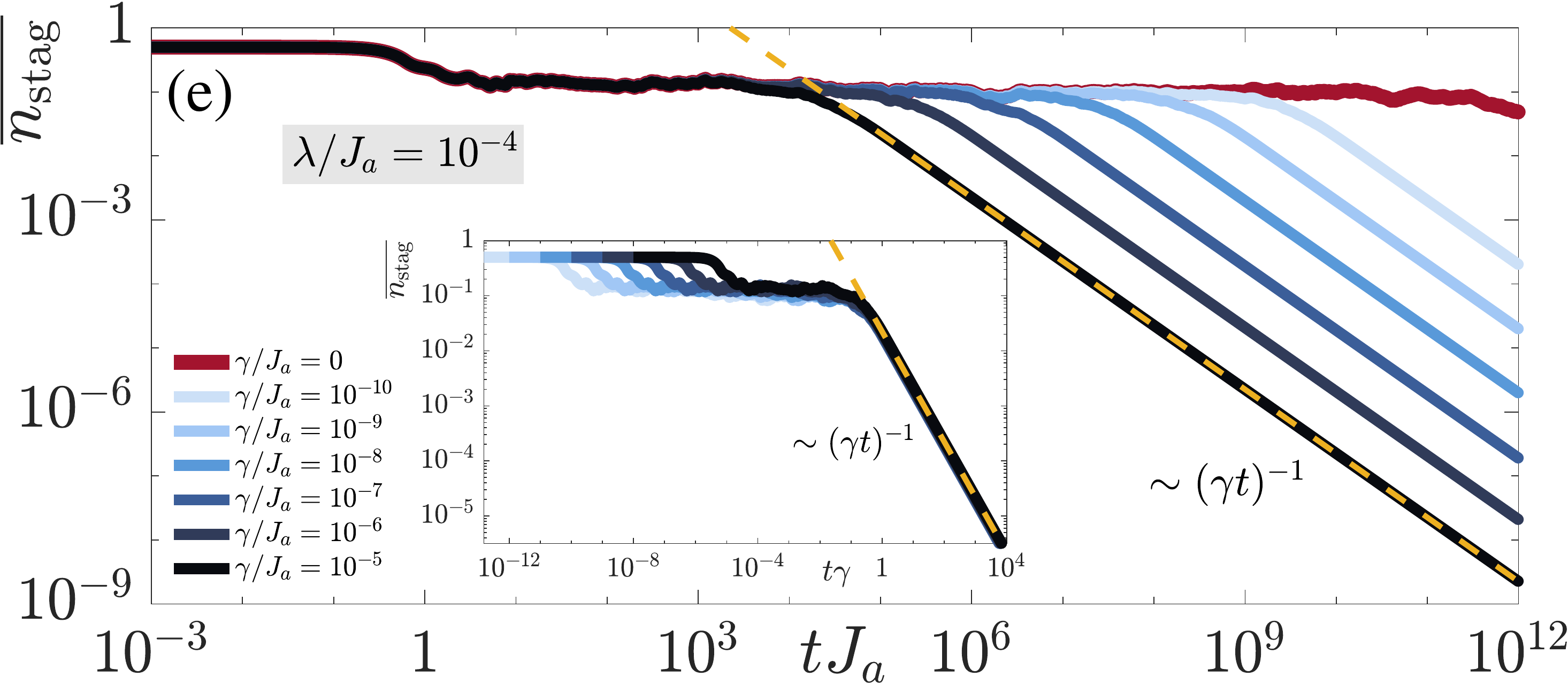}\quad
	\includegraphics[width=.48\textwidth]{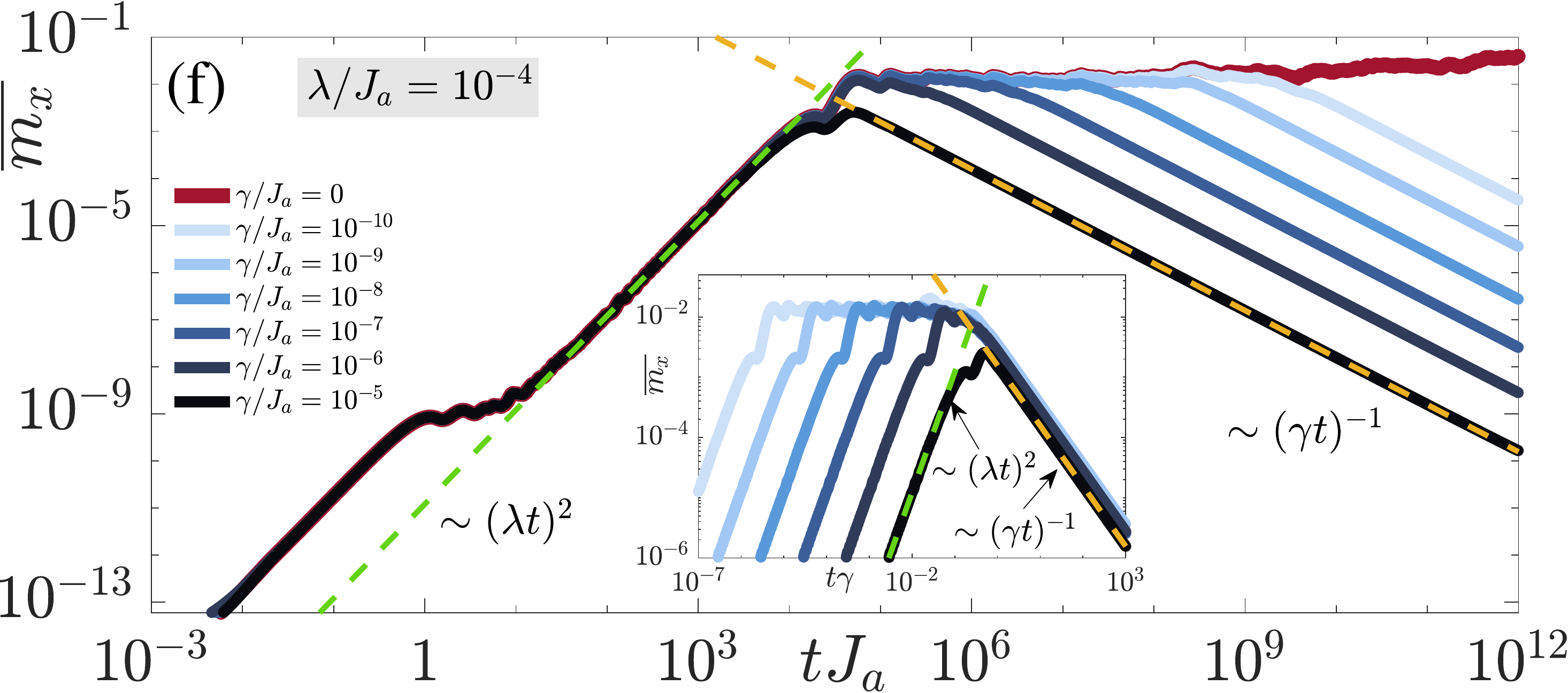}
	\caption{(Color online). (a) Full unitary dynamics of the quench scenario illustrated in Fig.~\ref{fig:Z2LGT_InitialStates}(a). Two prethermal plateaus at timescales $\lambda^{-1}$ and $J_a\lambda^{-2}$ (see insets) appear as explained numerically and analytically through a Magnus expansion in Refs.~\onlinecite{Halimeh2020b} and~\onlinecite{Halimeh2020c}. (b) Complementary results to those of the joint submission Ref.~\onlinecite{Halimeh2020f}, where here we fix $\gamma$ and scan $\lambda$. The conclusion remains the same, with the gauge violation exhibiting diffusive scaling $\varepsilon\sim\gamma t$ at short times for sufficiently small $\lambda$, and ballistic scaling $\sim(\lambda t)^2$ at sufficiently large $\lambda$, with the timescale of the crossover from the former to the latter being $t\propto\gamma/\lambda^2$. Upper inset shows the second (and final in the case of $N=4$ matter sites) prethermal timescale getting compromised at smaller values of $\lambda$ at which $\gamma=10^{-6}$ dominates. Lower inset shows that the first prethermal timescale is more resilient than the second, as its timescale persists for smaller values of $\lambda$. (c,d) Same as (b) but for the supersector projectors $\mathcal{P}_2$ and $\mathcal{P}_4$. As we can see, the crossover from the diffusive to ballistic regime is again at $t\propto\gamma/\lambda^2$, but whereas $\mathcal{P}_2$ shows the same scaling orders as $\varepsilon$, $\mathcal{P}_4\sim\gamma^2t^2$ in the diffusive regime and $\mathcal{P}_4\sim\lambda^4t^4$ in the ballistic regime. (e,f) Influence of gauge violation on local observables. As exemplified by (e) the staggered particle density and (f) the electric field, the dynamics is practically indistinguishable from the decoherence-free model before the timescale $\propto \gamma^{-1}$. Note how the electric field shows no diffusive behavior at the onset as the gauge violation and supersector projectors do. This is because the corresponding correction $\gamma t\mathcal{L}\rho_0$ to the unitary part of the density matrix makes a vanishing contribution to the electric field, as explained in Sec.~\ref{sec:TDPT_leadingIncoherent}.
	}
	\label{fig:gammaFixed} 
\end{figure*}

In the implementation of the $\mathrm{Z}_2$ LGT without unrealistic fine-tuning, coherent error terms emerge with $[H_1,G_j]\neq0$. Here, inspired by the effective coherent errors of the building block of Ref.~\onlinecite{Schweizer2019}, we assume the errors to have the form of unassisted matter tunneling and gauge flipping, which can be formalized as
\begin{align}\nonumber
\lambda H_1=&\,\lambda \sum_{j=1}^N\Big[\big(c_1a_j^\dagger \tau^-_{j,j+1} a_{j+1}+c_2a_j^\dagger\tau^+_{j,j+1} a_{j+1}+\text{H.c.}\big)\\\label{eq:H1}
&+a_j^\dagger a_j\big(c_3\tau^z_{j,j+1}-c_4\tau^z_{j-1,j}\big)\Big].
\end{align}
The strength of these errors is given by $\lambda$, and the coefficients $c_{1\ldots4}$ depend on a dimensionless driving parameter $\chi$ that is tunable in the experiment of Ref.~\onlinecite{Schweizer2019}. The specific expressions for these coefficients can be found in Appendix~\ref{sec:NumSpec}. Just as in the joint submission,\cite{Halimeh2020f} we show here results for $\chi=1.84$, but we have also checked that our results hold for various values of $\chi$ within the range found in the Floquet setup of Ref.~\onlinecite{Schweizer2019}.

As for the case of a global symmetry discussed above, the system is prepared in an initial state $\rho_0$, which at $t=0$ is quenched by $H_0+\lambda H_1$ and decoherence is turned on. The subsequent dynamics is computed using the Lindblad master equation
\begin{align}\nonumber
\dot{\rho}=&-i[H_0+\lambda H_1,\rho]\\\nonumber
&+\gamma\sum_{j=1}^N\Big(L^\text{m}_j\rho L^{\text{m}\dagger}_j+L^\text{g}_{j,j+1}\rho L^{\text{g}\dagger}_{j,j+1}\\\label{eq:EOM}
&-\frac{1}{2}\big\{L^{\text{m}\dagger }_jL^\text{m}_j+L^{\text{g}\dagger }_{j,j+1}L^\text{g}_{j,j+1},\rho\big\}\Big),
\end{align}
where $L^\text{m}_j$ and $L^\text{g}_{j,j+1}$ are the jump operators coupling the matter and gauge fields, respectively, to the environment at strength $\gamma$. We are interested in dynamics of the gauge violation, supersector projector, electric field, and staggered boson number, given respectively by,
\begin{align}\label{eq:Measure}
\varepsilon(t)&=\Tr\big\{\mathcal{G}\rho(t)\big\},\,\,\,\mathcal{G}=\frac{1}{N}\sum_j\big[G_j-g_j^\text{tar}\big],\\\label{eq:SupProj}
\langle\mathcal{P}_M(t)\rangle&=\Tr\big\{\rho(t)\mathcal{P}_M\big\},\,\,\,\mathcal{P}_M=\sum_{\mathbf{g};\,\sum_jg_j=2M}P_\mathbf{g},\\
m_x(t)&=\frac{1}{N}\Big|\Tr\Big\{\rho(t)\sum_j\tau_{j,j+1}^x\Big\}\Big|,\\
n_\text{stag}(t)&=\frac{1}{N}\Big|\Tr\Big\{\rho(t)\sum_j(-1)^ja_j^\dagger a_j\Big\}\Big|.
\end{align}
Note that the form of the gauge violation in Eq.~\eqref{eq:Measure} is specific for the $\mathrm{Z}_2$ LGT. It is a special case of Eq.~\eqref{eq:MeasureGeneral} using $g_j^2=2g_j$ and dropping an irrelevant factor of $2$.

\subsection{Quench dynamics}

As shown in Ref.~\onlinecite{Halimeh2020f}, the coexistence of unitary and incoherent gauge-breaking processes leads to competing timescales due to $\lambda>0$ and $\gamma>0$ in the $\mathrm{Z}_2$ LGT. While incoherent gauge-breaking processes yield a single timescale $1/\gamma$, coherent errors can generate a sequence or \textit{staircase} of prethermal plateaus with timescales $J_a^{s-1}/\lambda^s$ with $s=0,1,2,\ldots,N/2$.\cite{Halimeh2020b,Halimeh2020c} These coherent timescales arise due to unitary dynamics in a gauge theory as a result of resonances between different gauge-invariant sectors coupled through $H_1$, as can be shown through a Magnus expansion.\cite{Halimeh2020c} In the case of the initial states shown in Fig.~\ref{fig:Z2LGT_InitialStates} with $N=4$ matter sites, this means two plateaus at timescales $1/\lambda$ and $J_a/\lambda^2$ (the one at timescale $\propto\lambda^0$ does not appear in this case\cite{Halimeh2020b}), at which maximal violation is attained; see Fig.~\ref{fig:gammaFixed}(a) for the \textit{staggered} initial product state shown in Fig.~\ref{fig:Z2LGT_InitialStates}(a).

The picture changes significantly when $\gamma>0$; see Fig.~\ref{fig:gammaFixed}(b). When $\lambda=0$ in this case, the gauge violation accumulates diffusively as $\varepsilon\sim\gamma t$ until it reaches a maximal value of unity at $t\approx1/\gamma$. This gauge violation due to purely incoherent gauge-breaking processes shows no signatures of prethermalization. The picture starts to change for $\lambda>\gamma$, as then the prethermal plateau at timescale $\propto1/\lambda$ can still appear unaffected by the decoherence, which becomes dominant for $t\gtrsim1/\gamma$; see lower inset of Fig.~\ref{fig:gammaFixed}(b). The effects of decoherence on the second plateau, which in the purely unitary case appears at timescale $\propto J_a/\lambda^2$, are more prominent as can be seen in the upper panel of Fig.~\ref{fig:gammaFixed}(b). This plateau survives only when $\lambda^2/J_a\gtrsim\gamma$ as then the final prethermal timescale $\propto J_a/\lambda^2$ appears earlier than the decoherence timescale of $/\gamma$.

It is interesting to examine again the short-time scaling of the gauge violation for finite $\lambda$ in Fig.~\ref{fig:gammaFixed}(b). The behavior concurs with the conclusions of Ref.~\onlinecite{Halimeh2020f}, where we observe the diffusive scaling $\varepsilon\sim\gamma t$ for $t\lesssim \gamma/\lambda^2$, while for later times $t\gtrsim\gamma/\lambda^2$ the violation is dominated by coherent errors and $\varepsilon\sim(\lambda t)^2$. Intriguingly, we thus find in general two regimes where incoherent errors dominate at finite $\lambda$: the first at evolution times $t\lesssim\gamma/\lambda^2$ and the second for $t\gtrsim1/\gamma$. At intermediate times, the coherent gauge-breaking processes dominate when $\lambda>\gamma$, and both prethermal plateaus appear for $N=4$ matter sites when $\lambda^2>\gamma J_a$. 

Let us now again look at the dynamics of the supersector projectors in the presence of decoherence. This is shown for the projectors onto the supersectors $M=2$ and $M=4$ in Fig.~\ref{fig:gammaFixed}(c,d). Congruent to the conclusions of the joint submission,\cite{Halimeh2020f} we get the same short-time scaling for the supersector projectors $\mathcal{P}_2$ and $\mathcal{P}_4$, with the crossover from the diffusive regime where $\langle\mathcal{P}_2\rangle\sim\gamma t$ and $\langle\mathcal{P}_4\rangle\sim\gamma^2 t^2$ at $t\lesssim\gamma/\lambda^2$ to the ballistic regime where $\langle\mathcal{P}_2\rangle\sim\lambda^2 t^2$ and $\langle\mathcal{P}_4\rangle\sim\lambda^4 t^4$ at $t\gtrsim\gamma/\lambda^2$. The deterioration of the first prethermal timescale can also be observed in these quantities, and that of the second prethermal plateau is observed in $\langle\mathcal{P}_4\rangle$. The larger $\lambda$ is, the greater the integrity of the prethermal plateaus, with the first plateau exhibiting greater resilience as it survives smaller values of $\lambda$ than its second counterpart. Interestingly, at long times $t\gtrsim1/\gamma$, both projectors relax to the values $\langle\mathcal{P}_2\rangle\approx0.125$ and $\langle\mathcal{P}_4\rangle=\langle\mathcal{P}_0\rangle\approx0.75$, meaning that $\langle\mathcal{P}_2\rangle/\langle\mathcal{P}_4\rangle=\langle\mathcal{P}_2\rangle/\langle\mathcal{P}_0\rangle=6$, which is equal to the ratio of number of gauge sectors in each supersector. This generally does not happen in the case of no decoherence ($\gamma=0$), but in the presence of decoherence at any $\gamma>0$, the long-time limit will ascribe to this behavior. This is due to the fact that the gauge violation has fully diffused in the space of gauge sectors, occupying an equal distribution among all of them.

We also include the dynamics of the staggered boson number in Fig.~\ref{fig:gammaFixed}(e) [recall that the total boson number is conserved since $H_0$ and $H_1$ both have global $\mathrm{U}(1)$ symmetry, and there is only dephasing on the matter fields] and the electric field in Fig.~\ref{fig:gammaFixed}(f). One cannot discern any diffusive behavior at early times in these observables (even by looking at the deviation from the fully unitary case). A deeper reason may be that these local observables are not related to a divergence measure through the gauge sectors, contrary to the gauge violation (see Sec.~\ref{sec:divergencemeasure}).
The leading-order (in $\gamma$) correction to the unitary part of the density matrix is $\gamma t\mathcal{L}\rho_0$ makes a vanishing contribution to both of these observables as discussed in Sec.~\ref{sec:TDPT_leadingIncoherent}.

\begin{figure}[htp]
	\centering
	\includegraphics[width=.48\textwidth]{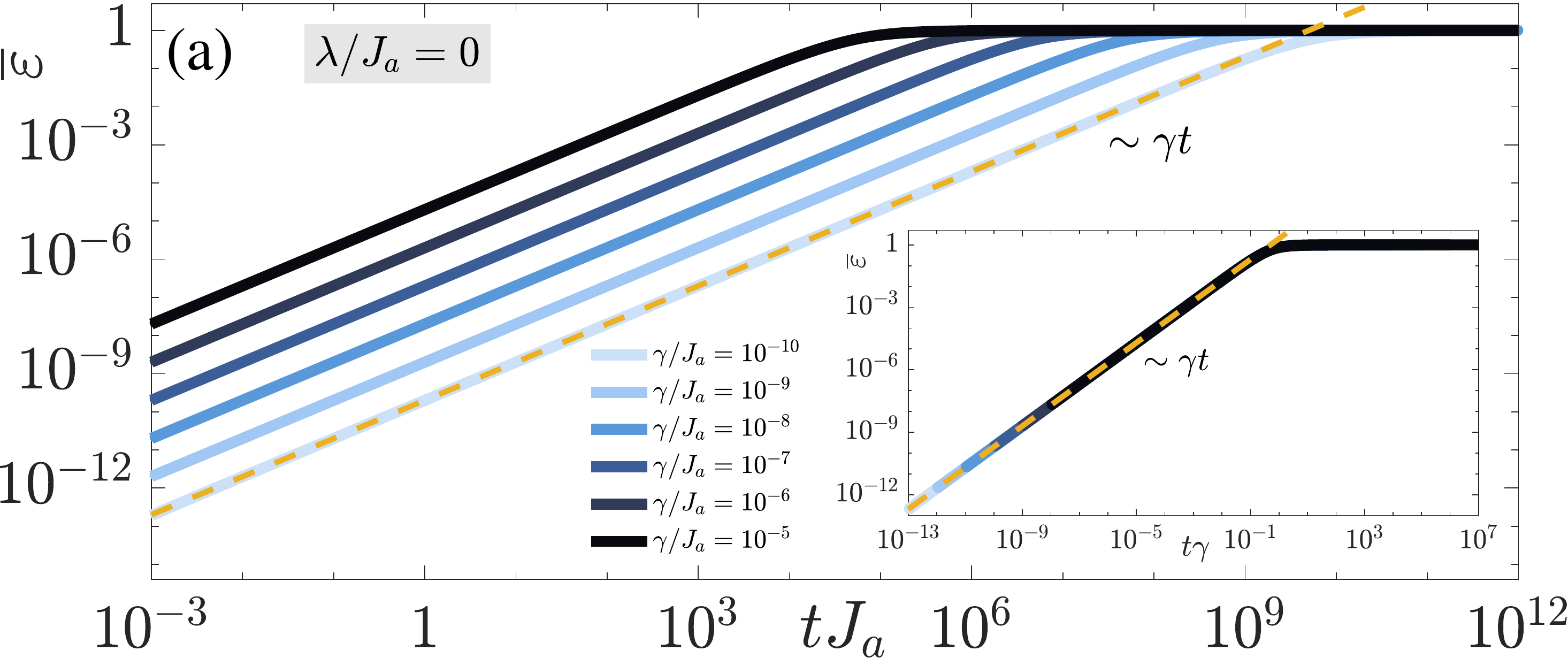}\\
	\includegraphics[width=.48\textwidth]{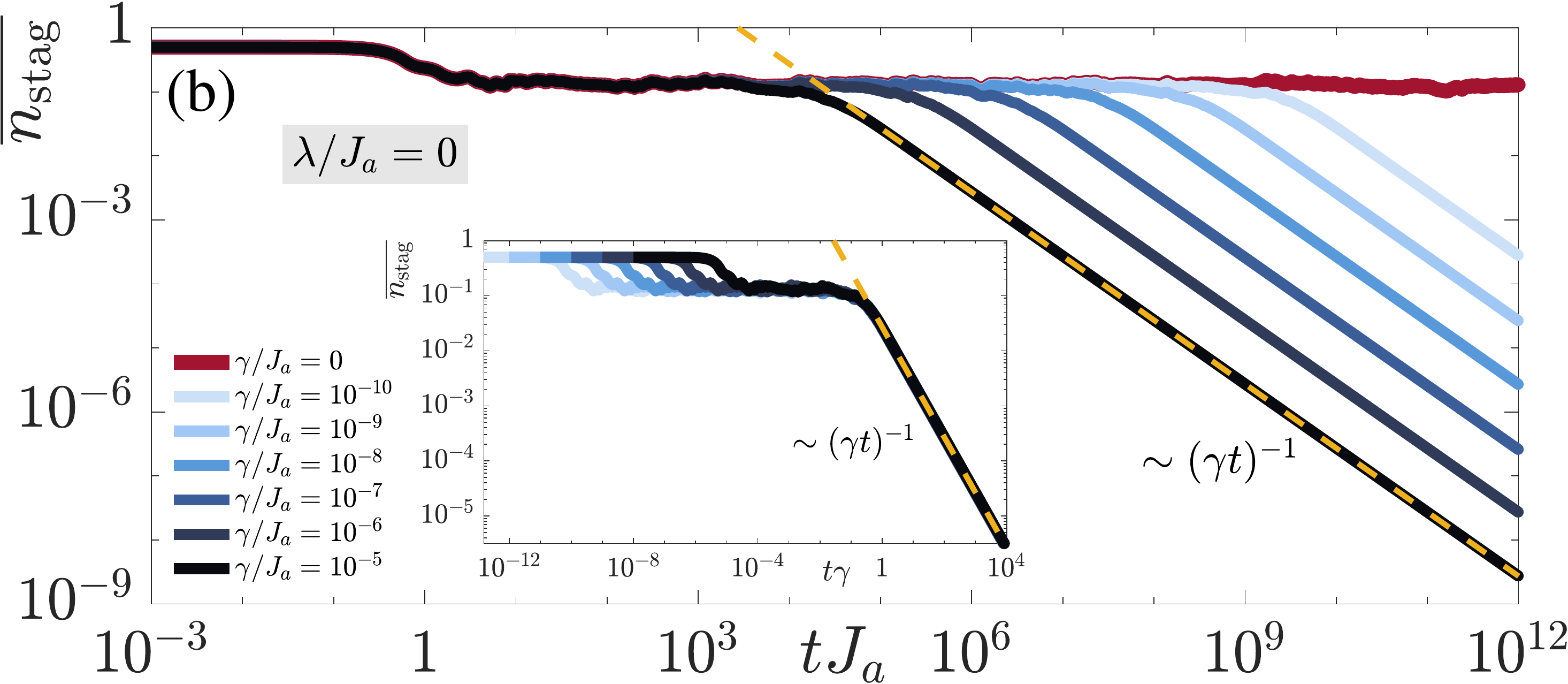}
	\caption{(Color online). Dynamics of (a) the gauge violation and (b) the staggered matter field for $\lambda=0$ at various values of the environment-coupling strength $\gamma$ (see legend). The behavior is qualitatively similar to that of Figs.~\ref{fig:gammaFixed}(a,e), respectively, albeit here there is no signature of the prethermal plateaus in the gauge violation since $\lambda=0$.}
	\label{fig:obs} 
\end{figure}

In the quench dynamics of the joint submission\cite{Halimeh2020f} and Fig.~\ref{fig:gammaFixed}, we have focused on $\lambda>0$. For completeness, we show in Fig.~\ref{fig:obs} the effect of $\gamma$ on the dynamics of gauge violation and staggered boson number, considering quench dynamics for the same initial state as in Fig.~\ref{fig:Z2LGT_InitialStates}(a) but for $\lambda=0$. The gauge violation [Fig.~\ref{fig:obs}(a)] spreads diffusively in the gauge sectors scaling as $\varepsilon\sim\gamma t$ at short times before settling into a maximal-violation steady state at $t\approx1/\gamma$. The staggered boson number [Fig.~\ref{fig:obs}(b)] behaves much the same way as in the case of $\lambda>0$ in Fig.~\ref{fig:gammaFixed}(e): it deviates from the purely coherent dynamics at $t\approx1/\gamma$, with its temporal average decaying $\sim(\gamma t)^{-1}$ at late times.

\subsection{Variations of jump operator}

\begin{figure}[htp]
	\centering
	\includegraphics[width=.48\textwidth]{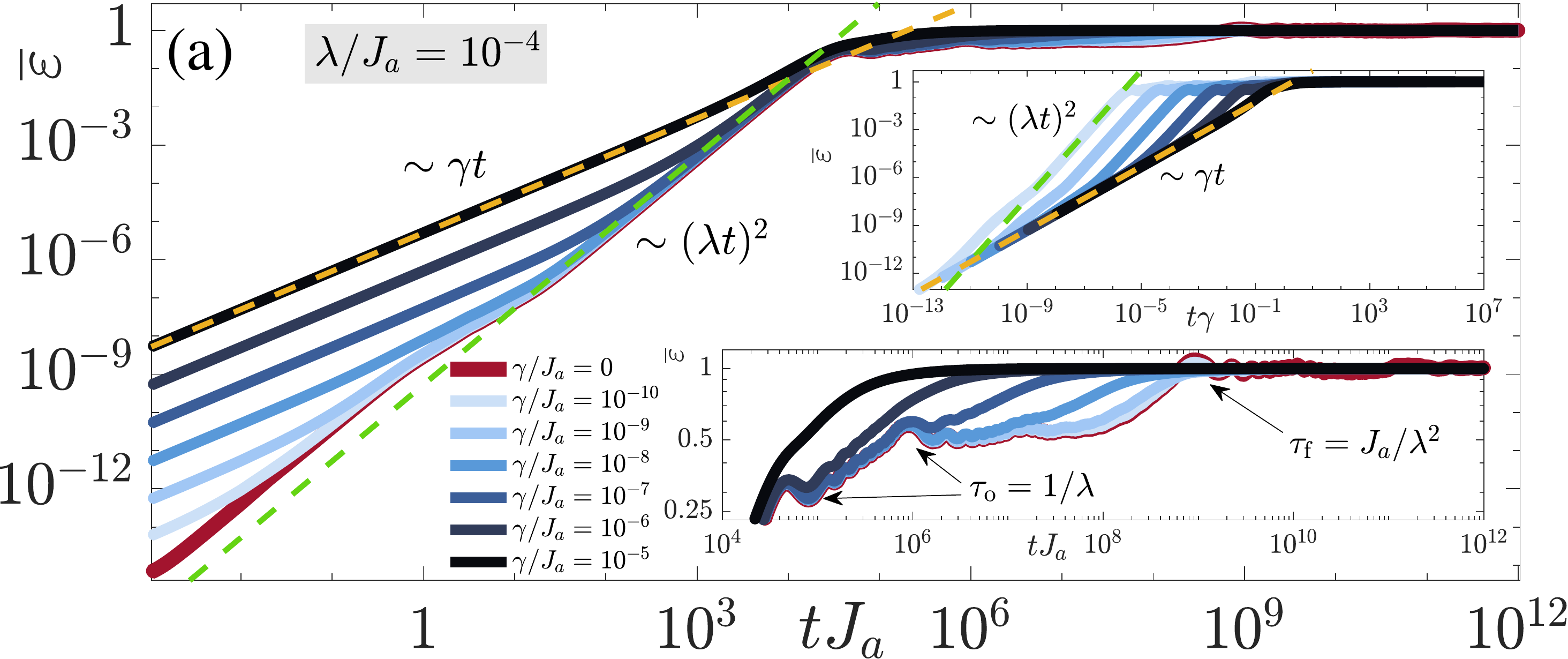}\\
	\includegraphics[width=.48\textwidth]{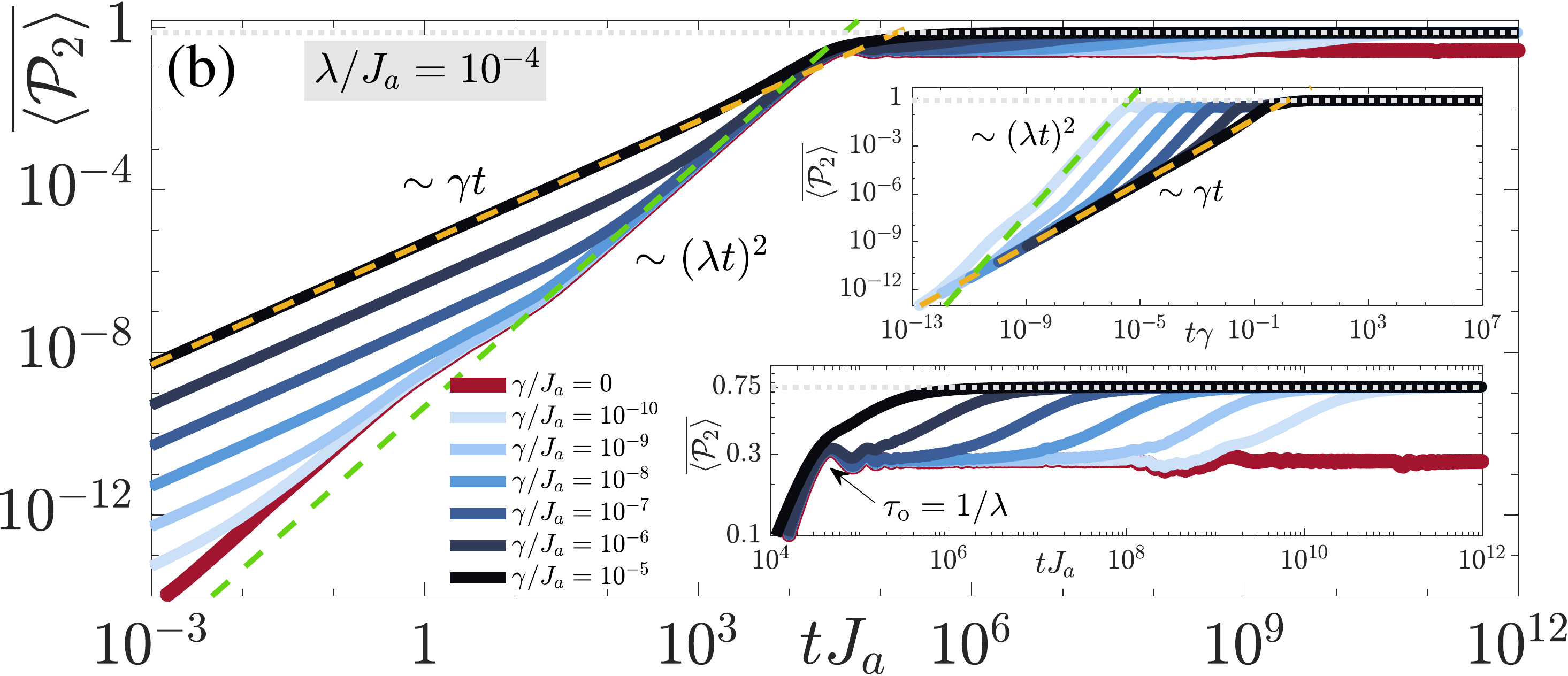}\\
	\includegraphics[width=.48\textwidth]{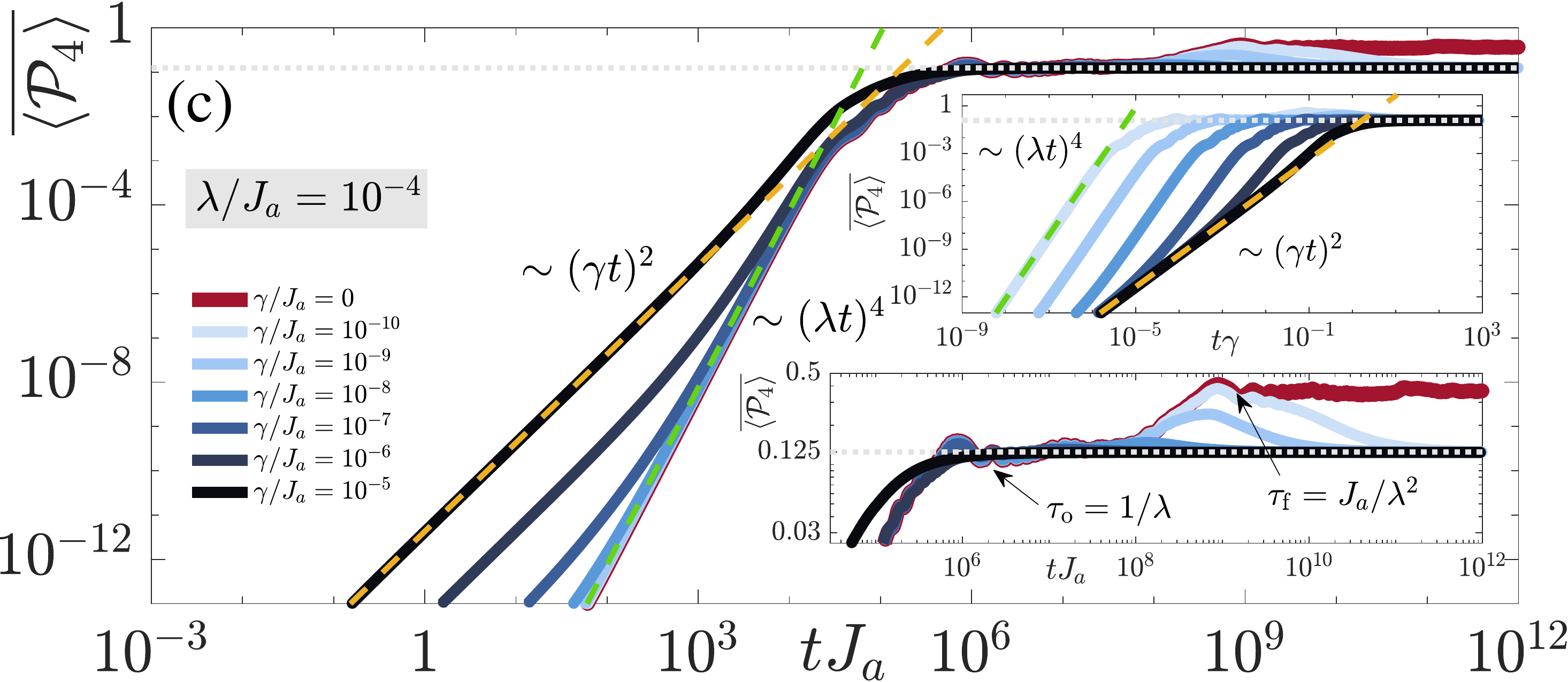}
	\caption{(Color online). Same as Fig.~1 of Ref.~\onlinecite{Halimeh2020f} but with a different jump operator on the gauge links. Quench dynamics of (a) the gauge violation, (b) supersector projector $\mathcal{P}_2$, and (c) supersector projector $\mathcal{P}_4$ in the open $\mathrm{Z}_2$ LGT starting in the gauge-invariant initial state of Fig.~\ref{fig:eBHM_InitialStates}(a), in the presence of the unitary gauge-breaking errors of Eq.~\eqref{eq:H1} at strength $\lambda=10^{-4}J_a$, and coupling to the environment at strength $\gamma$ with the jump operators $L^\mathrm{m}_j=a_j^\dagger a_j$ and $L^\mathrm{g}_{j,j+1}=\tau^-_{j,j+1}$ on matter sites and gauge links, respectively. The diffusive-to-ballistic crossover is again at $t\propto\gamma/\lambda^2$.}
	\label{fig:DiffDiss} 
\end{figure}

So far, we have included dissipation in the gauge fields as governed by the jump operator $L_{j,j+1}^\mathrm{g}=\tau^z_{j,j+1}$. To corroborate the generality of our results, we study the effect of a different dissipative jump operator $L_{j,j+1}^\mathrm{g}=\tau^-_{j,j+1}$ at various values of $\gamma$ in the presence of coherent gauge-breaking terms at strength $\lambda=10^{-4}J_a$. We show the associated gauge-violation and supersector-projector dynamics in Fig.~\ref{fig:DiffDiss}. The qualitative picture is unchanged, and we see that the diffusive-to-ballistic crossover is also at $t\propto\gamma/\lambda^2$, with scaling $\sim\gamma t$ ($\sim\gamma^2t^2$) in the diffusive regime and scaling $\sim\lambda^2t^2$ ($\sim\lambda^4t^4$) in the ballistic regime in the short-time dynamics of the gauge violation and $\langle\mathcal{P}_2\rangle$ ($\langle\mathcal{P}_4\rangle$).

\subsection{Other initial states}

\begin{figure}[htp]
	\centering
	\includegraphics[width=.48\textwidth]{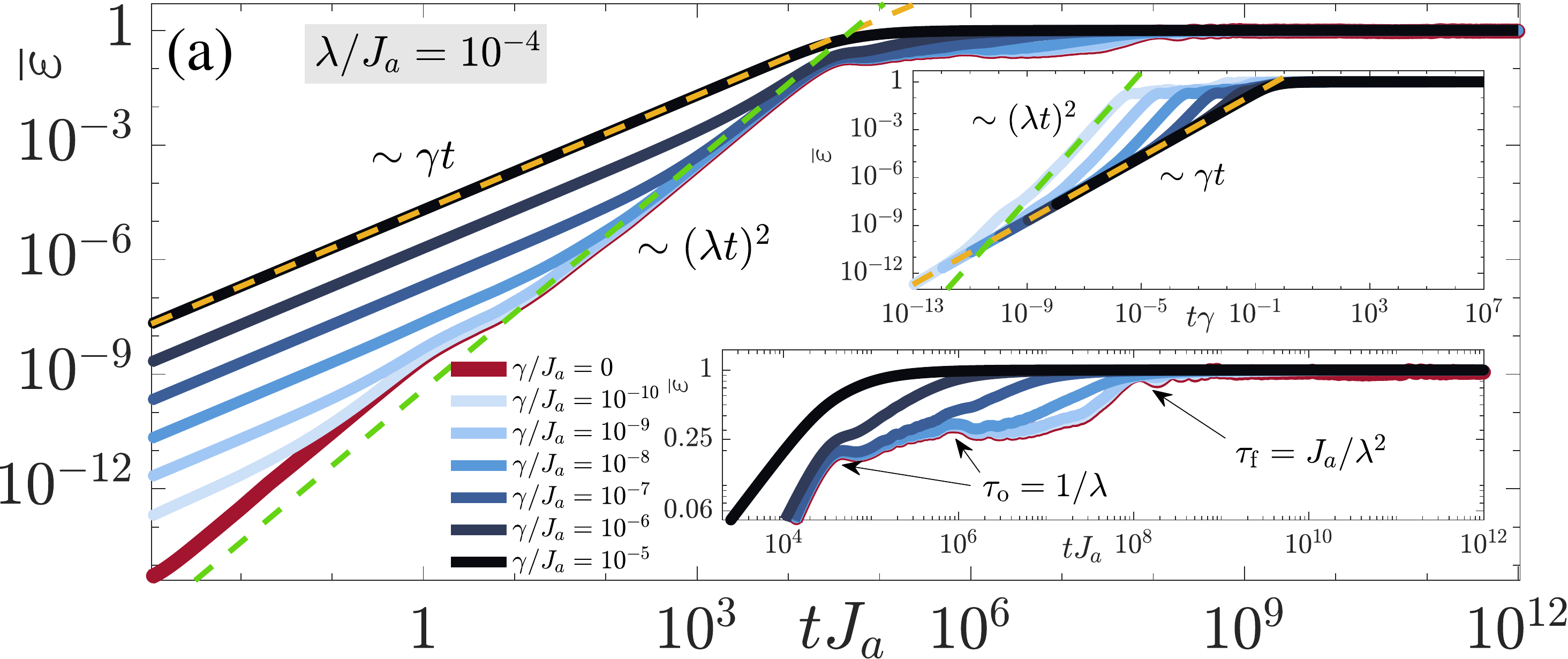}\\
	\includegraphics[width=.48\textwidth]{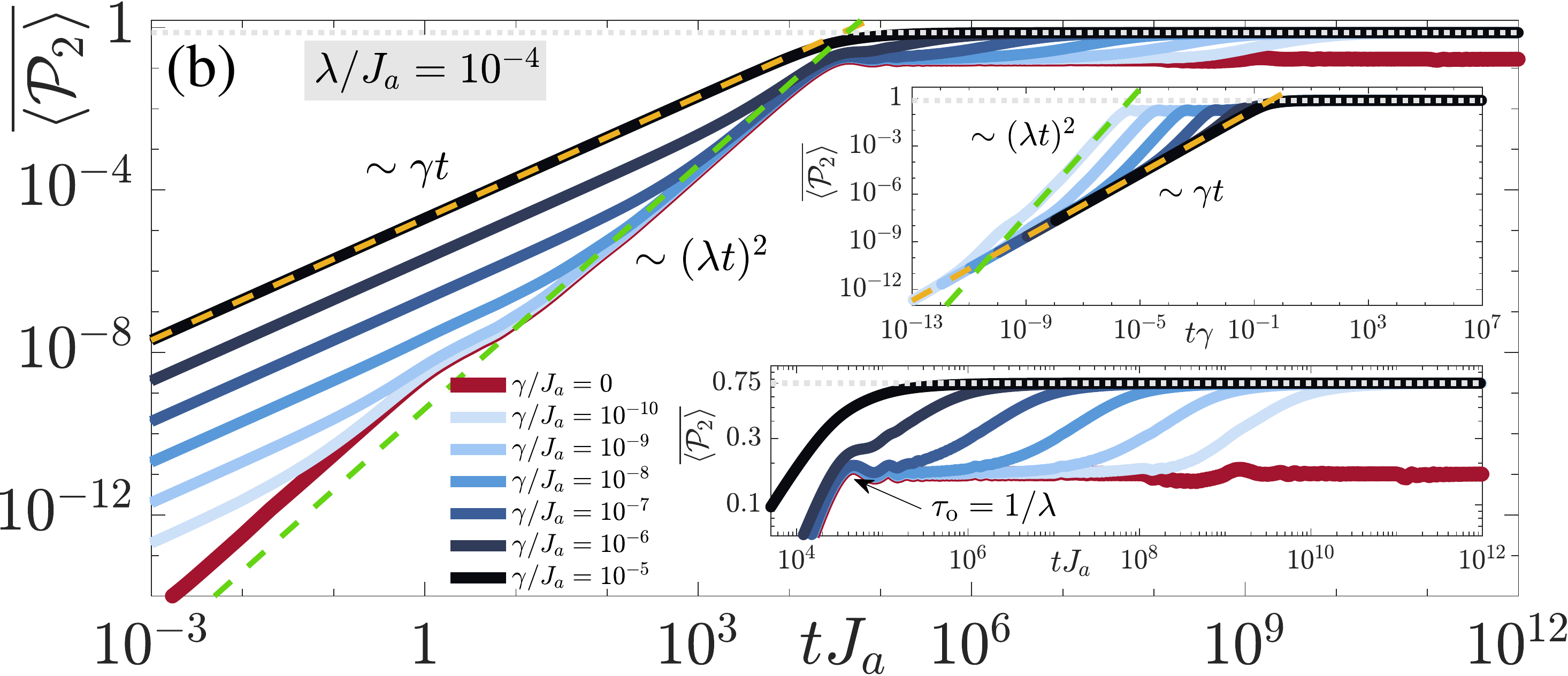}\\
	\includegraphics[width=.48\textwidth]{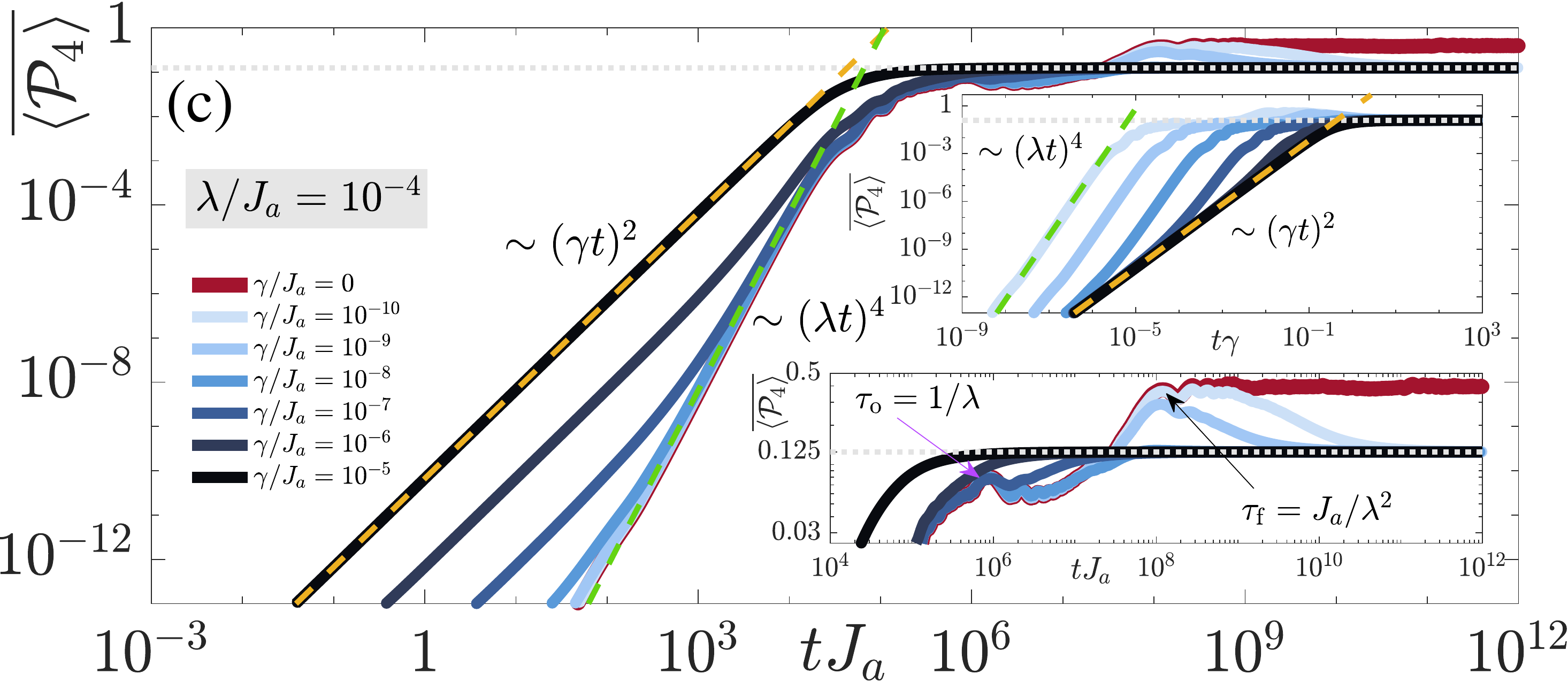}
	\caption{(Color online). Same as Fig.~\ref{fig:DiffDiss} but starting in the gauge-invariant ``domain-wall'' initial product state of Fig.~\ref{fig:Z2LGT_InitialStates}(b) and with $L^\mathrm{g}_{j,j+1}=\tau^z_{j,j+1}$.}
	\label{fig:DiffInitState} 
\end{figure}

Furthermore, our conclusions remain unaltered for other initial states. Whereas in the joint submission Ref.~\onlinecite{Halimeh2020f} and hitherto in this paper our initial state has been the staggered product state in Fig.~\ref{fig:Z2LGT_InitialStates}(a), in Fig.~\ref{fig:DiffInitState} we show the gauge-violation and supersector-projector dynamics for the ``domain-wall'' product state in Fig.~\ref{fig:Z2LGT_InitialStates}(b). Again, here we include coherent gauge-breaking terms at strength $\lambda=10^{-4}J_a$ and study the effects of decoherence at various values of the environment coupling $\gamma$, with the jump operators $L^\mathrm{m}_j=a_j^\dagger a_j$ and $L^\mathrm{g}_{j,j+1}=\tau^z_{j,j+1}$. The qualitative picture is identical to that of Fig.~1 in Ref.~\onlinecite{Halimeh2020f} and Fig.~\ref{fig:DiffInitState}.

\subsection{Multiple quantum coherences}
\begin{figure}[htp]
	\centering
	\includegraphics[width=.48\textwidth]{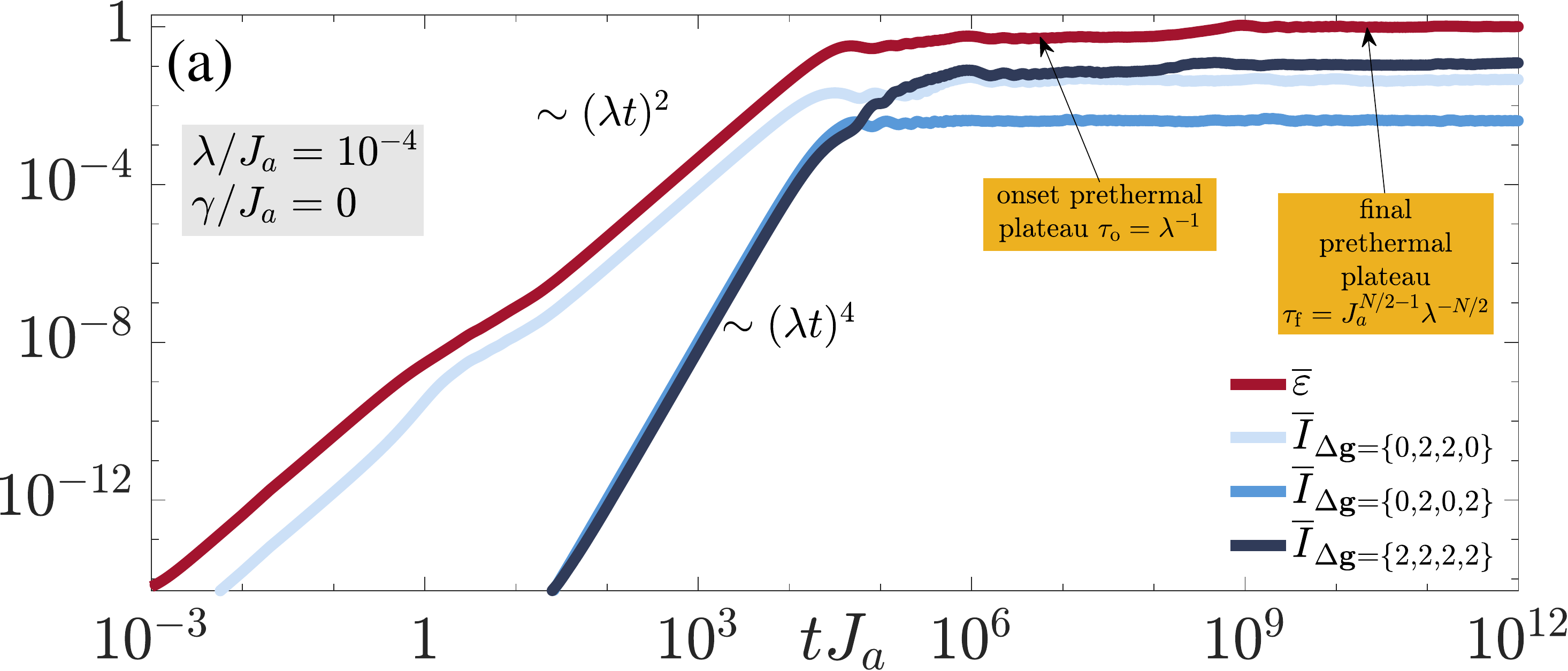}\\
	\includegraphics[width=0.32\linewidth]{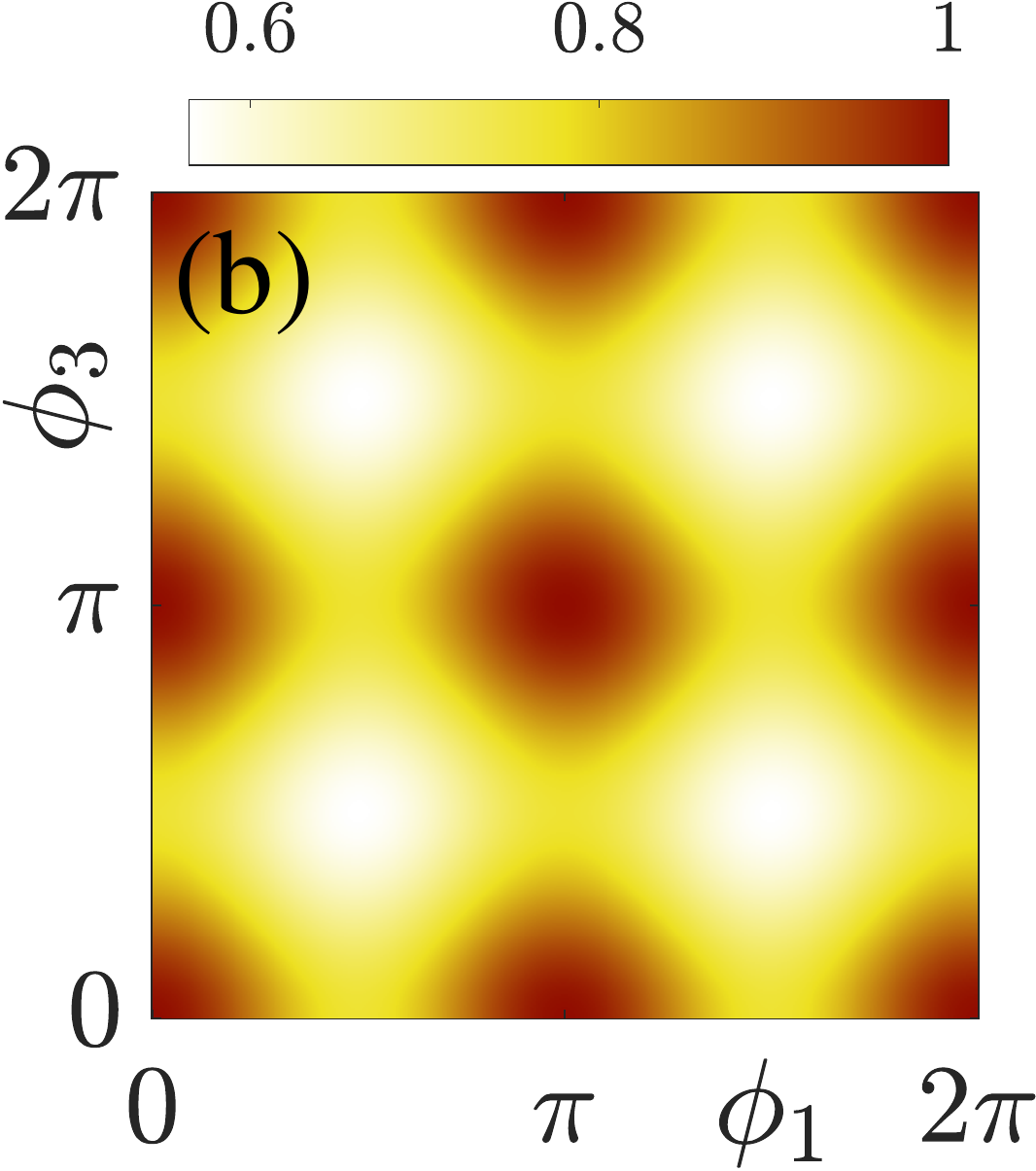}
	\includegraphics[width=0.32\linewidth]{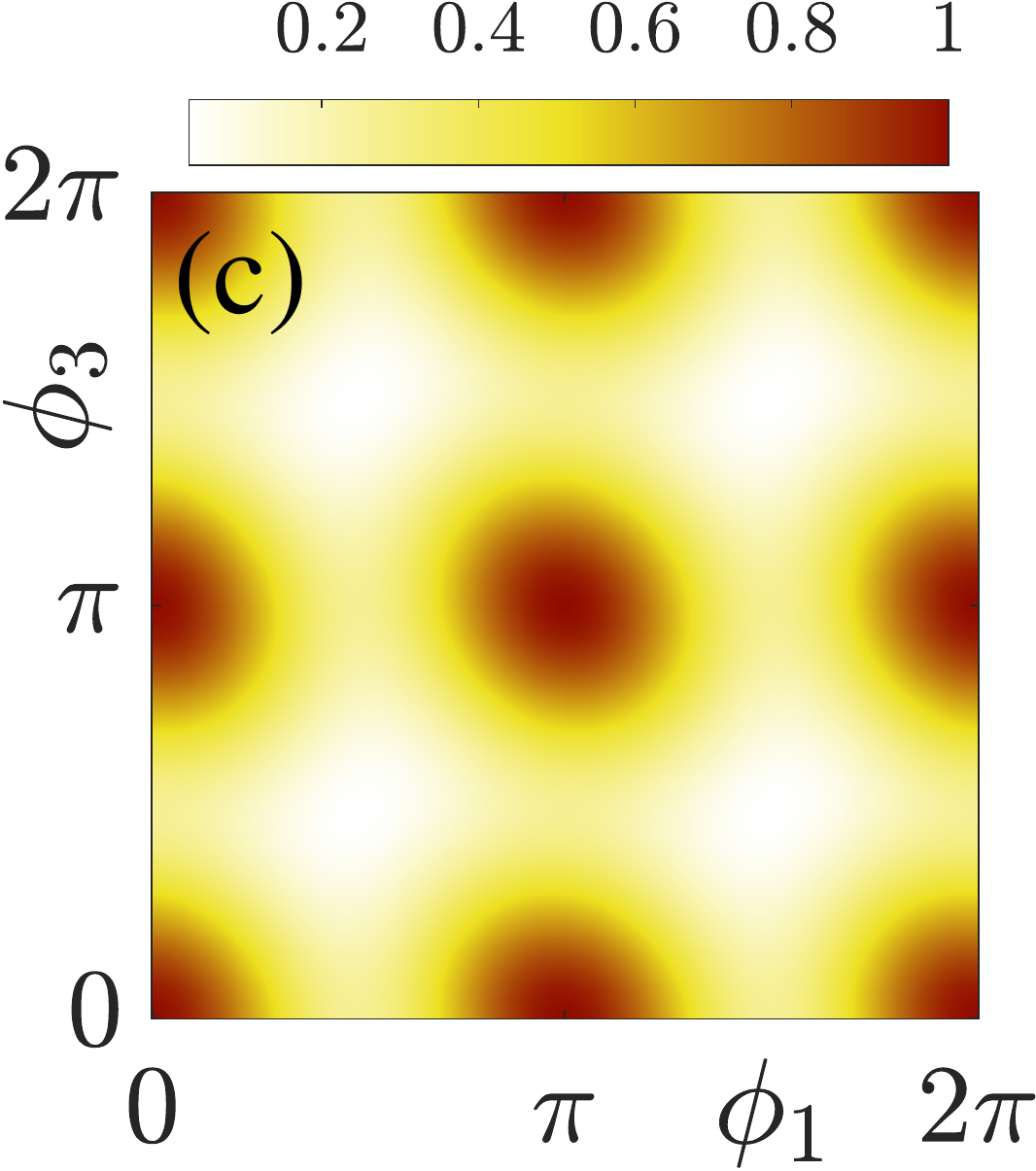}
	\includegraphics[width=0.32\linewidth]{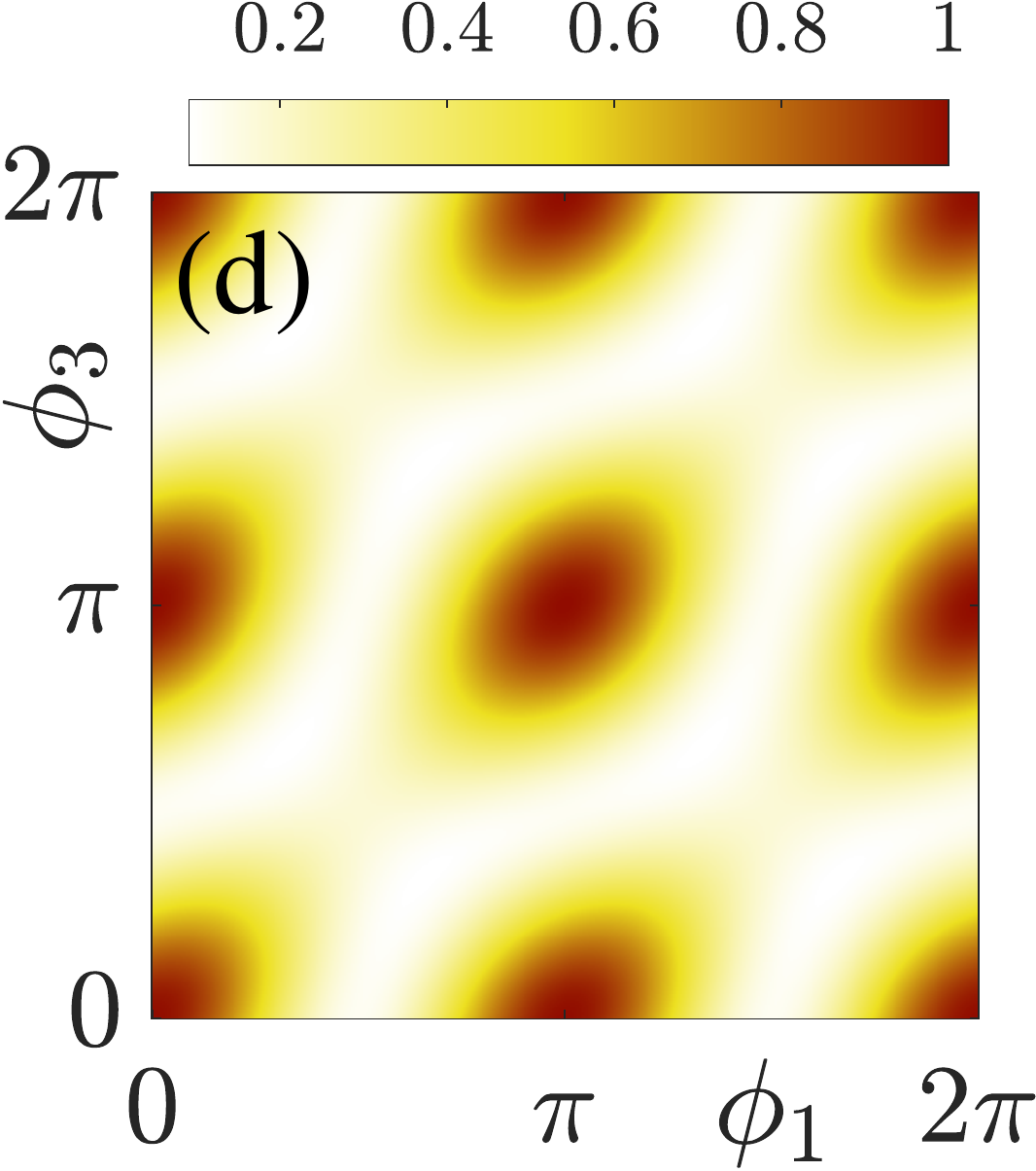}\\
	\includegraphics[width=0.32\linewidth]{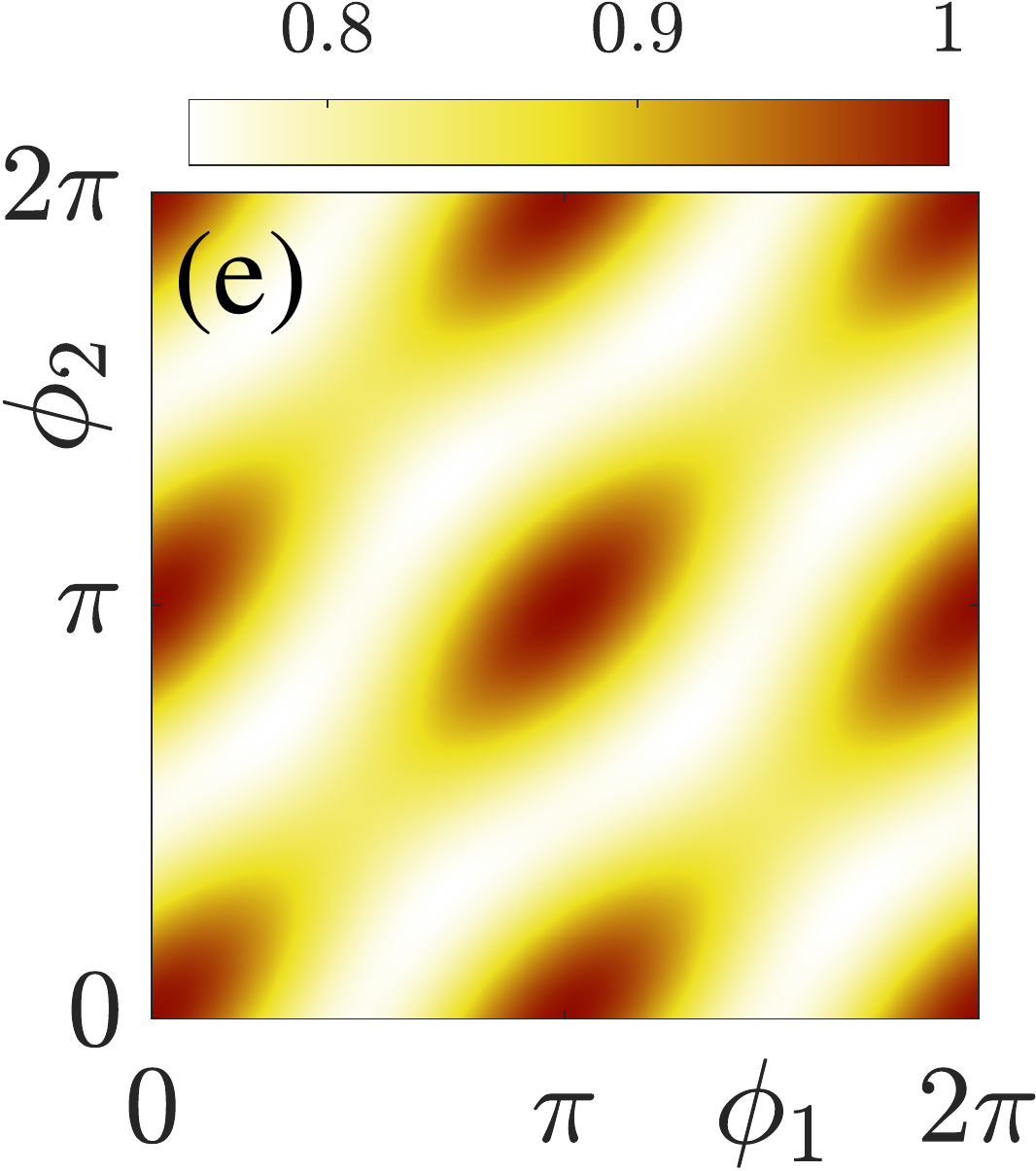}
	\includegraphics[width=0.32\linewidth]{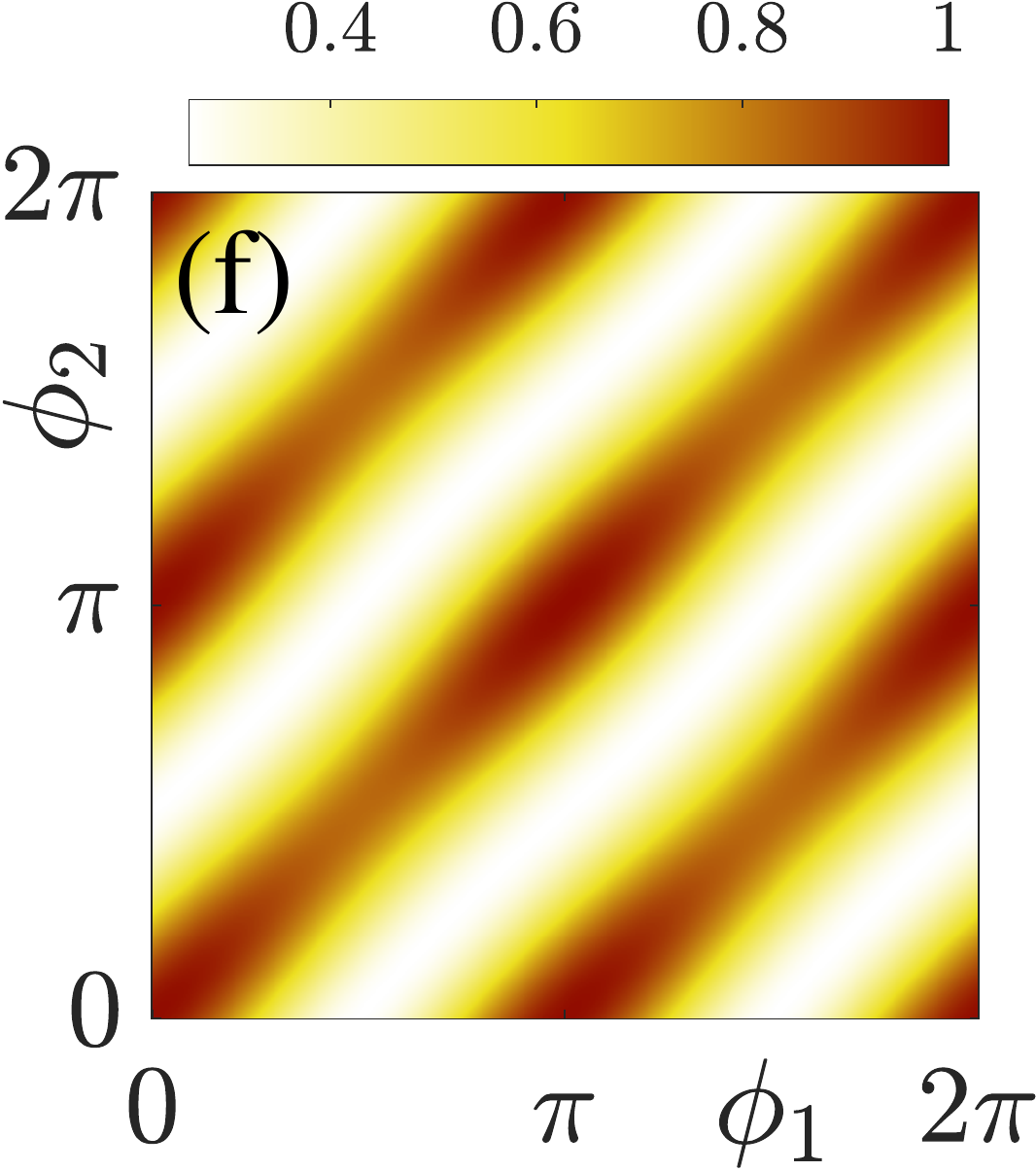}
	\includegraphics[width=0.32\linewidth]{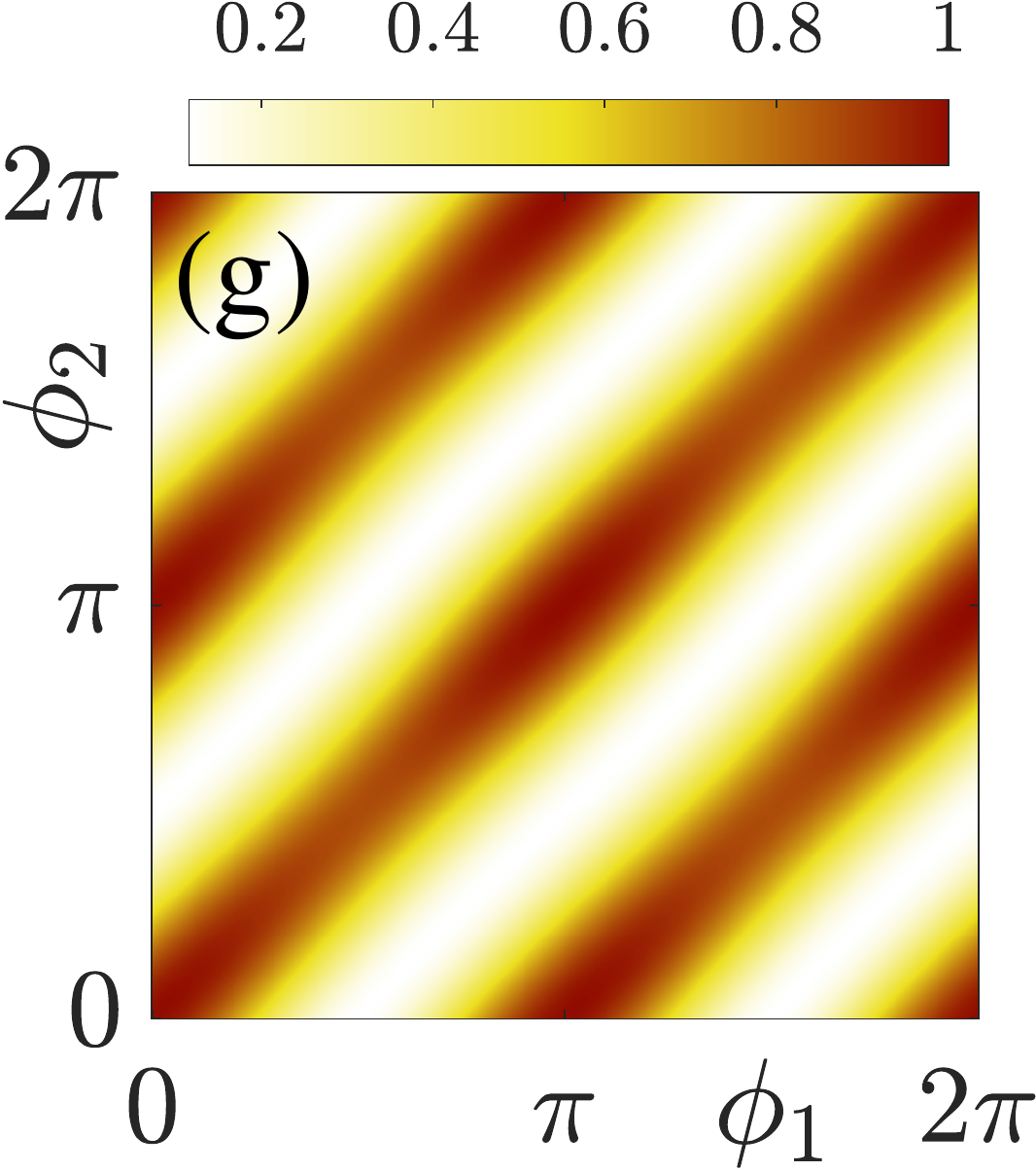}\\
	\includegraphics[width=0.32\linewidth]{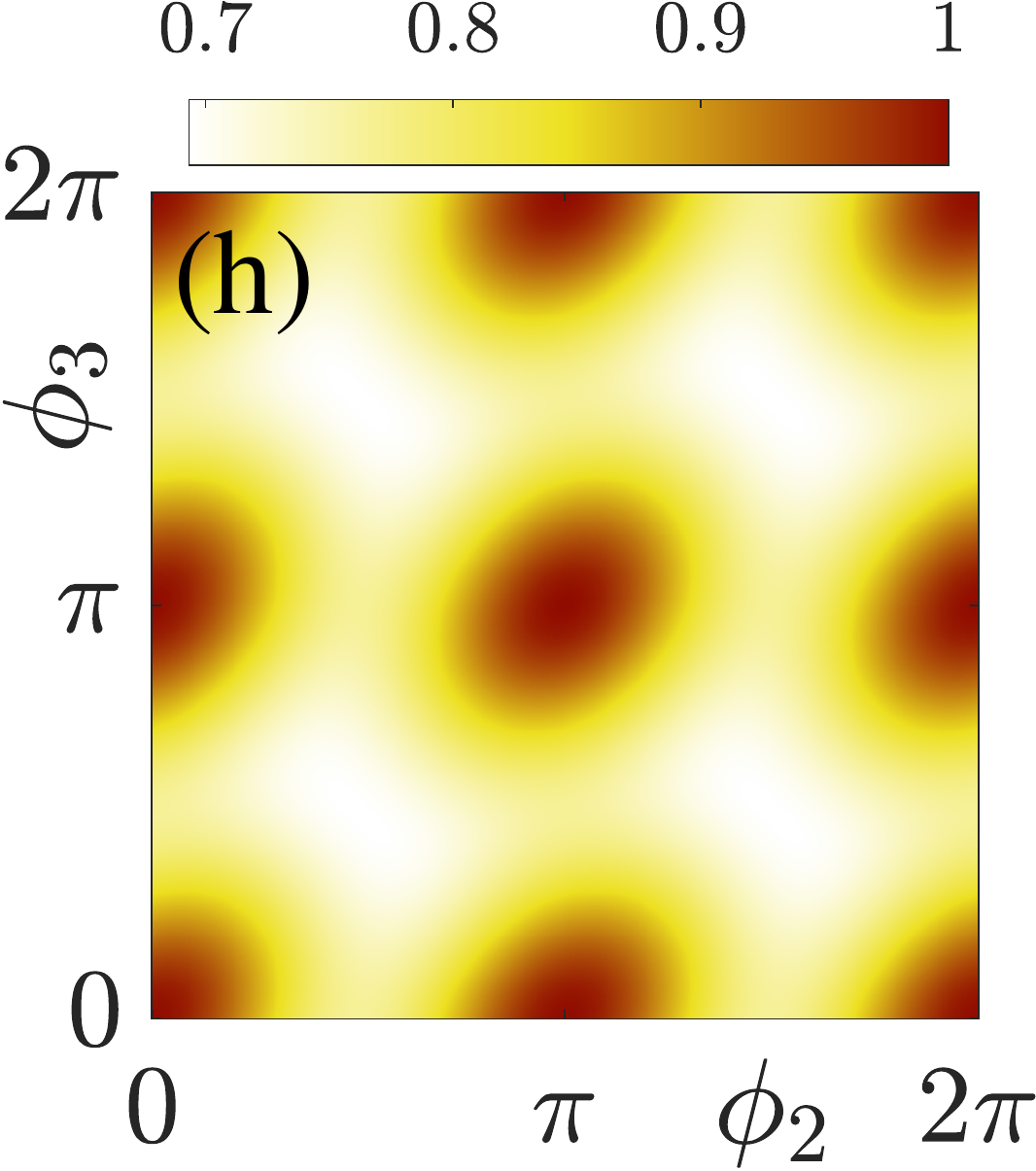}
	\includegraphics[width=0.32\linewidth]{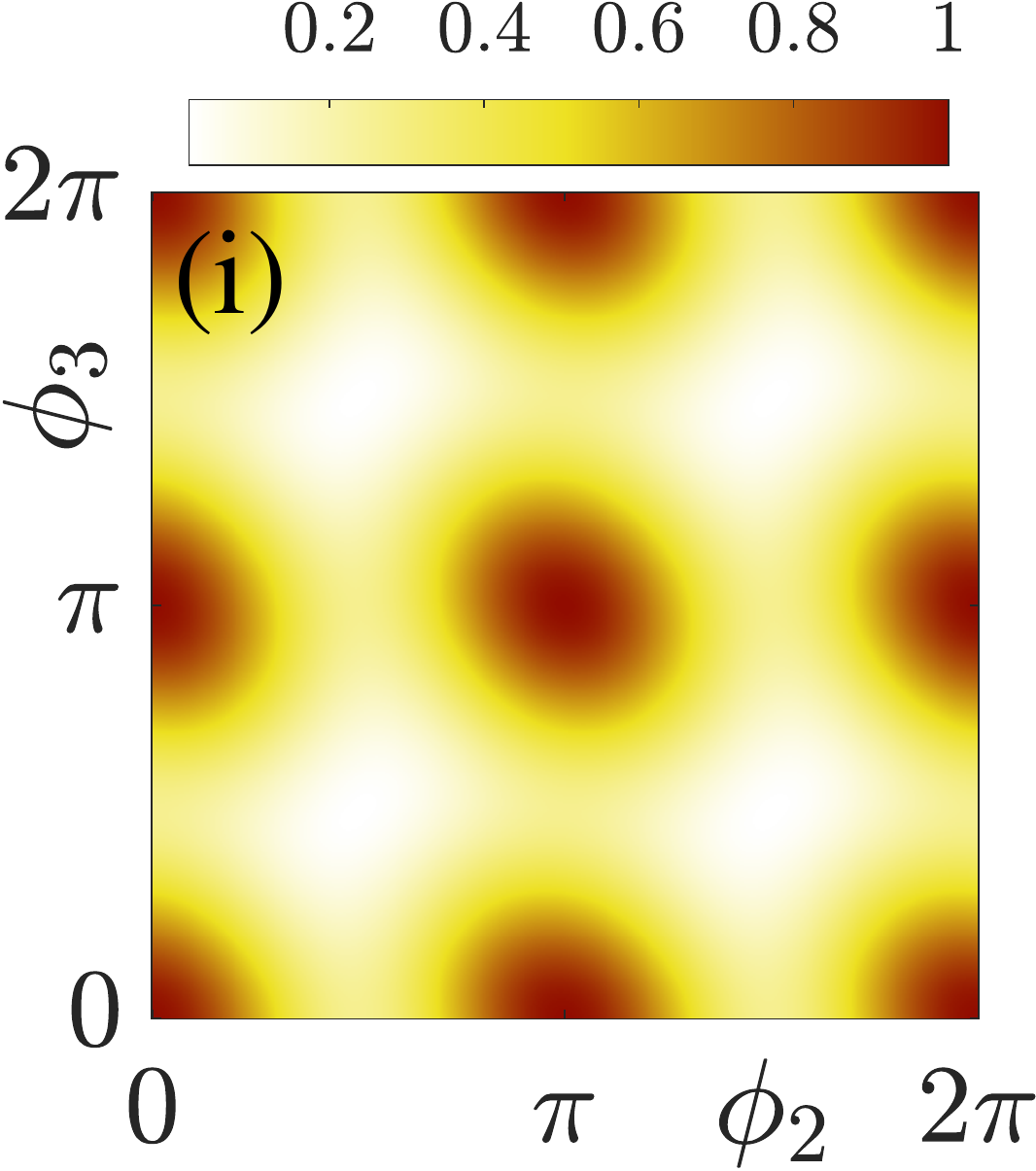}
	\includegraphics[width=0.32\linewidth]{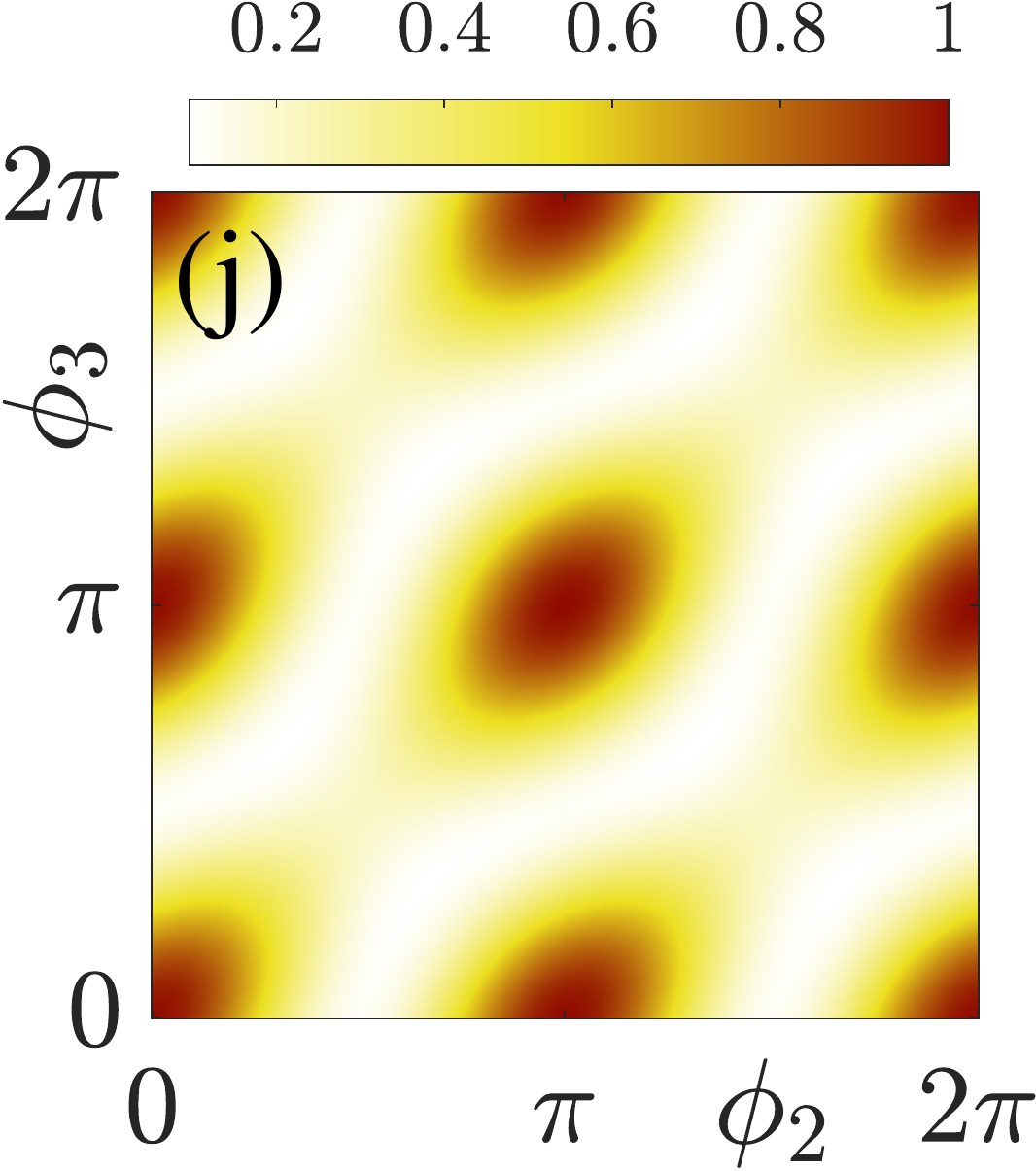}
	\caption{(Color online). Same as Fig.~2 of Ref.~\onlinecite{Halimeh2020f} but with no decoherence, and additionally showing the MQC spectra at different angles. (a) Dominant MQC. (b-j) MQC spectra, with rows from top to bottom $F(\phi_1,0,\phi_3,0)$, $F(\phi_1,\phi_2,0,0)$, and $F(0,\phi_2,\phi_3,0)$, and columns from left to right $tJ_a=10^4$, $tJ_a=10^5$, and $tJ_a=10^6$. Unsurprisingly, in the absence of decoherence, there is no decay of the intensities, and the bandwidth of the MQC spectra increases with time, in contrast to the case of decoherence in Fig.~2 of Ref.~\onlinecite{Halimeh2020f}, where decoherence diminishes the spectrum.}
	\label{fig:Fig2SM} 
\end{figure}

In the $\mathrm{Z}_2$ LGT, the MQC are a generalization of those given in Eq.~\eqref{eq:eBHM_MQC}, and read
\begin{subequations}
	\begin{align}
	I_{\Delta \mathbf{g}}&=\Tr\big\{\rho_{\Delta \mathbf{g}}^\dagger\rho_{\Delta \mathbf{g}}\big\},\\
	\rho_{\Delta \mathbf{g}}&=\sum_g P_{\mathbf{g}+\Delta \mathbf{g}}\rho P_\mathbf{g},
	\end{align}
\end{subequations}
where now they measure quantum coherences between gauge-invariant \textit{sectors} and not just supersectors as in the case of the eBHM (see Sec.~\ref{sec:eBHM_MQC}). Since the deviations between sectors are vectors, the MQC spectra
\begin{align}
F_{\boldsymbol{\phi}}=\sum_{\Delta\mathbf{g}}I_{\Delta\mathbf{g}}e^{-i\sum_j\phi_j\Delta g_j}
\end{align}
depend on $N$ angles $\boldsymbol{\phi}=\{\phi_1,\phi_2,\ldots,\phi_N\}$. In the joint submission Ref.~\onlinecite{Halimeh2020f}, we provide results for the MQC and their spectra, for a given choice of angles, at $\lambda=10^{-4}J_a$ and $\gamma=10^{-6}J_a$. Let us now look at these results but with no decoherence, i.e., $\lambda=10^{-4}J_a$ and $\gamma=10^{-6}J_a$. The corresponding results are shown in Fig.~\ref{fig:Fig2SM}. In the absence of decoherence, the MQC over evolution time and settle at a steady-state value at long times [Fig.~\ref{fig:Fig2SM}(a)], in contrast to the case with decoherence of Ref.~\onlinecite{Halimeh2020f}, where their temporal averages decay $\sim(\gamma t)^{-1}$ at $t>\gamma^{-1}$. The spectrum also behaves fundamentally differently. Whereas in the case with decoherence the spectrum almost vanishes for $t\gtrsim\gamma^{-1}$, in the closed-system case it is maximal in this temporal regime.

\begin{figure}[htp]
\centering
\includegraphics[width=0.32\linewidth]{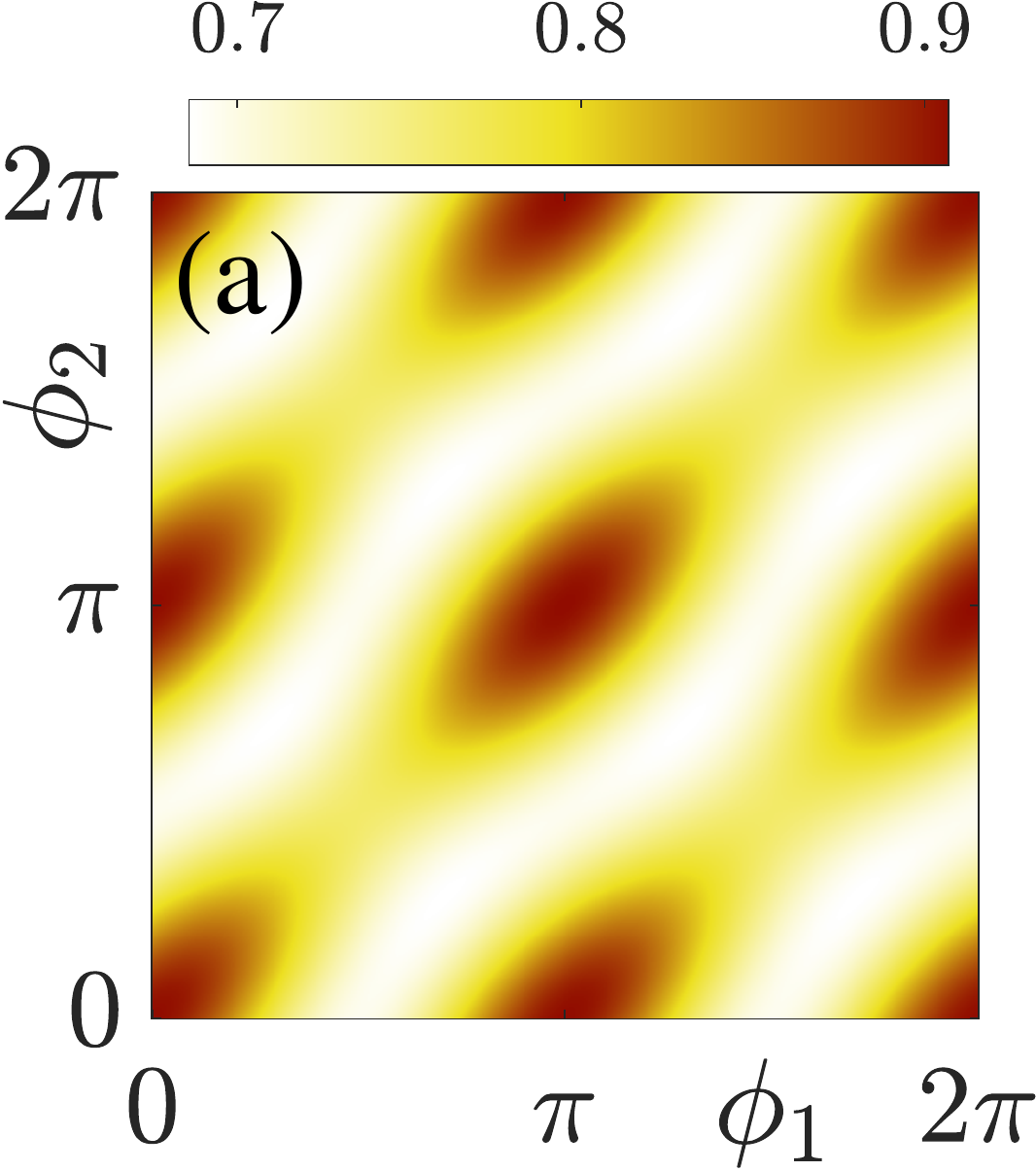}
\includegraphics[width=0.32\linewidth]{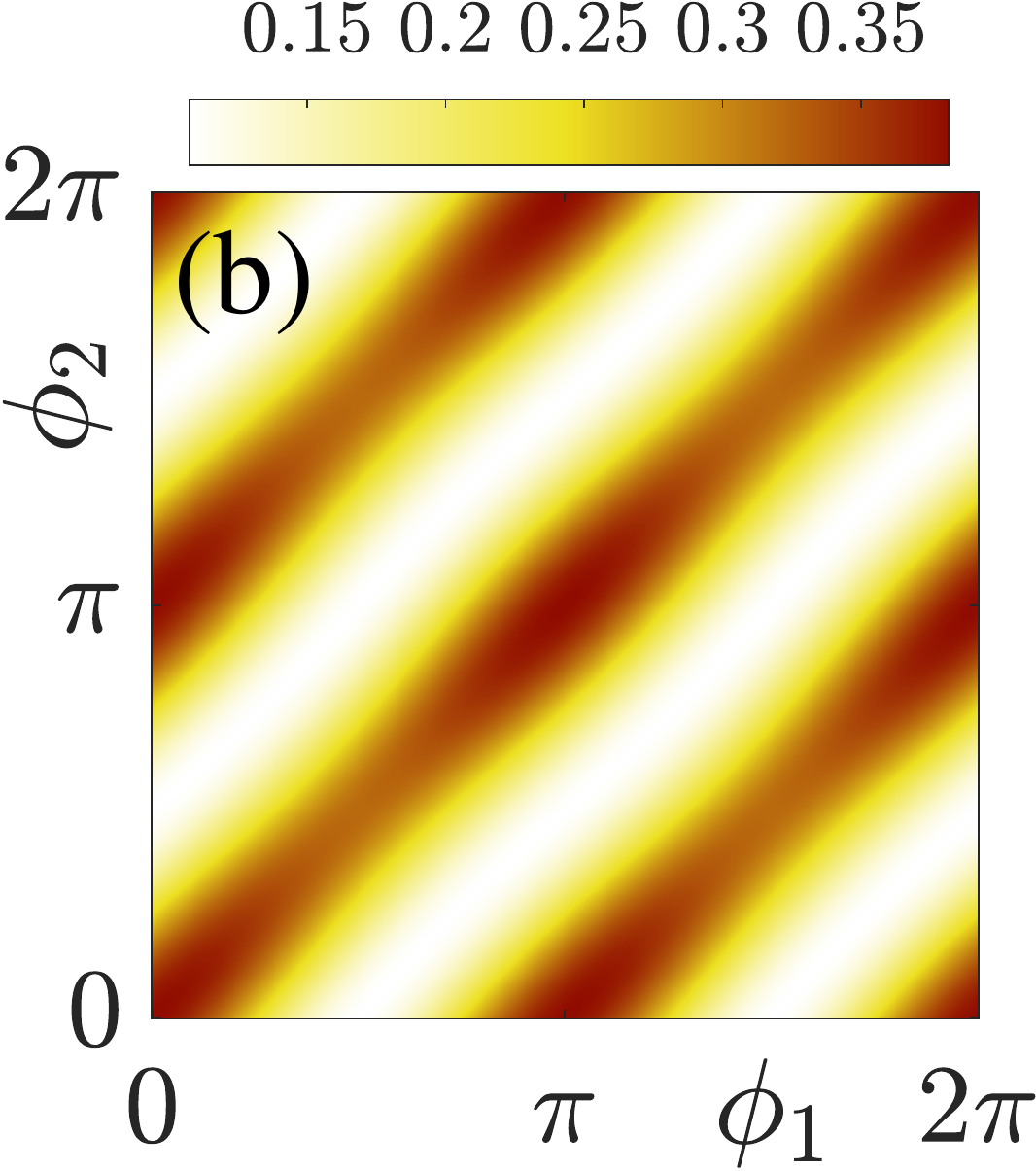}
\includegraphics[width=0.32\linewidth]{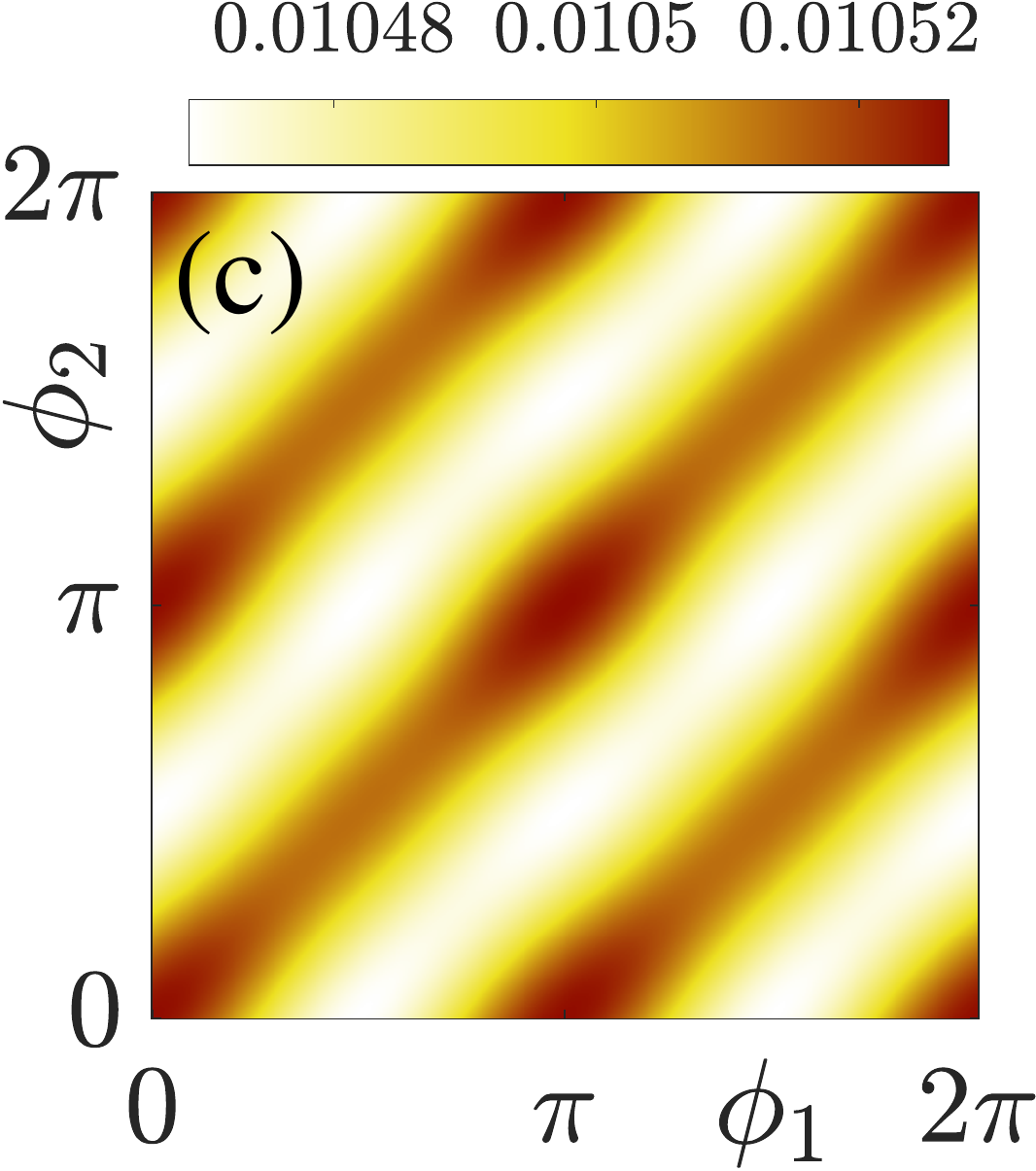}\\
\includegraphics[width=0.32\linewidth]{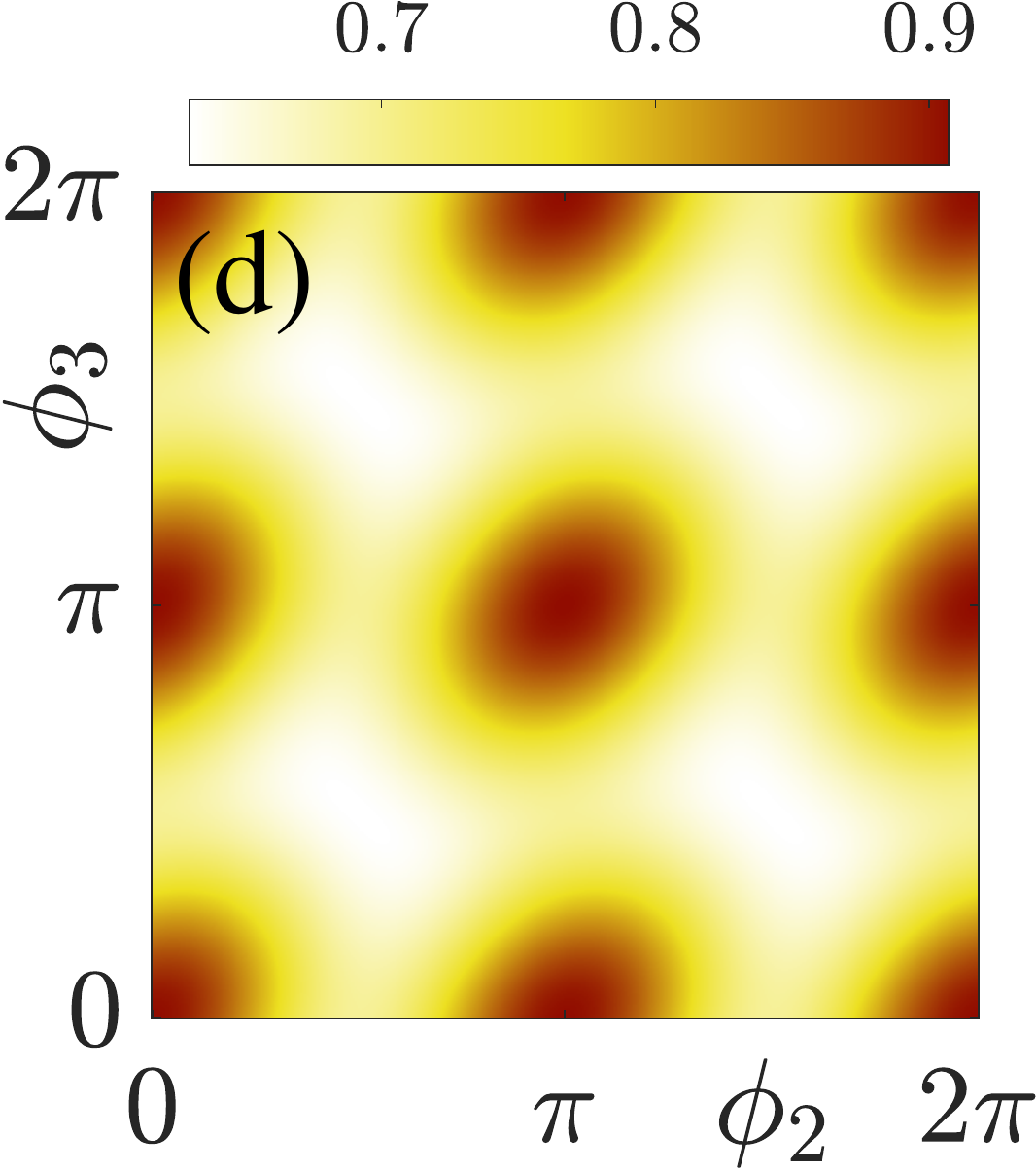}
\includegraphics[width=0.32\linewidth]{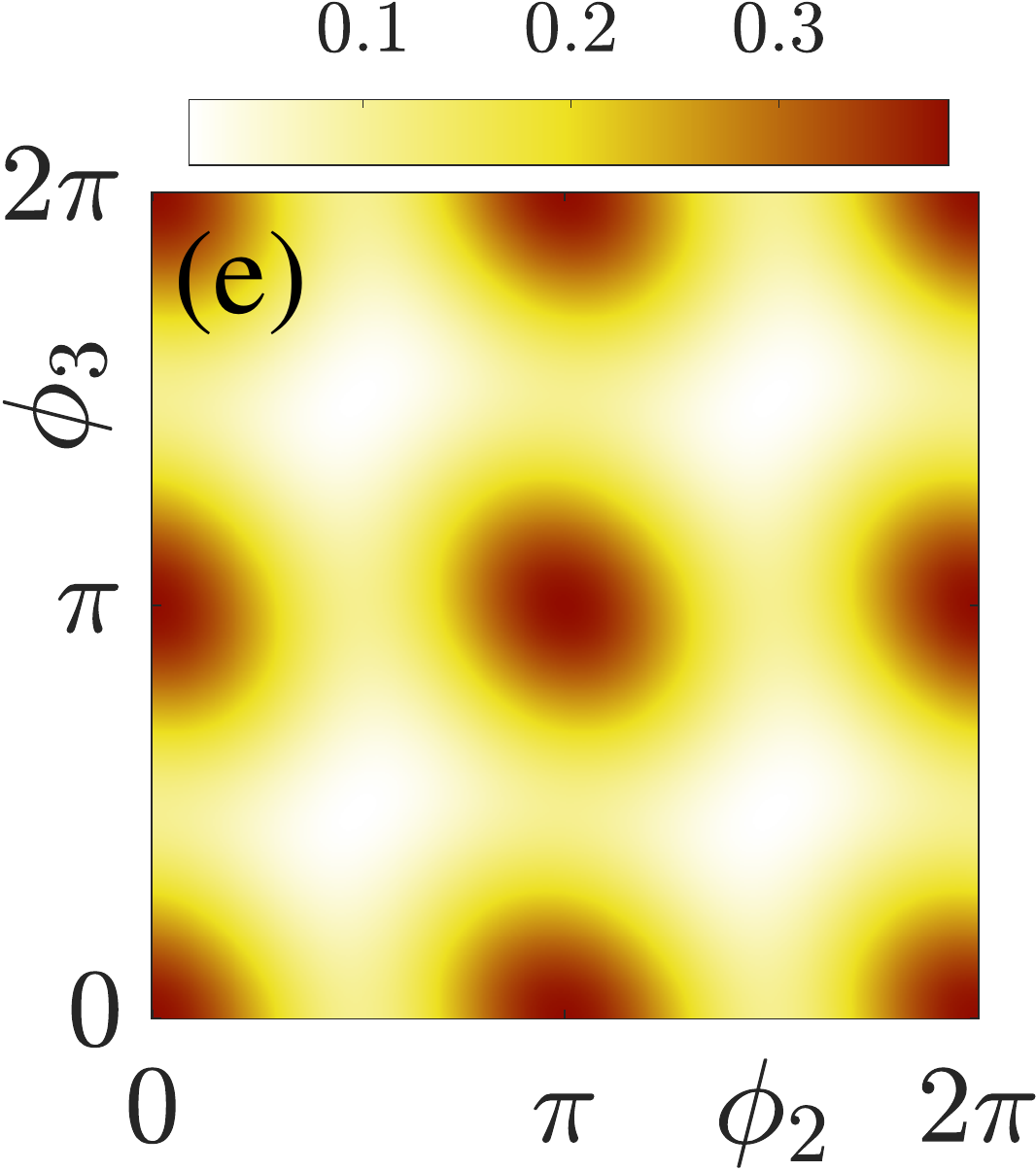}
\includegraphics[width=0.32\linewidth]{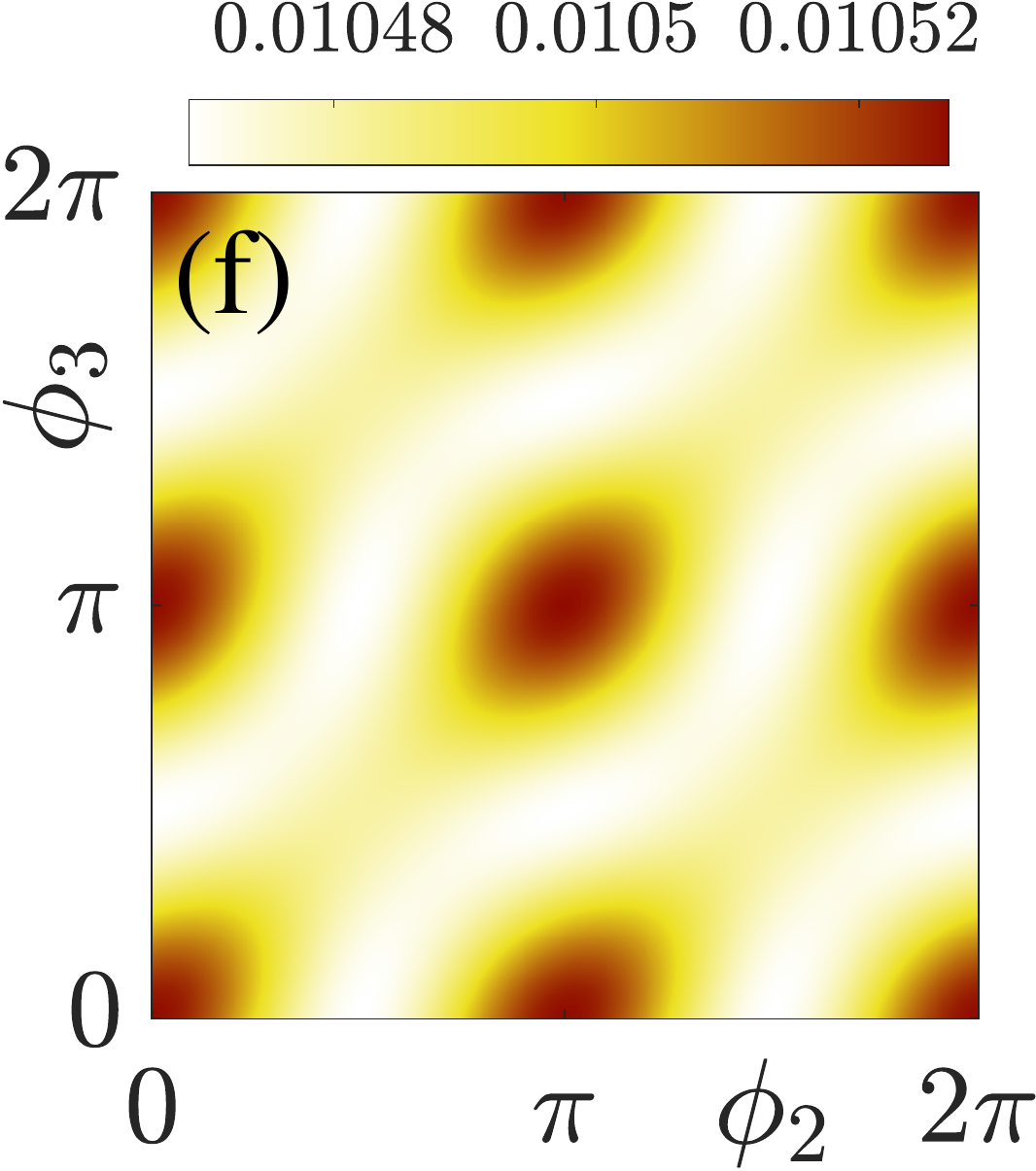}
\caption{(Color online). Same as Fig.~2(c-e) of Ref.~\onlinecite{Halimeh2020f} but for different angles of the MQC spectrum, where again $\lambda=10^{-4}J_a$ and $\gamma=10^{-6}J_a$. We show $F(\phi_1,\phi_2,0,0)$ at evolution times (a) $t=10^4/J_a$, (b) $t=10^5/J_a$, and (c) $t=10^6/J_a$, and we show $F(0,\phi_2,\phi_3,0)$ at evolution times (d) $t=10^4/J_a$, (e) $t=10^5/J_a$, and (f) $t=10^6/J_a$. The results show that decoherence diminishes the spectrum, in agreement with the conclusion from Fig.~2(c-e) of the joint submission.}
\label{fig:MQC_SM} 
\end{figure}

\begin{figure}[htp]
\centering
\includegraphics[width=.48\textwidth]{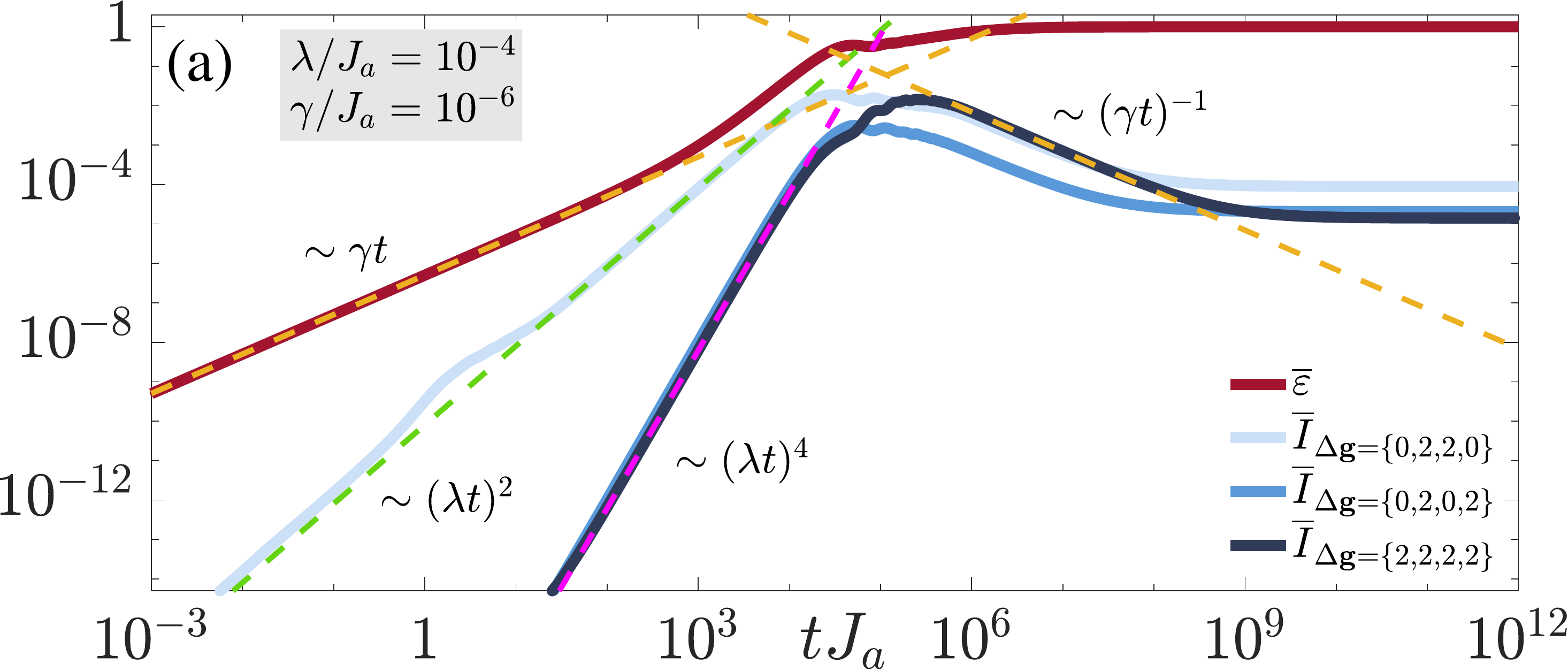}\\
\includegraphics[width=0.32\linewidth]{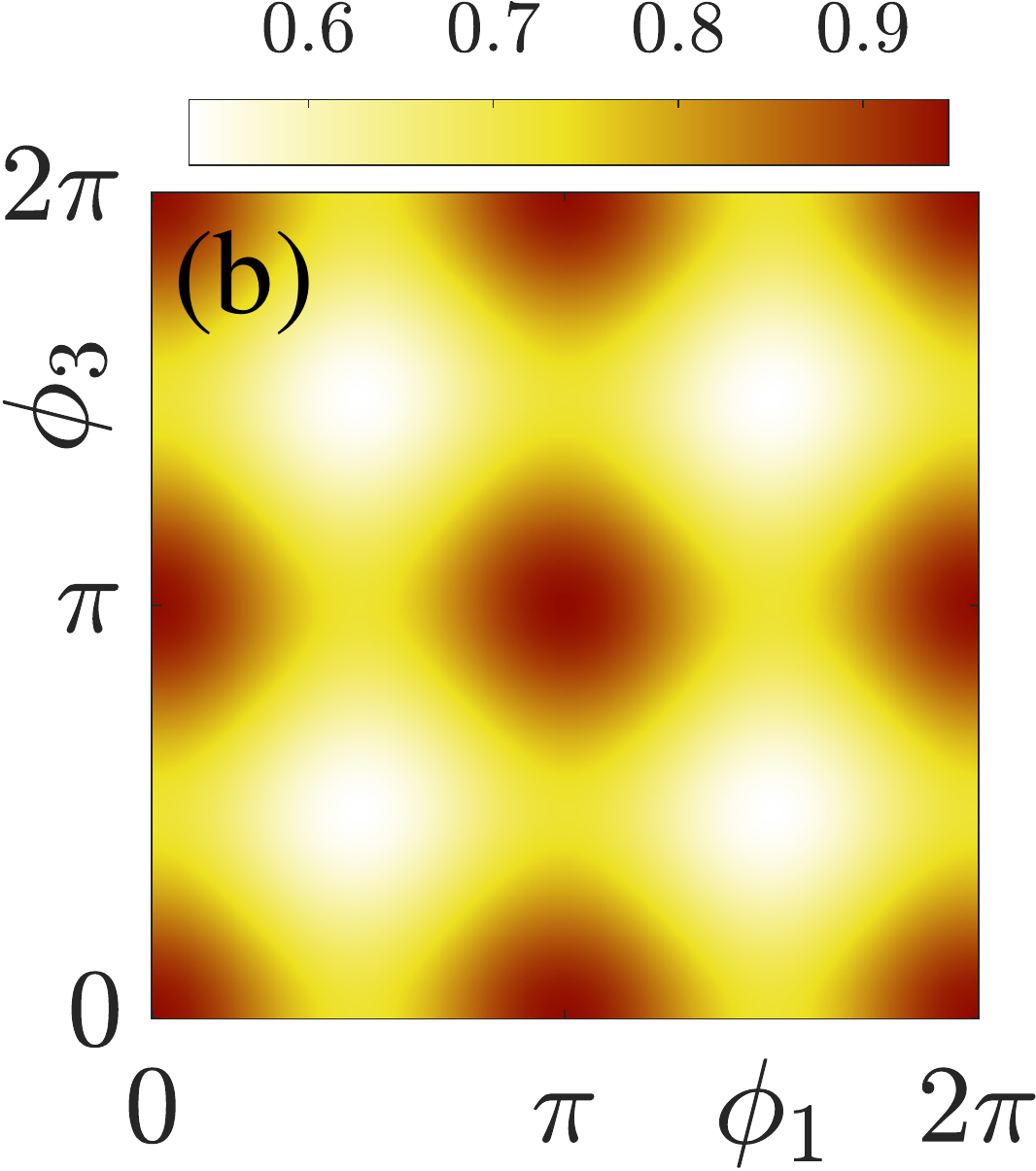}
\includegraphics[width=0.32\linewidth]{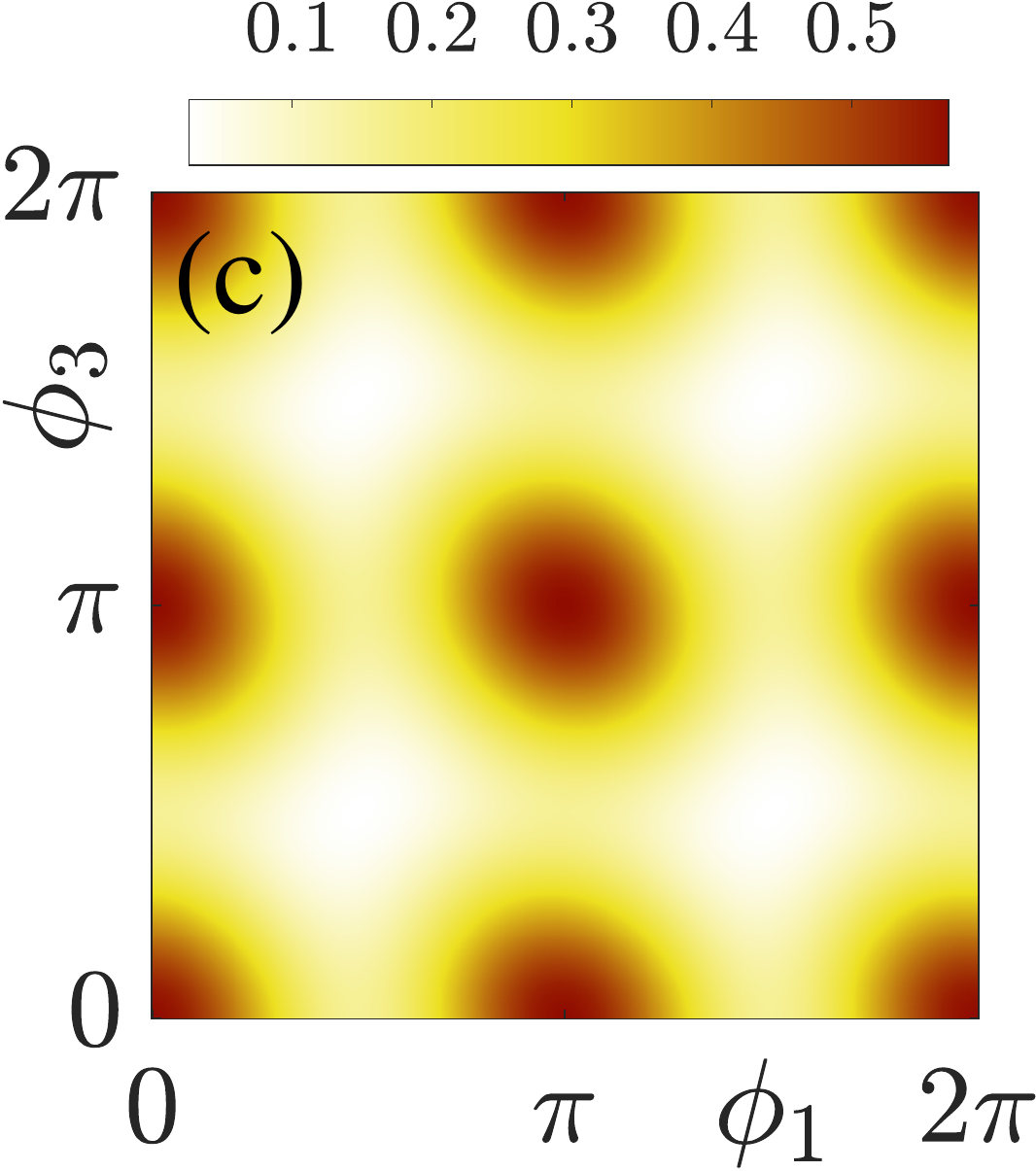}
\includegraphics[width=0.32\linewidth]{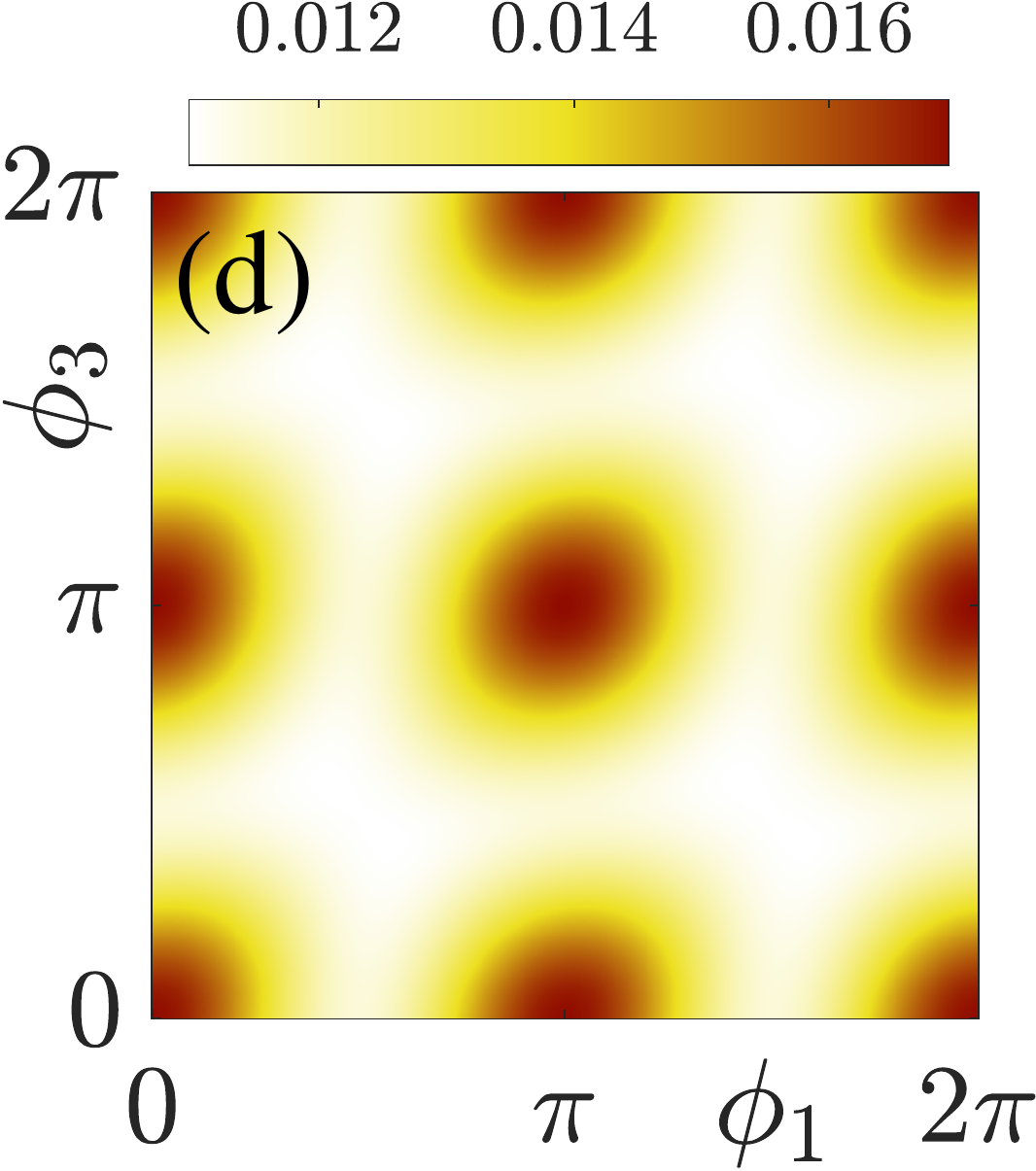}\\
\includegraphics[width=0.32\linewidth]{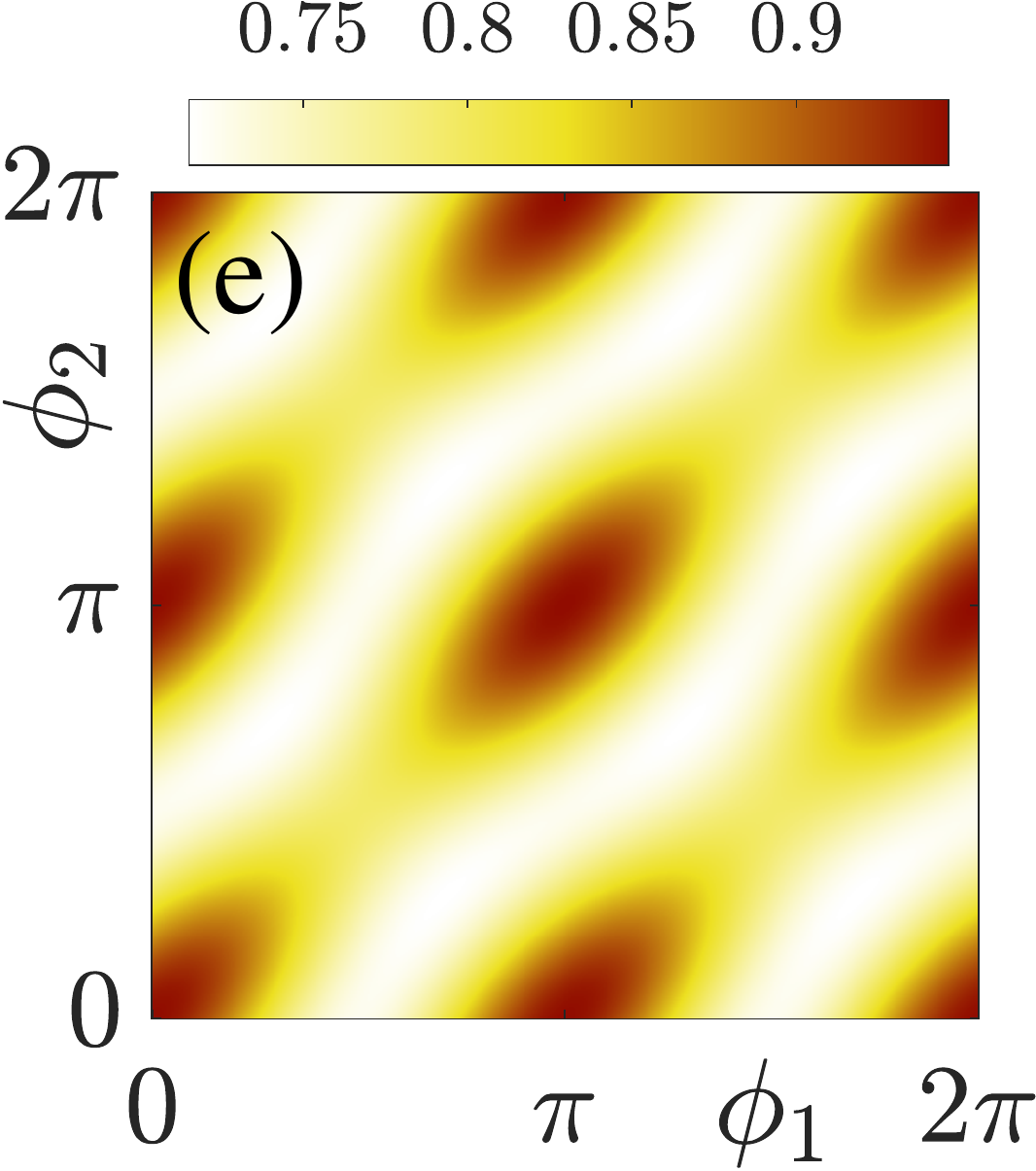}
\includegraphics[width=0.32\linewidth]{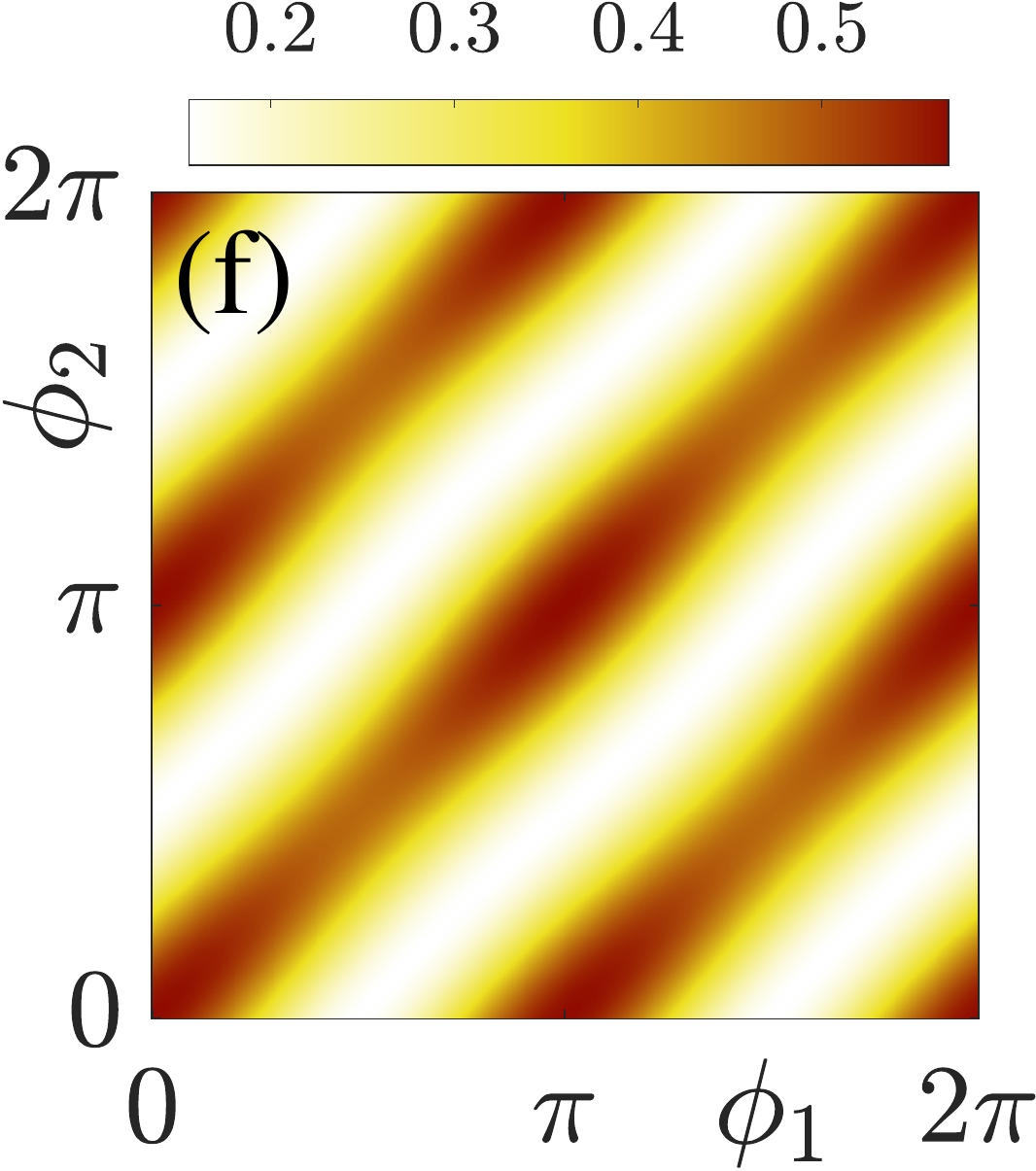}
\includegraphics[width=0.32\linewidth]{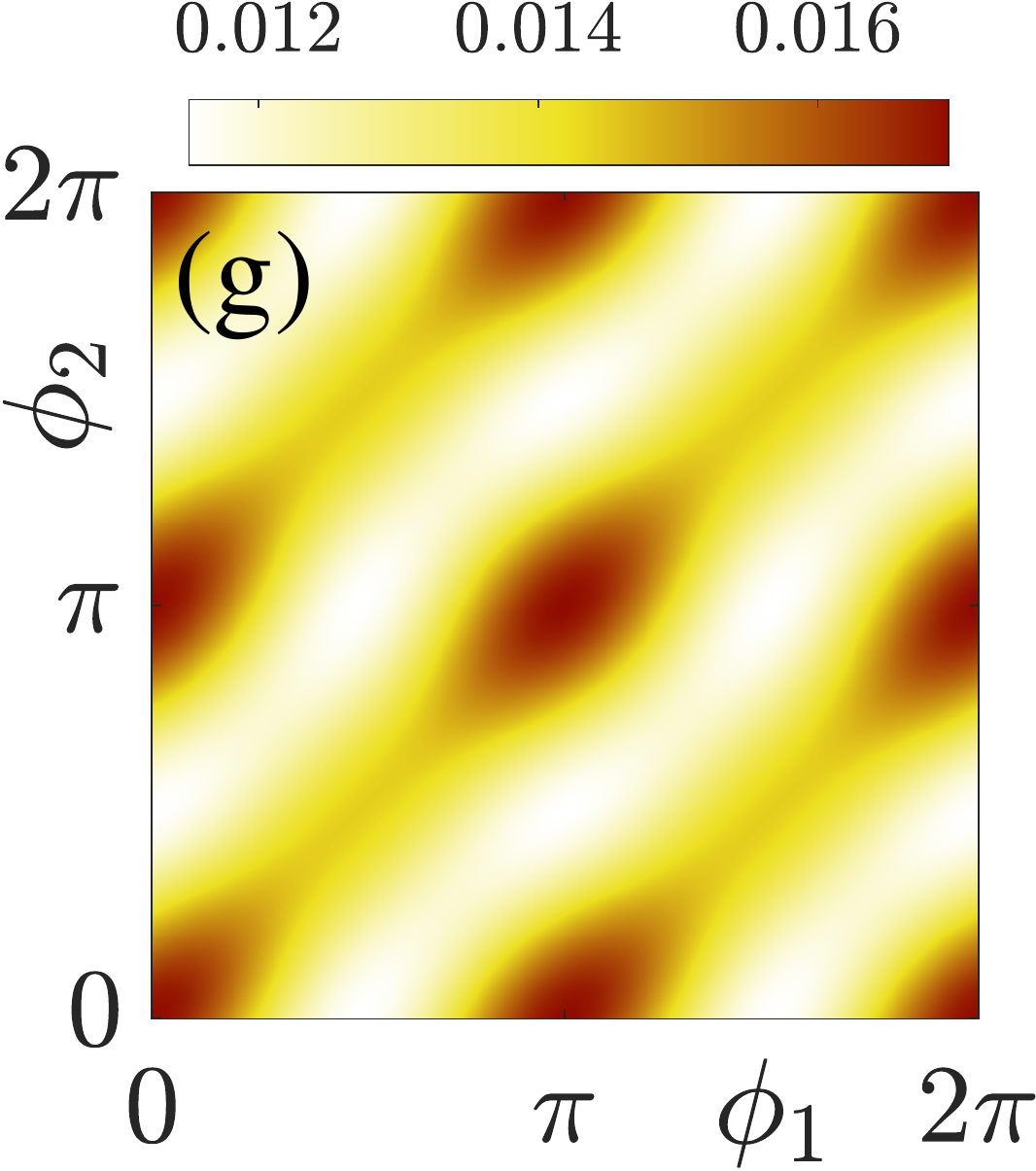}\\
\includegraphics[width=0.32\linewidth]{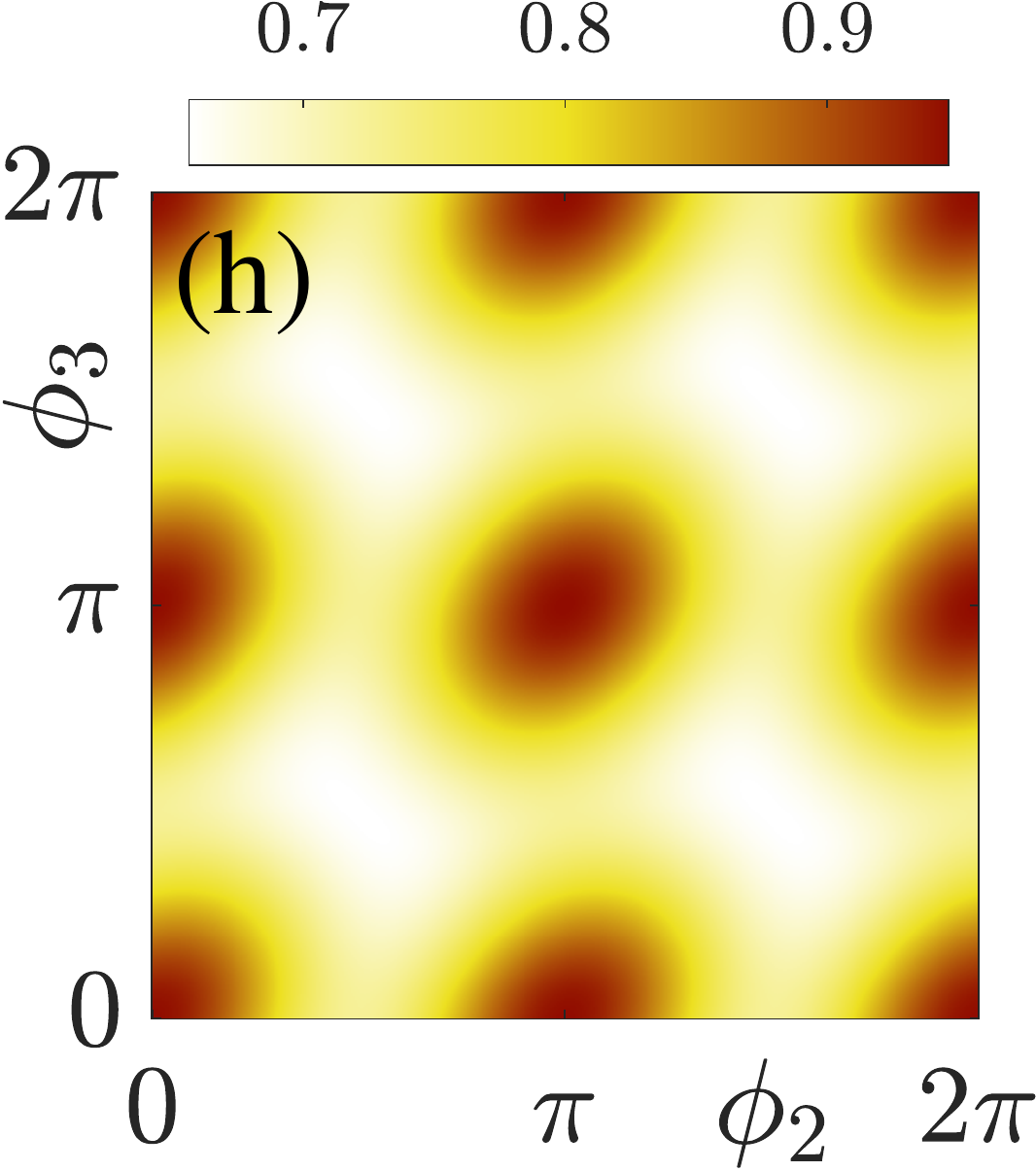}
\includegraphics[width=0.32\linewidth]{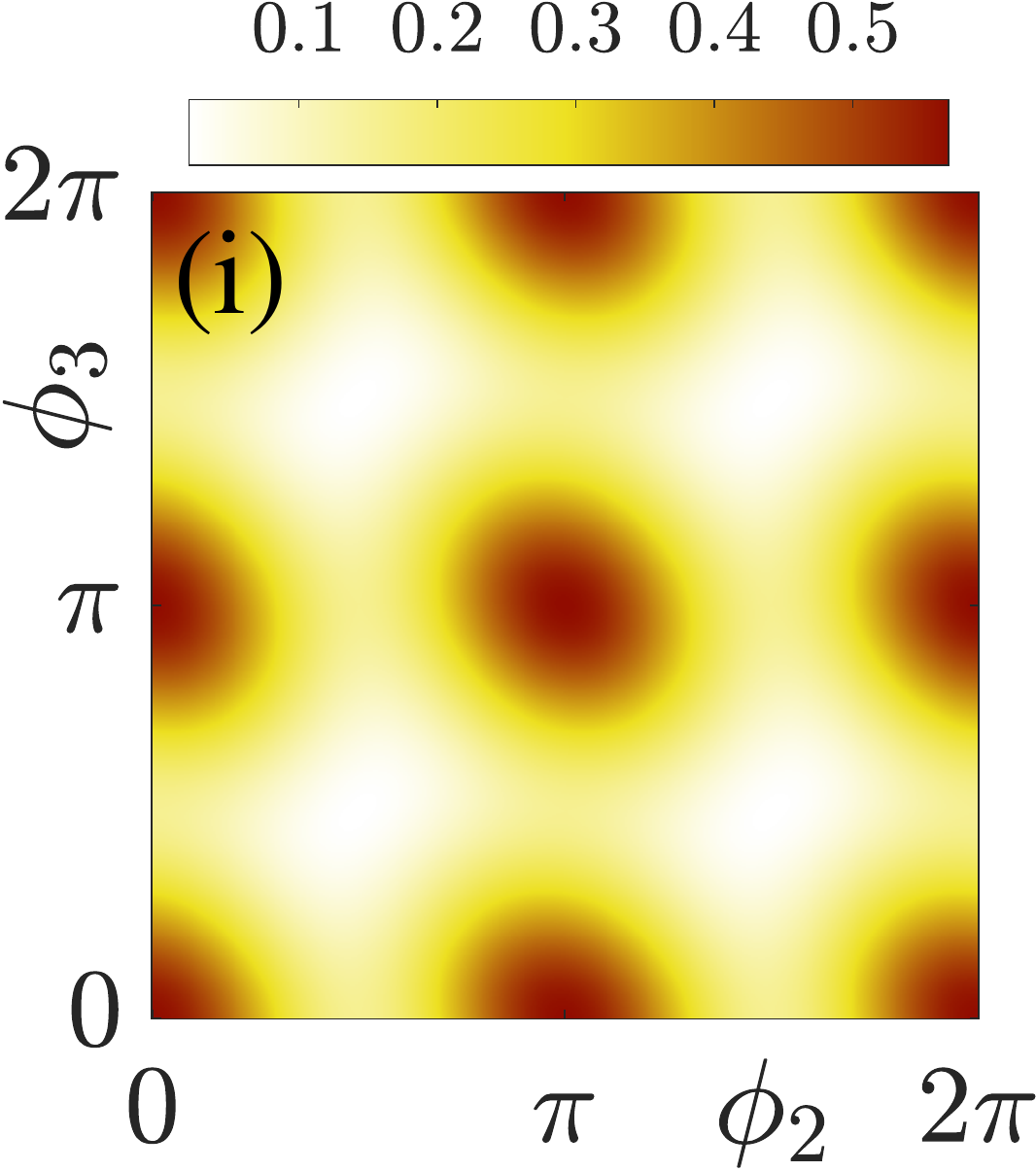}
\includegraphics[width=0.32\linewidth]{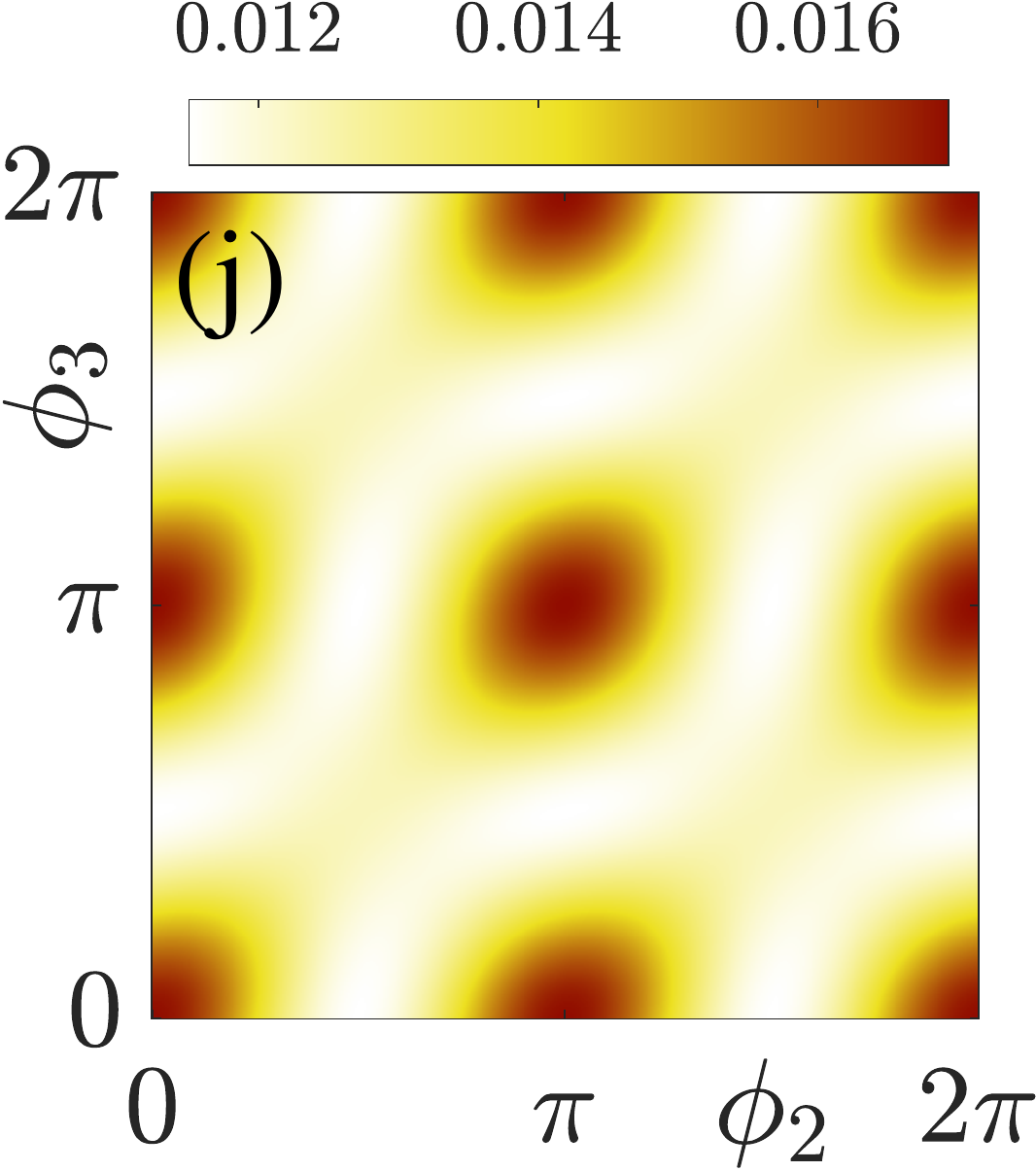}
\caption{(Color online). Same as Fig.~2 of Ref.~\onlinecite{Halimeh2020f} but with $L^\mathrm{g}_{j,j+1}=\tau^-_{j,j+1}$ and including additional angles for the MQC spectra. (a) Dominant MQC. (b-j) MQC spectra, with rows from left to right $F(\phi_1,0,\phi_3,0)$, $F(\phi_1,\phi_2,0,0)$, and $F(0,\phi_2,\phi_3,0)$ and columns from left to right $tJ_a=10^4$, $tJ_a=10^5$, and $tJ_a=10^6$. In contrast to the case of $L^\mathrm{g}_{j,j+1}=\tau^z_{j,j+1}$ as in Fig.~2 of Ref.~\onlinecite{Halimeh2020f}, here the MQC do not decay to zero, but rather saturate at finite steady-state values. This behavior depends on the fixed points of the Liouvillian superoperator.}
\label{fig:MQC_DiffDiss} 
\end{figure}

Our choice of the MQC-spectrum angles in the main text is neither special nor unique. Due to symmetry, we have
\begin{subequations}
\begin{align}
&F(\phi_1,0,\phi_3,0)=F(0,\phi_2,0,\phi_4),\\
&F(\phi_1,\phi_2,0,0)=F(0,0,\phi_3,\phi_4),\\
&F(0,\phi_2,\phi_3,0)=F(\phi_1,0,0,\phi_4).
\end{align}
\end{subequations}
For completeness, we show in Fig.~\ref{fig:MQC_SM} for the MQC spectra $F(\phi_1,\phi_2,0,0)$ and $F(0,\phi_2,\phi_3,0)$ in the presence of gauge-breaking coherent and incoherent errors at strengths $\lambda=10^{-4}J_a$ and $\gamma=10^{-6}$, where the initial state is that of Fig.~\ref{fig:Z2LGT_InitialStates}(a) and the dynamics is governed by Eq.~\eqref{eq:EOM}. Similarly to their counterpart $F(\phi_1,0,\phi_3,0)$, both $F(\phi_1,\phi_2,0,0)$ and $F(0,\phi_2,\phi_3,0)$ diminish in the presence of decoherence.

It is interesting to compare these findings to the dynamics of the MQC with a different jump operator. As such, we again start in the staggered initial state of Fig.~\ref{fig:Z2LGT_InitialStates}(a) and set the jump operator $L^\mathrm{g}_{j,j+1}=\tau^-_{j,j+1}$. Once again, we set $\lambda=10^{-4}J_a$ and $\gamma=10^{-6}J_a$, with $L^\mathrm{m}_j=a_j^\dagger a_j$. (The dynamics of the gauge violation and supersector projectors for this quench protocol are shown in Fig.~\ref{fig:DiffDiss}.) The corresponding MQC results are shown in Fig.~\ref{fig:MQC_DiffDiss}. Unlike the case of $L^\mathrm{g}_{j,j+1}=\tau^z_{j,j+1}$ (see Fig.~2 of Ref.~\onlinecite{Halimeh2020f}), the MQC do not decay to zero when $L^\mathrm{g}_{j,j+1}=\tau^-_{j,j+1}$, but rather saturate at finite steady-state values. This behavior depends on the fixed point of the Liouvillian superoperator. Indeed, we have checked (not shown) that when the ``domain-wall'' product state shown in Fig.~\ref{fig:Z2LGT_InitialStates}(b) is the initial state, the MQC take on the same steady-state values as those shown in Fig.~\ref{fig:MQC_DiffDiss}.

\subsection{Variations of relative strength of incoherent errors}

\begin{figure}[htp]
	\centering
	\includegraphics[width=.48\textwidth]{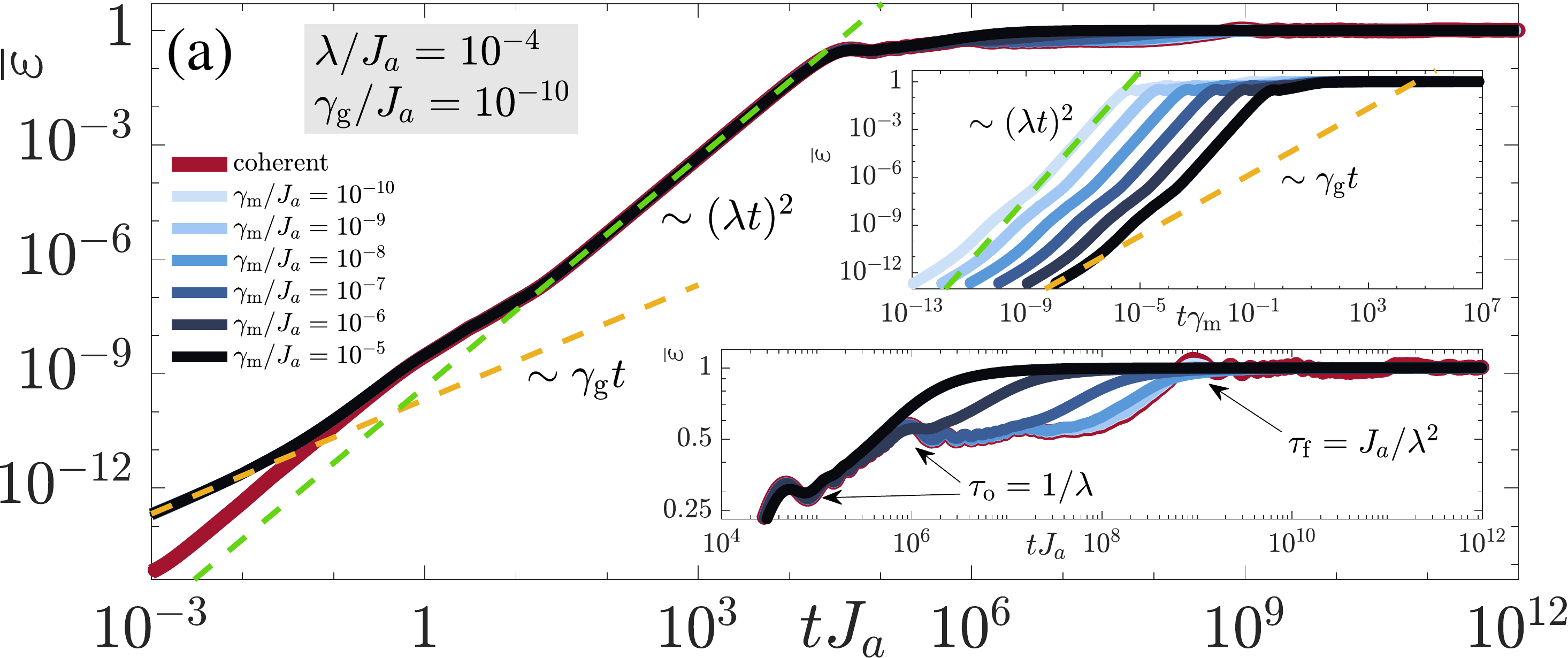}\quad
	\includegraphics[width=.48\textwidth]{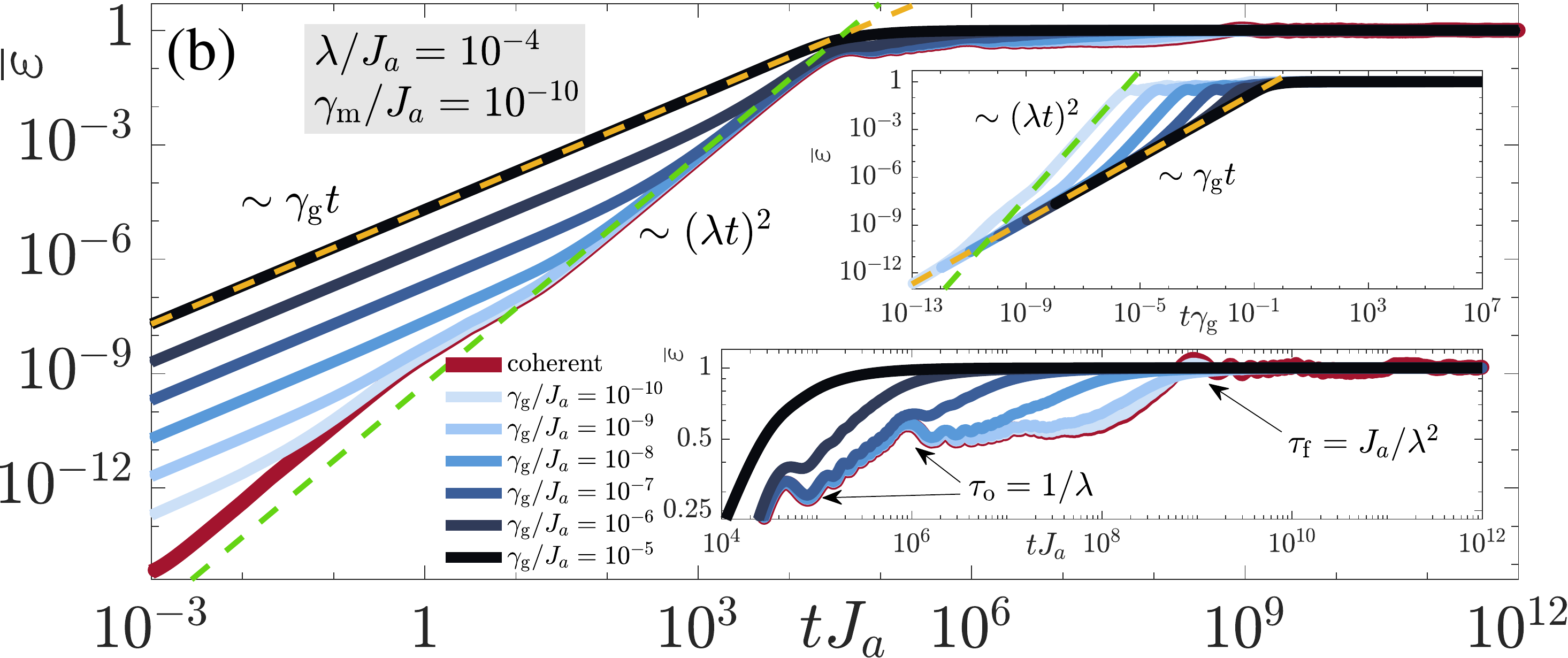}
	\caption{(Color online). Dynamics of the gauge-invariance violation at different environment-coupling strengths $\gamma_\mathrm{m}$ for the dephasing on matter fields and $\gamma_\mathrm{g}$ for the dissipation on gauge links, with fixed strength $\lambda=10^{-4}J_a$ of coherent gauge-breaking processes. (a) Gauge violation at various values of $\gamma_\mathrm{m}$ for fixed value of $\gamma_\mathrm{g}=10^{-10}J_a$. (b) Gauge violation at various values of $\gamma_\mathrm{g}$ for a fixed value of $\gamma_\mathrm{m}=10^{-10}J_a$. Dissipation clearly shows an effect on the gauge violation at short times, whereas dephasing does not. However, at late times dephasing also has a clear effect on the prethermal plateaus, as shown in the lower inset of (a).
	}
	\label{fig:relativeGamma} 
\end{figure}

Let us now investigate the effect of turning on the dephasing and dissipation at different strengths $\gamma_\mathrm{m}$ and $\gamma_\mathrm{g}$, respectively. The Lindblad master equation generalizes to
\begin{align}\nonumber
\dot{\rho}=&-i[H_0+\lambda H_1,\rho]\\\nonumber
&+\sum_{j=1}^N\Big[\gamma_\mathrm{m}\Big(L^\text{m}_j\rho L^{\text{m}\dagger}_j-\frac{1}{2}\big\{L^{\text{m}\dagger }_jL^\text{m}_j,\rho\big\}\Big)\\\label{eq:EOM_relative}
&+\gamma_\mathrm{g}\Big(L^\text{g}_{j,j+1}\rho L^{\text{g}\dagger}_{j,j+1}-\frac{1}{2}\big\{L^{\text{g}\dagger }_{j,j+1}L^\text{g}_{j,j+1},\rho\big\}\Big)\Big].
\end{align}
Interestingly, the dephasing strength $\gamma_\mathrm{m}$ has little effect on the short-time dynamics of the gauge violation, which at short times $t\lesssim\gamma_\mathrm{g}/\lambda^2$ scales as $\varepsilon\sim\gamma_\mathrm{g}t$, and at intermediate times $\gamma_\mathrm{g}/\lambda^2<t\lesssim1/\lambda$ scales as $\varepsilon\sim(\lambda t)^2$ due to the dominance of coherent gauge-breaking terms. In fact, it can be shown in TDPT (see Sec.~\ref{sec:TDPT_leadingIncoherent}) that the contribution to the gauge violation due to dephasing at short times vanishes. However, we find that both $\gamma_\mathrm{m}$ and $\gamma_\mathrm{g}$ have a significant effect on the later timescale at which decoherence dominates at maximal violation, with this timescale being roughly $1/\max\{\gamma_\mathrm{g},\gamma_\mathrm{m}\}$, as can be seen in the lower insets of Fig.~\ref{fig:relativeGamma}(a,b). In particular, dephasing incurs a nonperturbative effect on the prethermal plateaus, as shown in the lower inset of Fig.~\ref{fig:relativeGamma}(a).

\subsection{Dynamics under gauge protection}
\begin{figure}[htp]
	\centering
	\includegraphics[width=.48\textwidth]{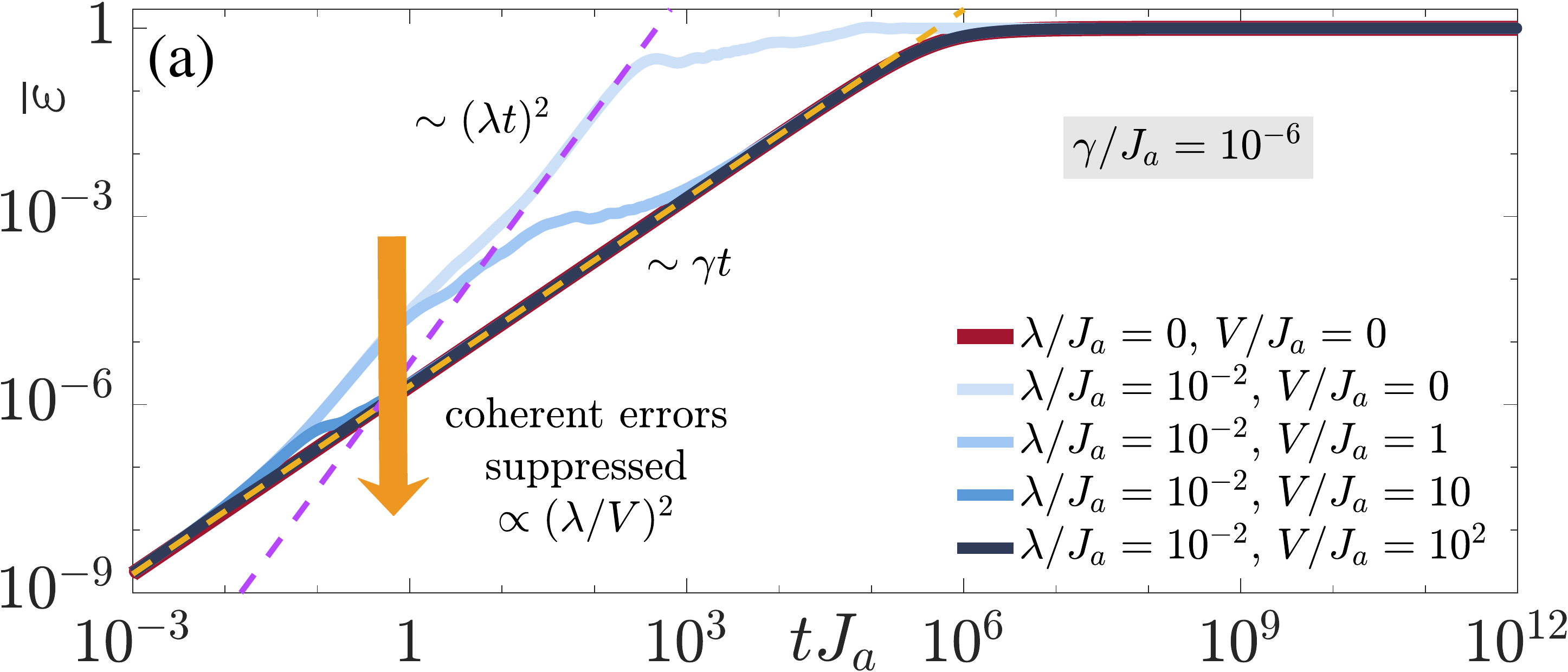}\quad
	\includegraphics[width=.48\textwidth]{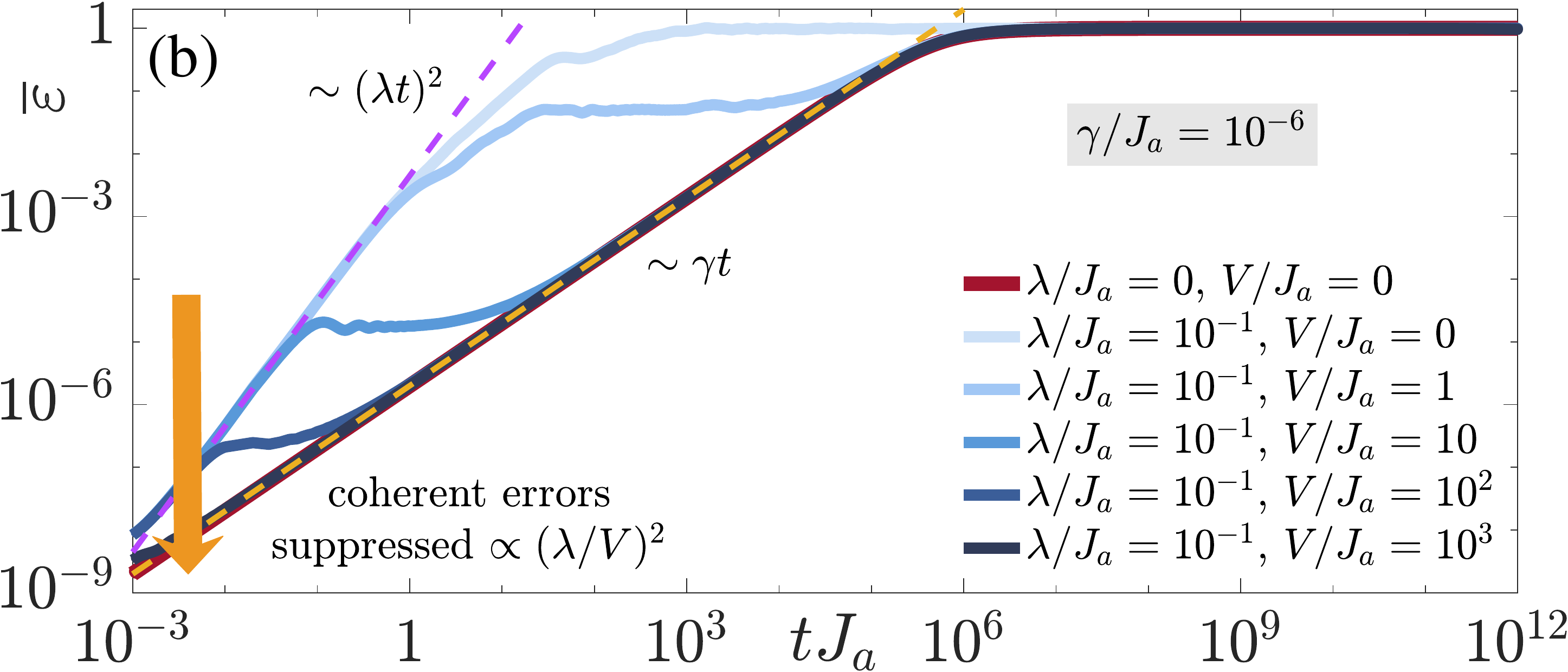}
	\caption{(Color online). Using the setup of Fig.~\ref{fig:Z2LGT_InitialStates}(a) under decoherence, we add a protection term \cite{Halimeh2020a} $V H_G=V\sum_jG_j$ such that the coherent quench is actuated by $H_0+\lambda H_1+VH_G$, which suppresses only the coherent gauge-breaking errors, but has no effect on the gauge violation due to Markovian decoherence. (a) At fixed $\lambda=10^{-2}J_a$ and $\gamma=10^{-6}J_a$ (blue curves), we see that at large enough protection strength $V$, the diffusive scaling $\varepsilon\sim\gamma t$ seen at short times emerges again before the maximal violation is reached, and, in some cases, after the ballistic scaling $\varepsilon\sim(\lambda t)^2$ has appeared at intermediate times. For reference, we also show the gauge violation for $\lambda=0$ and $\gamma=10^{-6}J_a$ (red curve), which exhibits only scaling $\varepsilon\sim\gamma t$ before saturating at its maximal value for $t\gtrsim1/\gamma$. At nonzero $\lambda$ and without energy protection, the gauge violation scales $\varepsilon\sim\gamma t$ only at early times $t\lesssim\gamma/\lambda^2$ before scaling $\varepsilon\sim(\lambda t)^2$ at intermediate times $t\gtrsim\gamma/\lambda^2$, and finally reaching the maximal violation at $t\propto\min\{\lambda^{-N/2},\gamma^{-1}\}$. (b) The same as panel (a) but for $\lambda=10^{-1}J_a$.}
	\label{fig:protection} 
\end{figure}

Recently, several theoretical works have proposed to use gauge protection to suppress processes driving the system out of its initial gauge-invariant sector,\cite{Zohar2011,Zohar2012,Banerjee2012,Zohar2013,Hauke2013,Kuehn2014,Kuno2015,Kuno2017,Negretti2017,Barros2019,Halimeh2020a,Halimeh2020e,Mathis2020,Lamm2020,Tran2020} and the principle has been demonstrated experimentally for a $\mathrm{U}(1)$ gauge theory.\cite{Yang2020} The basic idea is to introduce a suitable energy-penalty term, which for a $\mathrm{Z}_2$ gauge theory reads
\begin{align}
VH_G=V\sum_jG_j,
\end{align}
where $V$ controls the protection strength. For sufficiently large $V$, the associated gauge violation due to $H_1$ is suppressed by $(\lambda/V)^2$, and the ensuing dynamics is perturbatively close to a renormalized version of the ideal gauge theory.\cite{Halimeh2020a,Halimeh2020e} Using as quench Hamiltonian $H=H_0+\lambda H_1+VH_G$, and numerically solving the respective Lindblad master equation 
\begin{align}\nonumber
	\dot{\rho}=&-i[H_0+\lambda H_1+VH_G,\rho]\\\nonumber
	&+\gamma\sum_{j=1}^N\Big(L^\text{m}_j\rho L^{\text{m}\dagger}_j+L^\text{g}_{j,j+1}\rho L^{\text{g}\dagger}_{j,j+1}\\
	&-\frac{1}{2}\big\{L^{\text{m}\dagger }_jL^\text{m}_j+L^{\text{g}\dagger }_{j,j+1}L^\text{g}_{j,j+1},\rho\big\}\Big)
\end{align}
at fixed values of $\gamma$ and $\lambda$, we obtain the gauge-violation dynamics shown in Fig.~\ref{fig:protection}. We find that a finite $V$ suppresses only coherent contributions to the gauge violation, but not incoherent ones. This finding is not surprising as the dissipative errors in our work are modelled by a Markovian Lindblad master equation that couples states regardless of their energy differences, and which is thus oblivious to energy penalties.

Interestingly, for intermediate values of the protection strength ($V=J_a$ at $\lambda=10^{-2}J_a$), we see that after going from diffusive ($\varepsilon\sim\gamma t$) at $t\lesssim\gamma/\lambda^2$ to ballistic ($\varepsilon\sim\lambda^2 t^2$) dynamics at $t\gtrsim\gamma/\lambda^2$, the gauge violation again exhibits diffusive behavior before settling into a maximal-violation steady state at $t\approx1/\gamma$. The larger $V$ is, the shorter is the intermediate ballistic regime. At sufficiently large $V$, coherent errors are almost completely suppressed and the ballistic regime vanishes, with the gauge violation scaling as $\varepsilon\sim\gamma t$ for all times $t\lesssim1/\gamma$.

\subsection{Dynamics under particle loss}

\begin{figure}[htp]
	\centering
	\includegraphics[width=.48\textwidth]{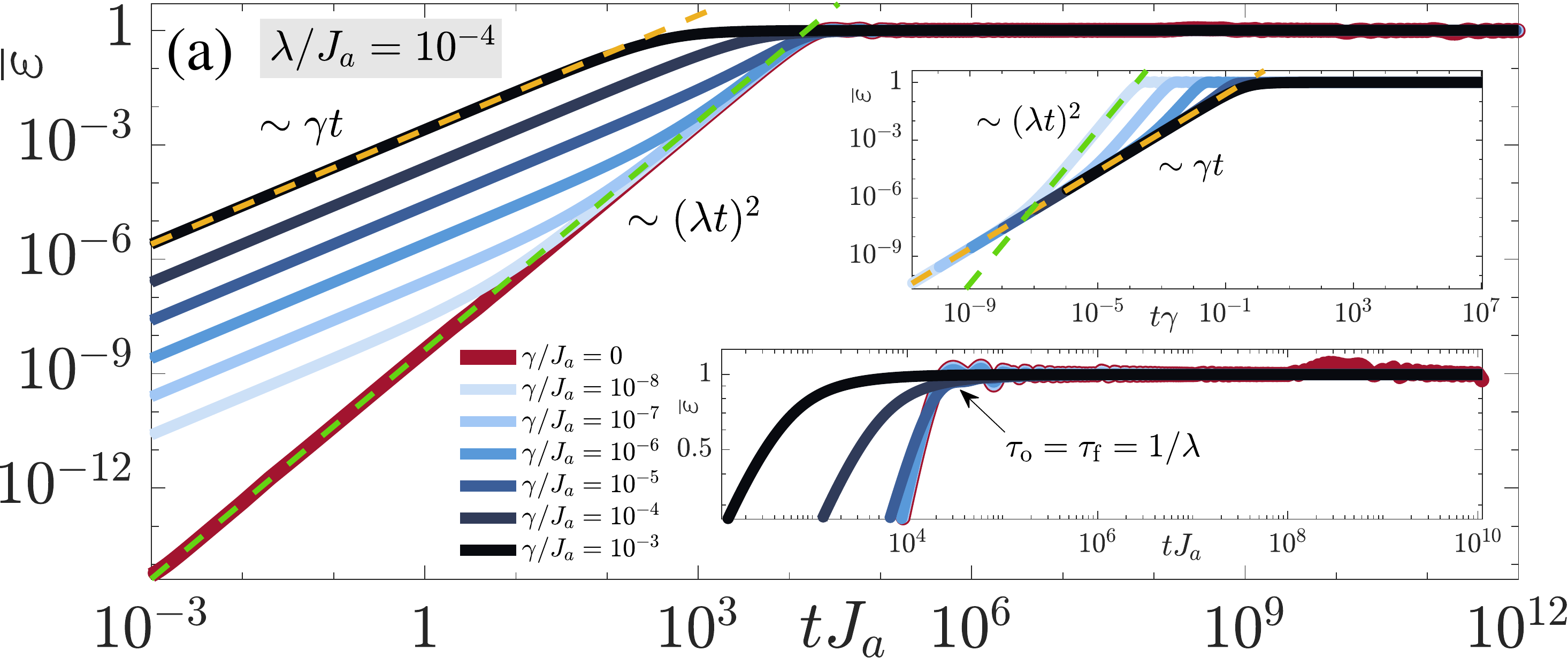}\quad
	\includegraphics[width=.48\textwidth]{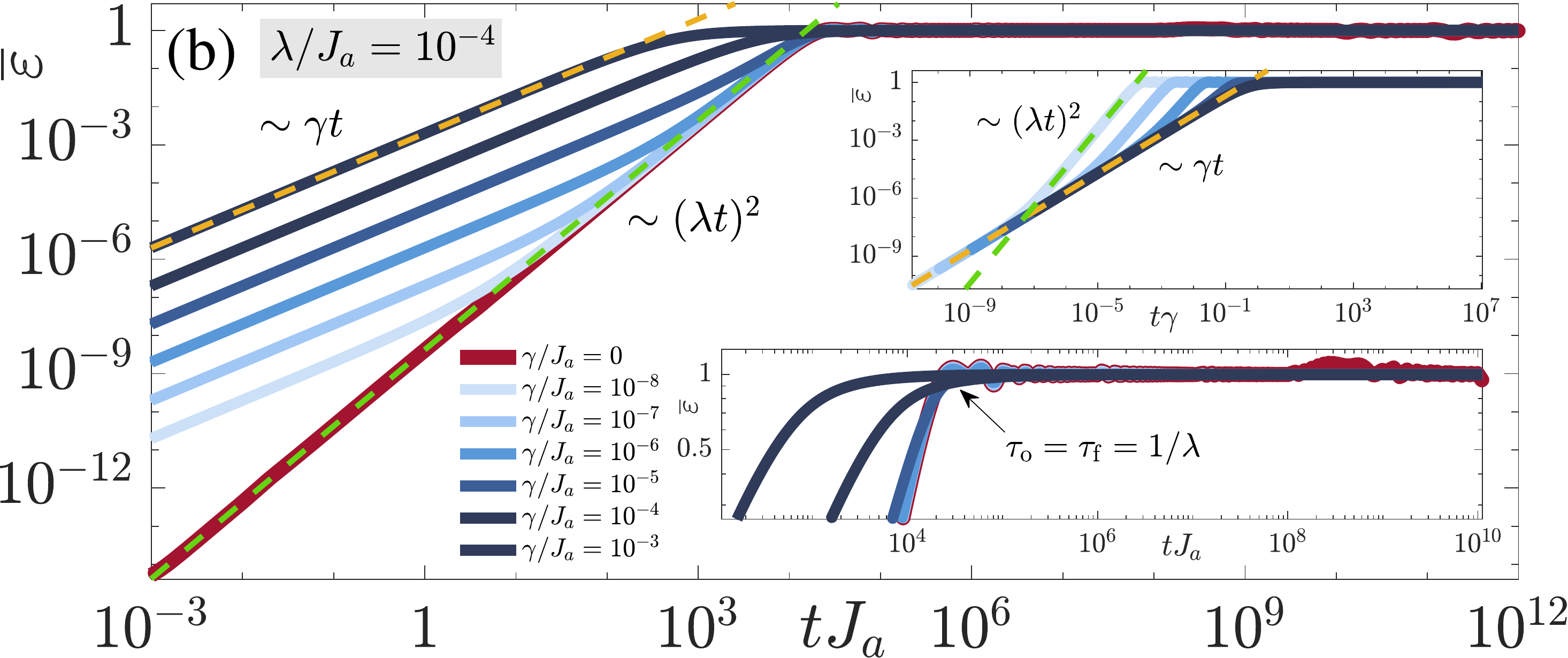}
	\caption{(Color online). Dynamics of gauge violation in the $\mathrm{Z}_2$ LGT with $N=2$ matter sites in the presence of dissipation in the gauge fields with jump operators $L^\mathrm{g}_{j,j+1}=\tau^z_{j,j+1}$ and (a) dissipation in the matter fields with  jump operators $L^\mathrm{m}_j=a_j$ and (b) dephasing in the matter fields with  jump operators $L^\mathrm{m}_j=a_j^\dagger a_j$, just as in all the results for the $\mathrm{Z}_2$ LGT before now. Coherent gauge breaking is at strength $\lambda=10^{-4}J_a$. The behavior is qualitatively identical whether the matter fields are subjected to dephasing or dissipation. The fact that we have only $N=2$ matter sites here brings about only a single prethermal plateau instead of two as in the case of $N=4$ matter sites.\cite{Halimeh2020b,Halimeh2020c}}
	\label{fig:ParticleLoss} 
\end{figure}

Until now, within Sec.~\ref{sec:Z2LGT} we have considered only dephasing in the matter fields in order to allow for the conservation of particle number, thereby enabling us to achieve larger system sizes (see Appendix~\ref{sec:NumSpec} for further details). Here, we consider also dissipation in the matter fields. In particular, we will choose $L^\mathrm{m}_j=a_j$ while also using $L^\mathrm{g}_{j,j+1}=\tau^z_{j,j+1}$ and fixing $\lambda=10^{-4}J_a$. Again, we consider the initial state in Fig.~\ref{fig:Z2LGT_InitialStates}(a) for $N=2$ matter sites (here, $N=4$ matter sites is numerically intractable for the evolution times we need to reach in our calculations), and solve Eq.~\eqref{eq:EOM} for $\lambda=10^{-4}J_a$ using several values of $\gamma$. The corresponding results for the gauge-violation dynamics are shown in Fig.~\ref{fig:ParticleLoss}(a). For ease of comparison, we repeat these results for the case of dephasing on matter fields with  jump operators $L^\mathrm{m}_j=a_j^\dagger a_j$ in Fig.~\ref{fig:ParticleLoss}(b), just as in all the results before this point. Aside from the absence of a second prethermal plateau due to the halved matter-site number, the results are very similar to those in Fig.~1(b) of Ref.~\onlinecite{Halimeh2020f}, and the qualitative behavior is identical: the gauge violation displays a crossover from diffusive scaling $\varepsilon\sim\gamma t$ to ballistic scaling $\varepsilon\sim\lambda^2t^2$ at $t\propto\gamma/\lambda^2$ when $\gamma<\lambda$. As such, we can conclude that our results in Ref.~\onlinecite{Halimeh2020f} are general, and are not restricted by considering only dephasing in the matter fields.

\subsection{Decoherence starting from equilibrium}

\begin{figure}[htp]
	\centering
	\includegraphics[width=.48\textwidth]{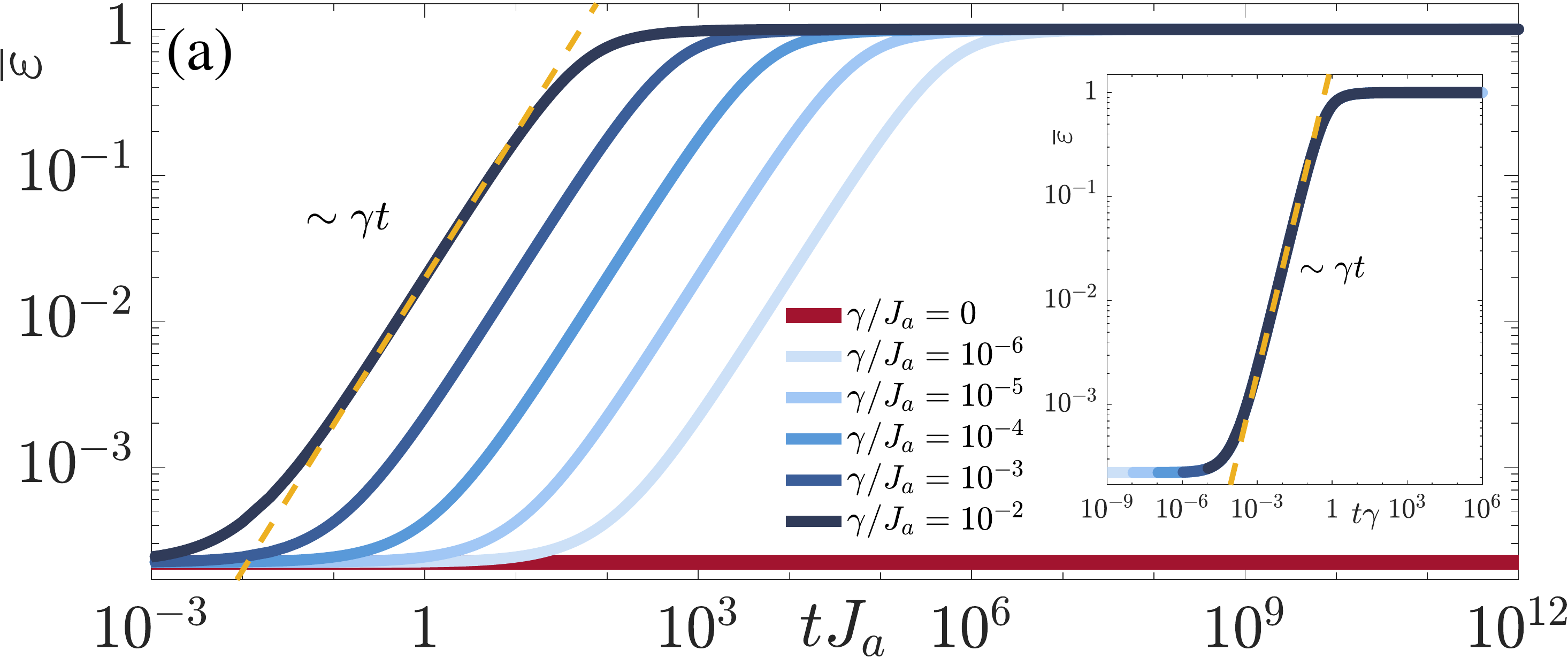}\\
	\includegraphics[width=.48\textwidth]{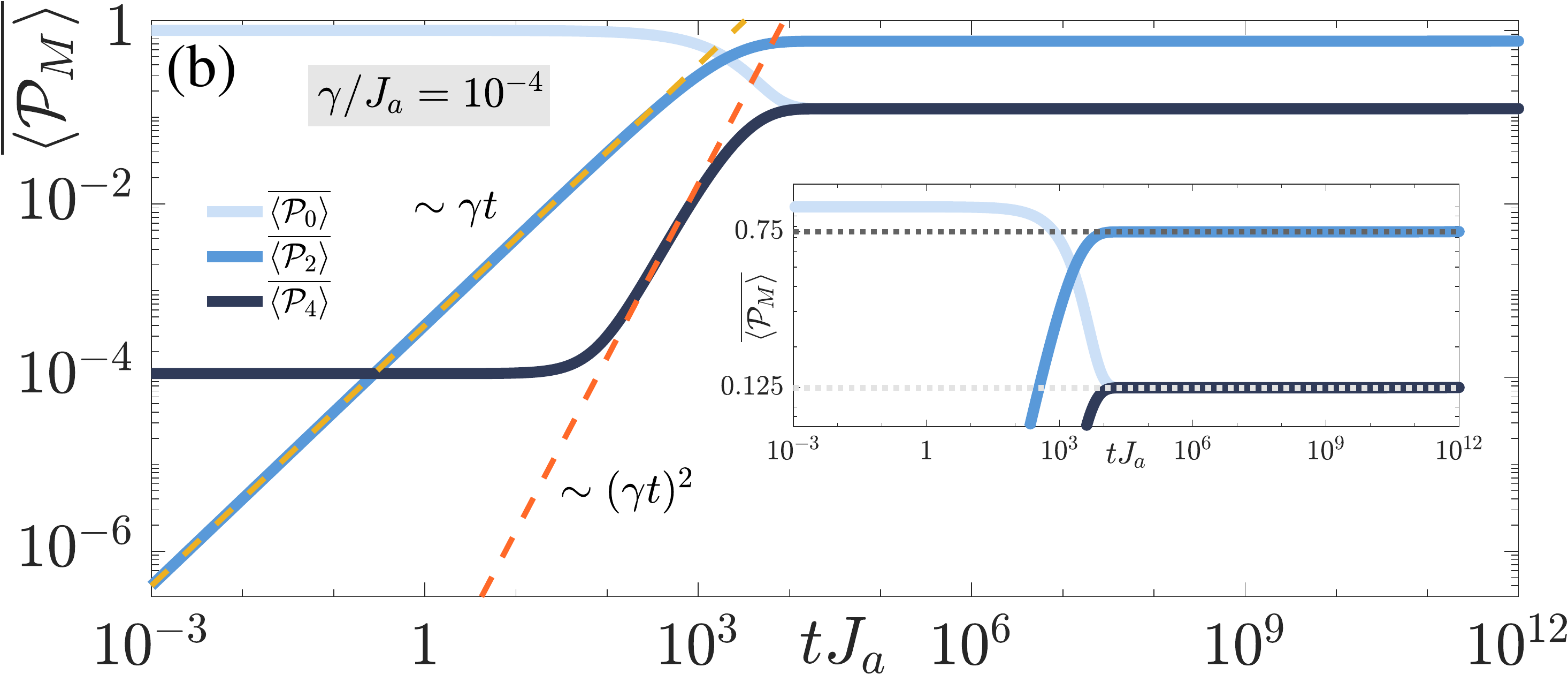}
	\caption{(Color online). (a) Time evolution of  gauge-invariance violation after starting in the ground state of the $\mathrm{Z}_2$ LGT and switching on decoherence with strength $\gamma$ at $t=0$. Note that the ground state of $H_0$ is not gauge-invariant, a consequence of the fact that different gauge-invariant sectors of $H_0$ have energy overlaps. Similarly to the case of quench dynamics, the gauge violation at short times scales $\sim\gamma t$ before plateauing at its maximal value.
		(b) Projectors onto the three accessible gauge-invariant supersectors. Their steady-state expectation values are proportional to the number of constituent gauge-invariant sectors.}
	\label{fig:Fig3} 
\end{figure}

\begin{figure}[htp]
	\centering
	\includegraphics[width=.48\textwidth]{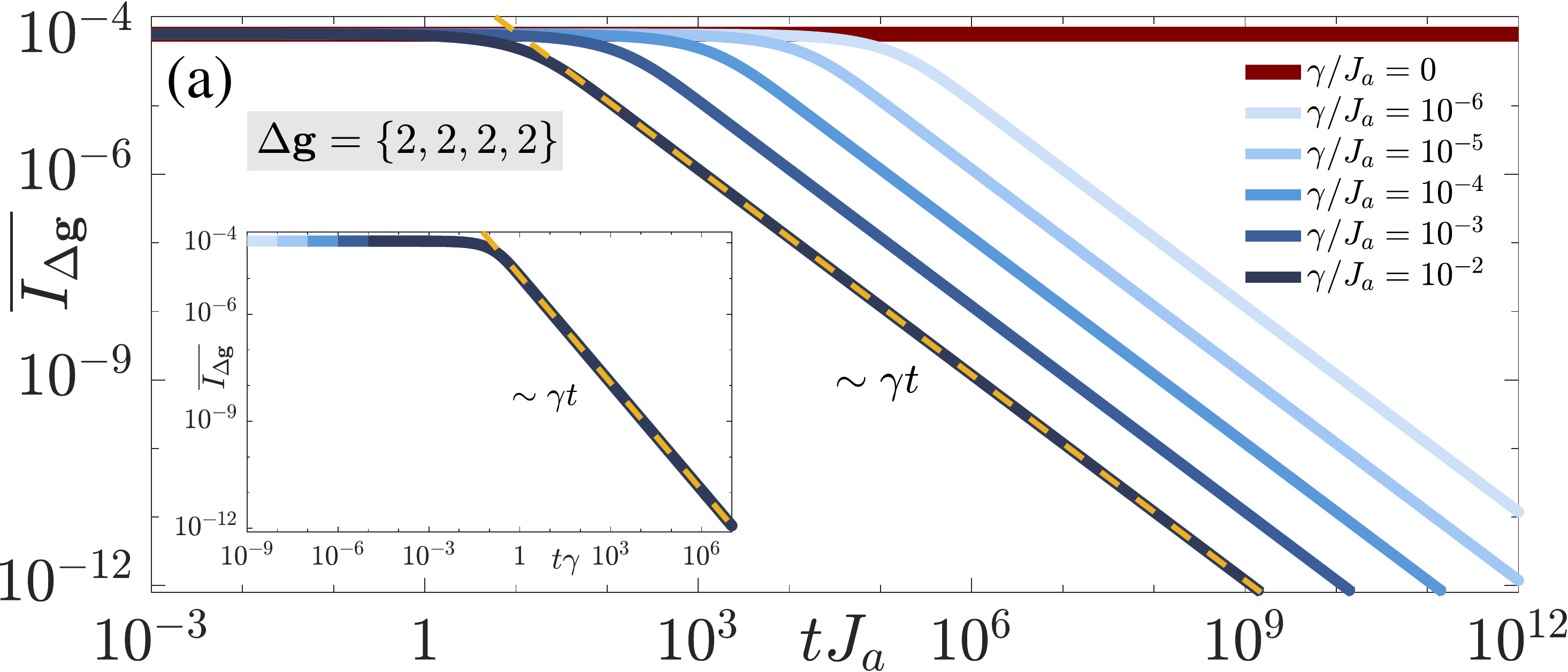}\quad
	\includegraphics[width=.48\textwidth]{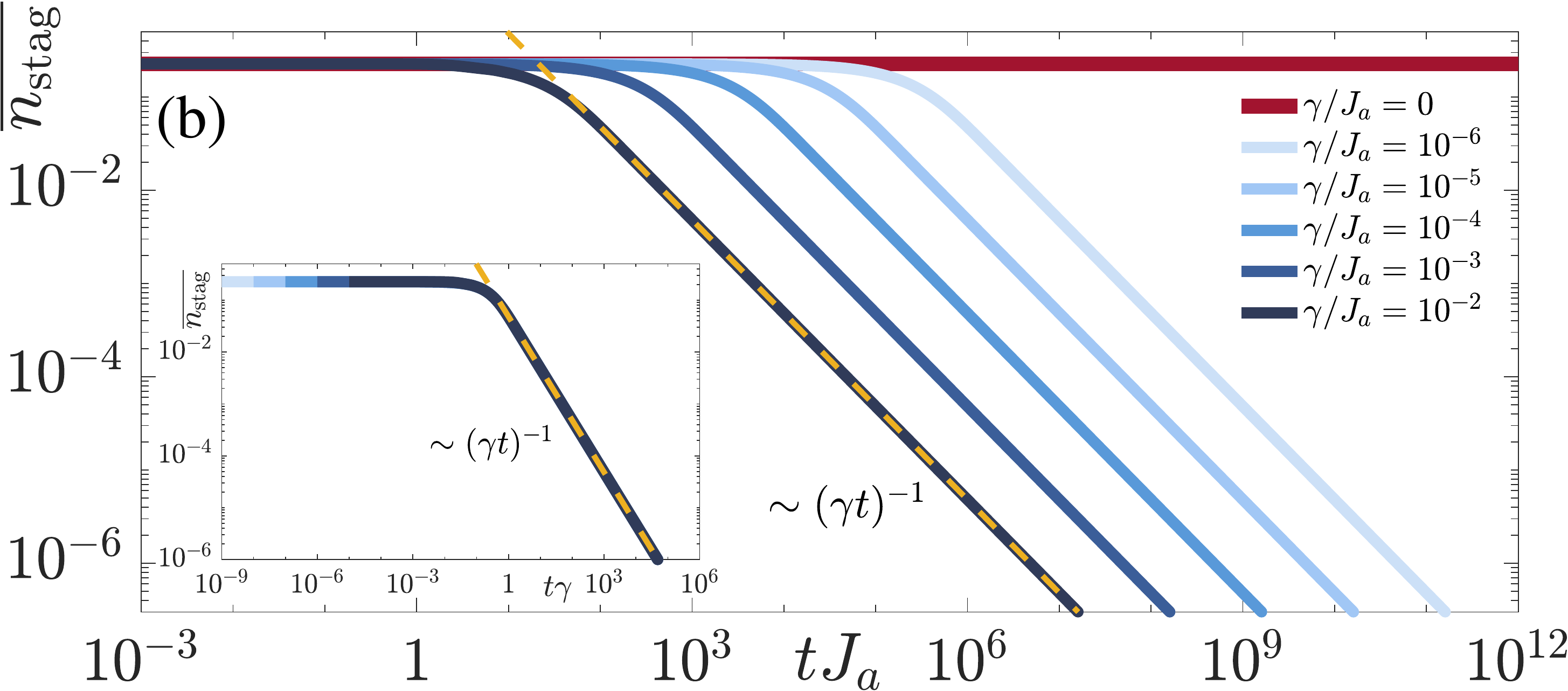}
	\caption{(Color online). Same scenario as in Fig.~\ref{fig:Fig3}: we start in the ground state of $H_0$ and switch on decoherence at $t=0$. Running averages of the (a) MQC intesity $I_{\Delta\mathbf{g}=\{2,2,2,2\}}$ and (b) staggered boson number are shown. Unlike the gauge violation and supersector projectors, these observables show no trace of diffusive behavior at short times, but they decay $\sim(\gamma t)^{-1}$ due to decoherence for $t\gtrsim1/\gamma$.
	}
	\label{fig:staticsObs} 
\end{figure}

\begin{figure}[htp]
	\centering
	\includegraphics[width=.48\textwidth]{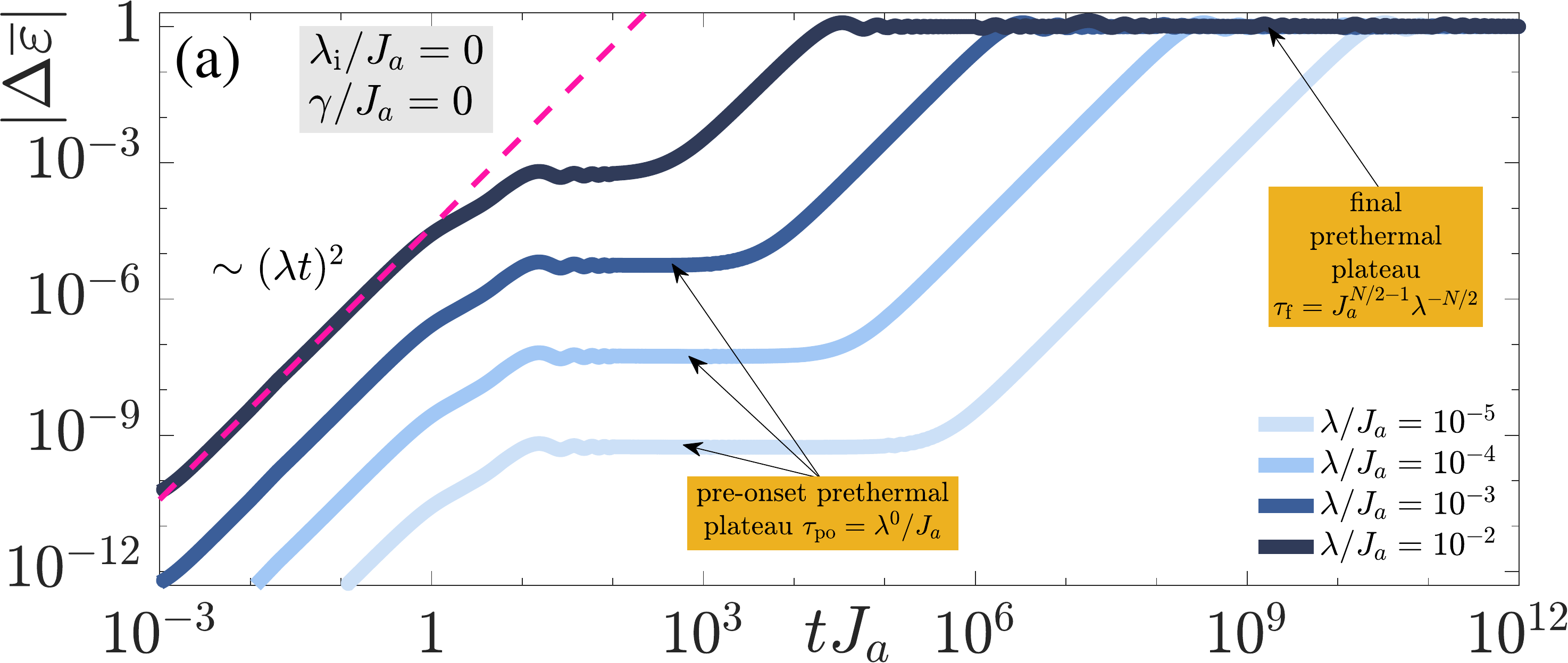}\quad
	\includegraphics[width=.48\textwidth]{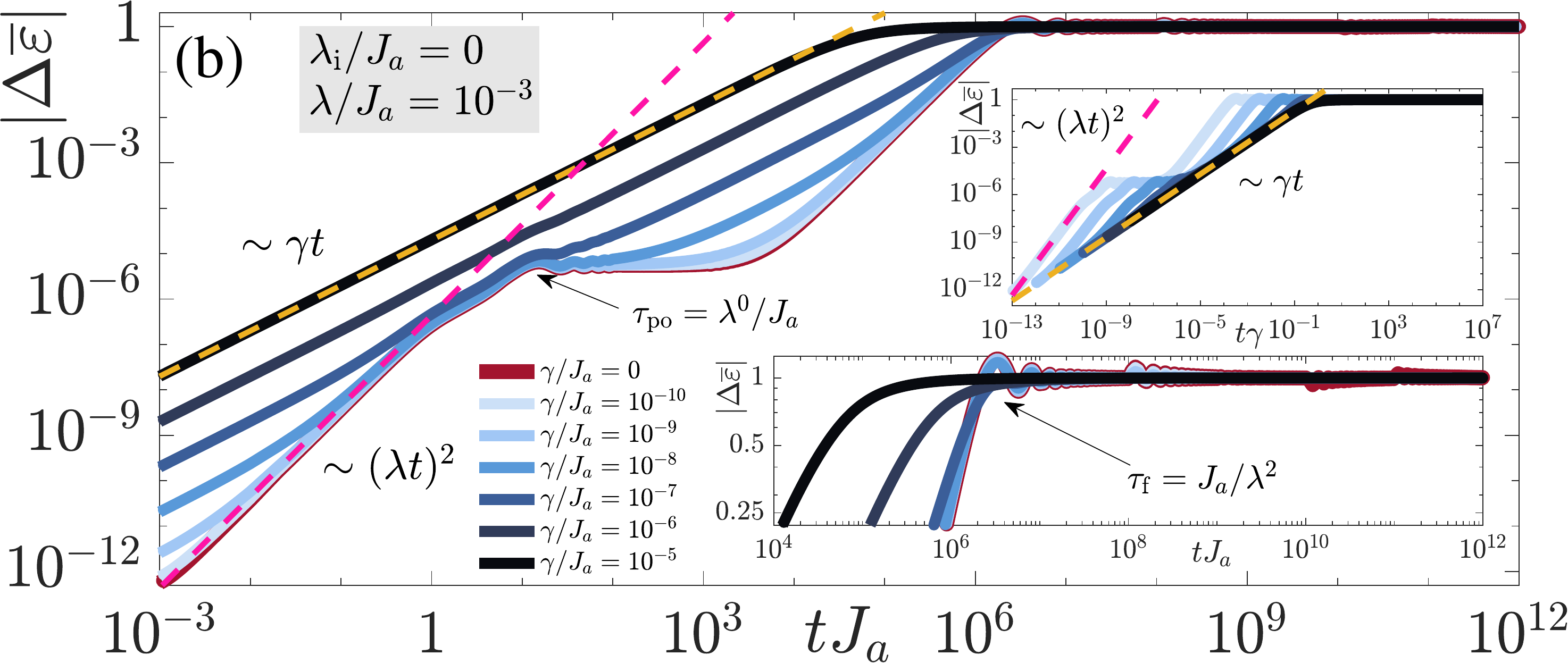}
	\caption{(Color online). Gauge violation over evolution time after starting in the (gauge-noninvariant) ground state of $H_0$ and quenching with $H_0+\lambda H_1$ with Markovian decoherence through jump operators $L^\mathrm{m}_j=a_j^\dagger a_j$ and $L^\mathrm{g}_{j,j+1}=\tau^z_{j,j+1}$. (a) In the case of $\gamma=0$ (no decoherence), the pre-onset and final prethermal plateaus appear and the violation at early times scales $\sim\lambda^2 t^2$, because $\rho_0$ is the ground state of $H_0$. (b) When decoherence is turned on, a diffusive-to-ballistic crossover appears at timescale $\propto\gamma/\lambda^2$ taking the violation from a diffusive spread $\sim\gamma t$ to a ballistic scaling $\sim\lambda^2 t^2$.
	}
	\label{fig:lami0} 
\end{figure}

\begin{figure}[htp]
	\centering
	\includegraphics[width=.48\textwidth]{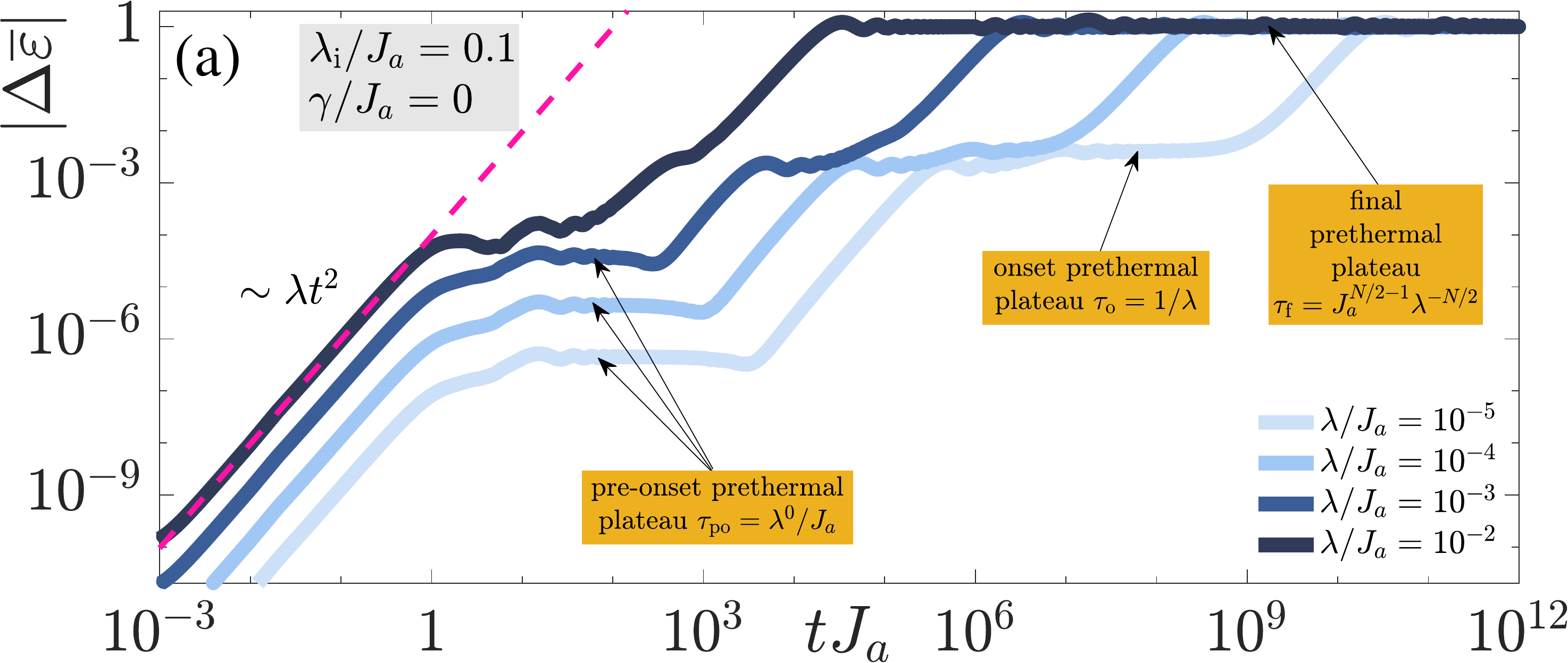}\quad
	\includegraphics[width=.48\textwidth]{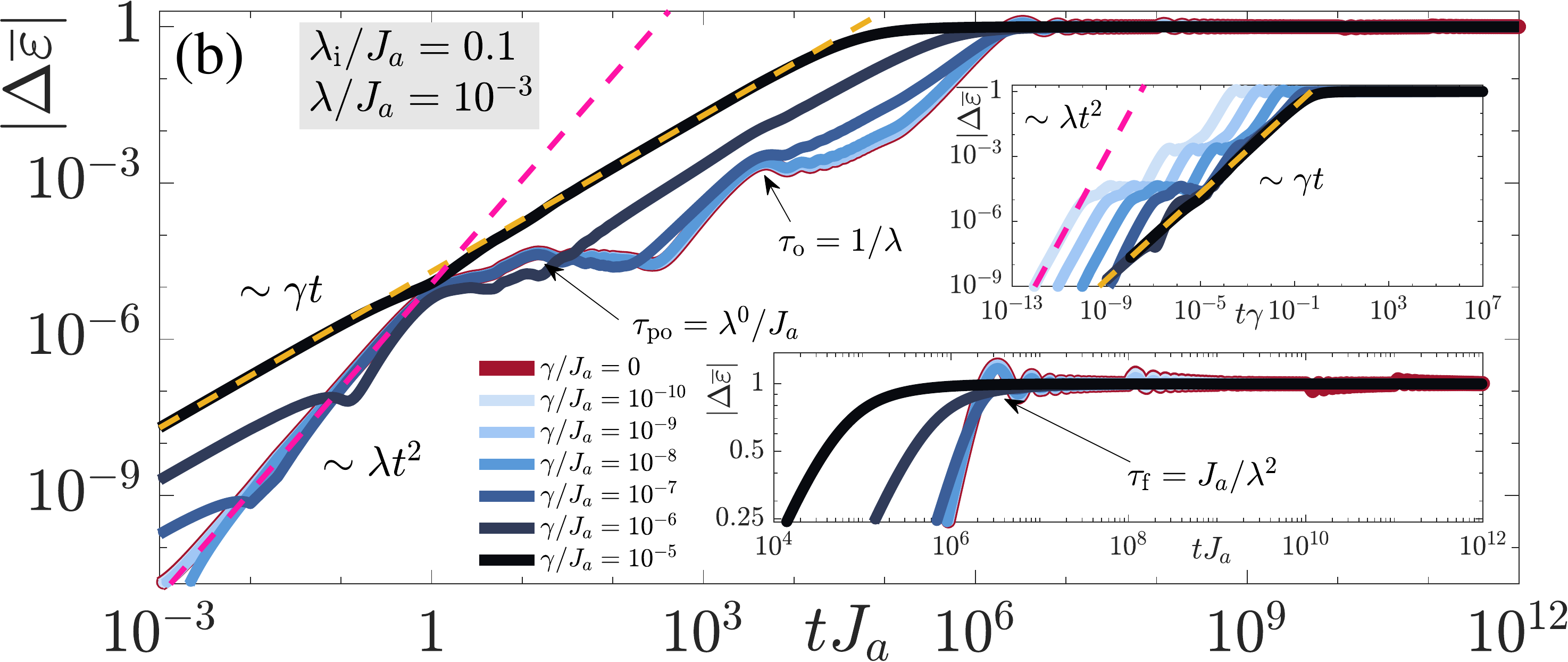}
	\caption{(Color online). Same as Fig.~\ref{fig:lami0} but with $\lambda_\text{i}=0.1$, and thus $\rho_0$ is the ground state of $H_0+\lambda_\text{i}H_1$. The initial state $\rho_0$ is again gauge-noninvariant, but not a ground state of $H_0$. This makes the leading order of coherent contribution $\propto\lambda t^2$, which identically vanishes in the case of $\lambda_\text{i}=0$ of Fig.~\ref{fig:lami0}. This therefore leads to a crossover from a diffusive spread $\sim\gamma t$ to a ballistic scaling $\sim\lambda t^2$ at an earlier timescale $t\propto\gamma/\lambda<\gamma/\lambda^2$. Note here in the purely coherent case (a) how there are three prethermal plateaus instead of just two as in the case of Fig.~\ref{fig:lami0}.
	}
	\label{fig:gaugenoninvariant} 
\end{figure}

Aside from quench dynamics, we also study the effect of decoherence through jump operators $L^\mathrm{m}_j=a_j^\dagger a_j$ and $L^\mathrm{g}_{j,j+1}=\tau^z_{j,j+1}$ on the ground state of an LGT with $N=4$ matter sites and periodic boundary conditions. For this aim, we prepare our system in the ground state of $H_0$, and switch on decoherence at $t=0$ according to the Lindblad master equation~\eqref{eq:EOM} with $\lambda=0$. Such a scenario may occur, e.g., when variational state preparation\cite{Kokail2019} is used to achieve a good approximation to a gauge-invariant initial ground state, which is then stored in the quantum computer and thus subjected to decoherence.
As we do not project the ground state into the target gauge-invariant sector $\mathbf{g}_\text{tar}$, the gauge violation at $t=0$ starts at a finite nonzero value. 
At $t>0$, it grows as $\varepsilon\sim\gamma t$ at short times until the system settles into a maximal-violation steady state at $t\approx1/\gamma$, as shown in Fig.~\ref{fig:Fig3}(a).

It is also instructive to investigate the projectors onto the three relevant gauge-invariant supersectors $\mathcal{P}_{M=0,2,4}$ (the supersector projectors $\mathcal{P}_M$ with odd $M$ are of zero norm in the half-filling global-symmetry sector of the $\mathrm{Z}_2$ LGT). Figure~\ref{fig:Fig3}(b) shows the three projectors that give rise to nonzero expectation values in the case of $N=4$ matter sites. 
Interestingly, the steady-state expectation values are proprotional to the number of gauge-invariant sectors within the associated supersector. Indeed, for $N=4$ matter sites, the supersector $\mathcal{P}_2$ contains six different gauge sectors, while the two supersectors represented by $\mathcal{P}_0$ and $\mathcal{P}_4$ each contains a single gauge-invariant sector. As shown in the inset, $\langle\mathcal{P}_2\rangle\approx0.75$, $\langle\mathcal{P}_0\rangle=\langle\mathcal{P}_4\rangle\approx0.125$, and thus $\langle\mathcal{P}_2\rangle/\langle\mathcal{P}_0\rangle=\langle\mathcal{P}_2\rangle/\langle\mathcal{P}_4\rangle=6$. This indicates that with decoherence the system evolves at late times into a steady state where the gauge violation has spread into an equal distribution over all gauge sectors. This is qualitatively and quantitatively identical to the behavior of these projectors at late times in the quench dynamics with decoherence shown in Fig.~1(c,d) of Ref.~\onlinecite{Halimeh2020f}, despite here the initial state and dynamic process being fundamentally different. Hence, in these results decoherence erases the memory of the initial state. For completeness, we additionally present the associated results for the MQC intensity $I_{\{2,2,2,2\}}$ in Fig.~\ref{fig:staticsObs}(a) and staggered particle density in Fig.~\ref{fig:staticsObs}(b). Starting at its ground-state value, each of these observables shows a decay $\sim(\gamma t)^{-1}$ in its temporal average at a time $t\approx1/\gamma$. The electric field (not shown) behaves also qualitatively the same.

A more interesting scenario is starting in ground state of $H_0+\lambda_\text{i} H_1$, which may become relevant in situations where a pre-quench state-preparation protocol is already subject to gauge-breaking errors. Not only is the initial state here gauge-noninvariant, the presence of $H_1$ allows for a competition between coherent errors and their incoherent counterparts due to decoherence. 

We first consider the case when $\lambda_\text{i}=0$, i.e., we have managed to prepare the system in the ground state of the ideal gauge theory without any coherent errors, but we shall assume that upon quenching, unitary errors $\lambda H_1$ will be present. This scenario may appear when the preparation follows one protocol, e.g., a variational principle,\cite{Kokail2019} while the quench dynamics is studied with another, e.g., an analog quantum-simulation scheme. The ensuing dynamics of the gauge-violation change is shown in Fig.~\ref{fig:lami0}(a) for the case without decoherence but with finite $\lambda>0$. The gauge violation grows $\sim\lambda^2t^2$ at early times, with all lower-order coherent contributions vanishing identically as rigorously explained in Sec.~\ref{sec:TDPT_coherent} through TDPT. The gauge violation exhibits the pre-onset and final plateaus at timescales $\propto1$ and $\propto J_a/\lambda^2$, respectively, but the onset plateau at timescale $\propto1/\lambda$, prominent in the case of gauge-invariant states, is missing here. Maximal violation occurs at the final prethermal timescale $\propto J_a/\lambda^2$. Upon introducing decoherence through jump operators $L^\mathrm{m}_j=a_j^\dagger a_j$ and $L^\mathrm{g}_{j,j+1}=\tau^z_{j,j+1}$ in Fig.~\ref{fig:lami0}(b) at fixed $\lambda$, a crossover emerges at $t\propto\gamma/\lambda^2$ from a diffusive spread $\sim\gamma t$ in the gauge violation to a ballistic spread $\sim\lambda^2t^2$, similarly to the generic behavior we find when starting in a gauge-invariant initial state. Moreover, decoherence compromises the prethermal plateaus, with its effect more apparent on the later plateaus.
	
On the other hand, when $\lambda_\text{i}\neq0$, i.e., when $\rho_0$ is gauge-noninvariant but also not the ground state of $H_0$, a lower-order coherent contribution $\propto\lambda t^2$ that vanishes in the case of $\lambda_\text{i}=0$, now becomes finite, and thus the gauge-violation change scales $\sim\lambda t^2$ at early times in the case of no decoherence, as shown in Fig.~\ref{fig:gaugenoninvariant}(a). Interestingly, here we find all three prethermal plateaus: pre-onset at timescale $t\propto1$, onset at $t\propto1/\lambda$, and final at $t\propto J_a/\lambda^2$. By switching on decoherence through jump operators $L^\mathrm{m}_j=a_j^\dagger a_j$ and $L^\mathrm{g}_{j,j+1}=\tau^z_{j,j+1}$ at a fixed values of $\lambda$, we see that a crossover appears where the gauge-violation difference goes from a diffusive scaling $\sim\gamma t$ to a ballistic spread $\sim\lambda t^2$ at the timescale $t\propto\gamma/\lambda$. As expected, decoherence also compromises the prethermal plateaus in this case.

\section{$\mathrm{U}(1)$ quantum link model}\label{sec:U1QLM}

\begin{figure}[htp]
	\centering
	\includegraphics[width=.48\textwidth]{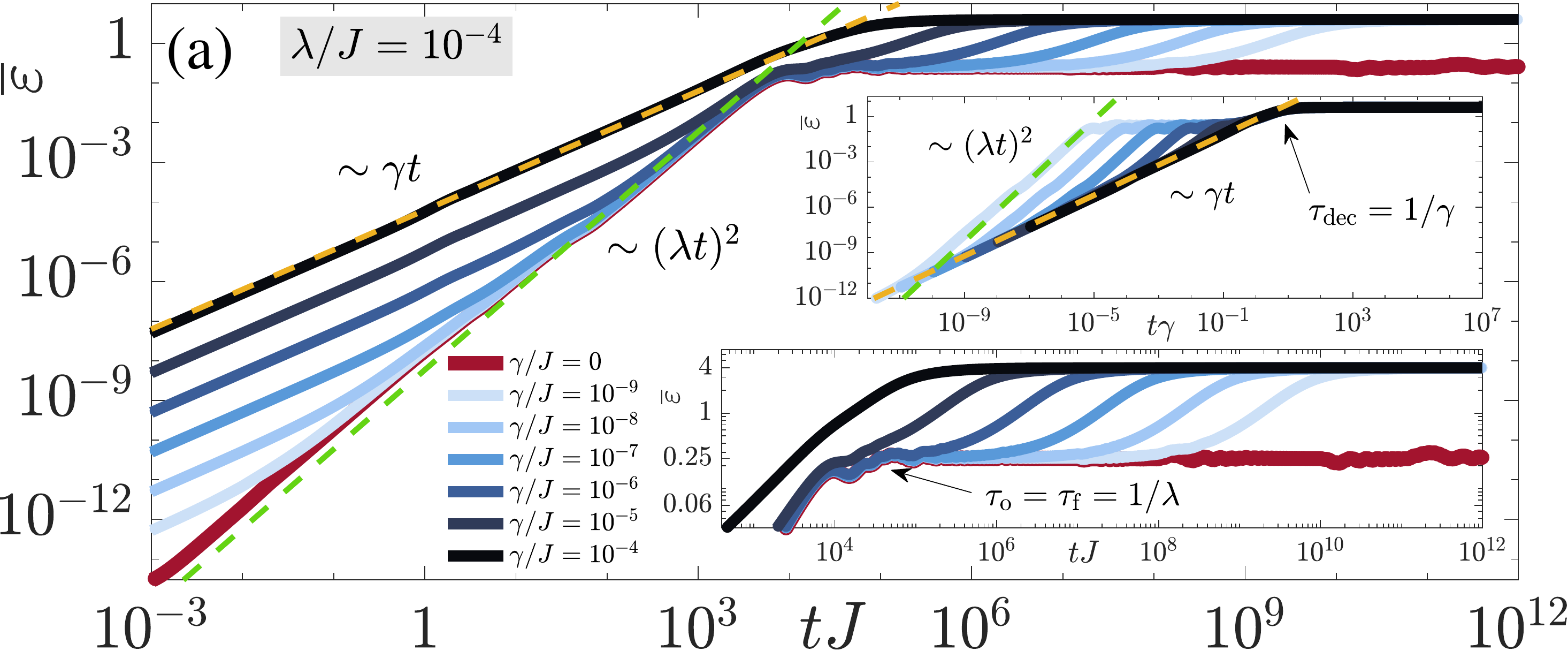}\\
	\includegraphics[width=.48\textwidth]{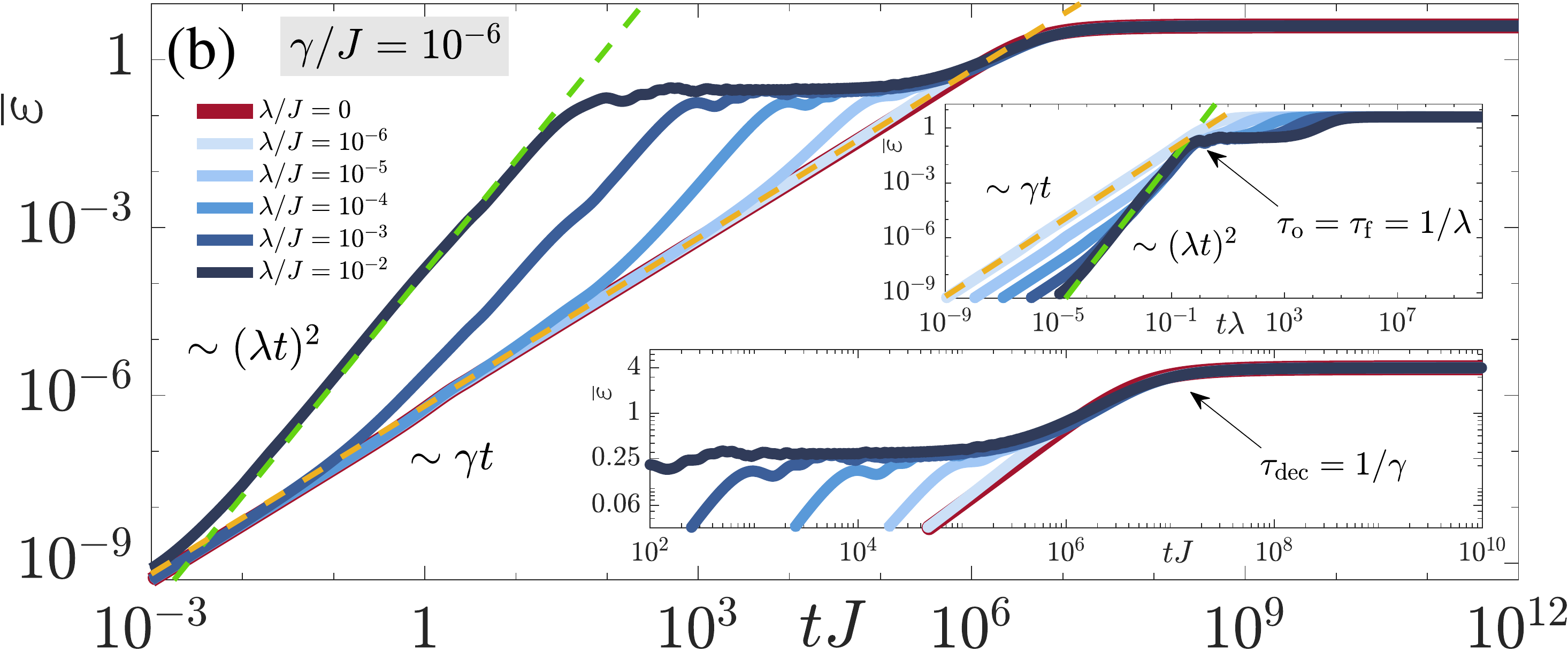}
	\caption{(Color online). A system with $N=2$ matter sites and periodic boundary conditions, prepared in a gauge-invariant initial state with zero bosons and a N\'eel configuration of the electric field is quenched with $H_0+\lambda H_1$  in the presence of decoherence through jump operators $L^\mathrm{m}_j=\sigma_j^z$ and $L^\mathrm{g}_{j,j+1}=\tau^-_{j,j+1}$ both at environment-coupling strength $\gamma$. Decoherence compromises the prethermal plateaus, and in fact leads the violation to a maximal value not attained by the coherent errors is alone. This is due to the absence of resonance between a few of the gauge-invariant sectors in the $\mathrm{U}(1)$ QLM, which does not have an analog in the $\mathrm{Z}_2$ LGT. This maximal value depends on the fixed points of the Liouvillian superoperator. The short-time dynamics is qualitatively the same as in the generic case of the $\mathrm{Z}_2$ LGT when the initial state is gauge-invariant: a diffusive-to-ballistic crossover arises at $t\propto\gamma/\lambda^2$ where the gauge violation goes from a diffusive spread $\sim\gamma t$ to a ballistic spread $\propto\lambda^2t^2$.}
	\label{fig:U1QLM} 
\end{figure}

To further demonstrate the generality of our findings, we now study the gauge-violation dynamics in the $\mathrm{U}(1)$ quantum link model (QLM) given by\cite{Wiese_review,Hauke2013,Yang2016,Yang2020}
\begin{align}
H_0=\sum_{j=1}^N\Big[-J\big(\sigma^-_j\tau^+_{j,j+1}\sigma^-_{j+1}+\text{H.c.}\big)+\frac{\mu}{2}\sigma^z_j\Big],
\end{align}
where the Pauli matrix $\sigma^+_j$ is the creation operator of a particle on site $j$, while the Pauli matrix $\tau^+_{j,j+1}$ ($\tau^z_{j,j+1}$) on link $j,j+1$ represents the gauge (electric) field. Here, we consider a lattice of $N=2$ matter sites and periodic boundary conditions. The Gauss's-law generator is
\begin{align}
G_j=\frac{(-1)^j}{2}\big(\tau^z_{j-1,j}+\sigma^z_j+\tau^z_{j,j+1}+1\big).
\end{align}
and has eigenvalues $g_j=-1,0,1,2$. This model has been the subject of recent ultracold-atom experiments.\cite{Mil2020,Yang2020} We calculate the quench dynamics of this model in the presence of decoherence through jump operators $L^\mathrm{m}_j=\sigma_j^z$ and $L^\mathrm{g}_{j,j+1}=\tau^-_{j,j+1}$, both at environment-coupling strength $\gamma$, by solving the Lindblad master equation~\eqref{eq:EOM} with 
\begin{align}
\lambda H_1=\lambda\sum_j\big(\sigma^-_j\sigma^-_{j+1}+\sigma^+_j\sigma^+_{j+1}+\tau^x_{j,j+1}\big),
\end{align}
which describes unassisted matter tunneling and gauge flipping, and we consider the jump operators $L^\mathrm{m}_j=\sigma^z_j$ and $L^\mathrm{g}_{j,j+1}=\tau^+_{j,j+1}$, although we have checked that other choices of the jump operators yield the same qualitative picture. Moreover, the gauge-invariant initial state $\rho_0$ has zero matter particles and a N\'eel configuration of the electric field. In our numerics, we have set $J=1$ and $\mu=0.05$, though we have checked that our conclusions are independent of this particular choice of parameters.

The ensuing time evolution of the gauge violation is shown in Fig.~\ref{fig:U1QLM}. Focusing first on the long-time dynamics, we again see how decoherence compromises the prethermal plateau, but, more dramatically than in the case of the $\mathrm{Z}_2$ LGT, it drives the gauge violation to a larger value. This is because the $\mathrm{U}(1)$ QLM has a larger number of gauge-invariant sectors (due to an eigenvalue $g_j$ having four possible values) some of which do not have resonances through $H_1$ with one another---unlike the $\mathrm{Z}_2$ LGT, here the value of $g_j$ restricts the possible values that $g_{j-1}$ and $g_{j+1}$ can take. In the presence of decoherence, resonances between these gauge-invariant sectors are facilitated, and this is what leads to a larger long-time violation compared to the purely unitary-error case. Furthermore, as can be seen in Fig.~\ref{fig:U1QLM}, the short-time dynamics and the diffusive-to-ballistic crossover are identical to those for the $\mathrm{Z}_2$ LGT and the eBHM, further validating the generality of our results. Indeed, as we rigorously show in Sec.~\ref{sec:TDPT} through TDPT, our conclusions are valid for any many-body system with local of global symmetries.

\section{Conclusions and outlook}\label{sec:conclusion}
In this work, we have considered the dynamics of quantum systems where a symmetry is slightly broken, with special focus on the interplay between coherent and dissipative errors. 
To obtain a general unifying picture, we have performed extensive numerical studies and analytical derivations, considering a broad range of scenarios including dynamics starting from initial product states and from ground states, as well as a variety of models with global symmetries [global $\mathrm{U}(1)$ symmetry in an extended Bose-Hubbard model corresponding to total particle-number conservation] and local symmetries [$\mathrm{Z}_2$ and $\mathrm{U}(1)$ gauge symmetries corresponding to Gauss's law]. 

From these, several generic features emerge. 
First, the symmetry violation---the expectation value of the symmetry generator---generically reveals a short-time crossover from diffusive to ballistic or even hyperballistic mean-square displacement across symmetry sectors. Second, for purely coherent errors interference effects can prevent the symmetry violation to reach its theoretical maximum, even in the long-time limit. Decoherence can lift these interference effects. As a consequence, the dynamics typically is dominated by decoherence at early and late times, while it is dominated by coherent errors in an intermediate time window. Third, the MQC is a powerful tool to reveal this complex interplay by quantifying the coherence between symmetry sectors encapsulated in the quantum state. Counterintuitively, we find situations where the addition of decoherence to coherent errors can increase the MQCs.

Our findings will be highly relevant for quantum simulation experiments on NISQ devices. They illuminate the general behavior with which the dynamics of a quantum many-body system under slight symmetry breaking deteriorates from the ideal model with intact conservation laws. These enable to estimate, e.g., target time scales that experimental technology needs to achieve in order to observe desired phenomena.

\section*{Acknowledgments}
This work is part of and supported by the Interdisciplinary Center Q@TN --- Quantum Science and Technologies at Trento, the DFG Collaborative Research Centre SFB 1225 (ISOQUANT), the Provincia Autonoma di Trento, and the ERC Starting Grant StrEnQTh (Project-ID 804305). 

\appendix
\begin{widetext}
	\section{Time-dependent perturbation theory}\label{sec:TDPT}
	In our results, we have seen that the symmetry violation transitions from diffusive behavior $\varepsilon\sim\gamma t$ at short times to ballistic behavior $\varepsilon\sim\lambda^m t^n$ with integers $m\geq1$ and even $n\geq2$ at intermediate times. To put these numerical results on an analytic footing, we rigorously derive here the corresponding scalings in time-dependent perturbation theory (TDPT).\cite{Altland_book,Halimeh2020a} Let us first rewrite Eqs.~\eqref{eq:EOM_eBHM} and~\eqref{eq:EOM} in the concise form
	\begin{subequations}
		\begin{align}\label{eq:EOMv2}
		\dot{\rho}=&\,(\mathcal{S}+\gamma\mathcal{L})\rho,\\
		\mathcal{S}\rho=&-i[H,\rho],\\
		\mathcal{L}\rho=&\sum_{j=1}^N\Big(L_j\rho L^\dagger_j-\frac{1}{2}\big\{L^\dagger_jL_j,\rho\big\}\Big).
		\end{align}
	\end{subequations}
	where, without loss of generality, we have used the jump operators $L_j$ without distinguishing between those acting on matter or gauge fields as that is inconsequential in the following derivations. Accordingly, the exact solution to Eq.~\eqref{eq:EOMv2} is
	\begin{align}
	\rho(t)=e^{(\mathcal{S}+\gamma\mathcal{L})t}\rho_0.
	\end{align}
	The Taylor expansion of this solution is
	\begin{align}\nonumber
	\rho(t)=&\sum_{n=0}^\infty(\mathcal{S}+\gamma\mathcal{L})^n\frac{t^n}{n!}\rho_0\\\label{eq:Taylor}
	=&\bigg\{1+\sum_{n=1}^\infty\bigg[\mathcal{S}^n+\gamma\sum_{m=0}^{n-1} \mathcal{S}^m\mathcal{L}\mathcal{S}^{n-m-1}+\gamma^2\sum_{m=1}^{n-1}\sum_{k=0}^{n-m-1}\mathcal{S}^{n-m-k-1}\mathcal{L}\mathcal{S}^k\mathcal{L}\mathcal{S}^{m-1}+\mathcal{O}(\gamma^3)\bigg]\frac{t^n}{n!}\bigg\}\rho_0.
	\end{align}
	
	\subsection{Coherent terms}\label{sec:TDPT_coherent}
	It is important to recall here that $\gamma$ is not the only perturbative parameter in this problem, because there are coherent errors at perturbative strength $\lambda$ encapsulated within the unitary processes of $\mathcal{S}$. In the case of $\gamma=0$, one can show that the leading order in gauge violation scales as $\varepsilon\sim(\lambda t)^2$ at short times $t\lesssim1/\lambda$ when starting in a symmetric state or a generic eigenstate of $H_0$, or as $\varepsilon\sim\lambda t^2$ when starting in an unsymmetric state that is also not an eigenstate of $H_0$. This can be seen by considering the first- and second-order (in $\mathcal{S}$) terms of Eq.~\eqref{eq:Taylor} for $\gamma=0$. If we were to write $\mathcal{S}=\mathcal{S}_0+\lambda\mathcal{S}_1$, where $\mathcal{S}_0$ contains all the processes due to $H_0$ while $\mathcal{S}_1$ all those due to $H_1$, the corresponding approximate density matrix would be
	\begin{align}\label{eq:UnitaryContribution}
	\rho(t)\approx\bigg[1+t\mathcal{S}_0+\frac{1}{2}t^2\mathcal{S}_0^2+\lambda t\mathcal{S}_1+\frac{1}{2}\lambda t^2\Big(\mathcal{S}_0\mathcal{S}_1+\mathcal{S}_1\mathcal{S}_0\Big)+\frac{1}{2}\lambda^2t^2\mathcal{S}_1^2\bigg]\rho_0.
	\end{align}
	Employing for convenience the \textit{symmetry-violation operator} $\mathcal{G}$ [see Eqs.~\eqref{eq:viol_eBHM} and~\eqref{eq:Measure}], it is straightforward to derive
	\begin{subequations}\label{eq:here}
		\begin{align}
		\Tr\big\{\mathcal{G}\rho_0\big\}&=0~\text{for all $\rho_0$ in target symmetry sector},\\\label{eq:coherent}
		t\Tr\big\{\mathcal{G}\mathcal{S}_0\rho_0\big\}&=-it\Tr\big\{\big[\mathcal{G},H_0\big]\rho_0\big\}=0,~\forall\rho_0,\\
		t^2\Tr\big\{\mathcal{G}\mathcal{S}_0^2\rho_0\big\}&=-t^2\Tr\big\{\mathcal{G}H_0H_0\rho_0-2\mathcal{G}H_0\rho_0H_0+\mathcal{G}\rho_0H_0H_0\big\}=0,~\forall\rho_0,\\\label{eq:lamt}
		\lambda t\Tr\big\{\mathcal{G}\mathcal{S}_1\rho_0\big\}&=-i\lambda t\Tr\big\{\big[\mathcal{G},H_1\big]\rho_0\big\} ,\\
		\lambda t^2\Tr\big\{\mathcal{G}\mathcal{S}_0\mathcal{S}_1\rho_0\big\}&=-\lambda t^2\Tr\big\{\mathcal{G}H_0\big[H_1,\rho_0\big]+\big[\rho_0,H_1\big]H_0\mathcal{G}\big\}=0,~\forall\rho_0,\\\label{eq:coherent_last}
		\lambda t^2\Tr\big\{\mathcal{G}\mathcal{S}_1\mathcal{S}_0\rho_0\big\}&=-\lambda t^2\Tr\big\{\mathcal{G}H_1\big[H_0,\rho_0\big]+\big[\rho_0,H_0\big]H_1\mathcal{G}\big\}.
		\end{align}
	\end{subequations}
In the Eqs.~\eqref{eq:here} that identically vanish for any initial state $\rho_0$, we have used the cyclic property of the trace, and the fact that $[H_0,\mathcal{G}]=0$. The remaining term in Eq.~\eqref{eq:UnitaryContribution} makes the following contribution to the symmetry violation:
\begin{align}\label{eq:coherent_remaining}
\frac{1}{2}\lambda^2t^2\Tr\big\{\mathcal{G}\mathcal{S}_1^2\rho_0\big\}=-\frac{1}{2}\lambda^2t^2\Tr\big\{\mathcal{G}H_1H_1\rho_0-2\mathcal{G}H_1\rho_0H_1+\mathcal{G}\rho_0H_1H_1\big\}\neq0\,\,\,\text{in general}.
\end{align}

In case $\rho_0$ is symmetric, then Eqs.~\eqref{eq:lamt} and~\eqref{eq:coherent_last} vanish since $\mathcal{G}\rho_0=\rho_0\mathcal{G}=\text{const.}\times\rho_0$, which means that the leading nonvanishing coherent contribution to the gauge violation is, at lowest order, $\propto\lambda^2 t^2$ due to Eq.~\eqref{eq:coherent_remaining}. This is shown numerically in Fig.~1(b) of Ref.~\onlinecite{Halimeh2020f}, and also in, e.g., Figs.~\ref{fig:gammaFixed},~\ref{fig:DiffDiss}, and~\ref{fig:DiffInitState} of this work for the $\mathrm{Z}_2$ LGT, Fig.~\ref{fig:U1QLM} for the $\mathrm{U}(1)$ QLM, and in Figs.~\ref{fig:eBHM_closed}(a,c) and~\ref{fig:eBHM_OneBodyError_open} for the eBHM. Note how in Figs.~\ref{fig:eBHM_closed}(b) and~\ref{fig:eBHM_TwoBodyError_open} the leading coherent contribution to the gauge violation is $\propto\lambda^2 t^4$ rather than $\propto\lambda^2 t^2$. This is due to a special interplay between the staggered symmetric initial state shown in Fig.~\ref{fig:eBHM_InitialStates}(a) and the two-body error term of Eq.~\eqref{eq:H1TwoBody}. In that case, since $\rho_0$ is symmetric, Eq.~\eqref{eq:coherent_remaining} reduces to $\lambda^2t^2\Tr\{\mathcal{G}H_1\rho_0H_1\}$. Moreover, $H_1\rho_0H_1$ vanishes as $H_1$ removes or adds bosons on two adjacent sites simultaneously, while $\rho_0$ has a staggered boson occupation. The next leading term in the Taylor expansion of the density matrix, $\propto\lambda^2 t^4 H_1H_0\rho_0H_0H_1$, does not vanish since $H_0$ actuates tunneling, leading to $H_0\rho_0H_0$ containing filled adjacent sites.

In case $\rho_0$ is unsymmetric, then theoretically Eq.~\eqref{eq:lamt} does not vanish in general. However, our numerical investigations suggest that this happens in one artificial and two pathological cases. The first pathological case is when $\rho_0$ is an eigenstate of a highly nonlocal Hamiltonian that does not commute with $\mathcal{G}$. The second is when $H_1$ is itself highly nonlocal and does not commute with either $\rho_0$ or $\mathcal{G}$. Usually in modern experimental setups the coherent gauge-breaking errors in the preparation of $\rho_0$ and during the dynamics are local. Our numerical checks reveal that in such realistic situations the coherent contribution $\propto\lambda t$ usually vanishes. Indeed, when the unitary errors in the preparation of $\rho_0$ and during the dynamics are due to the same term $H_1$, then Eq.~\eqref{eq:lamt} vanishes identically. Let $\rho_0$ be the ground state of $H_0+\lambda_\text{i}H_1$ with $\lambda_\text{i}\neq0$, while the unitary quench dynamics is actuated by $H_0+\lambda H_1$. One can derive

\begin{align}
\lambda t\Tr\big\{\big[\mathcal{G},H_1\big]\rho_0\big\}=\lambda t\Tr\big\{\mathcal{G}\big[H_1,\rho_0\big]\big\}=\frac{\lambda}{\lambda_\text{i}} t\Tr\big\{\mathcal{G}\big[H_0+\lambda_\text{i}H_1,\rho_0\big]\big\}=0,
\end{align}
where we have invoked Eq.~\eqref{eq:coherent}. This is why the leading coherent order in such a case is $\propto\lambda t^2$, as can be seen in Fig.~\ref{fig:eBHM_unsymmetric} for the eBHM and Fig.~\ref{fig:gaugenoninvariant} for the $\mathrm{Z}_2$ LGT. Thus, when $\rho_0$ is unsymmetric due to generic experimental preparation errors $H_1$, the nonvanishing leading coherent contribution to the symmetry violation  is $\propto\lambda t^2$ due to Eq.~\eqref{eq:coherent_remaining}. 

To illustrate the artificial case, let us assume we are in a common eigenbasis of $H_0$ and $\mathcal{G}$. Then if in this eigenbasis there are eigenstates that are degenerate with respect to $H_0$ but not to $\mathcal{G}$, then an arbitrary superposition of these eigenstates, itself still an eigenstate of $H_0$, is no longer an eigenstate of $\mathcal{G}$, thereby possibly leading to a finite contribution $\propto\lambda t$ due to Eq.~\eqref{eq:lamt}. Realistically, such a state seems difficult to prepare in experiment. Moreover, when $\rho_0$ is an eigenstate of $H_0$, no matter how artificially engineered, Eq.~\eqref{eq:coherent_remaining} completely vanishes by noting the cyclic property of the trace and that $[H_0,\rho_0]=0$. Therefore, when the initial state is a generic eigenstate of $H_0$, the leading coherent contribution to the gauge violation is $\propto\lambda^2t^2$, i.e., the same as that for a symmetric initial state. This is indeed what we see in Fig.~\ref{fig:lami0}.

In order to explain the short-time scalings of the supersector projectors in the ballistic regime in Figs.~\ref{fig:gammaFixed},~\ref{fig:DiffDiss}, and~\ref{fig:DiffInitState} of this work and Fig.~1(c,d) of Ref.~\onlinecite{Halimeh2020f} in the case of the $\mathrm{Z}_2$ LGT, let us focus on the case of a symmetric initial state, as is used in these aforementioned results. If we replace $\mathcal{G}$ with $\mathcal{P}_2$ in Eqs.~\eqref{eq:coherent}--\eqref{eq:coherent_remaining}, we will arrive at the same conclusions since $[\mathcal{G},\mathcal{P}_2]=[H_0,\mathcal{P}_2]=0$, and because $\mathcal{P}_2$ includes violations, with respect to the target gauge sector $\mathbf{g}_\text{tar}=\mathbf{0}$, due to first-order processes in $H_1$, meaning that $\lambda^2t^2\Tr\big\{\mathcal{P}_2H_1\rho_0H_1\big\}\neq0$ in general, which explains its ballistic behavior $\sim\lambda^2t^2$ at early times. 

For the same reasons, it is also found that Eqs.~\eqref{eq:coherent}--\eqref{eq:coherent_last} will all hold if we replace $\mathcal{G}$ with $\mathcal{P}_4$, but differently, we would get $\lambda^2t^2\Tr\big\{\mathcal{P}_4H_1\rho_0H_1\big\}=0$, because $H_1\rho_0H_1$ does not involve any second-order processes in $H_1$, and the (super)sector $M=4$ includes only such processes (this order could become nonzero for a less localized $H_1$ that breaks four local symmetry generators simultaneously). For similar reasons, coherent terms $\propto\lambda^3$ cannot involve second-order processes in $H_1$ on both sides of $\rho_0$ at the same time (the associated terms would be $H_1H_1H_1\rho_0$, $H_1H_1\rho_0H_1$, and their Hermitian conjugates), and their contribution to $\langle\mathcal{P}_4\rangle$ vanishes. Thus, for the error terms considered in this work the nonvanishing leading-order coherent contribution for $\mathcal{P}_4$ at early times is $\lambda^4t^4\Tr\big\{\mathcal{P}_4H_1H_1\rho_0H_1H_1\big\}/24\neq0\,\,\,\text{in general}$, which explains its scaling $\sim\lambda^4t^4$ in the ballistic regime.

\subsection{Leading incoherent terms}\label{sec:TDPT_leadingIncoherent}
In the presence of decoherence, the dominant correction to the unitary part of the density matrix at leading order of $\gamma$ is $\gamma t\mathcal{L}\rho_0$, as can be seen for $n=1$ (and, consequently, $m=0$) in Eq.~\eqref{eq:Taylor}. The contribution to the gauge violation $\varepsilon$ [see Eq.~\eqref{eq:Measure}] at short times due to the term $\gamma t\mathcal{L}\rho_0$ is
\begin{align}\label{eq:gammat}
	\gamma t\Tr\big\{\mathcal{G}\mathcal{L}\rho_0\big\}=\gamma t\sum_j\Tr\bigg\{\mathcal{G} L_j\rho_0 L_j^\dagger-\frac{1}{2}\mathcal{G} L_j^\dagger L_j\rho_0-\frac{1}{2}\mathcal{G}\rho_0L_j^\dagger L_j\bigg\}\neq0\,\,\,\text{in general},
\end{align}
regardless of whether $\rho_0$ is symmetric or not. Indeed, the term $\gamma t\mathcal{L}\rho_0$ involves only incoherent gauge-breaking processes, and its contribution will lead to diffusive scaling $\sim\gamma t$ in the gauge violation. This diffusive behavior will dominate over the leading-order coherent gauge breaking $\propto\lambda^2t^2$ for evolution times $t\lesssim\gamma/\lambda^2$ in case of a symmetric initial state or a generic eigenstate of $H_0$, as shown in Fig.~1(b) of Ref.~\onlinecite{Halimeh2020f} and Figs.~\ref{fig:gammaFixed},~\ref{fig:DiffDiss},~\ref{fig:DiffInitState},~\ref{fig:relativeGamma},~\ref{fig:ParticleLoss}, and~\ref{fig:lami0} in case of the $\mathrm{Z}_2$ LGT, Fig.~\ref{fig:U1QLM} for the $\mathrm{U}(1)$ QLM, and Fig.~\ref{fig:eBHM_OneBodyError_open} for the eBHM. In the case of a generic (i.e., not pathological or artificial; see discussion in Sec.~\ref{sec:TDPT_coherent}) unsymmetric initial state, the diffusive scaling $\sim\gamma t$ will dominate over the leading-order coherent contribution $\propto\lambda t^2$ for $t\lesssim\gamma/\lambda$, as seen in Fig.~\ref{fig:eBHM_unsymmetric} for the eBHM and Fig.~\ref{fig:gaugenoninvariant} for the $\mathrm{Z}_2$ LGT. However, this crossover time can be made even earlier, such as in the case of hyperballistic scaling $\propto\lambda^2t^4$ shown in Fig.~\ref{fig:eBHM_TwoBodyError_open}, where it becomes $t\propto(\gamma/\lambda^2)^{\frac{1}{3}}$. By replacing $\mathcal{G}$ with $\mathcal{P}_2$ in Eq.~\eqref{eq:gammat}, it is straightforward to see that the same dominant incoherent contribution to $\mathcal{P}_2$ is also $\propto\gamma t$, and it will therefore show diffusive scaling $\sim\gamma t$ at early times. Replacing $\mathcal{G}$ with the supersector projector $\mathcal{P}_4$ in Eq.~\eqref{eq:gammat}, we see that when $\rho_0$ is symmetric or a generic eigenstate of $H_0$, $\mathcal{P}_4$ cannot scale $\sim\gamma t$ in the diffusive regime. We will come back to this later. However, if $\rho_0$ is unsymmetric (but not a generic eigenstate of $H_0$ or another observable commuting with $\mathcal{G}$) with finite support in the supersector $M=2$, then $\mathcal{P}_4$ can show diffusive behavior $\sim\gamma t$ at early times.
	
We can also explain from Eq.~\eqref{eq:gammat} why when dissipation and dephasing have different environment-coupling strengths $\gamma_\mathrm{g}$ and $\gamma_\mathrm{m}$, respectively, the gauge violation at short times scales diffusively as $\varepsilon\sim\gamma_\mathrm{g}t$, with dephasing having no effect as shown in Fig.~\ref{fig:relativeGamma}. In the case of dephasing, $L_j^\mathrm{m}=a_j^\dagger a_j$ does not create a violation in the system because $\big[\mathcal{G},a_j^\dagger a_j\big]=\big[G_l,a_j^\dagger a_j\big]=0,\,\forall j,l$, and so the associated contribution $\gamma_\mathrm{m} t\sum_j\Tr\big\{\mathcal{G} a_j^\dagger a_j\rho_0 a_ja_j^\dagger\big\}=0$, where we recall that $\rho_0$ lies in the target symmetry sector $\mathbf{g}_\text{tar}=\mathbf{0}$. As such, the only remaining contribution from Eq.~\eqref{eq:gammat} is $\gamma_\mathrm{g} t\sum_j\Tr\big\{\mathcal{G} L^\mathrm{g}_{j,j+1}\rho_0 L^{\mathrm{g}\dagger}_{j,j+1}\big\}$, which does not vanish in general, because $\exists l:\,\big[G_l,L^\mathrm{g}_{j,j+1}\big]\neq0$ in the case of dissipation. The supersector projector $\langle\mathcal{P}_2\rangle$ exhibits the same behavior as the violation.
	
In contrast, the term $\gamma t\mathcal{L}\rho_0$ in the Taylor expansion of Eq.~\eqref{eq:EOM} leads to a vanishing contribution to $\langle\mathcal{P}_4\rangle$ because $\Tr\big\{\mathcal{P}_4L_j^{\mathrm{m}(\mathrm{g})}\rho_0L_j^{\mathrm{m}(\mathrm{g})\dagger}\big\}=0$ as the jump operators drive the system into the supersector $M=2$. As discussed in Sec.~\ref{sec:TDPT_coherent}, this is similar to the reason why $\Tr\big\{\mathcal{P}_4H_1\rho_0H_1\big\}=0$, as it involves first-order processes in $H_1$, which drive the system into the supersector $M=2$, and thus $\langle\mathcal{P}_4\rangle$ cannot show scaling $\sim\lambda^2t^2$ either.
	
It is important to note here that leading-order (in $\gamma$) corrections to the density matrix in Eq.~\eqref{eq:Taylor} also include terms that are quadratic in time, and involve the term $\gamma t^2(\mathcal{L}\mathcal{S}+\mathcal{S}\mathcal{L})\rho_0$, which can be rewritten as
\begin{align}\label{eq:Term0}
	\gamma t^2(\mathcal{L}\mathcal{S}+\mathcal{S}\mathcal{L})\rho_0=\big[\gamma t^2(\mathcal{L}\mathcal{S}_0+\mathcal{S}_0\mathcal{L})+\gamma\lambda t^2(\mathcal{L}\mathcal{S}_1+\mathcal{S}_1\mathcal{L})\big]\rho_0.
\end{align}
The purely incoherent gauge-breaking term $\gamma t^2(\mathcal{L}\mathcal{S}_0+\mathcal{S}_0\mathcal{L})\rho_0$ in Eq.~\eqref{eq:Term0} on the gauge violation $\varepsilon$ will always be dominated by that $\propto\gamma t$, which so far we have seen is a generic feature of the symmetry violation in presence of decoherence. As such, generically we will not see scaling $\sim\gamma t^2$ in the gauge violation.

\subsection{Mixed terms}
We now shift our attention to the component of Eq.~\eqref{eq:Term0} where unitary and incoherent gauge-breaking processes mix: $\gamma\lambda t^2(\mathcal{L}\mathcal{S}_1+\mathcal{S}_1\mathcal{L})\rho_0$. For this contribution to dominate over that $\propto\gamma t$, we must have $t>1/\lambda$, which is anyway beyond the perturbative regime as then prethermalization kicks in. Automatically this means that in generic situations the contribution $\propto\gamma\lambda t^2$ will not dominate over any of the (hyper)ballistic scalings that dominate over $\gamma t$ after the crossover time in the gauge violation.
	
\subsection{Higher-order incoherent terms}
We have thus far understood why the dominant scaling in the gauge violation is $\varepsilon\sim\gamma t$ at times $t\lesssim\gamma/\lambda^2$, beyond which we see in the ED results that the gauge violation scales as $\varepsilon\sim\lambda^2t^2$ up until evolution times $t\approx1/\lambda$ for a symmetric initial state. We also understand why the latter scale is not compromised by terms $\propto\gamma t^2$ or $\propto\gamma \lambda t^2$. One remaining term that merits investigation is $\gamma^2t^2\mathcal{L}^2\rho_0$ as it pertains to the supersector projector $\mathcal{P}_4$, to which its contribution is
\begin{align}\nonumber
	&\frac{1}{2}\gamma^2 t^2\Tr\big\{\mathcal{P}_4\mathcal{L}^2\rho_0\big\}\\\nonumber
	=&\,\frac{1}{2}\gamma^2 t^2\sum_j\Tr\bigg\{\mathcal{P}_4\mathcal{L}\bigg( L_j\rho_0 L_j^\dagger-\frac{1}{2} L_j^\dagger L_j\rho_0-\frac{1}{2}\rho_0 L_j^\dagger L_j\bigg)\bigg\}\\\nonumber
	=&\,\frac{1}{4}\gamma^2 t^2\sum_{j,l}\Tr\Big\{\mathcal{P}_4\Big( 2L_l L_j\rho_0 L_j^\dagger L_l^\dagger-L_l L_j^\dagger L_j\rho_0L_l^\dagger-L_l\rho_0  L_j^\dagger L_jL_l^\dagger-L_l^\dagger L_l L_j\rho_0 L_j^\dagger- L_j\rho_0 L_j^\dagger L_l^\dagger L_l+L_j^\dagger L_j\rho_0L_l^\dagger L_l  \Big)\Big\},
\end{align}
which is nonzero in general. This becomes the dominant incoherent contribution to $\mathcal{P}_4$, because it involves terms with second-order violating processes (quadratic in jump operators) on each side of $\rho_0$. This explains exactly why $\langle\mathcal{P}_4\rangle$ exhibits the scaling $\langle\mathcal{P}_4\rangle\sim\gamma^2t^2$ at times $t\lesssim\gamma/\lambda^2$ before scaling as $\langle\mathcal{P}_4\rangle\sim\lambda^4t^4$ for $t\gtrsim\gamma/\lambda^2$ for $\gamma\lesssim\lambda$, as shown in Fig.~1(d) of Ref.~\onlinecite{Halimeh2020f} and Figs.~\ref{fig:gammaFixed},~\ref{fig:DiffDiss}, and~\ref{fig:DiffInitState} of this work for the $\mathrm{Z}_2$ LGT.

Finally, we note that even though the contribution $\propto\gamma^2t^2$ to the gauge violation does not necessarily vanish, it will always be dominated by that $\propto\gamma t$ in generic situations.

\section{Numerics specifics}\label{sec:NumSpec}
In this Appendix we provide details pertaining to our numerical implementation. First we provide the exact expressions we used for the coefficients $c_n$ in $H_1$ of Eq.~\eqref{eq:H1}, which read
\begin{subequations}
	\begin{align}
	c_1=&\,\sum_{k>0}\frac{\mathcal{N}(\chi)}{k}\big[\mathcal{J}_{-k-1}(\chi)\mathcal{J}_{-k-2}(\chi)+\mathcal{J}_k(\chi)\mathcal{J}_{k+1}(\chi)-\mathcal{J}_{k-1}(\chi)\mathcal{J}_{k-2}(\chi)-\mathcal{J}_{-k}(\chi)\mathcal{J}_{-k+1}(\chi)\big],\\
	c_2=&\,\sum_{k>0}\frac{\mathcal{N}(\chi)}{k}\big[\mathcal{J}_{-k+1}(\chi)\mathcal{J}_{k-2}(\chi)+\mathcal{J}_{-k}(\chi)\mathcal{J}_{k-1}(\chi)-\mathcal{J}_{k+1}(\chi)\mathcal{J}_{-k-2}(\chi)-\mathcal{J}_{k}(\chi)\mathcal{J}_{-k-1}(\chi)\big],\\
	c_3=&\,\sum_{k>0}\frac{\mathcal{N}(\chi)}{k}\big[\mathcal{J}_{k-1}^2(\chi)+\mathcal{J}_{k-2}^2(\chi)-\mathcal{J}_{-k-1}^2(\chi)-\mathcal{J}_{-k-2}^2(\chi)\big],\\
	c_4=&\,\sum_{k>0}\frac{\mathcal{N}(\chi)}{k}\big[\mathcal{J}_{-k+1}^2(\chi)+\mathcal{J}_{-k}^2(\chi)-\mathcal{J}_{k+1}^2(\chi)-\mathcal{J}_{k}^2(\chi)\big],
	\end{align}
\end{subequations}
\end{widetext}
where $\mathcal{J}_q(\chi)$ is the $q^\text{th}$-order Bessel function of the first kind and we use a normalization factor $\mathcal{N}(\chi)$ to ensure that $\sum_{n=1}^4c_n=1$, in order to make the strength of the unitary gauge-breaking term independent of $\chi$, and solely dependent on $\lambda$.

\begin{figure}[htp]
	\centering
	\includegraphics[width=.48\textwidth]{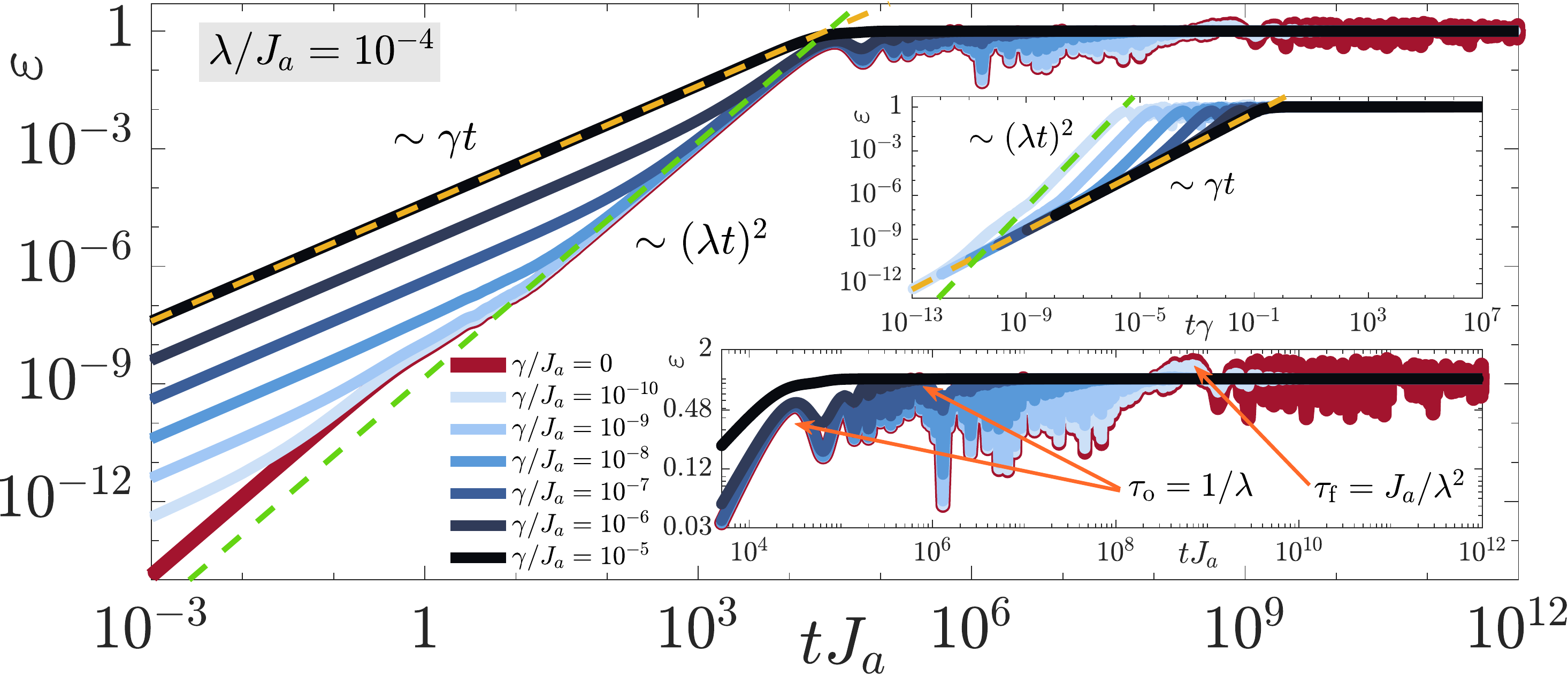}
	\caption{(Color online). Same as Fig.~1 of Ref.~\onlinecite{Halimeh2020f} but showing the raw signal rather than its temporal average. The qualitative picture is unaltered. The fluctuations in the raw signal are completely suppressed at times $t\gtrsim1/\gamma$.}
	\label{fig:raw} 
\end{figure}

\subsection{Implementational details}
All results presented in this work have been calculated using our in-house exact diagonalization toolkit LaGaDyn, \cite{LaGaDyn} where we have also performed benchmarks with QuTiP.\cite{Johansson2012,Johansson2013}

We solve Eq.~\eqref{eq:EOM} by rewriting it as
\begin{align}
\dot{\tilde{\rho}}=\mathcal{M}\tilde{\rho},
\end{align}
where we have matricized the equation of motion such that $\tilde{\rho}$ is a \textit{flattened} version of the density matrix $\rho$ where the latter's columns in left-to-right order are stacked on top of each other, and $\mathcal{M}$ is the corresponding \textit{Lindbladian superoperator} encapsulating all relevant unitary and incoherent processes from Eq.~\eqref{eq:EOM} in the resulting $\mathcal{H}\otimes\mathcal{H}$ space, where $\mathcal{H}$ is the Hilbert space of our model of interest.\cite{Havel2003} For the time evolution, we solve
\begin{align}\label{eq:EOM_numerics}
\tilde{\rho}(t)=e^{\mathcal{M}t}\tilde{\rho}_0,
\end{align}
using our exact exponentiation routine. We opt to use the latter instead of common methods based on iterative solutions of ordinary differential equations in order to be able to reliably achieve the large evolution times displayed in our results.

As mentioned in the main text, for numerical feasibility we have chosen in the main results to turn on only dephasing rather than dissipation on the matter fields. This leads to retaining a global $\mathrm{U}(1)$ symmetry in the form of particle-number conservation, because both $H_0$ and $H_1$ also conserve it. Taking into account the hard-core boson constraint and staying in the half-filling sector allows for reducing the number of states in the system's Hilbert space $\mathcal{H}$ from $2^{2N}$ to $\ell=2^N\binom{N}{N/2}$. However, our dynamics of Eq.~\eqref{eq:EOM_numerics} is not solved in $\mathcal{H}$ (whose size is $\ell\times\ell$), but rather in $\mathcal{H}\otimes\mathcal{H}$ (whose size is $\ell^2\times\ell^2$).

\subsection{Running average versus raw signal}
In this work and Ref.~\onlinecite{Halimeh2020f}, all results have shown the running temporal average $\overline A(t)=\int_0^t ds\, A(s)/t$ of all quantities $A(t)$. This is done in order to suppress fluctuations in the raw signal $A(t)$, which are prominent especially in the case of purely unitary dynamics. For comparison, we provide in Fig.~\ref{fig:raw} the raw data of the gauge-violation dynamics from Fig.~1(b) of Ref.~\onlinecite{Halimeh2020f} in the $\mathrm{Z}_2$ LGT with coherent and incoherent gauge-invariance breaking. As can be seen, the qualitative picture remains exactly the same. We see that decoherence itself behaves similar to the running average in that it fully suppresses fluctuations for times $t\gtrsim1/\gamma$. That may be seen as another instance of the effect of diffusion, in contrast to oscillations that are typical of coherent wave-like dynamics.

\bibliography{SymmViolBiblio}

\begin{thebibliography}{116}
\expandafter\ifx\csname natexlab\endcsname\relax\def\natexlab#1{#1}\fi
\expandafter\ifx\csname bibnamefont\endcsname\relax
  \def\bibnamefont#1{#1}\fi
\expandafter\ifx\csname bibfnamefont\endcsname\relax
  \def\bibfnamefont#1{#1}\fi
\expandafter\ifx\csname citenamefont\endcsname\relax
  \def\citenamefont#1{#1}\fi
\expandafter\ifx\csname url\endcsname\relax
  \def\url#1{\texttt{#1}}\fi
\expandafter\ifx\csname urlprefix\endcsname\relax\def\urlprefix{URL }\fi
\providecommand{\bibinfo}[2]{#2}
\providecommand{\eprint}[2][]{\url{#2}}

\bibitem[{\citenamefont{Sachdev}(2001)}]{Sachdev_book}
\bibinfo{author}{\bibfnamefont{S.}~\bibnamefont{Sachdev}},
  \emph{\bibinfo{title}{Quantum Phase Transitions}}
  (\bibinfo{publisher}{Cambridge University Press}, \bibinfo{year}{2001}), ISBN
  \bibinfo{isbn}{9780521004541},
  \urlprefix\url{https://books.google.de/books?id=Ih\_E05N5TZQC}.

\bibitem[{\citenamefont{Balents}(2010)}]{Balents_NatureReview}
\bibinfo{author}{\bibfnamefont{L.}~\bibnamefont{Balents}},
  \bibinfo{journal}{Nature} \textbf{\bibinfo{volume}{464}},
  \bibinfo{pages}{199} (\bibinfo{year}{2010}),
  \urlprefix\url{https://doi.org/10.1038/nature08917}.

\bibitem[{\citenamefont{Savary and Balents}(2016)}]{Savary2016}
\bibinfo{author}{\bibfnamefont{L.}~\bibnamefont{Savary}} \bibnamefont{and}
  \bibinfo{author}{\bibfnamefont{L.}~\bibnamefont{Balents}},
  \bibinfo{journal}{Reports on Progress in Physics}
  \textbf{\bibinfo{volume}{80}}, \bibinfo{pages}{016502}
  (\bibinfo{year}{2016}),
  \urlprefix\url{https://doi.org/10.1088%2F0034-4885%2F80%2F1%2F016502}.

\bibitem[{\citenamefont{Brading and Castellani}(2003)}]{Brading_book}
\bibinfo{author}{\bibfnamefont{K.}~\bibnamefont{Brading}} \bibnamefont{and}
  \bibinfo{author}{\bibfnamefont{E.}~\bibnamefont{Castellani}},
  \emph{\bibinfo{title}{Symmetries in Physics: Philosophical Reflections}}
  (\bibinfo{publisher}{Cambridge University Press}, \bibinfo{year}{2003}), ISBN
  \bibinfo{isbn}{9781139442022},
  \urlprefix\url{https://books.google.de/books?id=fB38Q9D-WdwC}.

\bibitem[{\citenamefont{Zee}(2003)}]{Zee_book}
\bibinfo{author}{\bibfnamefont{A.}~\bibnamefont{Zee}},
  \emph{\bibinfo{title}{Quantum Field Theory in a Nutshell}}
  (\bibinfo{publisher}{Princeton University Press}, \bibinfo{year}{2003}), ISBN
  \bibinfo{isbn}{9780691010199},
  \urlprefix\url{https://books.google.de/books?id=85G9QgAACAAJ}.

\bibitem[{\citenamefont{Tasaki}(1998)}]{Tasaki1998}
\bibinfo{author}{\bibfnamefont{H.}~\bibnamefont{Tasaki}},
  \bibinfo{journal}{Journal of Physics: Condensed Matter}
  \textbf{\bibinfo{volume}{10}}, \bibinfo{pages}{4353} (\bibinfo{year}{1998}),
  \urlprefix\url{https://doi.org/10.1088%2F0953-8984%2F10%2F20%2F004}.

\bibitem[{\citenamefont{Gersch and Knollman}(1963)}]{Gersch1963}
\bibinfo{author}{\bibfnamefont{H.~A.} \bibnamefont{Gersch}} \bibnamefont{and}
  \bibinfo{author}{\bibfnamefont{G.~C.} \bibnamefont{Knollman}},
  \bibinfo{journal}{Phys. Rev.} \textbf{\bibinfo{volume}{129}},
  \bibinfo{pages}{959} (\bibinfo{year}{1963}),
  \urlprefix\url{https://link.aps.org/doi/10.1103/PhysRev.129.959}.

\bibitem[{\citenamefont{Peskin and Schroeder}(2018)}]{Peskin2016}
\bibinfo{author}{\bibfnamefont{M.~E.} \bibnamefont{Peskin}} \bibnamefont{and}
  \bibinfo{author}{\bibfnamefont{D.~V.} \bibnamefont{Schroeder}},
  \emph{\bibinfo{title}{An Introduction To Quantum Field Theory}}
  (\bibinfo{publisher}{CRC Press}, \bibinfo{year}{2018}), ISBN
  \bibinfo{isbn}{9780429983184},
  \urlprefix\url{https://books.google.de/books?id=9EpnDwAAQBAJ}.

\bibitem[{\citenamefont{Jackson and Okun}(2001)}]{Jackson_review}
\bibinfo{author}{\bibfnamefont{J.~D.} \bibnamefont{Jackson}} \bibnamefont{and}
  \bibinfo{author}{\bibfnamefont{L.~B.} \bibnamefont{Okun}},
  \bibinfo{journal}{Rev. Mod. Phys.} \textbf{\bibinfo{volume}{73}},
  \bibinfo{pages}{663} (\bibinfo{year}{2001}),
  \urlprefix\url{https://link.aps.org/doi/10.1103/RevModPhys.73.663}.

\bibitem[{\citenamefont{Cheng and Li}(1984)}]{Cheng_book}
\bibinfo{author}{\bibfnamefont{T.}~\bibnamefont{Cheng}} \bibnamefont{and}
  \bibinfo{author}{\bibfnamefont{L.}~\bibnamefont{Li}},
  \emph{\bibinfo{title}{Gauge Theory of Elementary Particle Physics}}, Oxford
  science publications (\bibinfo{publisher}{Clarendon Press},
  \bibinfo{year}{1984}), ISBN \bibinfo{isbn}{9780198519614},
  \urlprefix\url{https://books.google.it/books?id=lk8GEzVNb10C}.

\bibitem[{\citenamefont{Mallayya et~al.}(2019)\citenamefont{Mallayya, Rigol,
  and De~Roeck}}]{Mallayya2019}
\bibinfo{author}{\bibfnamefont{K.}~\bibnamefont{Mallayya}},
  \bibinfo{author}{\bibfnamefont{M.}~\bibnamefont{Rigol}}, \bibnamefont{and}
  \bibinfo{author}{\bibfnamefont{W.}~\bibnamefont{De~Roeck}},
  \bibinfo{journal}{Phys. Rev. X} \textbf{\bibinfo{volume}{9}},
  \bibinfo{pages}{021027} (\bibinfo{year}{2019}),
  \urlprefix\url{https://link.aps.org/doi/10.1103/PhysRevX.9.021027}.

\bibitem[{\citenamefont{Ray et~al.}(2020)\citenamefont{Ray, Anglin, and
  Vardi}}]{Ray2020}
\bibinfo{author}{\bibfnamefont{S.}~\bibnamefont{Ray}},
  \bibinfo{author}{\bibfnamefont{J.~R.} \bibnamefont{Anglin}},
  \bibnamefont{and} \bibinfo{author}{\bibfnamefont{A.}~\bibnamefont{Vardi}}
  (\bibinfo{year}{2020}), \eprint{2009.01491},
  \urlprefix\url{https://arxiv.org/abs/2009.01491}.

\bibitem[{\citenamefont{Halimeh and Hauke}(2020{\natexlab{a}})}]{Halimeh2020b}
\bibinfo{author}{\bibfnamefont{J.~C.} \bibnamefont{Halimeh}} \bibnamefont{and}
  \bibinfo{author}{\bibfnamefont{P.}~\bibnamefont{Hauke}}
  (\bibinfo{year}{2020}{\natexlab{a}}), \eprint{2004.07248},
  \urlprefix\url{https://arxiv.org/abs/2004.07248}.

\bibitem[{\citenamefont{Halimeh and Hauke}(2020{\natexlab{b}})}]{Halimeh2020c}
\bibinfo{author}{\bibfnamefont{J.~C.} \bibnamefont{Halimeh}} \bibnamefont{and}
  \bibinfo{author}{\bibfnamefont{P.}~\bibnamefont{Hauke}}
  (\bibinfo{year}{2020}{\natexlab{b}}), \eprint{2004.07254},
  \urlprefix\url{https://arxiv.org/abs/2004.07254}.

\bibitem[{\citenamefont{Moeckel and Kehrein}(2008)}]{Moeckel2008}
\bibinfo{author}{\bibfnamefont{M.}~\bibnamefont{Moeckel}} \bibnamefont{and}
  \bibinfo{author}{\bibfnamefont{S.}~\bibnamefont{Kehrein}},
  \bibinfo{journal}{Phys. Rev. Lett.} \textbf{\bibinfo{volume}{100}},
  \bibinfo{pages}{175702} (\bibinfo{year}{2008}),
  \urlprefix\url{https://link.aps.org/doi/10.1103/PhysRevLett.100.175702}.

\bibitem[{\citenamefont{Eckstein et~al.}(2009)\citenamefont{Eckstein, Kollar,
  and Werner}}]{Eckstein2009}
\bibinfo{author}{\bibfnamefont{M.}~\bibnamefont{Eckstein}},
  \bibinfo{author}{\bibfnamefont{M.}~\bibnamefont{Kollar}}, \bibnamefont{and}
  \bibinfo{author}{\bibfnamefont{P.}~\bibnamefont{Werner}},
  \bibinfo{journal}{Phys. Rev. Lett.} \textbf{\bibinfo{volume}{103}},
  \bibinfo{pages}{056403} (\bibinfo{year}{2009}),
  \urlprefix\url{https://link.aps.org/doi/10.1103/PhysRevLett.103.056403}.

\bibitem[{\citenamefont{Kollar et~al.}(2011)\citenamefont{Kollar, Wolf, and
  Eckstein}}]{Kollar2011}
\bibinfo{author}{\bibfnamefont{M.}~\bibnamefont{Kollar}},
  \bibinfo{author}{\bibfnamefont{F.~A.} \bibnamefont{Wolf}}, \bibnamefont{and}
  \bibinfo{author}{\bibfnamefont{M.}~\bibnamefont{Eckstein}},
  \bibinfo{journal}{Phys. Rev. B} \textbf{\bibinfo{volume}{84}},
  \bibinfo{pages}{054304} (\bibinfo{year}{2011}),
  \urlprefix\url{https://link.aps.org/doi/10.1103/PhysRevB.84.054304}.

\bibitem[{\citenamefont{Tavora and Mitra}(2013)}]{Tavora2013}
\bibinfo{author}{\bibfnamefont{M.}~\bibnamefont{Tavora}} \bibnamefont{and}
  \bibinfo{author}{\bibfnamefont{A.}~\bibnamefont{Mitra}},
  \bibinfo{journal}{Phys. Rev. B} \textbf{\bibinfo{volume}{88}},
  \bibinfo{pages}{115144} (\bibinfo{year}{2013}),
  \urlprefix\url{https://link.aps.org/doi/10.1103/PhysRevB.88.115144}.

\bibitem[{\citenamefont{Nessi et~al.}(2014)\citenamefont{Nessi, Iucci, and
  Cazalilla}}]{Nessi2014}
\bibinfo{author}{\bibfnamefont{N.}~\bibnamefont{Nessi}},
  \bibinfo{author}{\bibfnamefont{A.}~\bibnamefont{Iucci}}, \bibnamefont{and}
  \bibinfo{author}{\bibfnamefont{M.~A.} \bibnamefont{Cazalilla}},
  \bibinfo{journal}{Phys. Rev. Lett.} \textbf{\bibinfo{volume}{113}},
  \bibinfo{pages}{210402} (\bibinfo{year}{2014}),
  \urlprefix\url{https://link.aps.org/doi/10.1103/PhysRevLett.113.210402}.

\bibitem[{\citenamefont{Essler et~al.}(2014)\citenamefont{Essler, Kehrein,
  Manmana, and Robinson}}]{Essler2014}
\bibinfo{author}{\bibfnamefont{F.~H.~L.} \bibnamefont{Essler}},
  \bibinfo{author}{\bibfnamefont{S.}~\bibnamefont{Kehrein}},
  \bibinfo{author}{\bibfnamefont{S.~R.} \bibnamefont{Manmana}},
  \bibnamefont{and} \bibinfo{author}{\bibfnamefont{N.~J.}
  \bibnamefont{Robinson}}, \bibinfo{journal}{Phys. Rev. B}
  \textbf{\bibinfo{volume}{89}}, \bibinfo{pages}{165104}
  (\bibinfo{year}{2014}),
  \urlprefix\url{https://link.aps.org/doi/10.1103/PhysRevB.89.165104}.

\bibitem[{\citenamefont{Bertini et~al.}(2016)\citenamefont{Bertini, Essler,
  Groha, and Robinson}}]{Bertini2016}
\bibinfo{author}{\bibfnamefont{B.}~\bibnamefont{Bertini}},
  \bibinfo{author}{\bibfnamefont{F.~H.~L.} \bibnamefont{Essler}},
  \bibinfo{author}{\bibfnamefont{S.}~\bibnamefont{Groha}}, \bibnamefont{and}
  \bibinfo{author}{\bibfnamefont{N.~J.} \bibnamefont{Robinson}},
  \bibinfo{journal}{Phys. Rev. B} \textbf{\bibinfo{volume}{94}},
  \bibinfo{pages}{245117} (\bibinfo{year}{2016}),
  \urlprefix\url{https://link.aps.org/doi/10.1103/PhysRevB.94.245117}.

\bibitem[{\citenamefont{Fagotti and Collura}(2015)}]{Fagotti2015}
\bibinfo{author}{\bibfnamefont{M.}~\bibnamefont{Fagotti}} \bibnamefont{and}
  \bibinfo{author}{\bibfnamefont{M.}~\bibnamefont{Collura}}
  (\bibinfo{year}{2015}), \eprint{1507.02678}.

\bibitem[{\citenamefont{Reimann and Dabelow}(2019)}]{Reimann2019}
\bibinfo{author}{\bibfnamefont{P.}~\bibnamefont{Reimann}} \bibnamefont{and}
  \bibinfo{author}{\bibfnamefont{L.}~\bibnamefont{Dabelow}},
  \bibinfo{journal}{Phys. Rev. Lett.} \textbf{\bibinfo{volume}{122}},
  \bibinfo{pages}{080603} (\bibinfo{year}{2019}),
  \urlprefix\url{https://link.aps.org/doi/10.1103/PhysRevLett.122.080603}.

\bibitem[{\citenamefont{Rigol et~al.}(2007)\citenamefont{Rigol, Dunjko,
  Yurovsky, and Olshanii}}]{Rigol2007}
\bibinfo{author}{\bibfnamefont{M.}~\bibnamefont{Rigol}},
  \bibinfo{author}{\bibfnamefont{V.}~\bibnamefont{Dunjko}},
  \bibinfo{author}{\bibfnamefont{V.}~\bibnamefont{Yurovsky}}, \bibnamefont{and}
  \bibinfo{author}{\bibfnamefont{M.}~\bibnamefont{Olshanii}},
  \bibinfo{journal}{Phys. Rev. Lett.} \textbf{\bibinfo{volume}{98}},
  \bibinfo{pages}{050405} (\bibinfo{year}{2007}),
  \urlprefix\url{https://link.aps.org/doi/10.1103/PhysRevLett.98.050405}.

\bibitem[{\citenamefont{Rigol}(2009{\natexlab{a}})}]{Rigol2009a}
\bibinfo{author}{\bibfnamefont{M.}~\bibnamefont{Rigol}},
  \bibinfo{journal}{Phys. Rev. Lett.} \textbf{\bibinfo{volume}{103}},
  \bibinfo{pages}{100403} (\bibinfo{year}{2009}{\natexlab{a}}),
  \urlprefix\url{https://link.aps.org/doi/10.1103/PhysRevLett.103.100403}.

\bibitem[{\citenamefont{Rigol}(2009{\natexlab{b}})}]{Rigol2009b}
\bibinfo{author}{\bibfnamefont{M.}~\bibnamefont{Rigol}},
  \bibinfo{journal}{Phys. Rev. A} \textbf{\bibinfo{volume}{80}},
  \bibinfo{pages}{053607} (\bibinfo{year}{2009}{\natexlab{b}}),
  \urlprefix\url{https://link.aps.org/doi/10.1103/PhysRevA.80.053607}.

\bibitem[{\citenamefont{Vidmar and Rigol}(2016)}]{Vidmar2016}
\bibinfo{author}{\bibfnamefont{L.}~\bibnamefont{Vidmar}} \bibnamefont{and}
  \bibinfo{author}{\bibfnamefont{M.}~\bibnamefont{Rigol}},
  \bibinfo{journal}{Journal of Statistical Mechanics: Theory and Experiment}
  \textbf{\bibinfo{volume}{2016}}, \bibinfo{pages}{064007}
  (\bibinfo{year}{2016}),
  \urlprefix\url{https://doi.org/10.1088%2F1742-5468%2F2016%2F06%2F064007}.

\bibitem[{\citenamefont{Essler and Fagotti}(2016)}]{Essler2016}
\bibinfo{author}{\bibfnamefont{F.~H.~L.} \bibnamefont{Essler}}
  \bibnamefont{and} \bibinfo{author}{\bibfnamefont{M.}~\bibnamefont{Fagotti}},
  \bibinfo{journal}{Journal of Statistical Mechanics: Theory and Experiment}
  \textbf{\bibinfo{volume}{2016}}, \bibinfo{pages}{064002}
  (\bibinfo{year}{2016}),
  \urlprefix\url{https://doi.org/10.1088%2F1742-5468%2F2016%2F06%2F064002}.

\bibitem[{\citenamefont{Cazalilla and Chung}(2016)}]{Cazalilla2016}
\bibinfo{author}{\bibfnamefont{M.~A.} \bibnamefont{Cazalilla}}
  \bibnamefont{and} \bibinfo{author}{\bibfnamefont{M.-C.} \bibnamefont{Chung}},
  \bibinfo{journal}{Journal of Statistical Mechanics: Theory and Experiment}
  \textbf{\bibinfo{volume}{2016}}, \bibinfo{pages}{064004}
  (\bibinfo{year}{2016}),
  \urlprefix\url{https://doi.org/10.1088%2F1742-5468%2F2016%2F06%2F064004}.

\bibitem[{\citenamefont{Caux}(2016)}]{Caux2016}
\bibinfo{author}{\bibfnamefont{J.-S.} \bibnamefont{Caux}},
  \bibinfo{journal}{Journal of Statistical Mechanics: Theory and Experiment}
  \textbf{\bibinfo{volume}{2016}}, \bibinfo{pages}{064006}
  (\bibinfo{year}{2016}),
  \urlprefix\url{https://doi.org/10.1088%2F1742-5468%2F2016%2F06%2F064006}.

\bibitem[{\citenamefont{Mallayya and Rigol}(2018)}]{Mallayya2018}
\bibinfo{author}{\bibfnamefont{K.}~\bibnamefont{Mallayya}} \bibnamefont{and}
  \bibinfo{author}{\bibfnamefont{M.}~\bibnamefont{Rigol}},
  \bibinfo{journal}{Phys. Rev. Lett.} \textbf{\bibinfo{volume}{120}},
  \bibinfo{pages}{070603} (\bibinfo{year}{2018}),
  \urlprefix\url{https://link.aps.org/doi/10.1103/PhysRevLett.120.070603}.

\bibitem[{\citenamefont{Stark and Kollar}(2013)}]{Stark2013}
\bibinfo{author}{\bibfnamefont{M.}~\bibnamefont{Stark}} \bibnamefont{and}
  \bibinfo{author}{\bibfnamefont{M.}~\bibnamefont{Kollar}}
  (\bibinfo{year}{2013}), \eprint{1308.1610}.

\bibitem[{\citenamefont{D'Alessio et~al.}(2016)\citenamefont{D'Alessio, Kafri,
  Polkovnikov, and Rigol}}]{DAlessio_review}
\bibinfo{author}{\bibfnamefont{L.}~\bibnamefont{D'Alessio}},
  \bibinfo{author}{\bibfnamefont{Y.}~\bibnamefont{Kafri}},
  \bibinfo{author}{\bibfnamefont{A.}~\bibnamefont{Polkovnikov}},
  \bibnamefont{and} \bibinfo{author}{\bibfnamefont{M.}~\bibnamefont{Rigol}},
  \bibinfo{journal}{Advances in Physics} \textbf{\bibinfo{volume}{65}},
  \bibinfo{pages}{239} (\bibinfo{year}{2016}),
  \eprint{https://doi.org/10.1080/00018732.2016.1198134},
  \urlprefix\url{https://doi.org/10.1080/00018732.2016.1198134}.

\bibitem[{\citenamefont{Lazarides et~al.}(2014)\citenamefont{Lazarides, Das,
  and Moessner}}]{Lazarides2014}
\bibinfo{author}{\bibfnamefont{A.}~\bibnamefont{Lazarides}},
  \bibinfo{author}{\bibfnamefont{A.}~\bibnamefont{Das}}, \bibnamefont{and}
  \bibinfo{author}{\bibfnamefont{R.}~\bibnamefont{Moessner}},
  \bibinfo{journal}{Phys. Rev. Lett.} \textbf{\bibinfo{volume}{112}},
  \bibinfo{pages}{150401} (\bibinfo{year}{2014}),
  \urlprefix\url{https://link.aps.org/doi/10.1103/PhysRevLett.112.150401}.

\bibitem[{\citenamefont{Canovi et~al.}(2016)\citenamefont{Canovi, Kollar, and
  Eckstein}}]{Canovi2016}
\bibinfo{author}{\bibfnamefont{E.}~\bibnamefont{Canovi}},
  \bibinfo{author}{\bibfnamefont{M.}~\bibnamefont{Kollar}}, \bibnamefont{and}
  \bibinfo{author}{\bibfnamefont{M.}~\bibnamefont{Eckstein}},
  \bibinfo{journal}{Phys. Rev. E} \textbf{\bibinfo{volume}{93}},
  \bibinfo{pages}{012130} (\bibinfo{year}{2016}),
  \urlprefix\url{https://link.aps.org/doi/10.1103/PhysRevE.93.012130}.

\bibitem[{\citenamefont{Zeh}(1970)}]{Zeh1970}
\bibinfo{author}{\bibfnamefont{H.~D.} \bibnamefont{Zeh}},
  \bibinfo{journal}{Foundations of Physics} \textbf{\bibinfo{volume}{1}},
  \bibinfo{pages}{69} (\bibinfo{year}{1970}),
  \urlprefix\url{https://doi.org/10.1007/BF00708656}.

\bibitem[{\citenamefont{Schlosshauer}(2005)}]{Schlosshauer2005}
\bibinfo{author}{\bibfnamefont{M.}~\bibnamefont{Schlosshauer}},
  \bibinfo{journal}{Rev. Mod. Phys.} \textbf{\bibinfo{volume}{76}},
  \bibinfo{pages}{1267} (\bibinfo{year}{2005}),
  \urlprefix\url{https://link.aps.org/doi/10.1103/RevModPhys.76.1267}.

\bibitem[{\citenamefont{Yoshihara et~al.}(2006)\citenamefont{Yoshihara,
  Harrabi, Niskanen, Nakamura, and Tsai}}]{Yoshihara2006}
\bibinfo{author}{\bibfnamefont{F.}~\bibnamefont{Yoshihara}},
  \bibinfo{author}{\bibfnamefont{K.}~\bibnamefont{Harrabi}},
  \bibinfo{author}{\bibfnamefont{A.~O.} \bibnamefont{Niskanen}},
  \bibinfo{author}{\bibfnamefont{Y.}~\bibnamefont{Nakamura}}, \bibnamefont{and}
  \bibinfo{author}{\bibfnamefont{J.~S.} \bibnamefont{Tsai}},
  \bibinfo{journal}{Phys. Rev. Lett.} \textbf{\bibinfo{volume}{97}},
  \bibinfo{pages}{167001} (\bibinfo{year}{2006}),
  \urlprefix\url{https://link.aps.org/doi/10.1103/PhysRevLett.97.167001}.

\bibitem[{\citenamefont{Kakuyanagi et~al.}(2007)\citenamefont{Kakuyanagi, Meno,
  Saito, Nakano, Semba, Takayanagi, Deppe, and Shnirman}}]{Kakuyanagi2007}
\bibinfo{author}{\bibfnamefont{K.}~\bibnamefont{Kakuyanagi}},
  \bibinfo{author}{\bibfnamefont{T.}~\bibnamefont{Meno}},
  \bibinfo{author}{\bibfnamefont{S.}~\bibnamefont{Saito}},
  \bibinfo{author}{\bibfnamefont{H.}~\bibnamefont{Nakano}},
  \bibinfo{author}{\bibfnamefont{K.}~\bibnamefont{Semba}},
  \bibinfo{author}{\bibfnamefont{H.}~\bibnamefont{Takayanagi}},
  \bibinfo{author}{\bibfnamefont{F.}~\bibnamefont{Deppe}}, \bibnamefont{and}
  \bibinfo{author}{\bibfnamefont{A.}~\bibnamefont{Shnirman}},
  \bibinfo{journal}{Phys. Rev. Lett.} \textbf{\bibinfo{volume}{98}},
  \bibinfo{pages}{047004} (\bibinfo{year}{2007}),
  \urlprefix\url{https://link.aps.org/doi/10.1103/PhysRevLett.98.047004}.

\bibitem[{\citenamefont{Bialczak et~al.}(2007)\citenamefont{Bialczak,
  McDermott, Ansmann, Hofheinz, Katz, Lucero, Neeley, O'Connell, Wang, Cleland
  et~al.}}]{Bialczak2007}
\bibinfo{author}{\bibfnamefont{R.~C.} \bibnamefont{Bialczak}},
  \bibinfo{author}{\bibfnamefont{R.}~\bibnamefont{McDermott}},
  \bibinfo{author}{\bibfnamefont{M.}~\bibnamefont{Ansmann}},
  \bibinfo{author}{\bibfnamefont{M.}~\bibnamefont{Hofheinz}},
  \bibinfo{author}{\bibfnamefont{N.}~\bibnamefont{Katz}},
  \bibinfo{author}{\bibfnamefont{E.}~\bibnamefont{Lucero}},
  \bibinfo{author}{\bibfnamefont{M.}~\bibnamefont{Neeley}},
  \bibinfo{author}{\bibfnamefont{A.~D.} \bibnamefont{O'Connell}},
  \bibinfo{author}{\bibfnamefont{H.}~\bibnamefont{Wang}},
  \bibinfo{author}{\bibfnamefont{A.~N.} \bibnamefont{Cleland}},
  \bibnamefont{et~al.}, \bibinfo{journal}{Phys. Rev. Lett.}
  \textbf{\bibinfo{volume}{99}}, \bibinfo{pages}{187006}
  (\bibinfo{year}{2007}),
  \urlprefix\url{https://link.aps.org/doi/10.1103/PhysRevLett.99.187006}.

\bibitem[{\citenamefont{Bylander et~al.}(2011)\citenamefont{Bylander,
  Gustavsson, Yan, Yoshihara, Harrabi, Fitch, Cory, Nakamura, Tsai, and
  Oliver}}]{Bylander2011}
\bibinfo{author}{\bibfnamefont{J.}~\bibnamefont{Bylander}},
  \bibinfo{author}{\bibfnamefont{S.}~\bibnamefont{Gustavsson}},
  \bibinfo{author}{\bibfnamefont{F.}~\bibnamefont{Yan}},
  \bibinfo{author}{\bibfnamefont{F.}~\bibnamefont{Yoshihara}},
  \bibinfo{author}{\bibfnamefont{K.}~\bibnamefont{Harrabi}},
  \bibinfo{author}{\bibfnamefont{G.}~\bibnamefont{Fitch}},
  \bibinfo{author}{\bibfnamefont{D.~G.} \bibnamefont{Cory}},
  \bibinfo{author}{\bibfnamefont{Y.}~\bibnamefont{Nakamura}},
  \bibinfo{author}{\bibfnamefont{J.-S.} \bibnamefont{Tsai}}, \bibnamefont{and}
  \bibinfo{author}{\bibfnamefont{W.~D.} \bibnamefont{Oliver}},
  \bibinfo{journal}{Nature Physics} \textbf{\bibinfo{volume}{7}},
  \bibinfo{pages}{565} (\bibinfo{year}{2011}),
  \urlprefix\url{https://doi.org/10.1038/nphys1994}.

\bibitem[{\citenamefont{Wang et~al.}(2015)\citenamefont{Wang, Shi, Hu, Han, Yu,
  and Wu}}]{Wang2015}
\bibinfo{author}{\bibfnamefont{H.}~\bibnamefont{Wang}},
  \bibinfo{author}{\bibfnamefont{C.}~\bibnamefont{Shi}},
  \bibinfo{author}{\bibfnamefont{J.}~\bibnamefont{Hu}},
  \bibinfo{author}{\bibfnamefont{S.}~\bibnamefont{Han}},
  \bibinfo{author}{\bibfnamefont{C.~C.} \bibnamefont{Yu}}, \bibnamefont{and}
  \bibinfo{author}{\bibfnamefont{R.~Q.} \bibnamefont{Wu}},
  \bibinfo{journal}{Phys. Rev. Lett.} \textbf{\bibinfo{volume}{115}},
  \bibinfo{pages}{077002} (\bibinfo{year}{2015}),
  \urlprefix\url{https://link.aps.org/doi/10.1103/PhysRevLett.115.077002}.

\bibitem[{\citenamefont{Kumar et~al.}(2016)\citenamefont{Kumar, Sendelbach,
  Beck, Freeland, Wang, Wang, Yu, Wu, Pappas, and McDermott}}]{Kumar2016}
\bibinfo{author}{\bibfnamefont{P.}~\bibnamefont{Kumar}},
  \bibinfo{author}{\bibfnamefont{S.}~\bibnamefont{Sendelbach}},
  \bibinfo{author}{\bibfnamefont{M.~A.} \bibnamefont{Beck}},
  \bibinfo{author}{\bibfnamefont{J.~W.} \bibnamefont{Freeland}},
  \bibinfo{author}{\bibfnamefont{Z.}~\bibnamefont{Wang}},
  \bibinfo{author}{\bibfnamefont{H.}~\bibnamefont{Wang}},
  \bibinfo{author}{\bibfnamefont{C.~C.} \bibnamefont{Yu}},
  \bibinfo{author}{\bibfnamefont{R.~Q.} \bibnamefont{Wu}},
  \bibinfo{author}{\bibfnamefont{D.~P.} \bibnamefont{Pappas}},
  \bibnamefont{and}
  \bibinfo{author}{\bibfnamefont{R.}~\bibnamefont{McDermott}},
  \bibinfo{journal}{Phys. Rev. Applied} \textbf{\bibinfo{volume}{6}},
  \bibinfo{pages}{041001} (\bibinfo{year}{2016}),
  \urlprefix\url{https://link.aps.org/doi/10.1103/PhysRevApplied.6.041001}.

\bibitem[{\citenamefont{Day et~al.}(2003)\citenamefont{Day, LeDuc, Mazin,
  Vayonakis, and Zmuidzinas}}]{Day2003}
\bibinfo{author}{\bibfnamefont{P.~K.} \bibnamefont{Day}},
  \bibinfo{author}{\bibfnamefont{H.~G.} \bibnamefont{LeDuc}},
  \bibinfo{author}{\bibfnamefont{B.~A.} \bibnamefont{Mazin}},
  \bibinfo{author}{\bibfnamefont{A.}~\bibnamefont{Vayonakis}},
  \bibnamefont{and}
  \bibinfo{author}{\bibfnamefont{J.}~\bibnamefont{Zmuidzinas}},
  \bibinfo{journal}{Nature} \textbf{\bibinfo{volume}{425}},
  \bibinfo{pages}{817} (\bibinfo{year}{2003}),
  \urlprefix\url{https://doi.org/10.1038/nature02037}.

\bibitem[{\citenamefont{Regal et~al.}(2008)\citenamefont{Regal, Teufel, and
  Lehnert}}]{Regal2008}
\bibinfo{author}{\bibfnamefont{C.~A.} \bibnamefont{Regal}},
  \bibinfo{author}{\bibfnamefont{J.~D.} \bibnamefont{Teufel}},
  \bibnamefont{and} \bibinfo{author}{\bibfnamefont{K.~W.}
  \bibnamefont{Lehnert}}, \bibinfo{journal}{Nature Physics}
  \textbf{\bibinfo{volume}{4}}, \bibinfo{pages}{555} (\bibinfo{year}{2008}),
  \urlprefix\url{https://doi.org/10.1038/nphys974}.

\bibitem[{\citenamefont{Poulin}(2010)}]{Poulin2010}
\bibinfo{author}{\bibfnamefont{D.}~\bibnamefont{Poulin}},
  \bibinfo{journal}{Phys. Rev. Lett.} \textbf{\bibinfo{volume}{104}},
  \bibinfo{pages}{190401} (\bibinfo{year}{2010}),
  \urlprefix\url{https://link.aps.org/doi/10.1103/PhysRevLett.104.190401}.

\bibitem[{\citenamefont{Marino and Silva}(2012)}]{Marino2012}
\bibinfo{author}{\bibfnamefont{J.}~\bibnamefont{Marino}} \bibnamefont{and}
  \bibinfo{author}{\bibfnamefont{A.}~\bibnamefont{Silva}},
  \bibinfo{journal}{Phys. Rev. B} \textbf{\bibinfo{volume}{86}},
  \bibinfo{pages}{060408} (\bibinfo{year}{2012}),
  \urlprefix\url{https://link.aps.org/doi/10.1103/PhysRevB.86.060408}.

\bibitem[{\citenamefont{Descamps}(2013)}]{Descamps2013}
\bibinfo{author}{\bibfnamefont{B.}~\bibnamefont{Descamps}},
  \bibinfo{journal}{Journal of Mathematical Physics}
  \textbf{\bibinfo{volume}{54}}, \bibinfo{pages}{092202}
  (\bibinfo{year}{2013}), \eprint{https://doi.org/10.1063/1.4820785},
  \urlprefix\url{https://doi.org/10.1063/1.4820785}.

\bibitem[{\citenamefont{Bernier et~al.}(2013)\citenamefont{Bernier, Barmettler,
  Poletti, and Kollath}}]{Bernier2013}
\bibinfo{author}{\bibfnamefont{J.-S.} \bibnamefont{Bernier}},
  \bibinfo{author}{\bibfnamefont{P.}~\bibnamefont{Barmettler}},
  \bibinfo{author}{\bibfnamefont{D.}~\bibnamefont{Poletti}}, \bibnamefont{and}
  \bibinfo{author}{\bibfnamefont{C.}~\bibnamefont{Kollath}},
  \bibinfo{journal}{Phys. Rev. A} \textbf{\bibinfo{volume}{87}},
  \bibinfo{pages}{063608} (\bibinfo{year}{2013}),
  \urlprefix\url{https://link.aps.org/doi/10.1103/PhysRevA.87.063608}.

\bibitem[{\citenamefont{Halimeh et~al.}(2018)\citenamefont{Halimeh, Punk, and
  Piazza}}]{Halimeh2018}
\bibinfo{author}{\bibfnamefont{J.~C.} \bibnamefont{Halimeh}},
  \bibinfo{author}{\bibfnamefont{M.}~\bibnamefont{Punk}}, \bibnamefont{and}
  \bibinfo{author}{\bibfnamefont{F.}~\bibnamefont{Piazza}},
  \bibinfo{journal}{Phys. Rev. B} \textbf{\bibinfo{volume}{98}},
  \bibinfo{pages}{045111} (\bibinfo{year}{2018}),
  \urlprefix\url{https://link.aps.org/doi/10.1103/PhysRevB.98.045111}.

\bibitem[{\citenamefont{Bernier et~al.}(2018)\citenamefont{Bernier, Tan,
  Bonnes, Guo, Poletti, and Kollath}}]{Bernier2018}
\bibinfo{author}{\bibfnamefont{J.-S.} \bibnamefont{Bernier}},
  \bibinfo{author}{\bibfnamefont{R.}~\bibnamefont{Tan}},
  \bibinfo{author}{\bibfnamefont{L.}~\bibnamefont{Bonnes}},
  \bibinfo{author}{\bibfnamefont{C.}~\bibnamefont{Guo}},
  \bibinfo{author}{\bibfnamefont{D.}~\bibnamefont{Poletti}}, \bibnamefont{and}
  \bibinfo{author}{\bibfnamefont{C.}~\bibnamefont{Kollath}},
  \bibinfo{journal}{Phys. Rev. Lett.} \textbf{\bibinfo{volume}{120}},
  \bibinfo{pages}{020401} (\bibinfo{year}{2018}),
  \urlprefix\url{https://link.aps.org/doi/10.1103/PhysRevLett.120.020401}.

\bibitem[{\citenamefont{Maier et~al.}(2019)\citenamefont{Maier, Brydges,
  Jurcevic, Trautmann, Hempel, Lanyon, Hauke, Blatt, and Roos}}]{Maier2019}
\bibinfo{author}{\bibfnamefont{C.}~\bibnamefont{Maier}},
  \bibinfo{author}{\bibfnamefont{T.}~\bibnamefont{Brydges}},
  \bibinfo{author}{\bibfnamefont{P.}~\bibnamefont{Jurcevic}},
  \bibinfo{author}{\bibfnamefont{N.}~\bibnamefont{Trautmann}},
  \bibinfo{author}{\bibfnamefont{C.}~\bibnamefont{Hempel}},
  \bibinfo{author}{\bibfnamefont{B.~P.} \bibnamefont{Lanyon}},
  \bibinfo{author}{\bibfnamefont{P.}~\bibnamefont{Hauke}},
  \bibinfo{author}{\bibfnamefont{R.}~\bibnamefont{Blatt}}, \bibnamefont{and}
  \bibinfo{author}{\bibfnamefont{C.~F.} \bibnamefont{Roos}},
  \bibinfo{journal}{Phys. Rev. Lett.} \textbf{\bibinfo{volume}{122}},
  \bibinfo{pages}{050501} (\bibinfo{year}{2019}),
  \urlprefix\url{https://link.aps.org/doi/10.1103/PhysRevLett.122.050501}.

\bibitem[{\citenamefont{Lenar\ifmmode \check{c}\else
  \v{c}\fi{}i\ifmmode~\check{c}\else \v{c}\fi{}
  et~al.}(2018)\citenamefont{Lenar\ifmmode \check{c}\else
  \v{c}\fi{}i\ifmmode~\check{c}\else \v{c}\fi{}, Lange, and
  Rosch}}]{Lenarcic2018}
\bibinfo{author}{\bibfnamefont{Z.}~\bibnamefont{Lenar\ifmmode \check{c}\else
  \v{c}\fi{}i\ifmmode~\check{c}\else \v{c}\fi{}}},
  \bibinfo{author}{\bibfnamefont{F.}~\bibnamefont{Lange}}, \bibnamefont{and}
  \bibinfo{author}{\bibfnamefont{A.}~\bibnamefont{Rosch}},
  \bibinfo{journal}{Phys. Rev. B} \textbf{\bibinfo{volume}{97}},
  \bibinfo{pages}{024302} (\bibinfo{year}{2018}),
  \urlprefix\url{https://link.aps.org/doi/10.1103/PhysRevB.97.024302}.

\bibitem[{\citenamefont{Lange et~al.}(2018)\citenamefont{Lange, Lenar\ifmmode
  \check{c}\else \v{c}\fi{}i\ifmmode~\check{c}\else \v{c}\fi{}, and
  Rosch}}]{Lange2018}
\bibinfo{author}{\bibfnamefont{F.}~\bibnamefont{Lange}},
  \bibinfo{author}{\bibfnamefont{Z.}~\bibnamefont{Lenar\ifmmode \check{c}\else
  \v{c}\fi{}i\ifmmode~\check{c}\else \v{c}\fi{}}}, \bibnamefont{and}
  \bibinfo{author}{\bibfnamefont{A.}~\bibnamefont{Rosch}},
  \bibinfo{journal}{Phys. Rev. B} \textbf{\bibinfo{volume}{97}},
  \bibinfo{pages}{165138} (\bibinfo{year}{2018}),
  \urlprefix\url{https://link.aps.org/doi/10.1103/PhysRevB.97.165138}.

\bibitem[{\citenamefont{Gross and Bloch}(2017)}]{Gross_review}
\bibinfo{author}{\bibfnamefont{C.}~\bibnamefont{Gross}} \bibnamefont{and}
  \bibinfo{author}{\bibfnamefont{I.}~\bibnamefont{Bloch}},
  \bibinfo{journal}{Science} \textbf{\bibinfo{volume}{357}},
  \bibinfo{pages}{995} (\bibinfo{year}{2017}), ISSN \bibinfo{issn}{0036-8075},
  \eprint{https://science.sciencemag.org/content/357/6355/995.full.pdf},
  \urlprefix\url{https://science.sciencemag.org/content/357/6355/995}.

\bibitem[{\citenamefont{Lewenstein et~al.}(2012)\citenamefont{Lewenstein,
  Sanpera, and Ahufinger}}]{Lewenstein_book}
\bibinfo{author}{\bibfnamefont{M.}~\bibnamefont{Lewenstein}},
  \bibinfo{author}{\bibfnamefont{A.}~\bibnamefont{Sanpera}}, \bibnamefont{and}
  \bibinfo{author}{\bibfnamefont{V.}~\bibnamefont{Ahufinger}},
  \emph{\bibinfo{title}{Ultracold Atoms in Optical Lattices: Simulating quantum
  many-body systems}} (\bibinfo{publisher}{OUP Oxford}, \bibinfo{year}{2012}),
  ISBN \bibinfo{isbn}{9780191627439},
  \urlprefix\url{https://books.google.de/books?id=Wpl91RDxV5IC}.

\bibitem[{\citenamefont{Hauke et~al.}(2012)\citenamefont{Hauke, Cucchietti,
  Tagliacozzo, Deutsch, and Lewenstein}}]{Hauke2012}
\bibinfo{author}{\bibfnamefont{P.}~\bibnamefont{Hauke}},
  \bibinfo{author}{\bibfnamefont{F.~M.} \bibnamefont{Cucchietti}},
  \bibinfo{author}{\bibfnamefont{L.}~\bibnamefont{Tagliacozzo}},
  \bibinfo{author}{\bibfnamefont{I.}~\bibnamefont{Deutsch}}, \bibnamefont{and}
  \bibinfo{author}{\bibfnamefont{M.}~\bibnamefont{Lewenstein}},
  \bibinfo{journal}{Reports on Progress in Physics}
  \textbf{\bibinfo{volume}{75}}, \bibinfo{pages}{082401}
  (\bibinfo{year}{2012}).

\bibitem[{\citenamefont{Zhou and Wu}(2006)}]{Zhou2006}
\bibinfo{author}{\bibfnamefont{F.}~\bibnamefont{Zhou}} \bibnamefont{and}
  \bibinfo{author}{\bibfnamefont{C.}~\bibnamefont{Wu}}, \bibinfo{journal}{New
  Journal of Physics} \textbf{\bibinfo{volume}{8}}, \bibinfo{pages}{166}
  (\bibinfo{year}{2006}),
  \urlprefix\url{https://doi.org/10.1088%2F1367-2630%2F8%2F8%2F166}.

\bibitem[{\citenamefont{Maruyama et~al.}(2007)\citenamefont{Maruyama, Koide,
  and Hatsugai}}]{Maruyama2007}
\bibinfo{author}{\bibfnamefont{I.}~\bibnamefont{Maruyama}},
  \bibinfo{author}{\bibfnamefont{T.}~\bibnamefont{Koide}}, \bibnamefont{and}
  \bibinfo{author}{\bibfnamefont{Y.}~\bibnamefont{Hatsugai}},
  \bibinfo{journal}{Phys. Rev. B} \textbf{\bibinfo{volume}{76}},
  \bibinfo{pages}{235105} (\bibinfo{year}{2007}),
  \urlprefix\url{https://link.aps.org/doi/10.1103/PhysRevB.76.235105}.

\bibitem[{\citenamefont{Roy and Saha}(2019)}]{Roy2019}
\bibinfo{author}{\bibfnamefont{A.}~\bibnamefont{Roy}} \bibnamefont{and}
  \bibinfo{author}{\bibfnamefont{K.}~\bibnamefont{Saha}}, \bibinfo{journal}{New
  Journal of Physics} \textbf{\bibinfo{volume}{21}}, \bibinfo{pages}{103050}
  (\bibinfo{year}{2019}),
  \urlprefix\url{https://doi.org/10.1088%2F1367-2630%2Fab4da0}.

\bibitem[{\citenamefont{Donatella et~al.}(2020)\citenamefont{Donatella, Biella,
  Boité, and Ciuti}}]{Donatella2020}
\bibinfo{author}{\bibfnamefont{K.}~\bibnamefont{Donatella}},
  \bibinfo{author}{\bibfnamefont{A.}~\bibnamefont{Biella}},
  \bibinfo{author}{\bibfnamefont{A.~L.} \bibnamefont{Boité}},
  \bibnamefont{and} \bibinfo{author}{\bibfnamefont{C.}~\bibnamefont{Ciuti}}
  (\bibinfo{year}{2020}), \eprint{2004.12883}.

\bibitem[{\citenamefont{Greiner et~al.}(2002)\citenamefont{Greiner, Mandel,
  Esslinger, H{\"a}nsch, and Bloch}}]{Greiner2002}
\bibinfo{author}{\bibfnamefont{M.}~\bibnamefont{Greiner}},
  \bibinfo{author}{\bibfnamefont{O.}~\bibnamefont{Mandel}},
  \bibinfo{author}{\bibfnamefont{T.}~\bibnamefont{Esslinger}},
  \bibinfo{author}{\bibfnamefont{T.~W.} \bibnamefont{H{\"a}nsch}},
  \bibnamefont{and} \bibinfo{author}{\bibfnamefont{I.}~\bibnamefont{Bloch}},
  \bibinfo{journal}{Nature} \textbf{\bibinfo{volume}{415}}, \bibinfo{pages}{39}
  (\bibinfo{year}{2002}), \urlprefix\url{https://doi.org/10.1038/415039a}.

\bibitem[{\citenamefont{Melnikov and Weinstein}(2000)}]{Melnikov2000}
\bibinfo{author}{\bibfnamefont{K.}~\bibnamefont{Melnikov}} \bibnamefont{and}
  \bibinfo{author}{\bibfnamefont{M.}~\bibnamefont{Weinstein}},
  \bibinfo{journal}{Phys. Rev. D} \textbf{\bibinfo{volume}{62}},
  \bibinfo{pages}{094504} (\bibinfo{year}{2000}),
  \urlprefix\url{https://link.aps.org/doi/10.1103/PhysRevD.62.094504}.

\bibitem[{\citenamefont{Tu et~al.}(2004)\citenamefont{Tu, Luo, and
  Gillies}}]{Tu2004}
\bibinfo{author}{\bibfnamefont{L.-C.} \bibnamefont{Tu}},
  \bibinfo{author}{\bibfnamefont{J.}~\bibnamefont{Luo}}, \bibnamefont{and}
  \bibinfo{author}{\bibfnamefont{G.~T.} \bibnamefont{Gillies}},
  \bibinfo{journal}{Reports on Progress in Physics}
  \textbf{\bibinfo{volume}{68}}, \bibinfo{pages}{77} (\bibinfo{year}{2004}),
  \urlprefix\url{https://doi.org/10.1088%2F0034-4885%2F68%2F1%2Fr02}.

\bibitem[{\citenamefont{Zohar and Reznik}(2011)}]{Zohar2011}
\bibinfo{author}{\bibfnamefont{E.}~\bibnamefont{Zohar}} \bibnamefont{and}
  \bibinfo{author}{\bibfnamefont{B.}~\bibnamefont{Reznik}},
  \bibinfo{journal}{Phys. Rev. Lett.} \textbf{\bibinfo{volume}{107}},
  \bibinfo{pages}{275301} (\bibinfo{year}{2011}),
  \urlprefix\url{https://link.aps.org/doi/10.1103/PhysRevLett.107.275301}.

\bibitem[{\citenamefont{Zohar et~al.}(2012)\citenamefont{Zohar, Cirac, and
  Reznik}}]{Zohar2012}
\bibinfo{author}{\bibfnamefont{E.}~\bibnamefont{Zohar}},
  \bibinfo{author}{\bibfnamefont{J.~I.} \bibnamefont{Cirac}}, \bibnamefont{and}
  \bibinfo{author}{\bibfnamefont{B.}~\bibnamefont{Reznik}},
  \bibinfo{journal}{Phys. Rev. Lett.} \textbf{\bibinfo{volume}{109}},
  \bibinfo{pages}{125302} (\bibinfo{year}{2012}),
  \urlprefix\url{https://link.aps.org/doi/10.1103/PhysRevLett.109.125302}.

\bibitem[{\citenamefont{Banerjee et~al.}(2012)\citenamefont{Banerjee, Dalmonte,
  M\"uller, Rico, Stebler, Wiese, and Zoller}}]{Banerjee2012}
\bibinfo{author}{\bibfnamefont{D.}~\bibnamefont{Banerjee}},
  \bibinfo{author}{\bibfnamefont{M.}~\bibnamefont{Dalmonte}},
  \bibinfo{author}{\bibfnamefont{M.}~\bibnamefont{M\"uller}},
  \bibinfo{author}{\bibfnamefont{E.}~\bibnamefont{Rico}},
  \bibinfo{author}{\bibfnamefont{P.}~\bibnamefont{Stebler}},
  \bibinfo{author}{\bibfnamefont{U.-J.} \bibnamefont{Wiese}}, \bibnamefont{and}
  \bibinfo{author}{\bibfnamefont{P.}~\bibnamefont{Zoller}},
  \bibinfo{journal}{Phys. Rev. Lett.} \textbf{\bibinfo{volume}{109}},
  \bibinfo{pages}{175302} (\bibinfo{year}{2012}),
  \urlprefix\url{https://link.aps.org/doi/10.1103/PhysRevLett.109.175302}.

\bibitem[{\citenamefont{Zohar et~al.}(2013)\citenamefont{Zohar, Cirac, and
  Reznik}}]{Zohar2013}
\bibinfo{author}{\bibfnamefont{E.}~\bibnamefont{Zohar}},
  \bibinfo{author}{\bibfnamefont{J.~I.} \bibnamefont{Cirac}}, \bibnamefont{and}
  \bibinfo{author}{\bibfnamefont{B.}~\bibnamefont{Reznik}},
  \bibinfo{journal}{Phys. Rev. Lett.} \textbf{\bibinfo{volume}{110}},
  \bibinfo{pages}{055302} (\bibinfo{year}{2013}),
  \urlprefix\url{https://link.aps.org/doi/10.1103/PhysRevLett.110.055302}.

\bibitem[{\citenamefont{Hauke et~al.}(2013)\citenamefont{Hauke, Marcos,
  Dalmonte, and Zoller}}]{Hauke2013}
\bibinfo{author}{\bibfnamefont{P.}~\bibnamefont{Hauke}},
  \bibinfo{author}{\bibfnamefont{D.}~\bibnamefont{Marcos}},
  \bibinfo{author}{\bibfnamefont{M.}~\bibnamefont{Dalmonte}}, \bibnamefont{and}
  \bibinfo{author}{\bibfnamefont{P.}~\bibnamefont{Zoller}},
  \bibinfo{journal}{Phys. Rev. X} \textbf{\bibinfo{volume}{3}},
  \bibinfo{pages}{041018} (\bibinfo{year}{2013}),
  \urlprefix\url{https://link.aps.org/doi/10.1103/PhysRevX.3.041018}.

\bibitem[{\citenamefont{Stannigel et~al.}(2014)\citenamefont{Stannigel, Hauke,
  Marcos, Hafezi, Diehl, Dalmonte, and Zoller}}]{Stannigel2014}
\bibinfo{author}{\bibfnamefont{K.}~\bibnamefont{Stannigel}},
  \bibinfo{author}{\bibfnamefont{P.}~\bibnamefont{Hauke}},
  \bibinfo{author}{\bibfnamefont{D.}~\bibnamefont{Marcos}},
  \bibinfo{author}{\bibfnamefont{M.}~\bibnamefont{Hafezi}},
  \bibinfo{author}{\bibfnamefont{S.}~\bibnamefont{Diehl}},
  \bibinfo{author}{\bibfnamefont{M.}~\bibnamefont{Dalmonte}}, \bibnamefont{and}
  \bibinfo{author}{\bibfnamefont{P.}~\bibnamefont{Zoller}},
  \bibinfo{journal}{Phys. Rev. Lett.} \textbf{\bibinfo{volume}{112}},
  \bibinfo{pages}{120406} (\bibinfo{year}{2014}),
  \urlprefix\url{https://link.aps.org/doi/10.1103/PhysRevLett.112.120406}.

\bibitem[{\citenamefont{K\"uhn et~al.}(2014)\citenamefont{K\"uhn, Cirac, and
  Ba\~nuls}}]{Kuehn2014}
\bibinfo{author}{\bibfnamefont{S.}~\bibnamefont{K\"uhn}},
  \bibinfo{author}{\bibfnamefont{J.~I.} \bibnamefont{Cirac}}, \bibnamefont{and}
  \bibinfo{author}{\bibfnamefont{M.-C.} \bibnamefont{Ba\~nuls}},
  \bibinfo{journal}{Phys. Rev. A} \textbf{\bibinfo{volume}{90}},
  \bibinfo{pages}{042305} (\bibinfo{year}{2014}),
  \urlprefix\url{https://link.aps.org/doi/10.1103/PhysRevA.90.042305}.

\bibitem[{\citenamefont{Kuno et~al.}(2015)\citenamefont{Kuno, Kasamatsu,
  Takahashi, Ichinose, and Matsui}}]{Kuno2015}
\bibinfo{author}{\bibfnamefont{Y.}~\bibnamefont{Kuno}},
  \bibinfo{author}{\bibfnamefont{K.}~\bibnamefont{Kasamatsu}},
  \bibinfo{author}{\bibfnamefont{Y.}~\bibnamefont{Takahashi}},
  \bibinfo{author}{\bibfnamefont{I.}~\bibnamefont{Ichinose}}, \bibnamefont{and}
  \bibinfo{author}{\bibfnamefont{T.}~\bibnamefont{Matsui}},
  \bibinfo{journal}{New Journal of Physics} \textbf{\bibinfo{volume}{17}},
  \bibinfo{pages}{063005} (\bibinfo{year}{2015}),
  \urlprefix\url{https://doi.org/10.1088%2F1367-2630%2F17%2F6%2F063005}.

\bibitem[{\citenamefont{Kuno et~al.}(2017)\citenamefont{Kuno, Sakane,
  Kasamatsu, Ichinose, and Matsui}}]{Kuno2017}
\bibinfo{author}{\bibfnamefont{Y.}~\bibnamefont{Kuno}},
  \bibinfo{author}{\bibfnamefont{S.}~\bibnamefont{Sakane}},
  \bibinfo{author}{\bibfnamefont{K.}~\bibnamefont{Kasamatsu}},
  \bibinfo{author}{\bibfnamefont{I.}~\bibnamefont{Ichinose}}, \bibnamefont{and}
  \bibinfo{author}{\bibfnamefont{T.}~\bibnamefont{Matsui}},
  \bibinfo{journal}{Phys. Rev. D} \textbf{\bibinfo{volume}{95}},
  \bibinfo{pages}{094507} (\bibinfo{year}{2017}),
  \urlprefix\url{https://link.aps.org/doi/10.1103/PhysRevD.95.094507}.

\bibitem[{\citenamefont{Dehkharghani et~al.}(2017)\citenamefont{Dehkharghani,
  Rico, Zinner, and Negretti}}]{Negretti2017}
\bibinfo{author}{\bibfnamefont{A.~S.} \bibnamefont{Dehkharghani}},
  \bibinfo{author}{\bibfnamefont{E.}~\bibnamefont{Rico}},
  \bibinfo{author}{\bibfnamefont{N.~T.} \bibnamefont{Zinner}},
  \bibnamefont{and} \bibinfo{author}{\bibfnamefont{A.}~\bibnamefont{Negretti}},
  \bibinfo{journal}{Phys. Rev. A} \textbf{\bibinfo{volume}{96}},
  \bibinfo{pages}{043611} (\bibinfo{year}{2017}),
  \urlprefix\url{https://link.aps.org/doi/10.1103/PhysRevA.96.043611}.

\bibitem[{\citenamefont{{Barros} et~al.}(2019)\citenamefont{{Barros},
  {Burrello}, and {Trombettoni}}}]{Barros2019}
\bibinfo{author}{\bibfnamefont{J.~C.~P.} \bibnamefont{{Barros}}},
  \bibinfo{author}{\bibfnamefont{M.}~\bibnamefont{{Burrello}}},
  \bibnamefont{and}
  \bibinfo{author}{\bibfnamefont{A.}~\bibnamefont{{Trombettoni}}},
  \bibinfo{journal}{ArXiv e-prints}  (\bibinfo{year}{2019}),
  \eprint{1911.06022}, \urlprefix\url{https://arxiv.org/abs/1911.06022}.

\bibitem[{\citenamefont{Schweizer et~al.}(2019)\citenamefont{Schweizer, Grusdt,
  Berngruber, Barbiero, Demler, Goldman, Bloch, and
  Aidelsburger}}]{Schweizer2019}
\bibinfo{author}{\bibfnamefont{C.}~\bibnamefont{Schweizer}},
  \bibinfo{author}{\bibfnamefont{F.}~\bibnamefont{Grusdt}},
  \bibinfo{author}{\bibfnamefont{M.}~\bibnamefont{Berngruber}},
  \bibinfo{author}{\bibfnamefont{L.}~\bibnamefont{Barbiero}},
  \bibinfo{author}{\bibfnamefont{E.}~\bibnamefont{Demler}},
  \bibinfo{author}{\bibfnamefont{N.}~\bibnamefont{Goldman}},
  \bibinfo{author}{\bibfnamefont{I.}~\bibnamefont{Bloch}}, \bibnamefont{and}
  \bibinfo{author}{\bibfnamefont{M.}~\bibnamefont{Aidelsburger}},
  \bibinfo{journal}{Nature Physics} \textbf{\bibinfo{volume}{15}},
  \bibinfo{pages}{1168} (\bibinfo{year}{2019}),
  \urlprefix\url{https://doi.org/10.1038/s41567-019-0649-7}.

\bibitem[{\citenamefont{Halimeh and Hauke}(2020{\natexlab{c}})}]{Halimeh2020a}
\bibinfo{author}{\bibfnamefont{J.~C.} \bibnamefont{Halimeh}} \bibnamefont{and}
  \bibinfo{author}{\bibfnamefont{P.}~\bibnamefont{Hauke}},
  \bibinfo{journal}{Phys. Rev. Lett.} \textbf{\bibinfo{volume}{125}},
  \bibinfo{pages}{030503} (\bibinfo{year}{2020}{\natexlab{c}}),
  \urlprefix\url{https://link.aps.org/doi/10.1103/PhysRevLett.125.030503}.

\bibitem[{\citenamefont{{Yang} et~al.}(2020)\citenamefont{{Yang}, {Sun}, {Ott},
  {Wang}, {Zache}, {Halimeh}, {Yuan}, {Hauke}, and {Pan}}}]{Yang2020}
\bibinfo{author}{\bibfnamefont{B.}~\bibnamefont{{Yang}}},
  \bibinfo{author}{\bibfnamefont{H.}~\bibnamefont{{Sun}}},
  \bibinfo{author}{\bibfnamefont{R.}~\bibnamefont{{Ott}}},
  \bibinfo{author}{\bibfnamefont{H.-Y.} \bibnamefont{{Wang}}},
  \bibinfo{author}{\bibfnamefont{T.~V.} \bibnamefont{{Zache}}},
  \bibinfo{author}{\bibfnamefont{J.~C.} \bibnamefont{{Halimeh}}},
  \bibinfo{author}{\bibfnamefont{Z.-S.} \bibnamefont{{Yuan}}},
  \bibinfo{author}{\bibfnamefont{P.}~\bibnamefont{{Hauke}}}, \bibnamefont{and}
  \bibinfo{author}{\bibfnamefont{J.-W.} \bibnamefont{{Pan}}},
  \bibinfo{journal}{ArXiv e-prints}  (\bibinfo{year}{2020}),
  \eprint{2003.08945}, \urlprefix\url{https://arxiv.org/abs/2003.08945}.

\bibitem[{\citenamefont{Halimeh
  et~al.}(2020{\natexlab{a}})\citenamefont{Halimeh, Ott, McCulloch, Yang, and
  Hauke}}]{Halimeh2020d}
\bibinfo{author}{\bibfnamefont{J.~C.} \bibnamefont{Halimeh}},
  \bibinfo{author}{\bibfnamefont{R.}~\bibnamefont{Ott}},
  \bibinfo{author}{\bibfnamefont{I.~P.} \bibnamefont{McCulloch}},
  \bibinfo{author}{\bibfnamefont{B.}~\bibnamefont{Yang}}, \bibnamefont{and}
  \bibinfo{author}{\bibfnamefont{P.}~\bibnamefont{Hauke}}
  (\bibinfo{year}{2020}{\natexlab{a}}), \eprint{2005.10249},
  \urlprefix\url{https://arxiv.org/abs/2005.10249}.

\bibitem[{\citenamefont{Halimeh
  et~al.}(2020{\natexlab{b}})\citenamefont{Halimeh, Lang, Mildenberger, Jiang,
  and Hauke}}]{Halimeh2020e}
\bibinfo{author}{\bibfnamefont{J.~C.} \bibnamefont{Halimeh}},
  \bibinfo{author}{\bibfnamefont{H.}~\bibnamefont{Lang}},
  \bibinfo{author}{\bibfnamefont{J.}~\bibnamefont{Mildenberger}},
  \bibinfo{author}{\bibfnamefont{Z.}~\bibnamefont{Jiang}}, \bibnamefont{and}
  \bibinfo{author}{\bibfnamefont{P.}~\bibnamefont{Hauke}}
  (\bibinfo{year}{2020}{\natexlab{b}}), \eprint{2007.00668},
  \urlprefix\url{https://arxiv.org/abs/2007.00668}.

\bibitem[{\citenamefont{Mathis et~al.}(2020)\citenamefont{Mathis, Mazzola, and
  Tavernelli}}]{Mathis2020}
\bibinfo{author}{\bibfnamefont{S.~V.} \bibnamefont{Mathis}},
  \bibinfo{author}{\bibfnamefont{G.}~\bibnamefont{Mazzola}}, \bibnamefont{and}
  \bibinfo{author}{\bibfnamefont{I.}~\bibnamefont{Tavernelli}}
  (\bibinfo{year}{2020}), \eprint{2005.10271},
  \urlprefix\url{https://arxiv.org/abs/2005.10271}.

\bibitem[{\citenamefont{Lamm et~al.}(2020)\citenamefont{Lamm, Lawrence, and
  Yamauchi}}]{Lamm2020}
\bibinfo{author}{\bibfnamefont{H.}~\bibnamefont{Lamm}},
  \bibinfo{author}{\bibfnamefont{S.}~\bibnamefont{Lawrence}}, \bibnamefont{and}
  \bibinfo{author}{\bibfnamefont{Y.}~\bibnamefont{Yamauchi}}
  (\bibinfo{year}{2020}), \eprint{2005.12688},
  \urlprefix\url{https://arxiv.org/abs/2005.12688}.

\bibitem[{\citenamefont{Tran et~al.}(2020)\citenamefont{Tran, Su, Carney, and
  Taylor}}]{Tran2020}
\bibinfo{author}{\bibfnamefont{M.~C.} \bibnamefont{Tran}},
  \bibinfo{author}{\bibfnamefont{Y.}~\bibnamefont{Su}},
  \bibinfo{author}{\bibfnamefont{D.}~\bibnamefont{Carney}}, \bibnamefont{and}
  \bibinfo{author}{\bibfnamefont{J.~M.} \bibnamefont{Taylor}}
  (\bibinfo{year}{2020}), \eprint{2006.16248},
  \urlprefix\url{https://arxiv.org/abs/2006.16248}.

\bibitem[{\citenamefont{Halimeh
  et~al.}(2020{\natexlab{c}})\citenamefont{Halimeh, Kasper, and
  Hauke}}]{Halimeh2020f}
\bibinfo{author}{\bibfnamefont{J.~C.} \bibnamefont{Halimeh}},
  \bibinfo{author}{\bibfnamefont{V.}~\bibnamefont{Kasper}}, \bibnamefont{and}
  \bibinfo{author}{\bibfnamefont{P.}~\bibnamefont{Hauke}}
  (\bibinfo{year}{2020}{\natexlab{c}}), \eprint{2009.07848},
  \urlprefix\url{https://arxiv.org/abs/2009.07848}.

\bibitem[{foo()}]{footnote}
\bibinfo{howpublished}{Note that continuous symmetries have Lie-algebra
  generators, but group theory does define generators for discrete symmetries
  as well, and these can be connected to Lie algebras in certain
  cases.\cite{Mariwalla1966} In this work, the term ``generator'' is used to
  denote the operator of the quantity conserved under the associated symmetry}.

\bibitem[{\citenamefont{Hauke and Heyl}(2015)}]{Hauke2015}
\bibinfo{author}{\bibfnamefont{P.}~\bibnamefont{Hauke}} \bibnamefont{and}
  \bibinfo{author}{\bibfnamefont{M.}~\bibnamefont{Heyl}},
  \bibinfo{journal}{Phys. Rev. B} \textbf{\bibinfo{volume}{92}},
  \bibinfo{pages}{134204} (\bibinfo{year}{2015}),
  \urlprefix\url{https://link.aps.org/doi/10.1103/PhysRevB.92.134204}.

\bibitem[{\citenamefont{Abanin et~al.}(2019)\citenamefont{Abanin, Altman,
  Bloch, and Serbyn}}]{Abanin2019}
\bibinfo{author}{\bibfnamefont{D.~A.} \bibnamefont{Abanin}},
  \bibinfo{author}{\bibfnamefont{E.}~\bibnamefont{Altman}},
  \bibinfo{author}{\bibfnamefont{I.}~\bibnamefont{Bloch}}, \bibnamefont{and}
  \bibinfo{author}{\bibfnamefont{M.}~\bibnamefont{Serbyn}},
  \bibinfo{journal}{Rev. Mod. Phys.} \textbf{\bibinfo{volume}{91}},
  \bibinfo{pages}{021001} (\bibinfo{year}{2019}),
  \urlprefix\url{https://link.aps.org/doi/10.1103/RevModPhys.91.021001}.

\bibitem[{\citenamefont{Hauke et~al.}(2016)\citenamefont{Hauke, Heyl,
  Tagliacozzo, and Zoller}}]{Hauke2016}
\bibinfo{author}{\bibfnamefont{P.}~\bibnamefont{Hauke}},
  \bibinfo{author}{\bibfnamefont{M.}~\bibnamefont{Heyl}},
  \bibinfo{author}{\bibfnamefont{L.}~\bibnamefont{Tagliacozzo}},
  \bibnamefont{and} \bibinfo{author}{\bibfnamefont{P.}~\bibnamefont{Zoller}},
  \bibinfo{journal}{Nature Physics} \textbf{\bibinfo{volume}{12}},
  \bibinfo{pages}{778} (\bibinfo{year}{2016}),
  \urlprefix\url{https://doi.org/10.1038/nphys3700}.

\bibitem[{\citenamefont{de~Almeida and Hauke}(2020)}]{Almeida2020}
\bibinfo{author}{\bibfnamefont{R.~C.} \bibnamefont{de~Almeida}}
  \bibnamefont{and} \bibinfo{author}{\bibfnamefont{P.}~\bibnamefont{Hauke}}
  (\bibinfo{year}{2020}), \eprint{2005.03049},
  \urlprefix\url{https://arxiv.org/abs/2005.03049}.

\bibitem[{\citenamefont{K\"uhner and Monien}(1998)}]{Kuehner1998}
\bibinfo{author}{\bibfnamefont{T.~D.} \bibnamefont{K\"uhner}} \bibnamefont{and}
  \bibinfo{author}{\bibfnamefont{H.}~\bibnamefont{Monien}},
  \bibinfo{journal}{Phys. Rev. B} \textbf{\bibinfo{volume}{58}},
  \bibinfo{pages}{R14741} (\bibinfo{year}{1998}),
  \urlprefix\url{https://link.aps.org/doi/10.1103/PhysRevB.58.R14741}.

\bibitem[{\citenamefont{K\"uhner et~al.}(2000)\citenamefont{K\"uhner, White,
  and Monien}}]{Kuehner2000}
\bibinfo{author}{\bibfnamefont{T.~D.} \bibnamefont{K\"uhner}},
  \bibinfo{author}{\bibfnamefont{S.~R.} \bibnamefont{White}}, \bibnamefont{and}
  \bibinfo{author}{\bibfnamefont{H.}~\bibnamefont{Monien}},
  \bibinfo{journal}{Phys. Rev. B} \textbf{\bibinfo{volume}{61}},
  \bibinfo{pages}{12474} (\bibinfo{year}{2000}),
  \urlprefix\url{https://link.aps.org/doi/10.1103/PhysRevB.61.12474}.

\bibitem[{\citenamefont{Dutta et~al.}(2015)\citenamefont{Dutta, Gajda, Hauke,
  Lewenstein, Lühmann, Malomed, Sowi{\'{n}}ski, and Zakrzewski}}]{Dutta2015}
\bibinfo{author}{\bibfnamefont{O.}~\bibnamefont{Dutta}},
  \bibinfo{author}{\bibfnamefont{M.}~\bibnamefont{Gajda}},
  \bibinfo{author}{\bibfnamefont{P.}~\bibnamefont{Hauke}},
  \bibinfo{author}{\bibfnamefont{M.}~\bibnamefont{Lewenstein}},
  \bibinfo{author}{\bibfnamefont{D.-S.} \bibnamefont{Lühmann}},
  \bibinfo{author}{\bibfnamefont{B.~A.} \bibnamefont{Malomed}},
  \bibinfo{author}{\bibfnamefont{T.}~\bibnamefont{Sowi{\'{n}}ski}},
  \bibnamefont{and}
  \bibinfo{author}{\bibfnamefont{J.}~\bibnamefont{Zakrzewski}},
  \bibinfo{journal}{Reports on Progress in Physics}
  \textbf{\bibinfo{volume}{78}}, \bibinfo{pages}{066001}
  (\bibinfo{year}{2015}),
  \urlprefix\url{https://doi.org/10.1088%2F0034-4885%2F78%2F6%2F066001}.

\bibitem[{\citenamefont{Kollath et~al.}(2010)\citenamefont{Kollath, Roux,
  Biroli, and Läuchli}}]{Kollath2010}
\bibinfo{author}{\bibfnamefont{C.}~\bibnamefont{Kollath}},
  \bibinfo{author}{\bibfnamefont{G.}~\bibnamefont{Roux}},
  \bibinfo{author}{\bibfnamefont{G.}~\bibnamefont{Biroli}}, \bibnamefont{and}
  \bibinfo{author}{\bibfnamefont{A.~M.} \bibnamefont{Läuchli}},
  \bibinfo{journal}{Journal of Statistical Mechanics: Theory and Experiment}
  \textbf{\bibinfo{volume}{2010}}, \bibinfo{pages}{P08011}
  (\bibinfo{year}{2010}),
  \urlprefix\url{https://doi.org/10.1088%2F1742-5468%2F2010%2F08%2Fp08011}.

\bibitem[{\citenamefont{Breuer and Petruccione}(2002)}]{Breuer_book}
\bibinfo{author}{\bibfnamefont{H.~P.} \bibnamefont{Breuer}} \bibnamefont{and}
  \bibinfo{author}{\bibfnamefont{F.}~\bibnamefont{Petruccione}},
  \emph{\bibinfo{title}{The Theory of Open Quantum Systems}}
  (\bibinfo{publisher}{Oxford University Press}, \bibinfo{year}{2002}), ISBN
  \bibinfo{isbn}{9780198520634},
  \urlprefix\url{https://books.google.de/books?id=0Yx5VzaMYm8C}.

\bibitem[{\citenamefont{Manzano}(2020)}]{Manzano2020}
\bibinfo{author}{\bibfnamefont{D.}~\bibnamefont{Manzano}},
  \bibinfo{journal}{AIP Advances} \textbf{\bibinfo{volume}{10}},
  \bibinfo{pages}{025106} (\bibinfo{year}{2020}),
  \eprint{https://doi.org/10.1063/1.5115323},
  \urlprefix\url{https://doi.org/10.1063/1.5115323}.

\bibitem[{\citenamefont{Baum et~al.}(1985)\citenamefont{Baum, Munowitz,
  Garroway, and Pines}}]{Pines1985}
\bibinfo{author}{\bibfnamefont{J.}~\bibnamefont{Baum}},
  \bibinfo{author}{\bibfnamefont{M.}~\bibnamefont{Munowitz}},
  \bibinfo{author}{\bibfnamefont{A.~N.} \bibnamefont{Garroway}},
  \bibnamefont{and} \bibinfo{author}{\bibfnamefont{A.}~\bibnamefont{Pines}},
  \bibinfo{journal}{The Journal of Chemical Physics}
  \textbf{\bibinfo{volume}{83}}, \bibinfo{pages}{2015} (\bibinfo{year}{1985}),
  \eprint{https://doi.org/10.1063/1.449344},
  \urlprefix\url{https://doi.org/10.1063/1.449344}.

\bibitem[{\citenamefont{Baum and Pines}(1986)}]{Pines1986}
\bibinfo{author}{\bibfnamefont{J.}~\bibnamefont{Baum}} \bibnamefont{and}
  \bibinfo{author}{\bibfnamefont{A.}~\bibnamefont{Pines}},
  \bibinfo{journal}{Journal of the American Chemical Society}
  \textbf{\bibinfo{volume}{108}}, \bibinfo{pages}{7447} (\bibinfo{year}{1986}),
  \urlprefix\url{https://pubs.acs.org/doi/abs/10.1021/ja00284a001}.

\bibitem[{\citenamefont{S\'anchez et~al.}(2014)\citenamefont{S\'anchez, Acosta,
  Levstein, Pastawski, and Chattah}}]{SpinDecoherence2014}
\bibinfo{author}{\bibfnamefont{C.~M.} \bibnamefont{S\'anchez}},
  \bibinfo{author}{\bibfnamefont{R.~H.} \bibnamefont{Acosta}},
  \bibinfo{author}{\bibfnamefont{P.~R.} \bibnamefont{Levstein}},
  \bibinfo{author}{\bibfnamefont{H.~M.} \bibnamefont{Pastawski}},
  \bibnamefont{and} \bibinfo{author}{\bibfnamefont{A.~K.}
  \bibnamefont{Chattah}}, \bibinfo{journal}{Phys. Rev. A}
  \textbf{\bibinfo{volume}{90}}, \bibinfo{pages}{042122}
  (\bibinfo{year}{2014}),
  \urlprefix\url{https://link.aps.org/doi/10.1103/PhysRevA.90.042122}.

\bibitem[{\citenamefont{\'Alvarez and Suter}(2010)}]{NMRLocalizationPRL2010}
\bibinfo{author}{\bibfnamefont{G.~A.} \bibnamefont{\'Alvarez}}
  \bibnamefont{and} \bibinfo{author}{\bibfnamefont{D.}~\bibnamefont{Suter}},
  \bibinfo{journal}{Phys. Rev. Lett.} \textbf{\bibinfo{volume}{104}},
  \bibinfo{pages}{230403} (\bibinfo{year}{2010}),
  \urlprefix\url{https://link.aps.org/doi/10.1103/PhysRevLett.104.230403}.

\bibitem[{\citenamefont{{\'A}lvarez et~al.}(2015)\citenamefont{{\'A}lvarez,
  Suter, and Kaiser}}]{NMRLocalizationScience2015}
\bibinfo{author}{\bibfnamefont{G.~A.} \bibnamefont{{\'A}lvarez}},
  \bibinfo{author}{\bibfnamefont{D.}~\bibnamefont{Suter}}, \bibnamefont{and}
  \bibinfo{author}{\bibfnamefont{R.}~\bibnamefont{Kaiser}},
  \bibinfo{journal}{Science} \textbf{\bibinfo{volume}{349}},
  \bibinfo{pages}{846} (\bibinfo{year}{2015}),
  \urlprefix\url{https://science.sciencemag.org/content/349/6250/846}.

\bibitem[{\citenamefont{G\"arttner et~al.}(2018)\citenamefont{G\"arttner,
  Hauke, and Rey}}]{Gaerttner2018}
\bibinfo{author}{\bibfnamefont{M.}~\bibnamefont{G\"arttner}},
  \bibinfo{author}{\bibfnamefont{P.}~\bibnamefont{Hauke}}, \bibnamefont{and}
  \bibinfo{author}{\bibfnamefont{A.~M.} \bibnamefont{Rey}},
  \bibinfo{journal}{Phys. Rev. Lett.} \textbf{\bibinfo{volume}{120}},
  \bibinfo{pages}{040402} (\bibinfo{year}{2018}),
  \urlprefix\url{https://link.aps.org/doi/10.1103/PhysRevLett.120.040402}.

\bibitem[{\citenamefont{Lewis-Swan et~al.}(2020)\citenamefont{Lewis-Swan,
  Muleady, and Rey}}]{LewisSwan2020}
\bibinfo{author}{\bibfnamefont{R.~J.} \bibnamefont{Lewis-Swan}},
  \bibinfo{author}{\bibfnamefont{S.~R.} \bibnamefont{Muleady}},
  \bibnamefont{and} \bibinfo{author}{\bibfnamefont{A.~M.} \bibnamefont{Rey}}
  (\bibinfo{year}{2020}), \eprint{2006.01313}.

\bibitem[{\citenamefont{G{\"{a}}rttner
  et~al.}(2017)\citenamefont{G{\"{a}}rttner, Bohnet, Safavi-Naini, Wall,
  Bollinger, and Rey}}]{Garttner2017}
\bibinfo{author}{\bibfnamefont{M.}~\bibnamefont{G{\"{a}}rttner}},
  \bibinfo{author}{\bibfnamefont{J.~G.} \bibnamefont{Bohnet}},
  \bibinfo{author}{\bibfnamefont{A.}~\bibnamefont{Safavi-Naini}},
  \bibinfo{author}{\bibfnamefont{M.~L.} \bibnamefont{Wall}},
  \bibinfo{author}{\bibfnamefont{J.~J.} \bibnamefont{Bollinger}},
  \bibnamefont{and} \bibinfo{author}{\bibfnamefont{A.~M.} \bibnamefont{Rey}},
  \bibinfo{journal}{Nature Physics} \textbf{\bibinfo{volume}{13}},
  \bibinfo{pages}{781} (\bibinfo{year}{2017}),
  \urlprefix\url{https://doi.org/10.1038/nphys4119}.

\bibitem[{\citenamefont{Zohar et~al.}(2017)\citenamefont{Zohar, Farace, Reznik,
  and Cirac}}]{Zohar2017}
\bibinfo{author}{\bibfnamefont{E.}~\bibnamefont{Zohar}},
  \bibinfo{author}{\bibfnamefont{A.}~\bibnamefont{Farace}},
  \bibinfo{author}{\bibfnamefont{B.}~\bibnamefont{Reznik}}, \bibnamefont{and}
  \bibinfo{author}{\bibfnamefont{J.~I.} \bibnamefont{Cirac}},
  \bibinfo{journal}{Phys. Rev. Lett.} \textbf{\bibinfo{volume}{118}},
  \bibinfo{pages}{070501} (\bibinfo{year}{2017}),
  \urlprefix\url{https://link.aps.org/doi/10.1103/PhysRevLett.118.070501}.

\bibitem[{\citenamefont{Barbiero et~al.}(2019)\citenamefont{Barbiero,
  Schweizer, Aidelsburger, Demler, Goldman, and Grusdt}}]{Barbiero2019}
\bibinfo{author}{\bibfnamefont{L.}~\bibnamefont{Barbiero}},
  \bibinfo{author}{\bibfnamefont{C.}~\bibnamefont{Schweizer}},
  \bibinfo{author}{\bibfnamefont{M.}~\bibnamefont{Aidelsburger}},
  \bibinfo{author}{\bibfnamefont{E.}~\bibnamefont{Demler}},
  \bibinfo{author}{\bibfnamefont{N.}~\bibnamefont{Goldman}}, \bibnamefont{and}
  \bibinfo{author}{\bibfnamefont{F.}~\bibnamefont{Grusdt}},
  \bibinfo{journal}{Science Advances} \textbf{\bibinfo{volume}{5}}
  (\bibinfo{year}{2019}),
  \urlprefix\url{https://advances.sciencemag.org/content/5/10/eaav7444}.

\bibitem[{\citenamefont{Borla et~al.}(2020)\citenamefont{Borla, Verresen,
  Grusdt, and Moroz}}]{Borla2019}
\bibinfo{author}{\bibfnamefont{U.}~\bibnamefont{Borla}},
  \bibinfo{author}{\bibfnamefont{R.}~\bibnamefont{Verresen}},
  \bibinfo{author}{\bibfnamefont{F.}~\bibnamefont{Grusdt}}, \bibnamefont{and}
  \bibinfo{author}{\bibfnamefont{S.}~\bibnamefont{Moroz}},
  \bibinfo{journal}{Phys. Rev. Lett.} \textbf{\bibinfo{volume}{124}},
  \bibinfo{pages}{120503} (\bibinfo{year}{2020}),
  \urlprefix\url{https://link.aps.org/doi/10.1103/PhysRevLett.124.120503}.

\bibitem[{\citenamefont{Kokail et~al.}(2019)\citenamefont{Kokail, Maier, van
  Bijnen, Brydges, Joshi, Jurcevic, Muschik, Silvi, Blatt, Roos
  et~al.}}]{Kokail2019}
\bibinfo{author}{\bibfnamefont{C.}~\bibnamefont{Kokail}},
  \bibinfo{author}{\bibfnamefont{C.}~\bibnamefont{Maier}},
  \bibinfo{author}{\bibfnamefont{R.}~\bibnamefont{van Bijnen}},
  \bibinfo{author}{\bibfnamefont{T.}~\bibnamefont{Brydges}},
  \bibinfo{author}{\bibfnamefont{M.~K.} \bibnamefont{Joshi}},
  \bibinfo{author}{\bibfnamefont{P.}~\bibnamefont{Jurcevic}},
  \bibinfo{author}{\bibfnamefont{C.~A.} \bibnamefont{Muschik}},
  \bibinfo{author}{\bibfnamefont{P.}~\bibnamefont{Silvi}},
  \bibinfo{author}{\bibfnamefont{R.}~\bibnamefont{Blatt}},
  \bibinfo{author}{\bibfnamefont{C.~F.} \bibnamefont{Roos}},
  \bibnamefont{et~al.}, \bibinfo{journal}{Nature}
  \textbf{\bibinfo{volume}{569}}, \bibinfo{pages}{355} (\bibinfo{year}{2019}),
  \urlprefix\url{https://doi.org/10.1038/s41586-019-1177-4}.

\bibitem[{\citenamefont{Wiese}(2013)}]{Wiese_review}
\bibinfo{author}{\bibfnamefont{U.-J.} \bibnamefont{Wiese}},
  \bibinfo{journal}{Annalen der Physik} \textbf{\bibinfo{volume}{525}},
  \bibinfo{pages}{777} (\bibinfo{year}{2013}),
  \eprint{https://onlinelibrary.wiley.com/doi/pdf/10.1002/andp.201300104},
  \urlprefix\url{https://onlinelibrary.wiley.com/doi/abs/10.1002/andp.201300104}.

\bibitem[{\citenamefont{Yang et~al.}(2016)\citenamefont{Yang, Giri, Johanning,
  Wunderlich, Zoller, and Hauke}}]{Yang2016}
\bibinfo{author}{\bibfnamefont{D.}~\bibnamefont{Yang}},
  \bibinfo{author}{\bibfnamefont{G.~S.} \bibnamefont{Giri}},
  \bibinfo{author}{\bibfnamefont{M.}~\bibnamefont{Johanning}},
  \bibinfo{author}{\bibfnamefont{C.}~\bibnamefont{Wunderlich}},
  \bibinfo{author}{\bibfnamefont{P.}~\bibnamefont{Zoller}}, \bibnamefont{and}
  \bibinfo{author}{\bibfnamefont{P.}~\bibnamefont{Hauke}},
  \bibinfo{journal}{Phys. Rev. A} \textbf{\bibinfo{volume}{94}},
  \bibinfo{pages}{052321} (\bibinfo{year}{2016}),
  \urlprefix\url{https://link.aps.org/doi/10.1103/PhysRevA.94.052321}.

\bibitem[{\citenamefont{Mil et~al.}(2020)\citenamefont{Mil, Zache, Hegde, Xia,
  Bhatt, Oberthaler, Hauke, Berges, and Jendrzejewski}}]{Mil2020}
\bibinfo{author}{\bibfnamefont{A.}~\bibnamefont{Mil}},
  \bibinfo{author}{\bibfnamefont{T.~V.} \bibnamefont{Zache}},
  \bibinfo{author}{\bibfnamefont{A.}~\bibnamefont{Hegde}},
  \bibinfo{author}{\bibfnamefont{A.}~\bibnamefont{Xia}},
  \bibinfo{author}{\bibfnamefont{R.~P.} \bibnamefont{Bhatt}},
  \bibinfo{author}{\bibfnamefont{M.~K.} \bibnamefont{Oberthaler}},
  \bibinfo{author}{\bibfnamefont{P.}~\bibnamefont{Hauke}},
  \bibinfo{author}{\bibfnamefont{J.}~\bibnamefont{Berges}}, \bibnamefont{and}
  \bibinfo{author}{\bibfnamefont{F.}~\bibnamefont{Jendrzejewski}},
  \bibinfo{journal}{Science} \textbf{\bibinfo{volume}{367}},
  \bibinfo{pages}{1128 LP } (\bibinfo{year}{2020}),
  \urlprefix\url{http://science.sciencemag.org/content/367/6482/1128.abstract}.

\bibitem[{\citenamefont{Altland and Simons}(2010)}]{Altland_book}
\bibinfo{author}{\bibfnamefont{A.}~\bibnamefont{Altland}} \bibnamefont{and}
  \bibinfo{author}{\bibfnamefont{B.}~\bibnamefont{Simons}},
  \emph{\bibinfo{title}{Condensed Matter Field Theory}}, Cambridge books online
  (\bibinfo{publisher}{Cambridge University Press}, \bibinfo{year}{2010}), ISBN
  \bibinfo{isbn}{9780521769754},
  \urlprefix\url{https://books.google.de/books?id=GpF0Pgo8CqAC}.

\bibitem[{LaG()}]{LaGaDyn}
\bibinfo{howpublished}{J. C. Halimeh \textit{et al.} (in preparation)}.

\bibitem[{\citenamefont{Johansson et~al.}(2012)\citenamefont{Johansson, Nation,
  and Nori}}]{Johansson2012}
\bibinfo{author}{\bibfnamefont{J.}~\bibnamefont{Johansson}},
  \bibinfo{author}{\bibfnamefont{P.}~\bibnamefont{Nation}}, \bibnamefont{and}
  \bibinfo{author}{\bibfnamefont{F.}~\bibnamefont{Nori}},
  \bibinfo{journal}{Computer Physics Communications}
  \textbf{\bibinfo{volume}{183}}, \bibinfo{pages}{1760 }
  (\bibinfo{year}{2012}),
  \urlprefix\url{http://www.sciencedirect.com/science/article/pii/S0010465512000835}.

\bibitem[{\citenamefont{Johansson et~al.}(2013)\citenamefont{Johansson, Nation,
  and Nori}}]{Johansson2013}
\bibinfo{author}{\bibfnamefont{J.}~\bibnamefont{Johansson}},
  \bibinfo{author}{\bibfnamefont{P.}~\bibnamefont{Nation}}, \bibnamefont{and}
  \bibinfo{author}{\bibfnamefont{F.}~\bibnamefont{Nori}},
  \bibinfo{journal}{Computer Physics Communications}
  \textbf{\bibinfo{volume}{184}}, \bibinfo{pages}{1234 }
  (\bibinfo{year}{2013}),
  \urlprefix\url{http://www.sciencedirect.com/science/article/pii/S0010465512003955}.

\bibitem[{\citenamefont{Havel}(2003)}]{Havel2003}
\bibinfo{author}{\bibfnamefont{T.~F.} \bibnamefont{Havel}},
  \bibinfo{journal}{Journal of Mathematical Physics}
  \textbf{\bibinfo{volume}{44}}, \bibinfo{pages}{534} (\bibinfo{year}{2003}),
  \eprint{https://aip.scitation.org/doi/pdf/10.1063/1.1518555},
  \urlprefix\url{https://aip.scitation.org/doi/abs/10.1063/1.1518555}.

\bibitem[{\citenamefont{Mariwalla}(1966)}]{Mariwalla1966}
\bibinfo{author}{\bibfnamefont{K.~H.} \bibnamefont{Mariwalla}},
  \bibinfo{journal}{Journal of Mathematical Physics}
  \textbf{\bibinfo{volume}{7}}, \bibinfo{pages}{114} (\bibinfo{year}{1966}),
  \eprint{https://doi.org/10.1063/1.1704797},
  \urlprefix\url{https://doi.org/10.1063/1.1704797}.

\end{thebibliography}
\end{document}